\documentclass[12pt]{article}

\RequirePackage{a4}
\usepackage{epsfig}
\usepackage[utf8x]{inputenc}
\usepackage{url}
\usepackage{here}
\usepackage{amssymb}
\usepackage{graphicx}
\usepackage{comment}
\usepackage{verbatim}
\usepackage{fancyvrb}

\usepackage{lineno}
\usepackage{hyperref}
\usepackage{breakurl}
\usepackage{etoolbox}
\gappto{\UrlBreaks}{\UrlOrds}

\usepackage{amsmath}   

\VerbatimFootnotes

\newcommand{\bm}[1]{\mbox{\boldmath$#1$}}

\newcommand{\Inf}{\mbox{Inf}}
\newcommand{\NoInf}{\mbox{NoInf}}
\newcommand{\Pos}{\mbox{Pos}}
\newcommand{\Neg}{\mbox{Neg}}
\newcommand{\Infs}{\mbox{\footnotesize Inf}}
\newcommand{\NoInfs}{\mbox{\footnotesize NoInf}}

\begin{document}

\makeatletter
\def\maketitle{%
\cleardoublepage 
\thispagestyle{empty}%
\begingroup 
\topskip\z@skip
\null\vfil
\begin{center}
\LARGE\centering
\openup\medskipamount
\vspace{-38pt}\@title\par\vspace{15pt}%
\mdseries\large\@author\par \vspace{15pt}%
\vfill
\end{center}}
\makeatother

\title{Checking individuals and sampling populations \\
  with  imperfect tests
}

\author{Giulio D'Agostini\footnote{
    Universit\`a ``La Sapienza'' and INFN, Roma, Italia,
    giulio.dagostini@roma1.infn.it
  }
 \  and Alfredo Esposito\footnote{
   Retired, alfespo@yahoo.it
  }
}

\date{}

\maketitle

\begin{abstract}
In the last months, due to the emergency of
Covid-19, questions related to the fact
of belonging or not to a particular class
of individuals (`infected or not infected'),
after being tagged as `positive' or `negative'
by a test, have never been so popular.
Similarly, there has been strong interest in
estimating the proportion of a population expected to hold
a given characteristics (`having or having had the virus'). 
Taking the cue from the many related discussions
on the media, in addition to those to which we took part, we
analyze these questions from a probabilistic perspective
(`Bayesian'), considering several effects that
play a role in evaluating the probabilities of interest. 
The resulting paper, written with didactic intent, 
is rather general and not strictly related to
pandemics: the basic ideas of Bayesian inference are
introduced and the uncertainties on the performances
of the tests are treated using the metrological concepts
of `systematics', and are propagated into
the quantities of interest following the rules of probability theory;
the separation of `statistical'
and  `systematic' contributions to the uncertainty  
on the inferred proportion of infectees allows to optimize
the sample size;
the role of `priors', often overlooked, is stressed,
however recommending the use of `flat priors', since
the resulting posterior distribution can be `reshaped'
by an `informative prior' in a later step;
details on the calculations are given, also deriving useful
approximated formulae, the tough work being however done
with the help of direct Monte Carlo simulations and
Markov Chain Monte Carlo,
implemented in R and JAGS (relevant code provided in appendix).
\end{abstract}
\mbox{}\vspace{-0.6cm}\mbox{}
{\footnotesize
\begin{flushright}
  {\small 
   {\sl   ``Grown-ups like numbers''}\\
   (The Little Prince)\\
   \mbox{}\\
  {\sl ``The theory of probabilities is basically\\
    just common sense reduced to calculus''}\\
  (Laplace)\\
  \mbox{}\\
  {\sl ``All models are wrong, but some are useful''\\
    (G. Box)
  }
  }
  \mbox{} \\  \mbox{} \\
\end{flushright}  
}

\tableofcontents
\mbox{}\\
\noindent
{\bf References} \  .  \  .  \  .  \  .  \  .  \  .  \  . \ . \ . \
\  .  \  .  \  .  \  .  \  .  \  .  \  . \ . \ . \
 \  .  \  .  \  .  \  .  \  .  \  .  \  . \ . \ . \  . \  {\bf 96} \\
{\bf Appendix A -- Some remarks on `{\em Bayes' formulae}'}
    \  .  \  .  \  .  \  .  \  .  \  .  \  . \ . \ . \ {\bf 99}\\
{\bf Appendix B -- R and JAGS code}
\  .  \  .  \  .  \  .  \  .  \  .  \  . \ . \ . \
 \  .  \  .  \  .  \  .  \  .  \  .  \  . \ . \  {\bf 103}\\
\setcounter{footnote}{0}

\section{Introduction}
The Covid-19 outbreak of these months
raised a new interest in data analysis,
especially among lay people, for long locked down and really flooded
by a tidal wave of numbers, whose meaning has often
been pretty unclear,
including that of the body counting, which should be in principle
the easiest to assess.
As practically anyone who has some experience in data
analysis, we were also tempted -- we have to confess --
to build up some models in order to
understand what was going on, and especially to forecast future numbers.
But we immediately gave up, and not only because
faced with numbers that were not really meaningful, without clear
conditions, within reasonable uncertainty, about how they were obtained.
The basic question is that, we realized soon,
we cannot treat a virus spreading in a
human population like a bacterial colony in a homogeneous medium,
or a continuous (or discretized) thermodynamic system.
People live -- fortunately! --
in far more complex communities (`clusters'), starting from the families,
villages and suburbs; then cities, regions, countries and
continents of different characteristics, population densities
and social behaviors. Then we would have to take into account 
`osmosis' of different kinds among the clusters,
due to local, intermediate and long distance movements
of individuals. Not to speak of the diffusion properties
of viruses in general and of this one in particular.

A related problem, which would complicate further the model,
was the fact that tests were applied,
at least at the beginning of pandemic,  
mainly to people showing evident symptoms
or at risk for several reasons, like personnel of the health system. 
We were then asking ourselves rather soon,
why tests were not also made on 
a possibly representative sample of the entire population,
independently of the presence of symptoms
or not.\footnote{For example we would have started
  choosing, in Italy,
  the families involved in the Auditel system\,\cite{Auditel}, 
  created with the purpose to infer the
  share of television programs, on the basis of which advertisers
  pay the TV channels. In general, in order to make sampling meaningful,
  the selection of individuals cannot be left to a voluntary choice
  that would inevitably bias the outcomes of the test campaign.}
This would be, in our opinion,
the best way to get an idea of the proportion of the
population affected  at a given
`instant' (to be understood as one or a few days)
and to take decisions accordingly.
It is quite obvious that surveys of this kind
would require rather fast and inexpensive tests,
to the detriment of their quality, thus unavoidably yielding
a not negligible fraction of so called {\em false positives} 
and {\em false negatives}. 

When we read in a newspaper\,\cite{FattoQuotidiano}
about a rather cheap
antibody blood test able to tag 
the individuals {\em being or having been infected}\,\footnote{In fact,
the test reported in Ref.\,\cite{FattoQuotidiano}
was claimed to be sensitive both to
Immunoglobulin M (IgM), the antibody related
to a current infection,
and  Immunoglobulin G (IgG) related to
a past infection \cite{IgM,IgG}.
Obviously, the effectiveness of these kind
of `serological tests' is not questioned here.
In particular, two kinds of immunoglobulins
will take some time to develop and they
are most likely characterized by decay times.
Therefore, the generic expression {\em infected individuals}
 (or in short {\em infectees}) 
has to be meant as the
{\em members of the population
which hold some `property' to which the test is sensitive
at the time in which it is performed.}
}
we decided to make some exercises in order to understand
whether such a `low quality' test would be adequate for the purpose
and what sample size would be required in order
to get `snapshots` of a population at regular times.
In fact Ref.\,\cite{FattoQuotidiano}
not only reported the relevant `probabilities',
namely 98\% to tag an Infected ({\em presently or previously}) as Positive
(`sensitivity') and 88\%  to tag a not-Infected as Negative
(`specificity'), but also the
numbers of tests from which these two numbers resulted.
This extra information is important to understand how
believable these two numbers are and how to propagate their uncertainty
into the other numbers of interest, together with other sources of uncertainty. 
This convinced us to go through the {\em exercise}
of understanding how the main uncertainties of the problem
would affect the conclusions:
\begin{itemize}
\item uncertainty due to sampling;
\item uncertainty due to the fact that the above
  probabilities differ from 1;
\item uncertainty about the exact values of these 
  `probabilities'.\footnote{If you are not used to attach a probability
  to numbers that might have by themselves the meaning of
probability, Ref.\,\cite{ProbProp} is recommended.}
\end{itemize}
Experts might argue that other sources of uncertainty should
be considered, but our point was that already clarifying some issues
related to the above contributions would have been of some interest.
From the probabilistic point of view, there is another
source of uncertainty to be taken into account, 
which is the {\em prior distribution}
of the proportion of infectees in the population,
however not as important as  when we have
to judge from a single test if an individual is infected or not.

The paper, written with didactic intent\footnote{The educational
  writing is an old idea that both the authors pursued in the past
  (see e.g. Refs.~\cite{Colombo, CEP, Streghe}),
  strongly believing in the necessity of
  making the management of uncertainty a basic tenet of scholastic
  (and not only) curricula.}
(and we have to admit that it was useful to clarify
some issues even to us), is organized in the following way.
\begin{itemize}
\item  Section ~\ref{sec:rough} shows some simple evaluations based on the
    nominal capabilities of the test, without entering in the
    probabilistic treatment of the problem. The limitations
    of such `rough reasoning' become immediately clear.
\item Then we move in Sec.~\ref{sec:Bayes1} to probabilistic
    reasoning, applied to the probability
    that a person tagged as positive/negative `is' (or `has been') really
    infected or not infected. The probabilistic tool needed
    to make this so called `probabilistic inversion' (Bayes' theorem)
    is then reminded and applied, showing the relevance
    of the probability that the individual is infected or not,
    based on other pieces of information/knowledge
    (`prior probability'), a fundamental ingredient
    of inference often overlooked.\footnote{This problem
    has been recently addressed by an article on Scientific
    American \cite{SCIAM}, with arguments similar
    to the simplistic one we are going to show in
    Sec.~\ref{sec:rough}, although complemented by a rather popular
    visualization of the question. But we have been surprised
    by the lack of
    any reference to probability theory and to the Bayes' rule
    in the paper.}
\item The effect of the uncertainty on sensititivity, specificity
      and proportion of infectees in the population  
     is discussed in Sec.~\ref{sec:uncertainty}.
  But, before doing that, we have to model the probability density
  function for these uncertain quantities. Hence an introduction
  to the application of Bayes' theorem
  to continuous quantities is required,
  including some notes on the use of {\em conjugate priors}.
\item  
  From Sec.~\ref{sec:fpN_from_ps} we switch our focus
  from  single individuals to  populations. Our aim, 
  that is inferring the {\em proportion of `infectees'}
  (meaning, let us repeat it once more,
  {\em `individuals being or having being infected'}) will be reached  
  in Secs.~\ref{sec:interring_p} and \ref{sec:direct}.  
  But, for didactic purposes,  we proceed by step, starting from
  the expected number of positives,  examining in depth the various
  sources of uncertainty. 
  In particular, in Sec.~\ref{sec:measurability_p}
  we study the measurability of $p$ and the dependence
  of its `resolution power' on the test performances and
  the sample size.  Most of the work is done using Monte Carlo methods,
  but some useful approximated formulae
  for the evaluation of uncertainty on the result are given as well.
\item
  The probabilistic inference of $p$, that is evaluating its probability
  density function $f(p)$, conditioned by data and well stated hypotheses,
  is finally done in Sec.~\ref{sec:interring_p}. Having to solve
  a multidimensional problem, in which $f(p)$ is finally obtained by
  marginalization, Markov Chain Monte Carlo (MCMC) methods
  become a must. In particular, we use JAGS~\cite{JAGS},
  interfaced with R~\cite{R} through the package
  rjags~\cite{rjags}. We also evaluate, by the help of JAGS, some joint
  probability distributions and the correlation coefficients among
  the variables of interest, thus showing the great power of 
  MCMC methods, that have given a decisive boost to Bayesian inference
  in the past decades.  
\item
  However, we show in  Sec.~\ref{sec:direct} how to solve
  the problem exactly, although not in closed form, and limiting ourselves
  to the pdf of $p$. A simple extension of
  the expression of the normalization constant allows to evaluate
  the first moments of the distribution, from which expected value,
  variance, skewness and kurtosis can be computed
  (and then an approximation of $f(p)$ can be `reconstructed').
\item
  An important issue, also of practical relevance,
  is the inference of the proportions of infectees in
  different populations, analyzed in Sec.~\ref{ss:inferring_Delta_p},
  after having been anticipated in Sec.~\ref{ss:predict_Delta_fP}.
  In fact, since the uncertainties about sensitivity and specificity
  act as systematic errors (hereafter `systematics'),
   the differences between these proportions
  can be determined better than each of them.
\item   
  The role of the prior in the inference of $p$, already
  analyzed in detail
  in Sec.~\ref{ss:priors_priors_priors}, is discussed again
  in Sec.~\ref{ss:more_on_priors}, with particular emphasis
  to the case in which priors are at odds `with the data'
  (in the sense specified there). The take away message
  will be to be very careful in taking literally 
  `comfortable' mathematical models, never forgetting
  the quotes by Laplace and Box reminded in the front page.
\end{itemize}
  Two appendixes complete the paper. Appendix A is
  a kind of summary of `Bayesian formulae', with emphasis
  on the practical importance of unnormalized posteriors obtained
  by a suitable choice of the so called {\em chain rule} of probability
  theory and on which
  most Monte Carlo methods to perform Bayesian inference are based.
  In Appendix B several R scripts are provided in order to allow the reader
  to reproduce most of the results presented in the paper.

\section{Rough reasoning based on expectations}\label{sec:rough}
\subsection{Setting up the problem}
Let us imagine we have a {\em population} of $N$ elements, 
a proportion $p$ of which shares a given character.
The simplest example is that of a box containing $N$ balls,
$n_1$  white and $n_2$ black.
Let $p$ be the proportion of white balls, i.e.  $p=n_1/N$.
If we extract at random $m$ balls, then we {\em roughly expect}
$m_1\approx n_1\times m/N = p\cdot m$ white and
$m_2\approx n_2\times m/N = (N-n_1)\times m/N = (1-p)\cdot m $ black.
A classical problem in probability theory is to {\em infer}
the proportion $p$ from the observed (`measured')
proportion $p_m = m_1/m$.

Obviously, if $m$ is equal to $N$,
i.e. if we completely empty the box, then
we acquire full knowledge of the box content and the solution
is trivial. However, in most cases 
we are unable to analyze the entire population
and we have to infer $p$ from a sample.
Therefore, although $p_m$ can be a reasonable {\em rough 
estimate} of $p$, we can never be sure about the true proportion.
At most, there are numerical values we shall believe
more (those around $p_m$) and others we shall believe less.
This problem was first tackled analytically
by Laplace in 1774\,\cite{Laplace_PC}.

Let us now complicate the problem, taking into account the fact that
we are not even sure about the characteristics of
each sampled individual, as, instead, it happens with 
black and white balls. This is exactly what happens with infections
of different kinds, unless the symptoms are so evident and unique
to rule out any other explanation.
We have then to rely on tests
that are typically not perfect,
especially if we have neither time nor money to inspect
in detail each individual in order to really {\em see} the
active {\em agent}. 
Sticking to tests providing only a binary response,\footnote{But
  we hardly believe that they only provide binary information,
  of the kind Yes/No, and we wonder why a
  (although slightly) more refined scale is not reported,
  even discretized in a few steps, like when we rank
  goods and services with stars.
  Anyway, we shall not touch this question in the present paper,
  but only wanted to express here our perplexity.}
as we hear and read in the media, and  
assuming that such testing devices and procedures
are planned to detect the infected individuals, we expect that 
if the answer is {\em positive} then
there should be a quite high chance that the individual is really infected,
and a small chance that she is not. Similarly,
if the answer is {\em negative}, there should be a
high chance that the individual is not infected. 
(The conditionals are due to the fact that there are other
pieces of information to take into account, as we shall see.)

We can characterize therefore the test by two {\em virtually continuous}
numbers  $\pi_1$ and $\pi_2$ 
in the range between 0 an 1  such that, depending on whether
the individual is infected or not,
the test procedure
provides positive and negative answers with
probabilities
\begin{eqnarray*}
  P(\Pos\,|\,\Inf) &=& \pi_1 \\
  P(\Neg\,|\,\Inf) &=& 1-\pi_1\,;\\
  && \\
  P(\Pos\,|\,\NoInf) &=& \pi_2 \\
  P(\Neg\,|\,\NoInf) &=& 1-\pi_2\,, 
\end{eqnarray*}
with self-evident meaning of the symbols
(we just remind that the `$|$'
indicates that what follows it plays the role of {\em conditions}
and therefore `$|$' should be read as ``under the condition'',
or ``conditioned by'').
More technically,
  $\pi_1$ is defined as test {\em sensitivity}, 
  while $(1-\pi_2)$ is the test {\em specificity}
  (see e.g. Ref.~\cite{Specificity}). Therefore,
  in order to fix the ideas, the
  test to which we are referring~\cite{FattoQuotidiano} has
  98\% sensitivity and 88\% specificity.

As it is easy to understand, the numerical quantities of $\pi_1$ and $\pi_2$
do not come from first principles, but result from previous
measurements. They are therefore affected by uncertainty as
all results in measurements typically are~\cite{ISO}.
Therefore, probability distributions have to be associated also to
the possible numerical values of these two test parameters.
Anyway, within this section we take the
freedom to use their {\em nominal values} of 0.98 and 0.12
for our first rough considerations.

\subsection{Fraction of sampled positives being really
  infected or not}\label{ss:NumberPositiveServa}
Putting all together, our {\em rough expectation} is that our sample
of $m$ individuals will contain $m_1\approx p\cdot m$ infected,
although we shall write it {\em within this section}
as an equality (`$m_1 = p\cdot m$'),
and ditto for other related numbers.
Out of these $m_1$ infected, 
$\pi_1 \cdot m_1$ will be tagged as positive and
$(1-\pi_1)\cdot m_1$ as negative.
Of the remaining $m_2= (1-p)\cdot m$, not infected,
$\pi_2\cdot m_2$ will be tagged as positive and
$(1-\pi_2)\cdot m_2$ as negative. In sum, the expected numbers of
positive and negative will be
\begin{eqnarray}
  n_P &=& \pi_1\cdot m_1 +  \pi_2\cdot m_2 \\
  n_N &=& (1-\pi_1)\cdot m_1 +  (1-\pi_2)\cdot m_2\,, 
\end{eqnarray}
which we can rewrite as
\begin{eqnarray}
  n_P &=& \pi_1\cdot p\cdot m +  \pi_2\cdot (1-p)\cdot m
  \label{eq:positivi_sample}\\
  n_N &=& (1-\pi_1)\cdot p\cdot m +  (1-\pi_2)\cdot (1-p)\cdot m\,,
  \label{eq:negativi_sample} 
\end{eqnarray}
So, just to fix the ideas with a numerical example
and sticking to $\pi_1=0.98$ and $\pi_2=0.12$
of Ref.\,\cite{FattoQuotidiano}, 
in the case we sample 10000 individuals
we get, {\em assuming} 10\% infected ($p=0.10$),
\begin{itemize}
\item number of infected in the sample: 1000
  (and hence 9000 not infected);  
\item infected tagged as positive: 980;
\item infected tagged as negative: 20;
\item not infected tagged as positive: 1080;
\item not infected tagged as negative: 7920;
\item total number of positive: 2060;
\item total number of negative: 7940;
\item fraction of the positives really infected: $980/2060 = 47.6\,\%$.
\end{itemize}
\subsection{Fraction of infectees in the positive sub-sample}
\label{ss:FractionInfected}
We see therefore that, contrary to naive intuition,
in spite of the apparent rather good quality of the test
($\pi_1=0.98$), the result is quite unreliable on the individual base:
a positive person has roughly 50\,\% chance of being really
infected.\footnote{This point is quite relevant
  when the so called {\em false positive} regards some disease
  with a strong social stigma (e.g. AIDS). Bad practices and negligence
  in dealing with test results and ignoring the population background
  caused genuine emotional suffering, heavy distress, up to suicide
  attempts \cite{negligence}. The same applies in forensics, where individual freedom and justice can be badly influenced
  by evidence mismanagement (See Ref.~\cite{Fenton, Koelher} and the references there). In a less tragic context,
  ignoring the role of the {\em priors} can cause bad decisions to be made
  (see e.g. Ref.~\cite{debunk} for an application
  concerning Information Security).} 
But this does not mean that the test was really useless.
It has indeed increased the probability of a {randomly
chosen individual} to be
infected from 10\% to 48\%.
On the contrary, the fraction of negatives really not infected
is $7920/7940 = 99.75\%$.
This result is also surprising on a first sight, being
the specificity
$(1-\pi_2)$ only 88\%, i.e. not `as good'
as the sensitivity $(\pi_1)$, as high as 98\%.
We shall see the reason in a while.
For the moment we just remark that
in this second case the probability of a randomly
chosen individual to be not infected has increased
from 90\,\% to 99.75\,\%.

The reason of these counter-intuitive results is due to the
role of the {\em prior probability} of being infected
or not, based on the best knowledge of the proportion of infected
individuals in the entire population.\footnote{We remind that
we are not taking into account symptoms or other reasons
that would increase or decrease the
probability of a particular individual to be
infected. For example, the journalist of Ref.~\cite{FattoQuotidiano}
  tells that he had `some suspicions' he could have been infected
  on a plane.
}
The easy explanation is that,
given the numbers we are playing with, the number of
positives is strongly `polluted' by the large {\em background}
of not infected individuals.

In order to see how the outcomes depend on $p$, let us lower
its value from 10\% to 1\%.
In this case our expectation will be of 1286 positives, out of which
only 98 infected and 1188 not infected (the details are left
as exercise).
The fraction of positives really infected becomes now only 7.6\,\%.
On the other hand the fraction of negatives really
not infected is as high as  99.98\,\%.
Figure \ref{fig:TruePosTrueNeg}
\begin{figure}[t]
  \begin{center}\epsfig{file=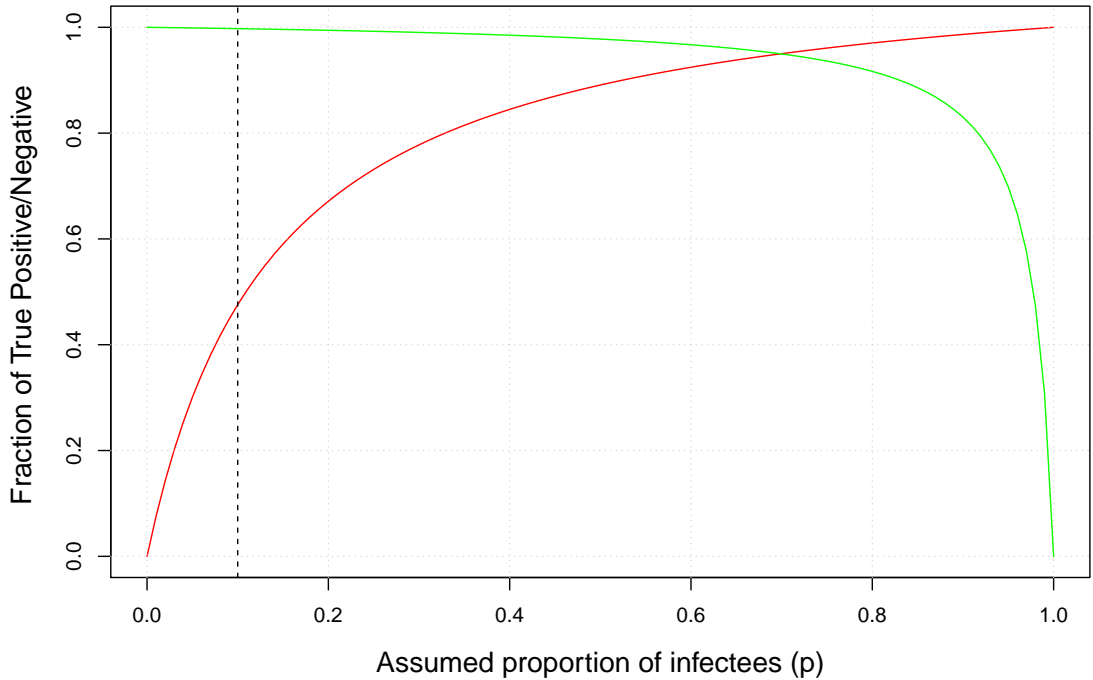,clip=,width=0.95\linewidth}
  \\  \mbox{} \vspace{-1.0cm} \mbox{}
  \end{center}
  \caption{\small \sf Fraction of `true positives' 
    (red line, starting at 0 for $p=0$) and
    `true negatives' (green line,
    starting at 1 for $p=0$) in the sample,
     as a function of the assumed proportion  $p$
    of infected individuals in the population, assuming
    $\pi_1=P(\Pos\,|\,\Inf)=0.98$
    and $\pi_2= P(\Pos\,|\,\NoInf) =0.12$\,. The results in correspondence
    of $p=0.1$, {\em arbitrarily} used as {\em reference value}
    in the numerical example of this section,    
    are marked by the vertical dashed line.
  }
   \label{fig:TruePosTrueNeg}
\end{figure}
shows how these numbers depend on the 
assumed proportion of infectees in the population
(and then in the sample, because of the rough reasoning we are following
in this section).

This should make definitively clear that
the probabilities of interest
not only depend, as trivially expected,
on the performances of the test, summarized here by
$\pi_1$ and $\pi_2$,  but also -- and quite strongly! -- 
 on the {\em assumed} proportion of infectees in the population.
More precisely, they depend on whether
the individual shows symptoms {\em possibly} related to the
searched for infection and on the probability that the same symptoms
could arise from other diseases. However 
we are not in the condition  to
enter into such  `details' in this paper and shall 
focus on {\em random samples of the population}.
Therefore, up to Sec~\ref{sec:uncertain_p}, in which
we deal with the probability that a tested individual
is infected or not on the basis of the test result, we shall refer
to $p$ as `proportion of infectees' in the population.
But {\em everything we are going to say is valid as well if
$p$ is our `prior' probability that a particular individual is 
infected}, based on our best knowledge of the case.

\subsection{Estimating the proportion of infectees in the population}
Now, after having seen what we can tell about a single
individual chosen at random and of which we have no information
about possible symptoms, contacts or behavior,
let us see what we can tell about
the proportion $p$ of infected in the population, based
on the tests performed on the sampled individuals.
The first idea is to solve
Eqs.\,(\ref{eq:positivi_sample}) and (\ref{eq:negativi_sample})
with respect to $p$,
from which it follows
\begin{eqnarray}
  p &=& \frac{n_P - \pi_2\cdot m}{(\pi_1-\pi_2)\cdot m}\,. \label{eq_naive_p}  
\end{eqnarray}
Applying this formula to the 2060 positives got in our numerical
example we re-obtain the input proportion of 10\%,
somehow getting reassured 
about the correctness of the reasoning.
If, instead, we get more positives, for example 2500, 3000 or 3500,
then the proportion would rise to 15.1\%, 20.1\% and 26.7\%, respectively,
which goes somehow in the `right direction'.
If, instead, we get less, for example 2000 or 1500, then the proportion
lowers to 9.3\% and 3.5\%, respectively, which also seems to go
into the right direction.

However, keeping lowering the number of positives something strange happens.
For $n_P= 1200$  Eq.\,(\ref{eq_naive_p}) vanishes
and it becomes even {\em negative} for smaller numbers of positives,
which is something concerning,
 indicating that the above formula
is not valid in general. But why did it nicely  give  the
exact result in the case of 2060 positives? And what is the reason  
why it yields {\em negative proportions} below  
1200 positives?
Moreover, Eq.\,(\ref{eq_naive_p}) has a worrying behavior of diverging
for $\pi_1=\pi_2$, even though irrelevant in practice, because such a test
would be ridiculous -- the same as tossing a coin to tag a person
Positive or Negative (but in such a case we would expect to learn
nothing from the `test', certainly
not that the real proposition of infectees diverges!). 

Let us see the limits of validity of the equation.
\begin{itemize}
\item The  lower limit  $p \ge 0$ implies, as we have
  already seen in the numerical
  example, $n_P \ge \pi_2 \cdot m$ and
    $\pi_1 > \pi_2$.\footnote{Mathematically,
    also negative numerator and denominator would yield
    a positive value of $p$, although this case makes no sense
    in practice, requiring $\pi_1$ smaller than  $\pi_2$.
    Moreover, the  mathematical divergence of Eq.\,(\ref{eq_naive_p})
    -- of no practical relevance, as we have already commented --
  for $\pi_1=\pi_2$ is indeed due to the fact Eq.\,(\ref{eq:positivi_sample})
  and (\ref{eq:negativi_sample}) become then $n_P = \pi_1\cdot m$ and
  $n_N = (1-\pi_1)\cdot m$, not depending any longer on $p$. 
  $[$In more detail, taking $\pi_2=\pi_1-\epsilon$, we get
    $p=(n_P-\pi_1\cdot m+\epsilon\cdot m)/(\epsilon\cdot m)$,
    diverging for $\epsilon\rightarrow 0$.$]$
  } 
\item The upper limit $p \le 1$  is reflected
  in the condition $n_P\le \pi_1\cdot  m$ (and $\pi_1 > \pi_2$).
  In our numeric example this would mean to have less than 9800 positives
  in our sample of 10000. But this ignores the fact that the
  proportion of infectees in the sample could be higher than that
  in the population.
\end{itemize}  
Anyway, it is clear that when the model contemplates
probabilistic effects we have to use sound
methods based on probability theory.

\subsection{Moving to probabilistic considerations}
Let us start seeing what is going on when there are no
infected individuals in the population, i.e. when $p=0$. 
In our rough reasoning none 
of the 10000 sampled individual will be infected.
But 12\% of them will be tagged as positive, exactly the critical value of
1200 we have seen above.
In reality we have neglected the fact that 1200 is an {\em expectation}, 
in the probabilistic meaning of {\em expected value}, but that
other values are also possible. In fact, given the assumed properties
of the test, the number of individuals which shall
result positive to the test is uncertain,
and precisely described by the well known binomial distribution
with `probability parameter' (see Ref.~\cite{ProbProp} for clarifications)
$\pi_2$. The expectation has therefore an uncertainty, that we quantify
with the {\em standard uncertainty} \cite{ISO}, i.e. the 
standard deviation of the related probability distribution.
Using the well known formula
resulting from the binomial distribution, which in our
case reads as $\sigma=\sqrt{\pi_2\cdot (1-\pi_2)\cdot m}$,
we get, using our numbers, $\sigma=32.5$. 
Since we are dealing with reasonably large numbers,
the Gaussian approximation holds and
we can easily calculate that there is about
16\% probability to get a number of positives
equal or below 1167, and so on. In particular we get 0.1\% probability
to observe a number equal or below 1100, which we could consider
a safe limit for practical purposes.

But, unfortunately, the story is a bit longer. In fact we don't have
to forget that  $\pi_2$ comes itself from measurements
and is therefore uncertain.
Therefore, although 0.12 is its `nominal value', also values below
0.10 are easily possible, yielding e.g. an expected number
of positives, among the not infected individuals, of $1000\pm 30$ for
$\pi_2=0.10$ and $800\pm 27$ for $\pi_2=0.08$
(hereafter, unless indicated otherwise,
we quote standard uncertainties).

Then there is the question that we  apply the tests
on the sample, and not on the entire population.
Therefore, unless the proportion of infectees in the population
is exactly 0 or 1, the proportion of infectees
in the sample (`$p_s$'), will differ from $p$.
For example, sticking to a reference $p=0.1$,
in the 10000 individuals sampled from a population
ten times larger
we do not expect exactly 1000 infected,
but $1000\pm 28$    
as we shall see in detail in Sec.~\ref{ss:binom_hg}
(we only anticipate, in answer to somebody who might
have quickly checked the numbers, that the standard uncertainty differs from
30, calculated from a binomial distribution, because this
kind of sampling belongs, contrary to the binomial, to the model
`extraction \underline{without} {\em reintroduction}').

\subsection{Summing up}
The simple reasoning based on {\em mean expectations}
leads to correct results only when all probabilistic effects
are negligible, an approximation which holds,
generally speaking, only for `large numbers'. 
Under this approximation
the numbers of individuals tagged as Positive
or Negative can be considered to
follow in a deterministic way from the assumptions,
one of which is the proportion of infectees. This number can
then be obtained inverting the deterministic relation, thus yielding
Eq.\,(\ref{eq_naive_p}). But when fluctuations around the mean expectations  
become important we need to use probability theory
in order to {\em infer} the parameter of interest.

As far as telling from a single test if a person tagged as Positive
is really infected, we have seen that the prior `assumed proportion' of infected
individuals in the entire populations plays a major role. 
We have seen how to get the probability of interest reasoning
on the fraction of positives really infected in the sample
of positives. In more general terms
this probability {\em has to} be calculated using
Bayes' theorem, that will be shortly reminded in the next section.


\section[Probability of infected in the light
  of the available information]{Probability of infected,
  in the light of the test result and of other relevant
  information}\label{sec:Bayes1}
Having seen the limitations of rough
reasoning in evaluating the probabilities of interest,
let us now start using consistently the rules
of probability theory. We begin focusing on the probability
of infected or not infected, given the test results
and the performances of the test.
We shall move to predict the number of positives in a sample
of tested individuals starting from Sec.~\ref{sec:fpN_from_ps}. 

 
\subsection{Bayes' rule at work}
The probability of Infected or Not Infected,
given the result of the test,
is easily calculated using a simple rule of probability theory
known as {\em Bayes' theorem}
(or {\em Bayes' rule}),\footnote{See Appendix A for details.}
thus obtaining, for the two probabilities to which we
are interested (the other two are obtained by complement),
\begin{eqnarray}
  P(\Inf\,|\,\Pos) & = & \frac{P(\Pos\,|\,\Inf)\cdot P_0(\Inf)}
  {P(\Pos) } \label{eq:P_Inf_Pos_first} \\
  &&\nonumber \\
  P(\NoInf\,|\,\Neg) & = & \frac{P(\Neg\,|\,\NoInf)\cdot P_0(\NoInf)}
  {P(\Neg)}\,, \label{eq:P_NoInf_Neg_first} 
\end{eqnarray}
where $P_0()$ stands for the {\em initial}, or {\em prior} probability,
i.e. `before'\footnote{This usual expression, regularly used in the
  literature together with the term {\em prior},
  could transmit the {\em wrong idea of time order} strictly needed,
  leading to the absurdity that the Bayes' theorem could not be applied
  if one did not `declare' (to a notary?) {\em in advance} her
  priors. What really matters, e.g. in this specific example,
 is the probability that the tested person could be infected
 or not, taking into account all other information but the test result.
 (We shall comment further on the meaning and the role of the priors,
 in particular in Sec.~\ref{ss:priors_priors_priors}.)
 \label{fn:notary}
}
the information of the test result is acquired, i.e.
the degree of belief we attach to the hypothesis that a person
could be e.g. infected, based on our best knowledge
of the person (including symptoms and habits)
and of the infection. 
As we have already said, if the person is chosen absolutely
at random, or we are unable to form our mind even having the person
in front of us, we can only use for $P_0(\Inf)$ the
proportion $p$ of infected individuals in the population,
or {\em assume} a value and provide probabilities
conditioned by that value, as we shall do in a while.
Therefore, hereafter the two `priors' will just be
$P_0(\Inf)= p$ and $P_0(\NoInf)= 1-p$.

Applying another well known theorem,
since the  hypotheses $\Inf$ and $\NoInf$
are exhaustive and mutually exclusive, we can rewrite the above
equations as
\begin{eqnarray}
  P(\Inf\,|\,\Pos) & = & \frac{P(\Pos\,|\,\Inf)\cdot P_0(\Inf)}
  {P(\Pos\,|\,\Inf)\cdot P_0(\Inf)+P(\Pos\,|\,\NoInf)\cdot P_0(\NoInf) }
  \label{eq:bayes0}  \\
  &&\nonumber \\
  P(\NoInf\,|\,\Neg) & = & \frac{P(\Neg\,|\,\NoInf)\cdot P_0(\NoInf)}
  {P(\Neg\,|\,\Inf)\cdot P_0(\Inf)+P(\Neg\,|\,\NoInf)\cdot P_0(\NoInf) }\,.
  \label{eq:bayes}  
\end{eqnarray}
In our model $P(\Pos\,|\,\Inf)$ and  $P(\Neg\,|\,\NoInf)$ depend
on our assumptions on the parameters $\pi_1$ and $\pi_2$, 
that is, including the other two probabilities of interest,
\begin{eqnarray}
  P(\Pos\,|\,\Inf, \pi_1) &=& \pi_1 \label{eq:P_given_pi1} \\
  P(\Pos\,|\,\NoInf, \pi_2) &=& \pi_2\,, \label{eq:P_given_pi2}\\
    P(\Neg\,|\,\Inf, \pi_1) &=& 1- \pi_1 \label{eq:N_given_pi1} \\
  P(\Neg\,|\,\NoInf, \pi_2) &=& 1- \pi_2\,, \label{eq:N_given_pi2}
\end{eqnarray}  
In the same way we can rewrite Eqs.\,(\ref{eq:bayes0}) and  (\ref{eq:bayes}),
adding, for completeness, also the other two probabilities of interest,
as 
\begin{eqnarray}
  P(\Inf\,|\,\Pos,\pi_1,\pi_2,p) & = &  \frac{\pi_1\cdot p}
  {\pi_1\cdot p + \pi_2\cdot (1-p)}
  \label{eq:P_Inf_Pos}\\
  P(\NoInf\,|\Neg,\pi_1,\pi_2,p) &=&  \frac{(1-\pi_2)\cdot (1-p)}
  {(1-\pi_1)\cdot p + (1-\pi_2)\cdot (1-p)}\,. \label{eq:P_NoInf_Neg} \\
  P(\NoInf\,|\,\Pos,\pi_1,\pi_2,p) & = &
  \frac{\pi_2\cdot (1-p)}{\pi_1\cdot p + \pi_2\cdot (1-p)}
  \label{eq:P_NoInf_Pos} \\
  P(\Inf\,|\,\Neg,\pi_1,\pi_2,p) &=&  \frac{(1-\pi_1)\cdot p}
  {(1-\pi_1)\cdot p + (1-\pi_2)\cdot (1-p)}\,. \label{eq:P_Inf_Neg} 
\end{eqnarray}
We also remind that the denominators have the meaning of `a priori
probabilities of the test results', being
\begin{eqnarray*}
  P(\Pos\,|\,\pi_1,\pi_2,p) &=&  \pi_1\cdot p + \pi_2\cdot (1-p) \\
  P(\Neg\,|\,\pi_1,\pi_2,p) &=&(1-\pi_1)\cdot p + (1-\pi_2)\cdot (1-p)\,.
\end{eqnarray*}
For example, taking the parameters of our numerical example
($p=0.1$, $\pi_1=0.98$ and $\pi_2=0.12$), an individual chosen
at random is expected to be tagged as positive or negative
with probabilities
20.6\% and 79.4\%, respectively.
\begin{figure}[t]
  \begin{center}
    \epsfig{file=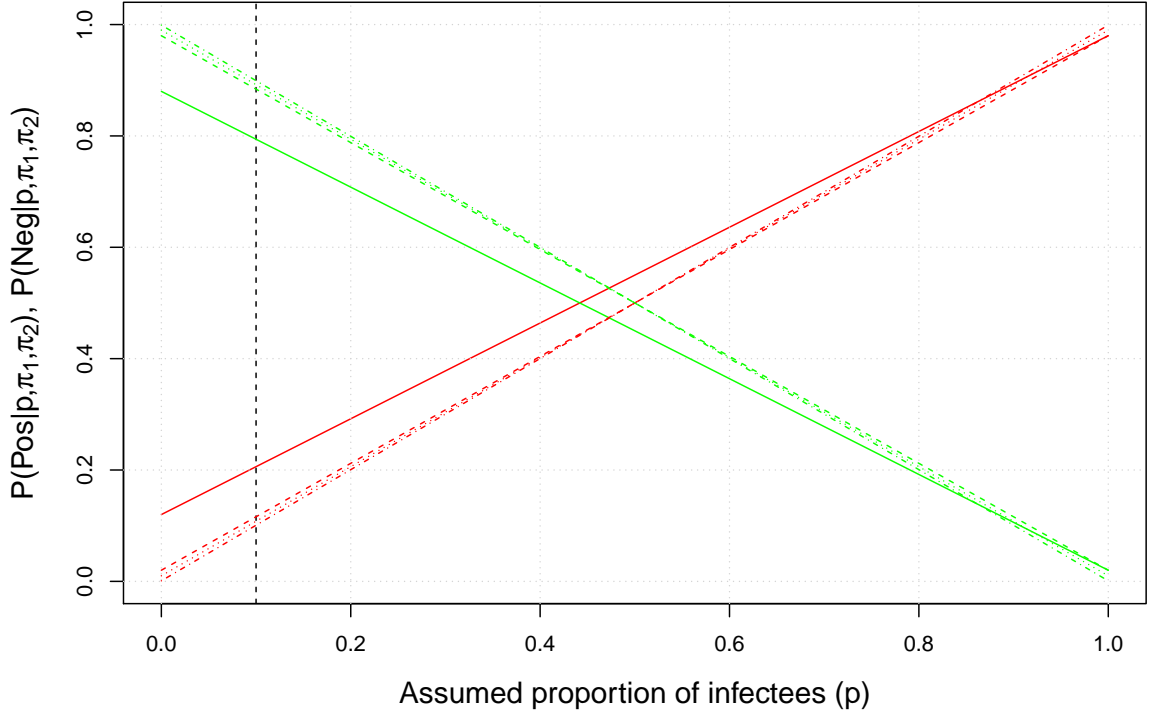,clip=,width=\linewidth}
      \\  \mbox{} \vspace{-1.0cm} \mbox{}
  \end{center}
  \caption{\small \sf Probability that an individual {\em chosen at random}
    will result Positive (red lines with positive slope) or Negative
    (green lines, negative slope)
    as a function of the assumed proportion of infectees
    in the population.
    Solid lines for $\pi_1=0.98$ and $\pi_2=0.12$;
    dashed for $\pi_1=0.98$ and $\pi_2=0.02$;
    dotted for  $\pi_1=0.99$ and $\pi_2=0.01$; dashed-dotted for
     $\pi_1=0.999$ and $\pi_2=0.001$.
  }
  \label{fig:Prob_Pos-Neg}
\end{figure}
Figure \ref{fig:Prob_Pos-Neg} shows these two probabilities
as a function of $p$ for some values of $\pi_1$ and $\pi_2$.

\begin{figure}[t]
  \begin{center}
    \epsfig{file=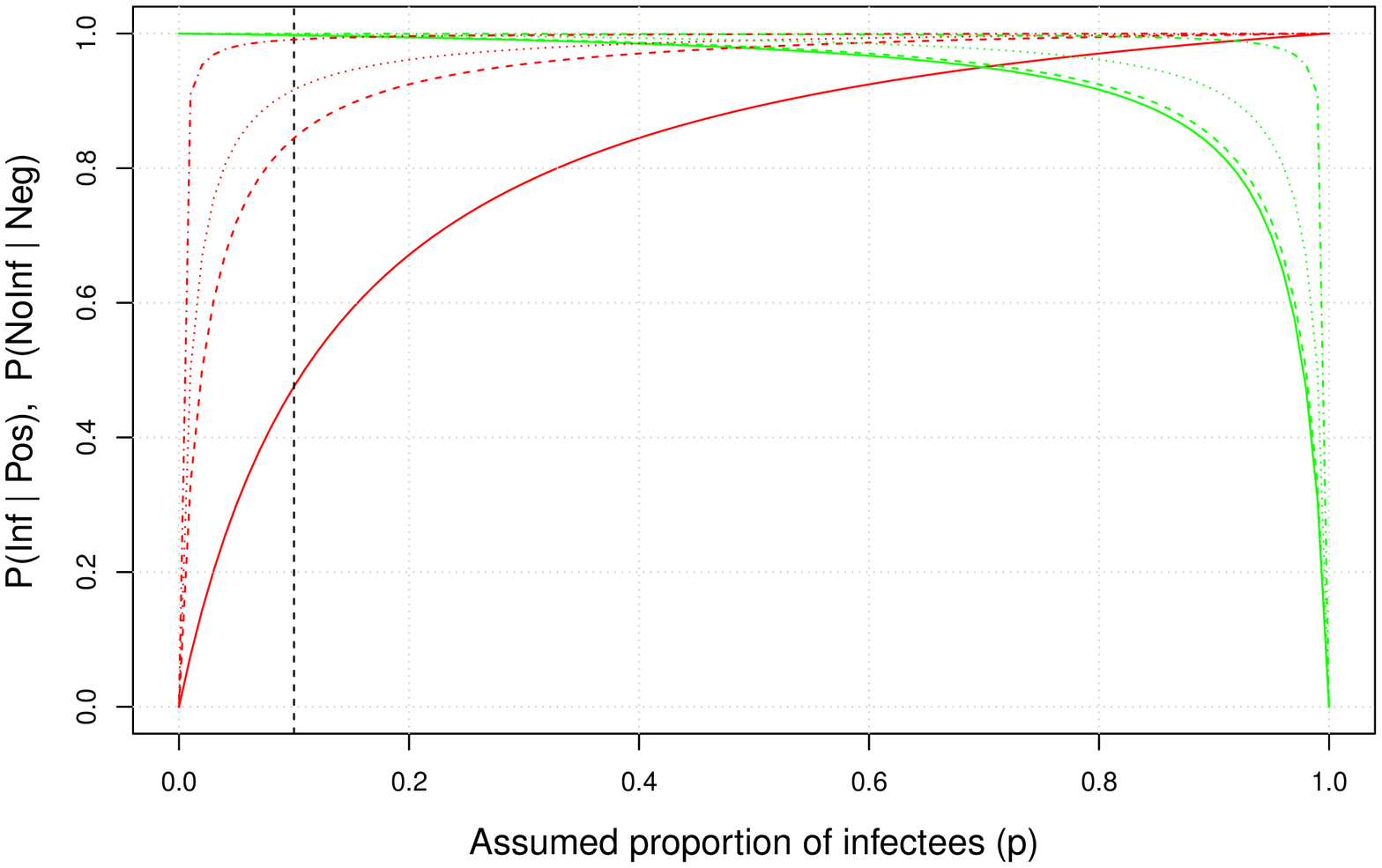,clip=,width=\linewidth}
    \\  \mbox{} \vspace{-1.0cm} \mbox{}
    \end{center}
  \caption{\small \sf Probability of `Infected if tagged as Positive'
    $[\,P(\Inf\,|\,\Pos)$, red line, null at $p=0\,]$
    and probability of `Not Infected if tagged as Negative'
    $[\,P(\NoInf\,|\,\Neg)$, green line, null at $p=1\,]$
    as a function of $p$, calculated from Eqs.\,(\ref{eq:P_Inf_Pos})
    and (\ref{eq:P_NoInf_Neg}) for $\pi_1=0.98$ and $\pi_2=0.12$
    (solid lines). For comparison, we have also included (dashed lines)
    the case of  $\pi_2$ reduced to 0.02, thus increasing the
    `specificity' to 0.98. Then there are the cases
    of a higher quality test $[\pi_1=(1-\pi_2)=0.99]$, shown by dotted lines
    and of an extremely good test $[\pi_1=(1-\pi_2)=0.999)]$
    shown by dotted-dashed lines.
    (The probabilities to tag an individual,
    chosen at random, as positive or negative,           
    for the same sets of parameters,
    were shown in Fig.~\ref{fig:Prob_Pos-Neg}.)
  }
  \label{fig:TruePosTrueNegBayes} 
\end{figure}
Figure \ref{fig:TruePosTrueNegBayes} shows,
by solid lines, $P(\Inf\,|\,\Pos,\pi_1,\pi_2,p)$
and  $P(\NoInf\,|\Neg,,\pi_1,\pi_2,p)$
as a function of $p$,
having fixed  $\pi_1$ and $\pi_2$ at our nominal values
0.98 and 0.12. They  are identical to those
of Fig.\,\ref{fig:TruePosTrueNeg},
the only difference being the label of the $y$ axis,
now expressed in terms of conditional probabilities.
In the same figure we have also added the results obtained
with other sets of parameters $\pi_1$ and $\pi_2$,
as indicated directly in the figure caption.\footnote{The reader might be surprised to
    see plots in which $p$ goes up to 1, but the reason is twofold:
    first, $p$ can be also interpreted in these plots as 
    the purely subjective degree of belief of the expert that
    the tested individual is infected, independently of the test result;
    second, the aim of this paper is rather general and,
    from a physicist's perspective, $p$ could have the meaning 
    of a detector efficiency, a branching ratio in particle decays,
    and whatever can be modeled by a binomial distribution.}

Analyzing the above four  formulae, besides the trivial ideal condition
obtained by $\pi_1=1$ and $\pi_2=0$, one can make
a risk analysis in order to optimize the parameters, depending
on the purpose of the test.
For example, we can rewrite Eq.\,(\ref{eq:P_Inf_Pos}) as
\begin{eqnarray}
  P(\Inf\,|\,\Pos,\pi_1,\pi_2,p) & = &
  \frac{1}{1+ \frac{\pi_2}{\pi_1}\cdot\frac{(1-p)}{p}}\,:
  \label{eq:P_Inf_Pos_1}
\end{eqnarray} 
if we want to be rather sure that a Positive is really infected, then
we need $\pi_2/\pi_1\ll 1$, unless $p\approx 1$.
Similarly, we can rewrite Eq.\,(\ref{eq:P_NoInf_Neg})
as
\begin{eqnarray*}
    P(\NoInf\,|\Neg,,\pi_1,\pi_2,p) &=& 
    \frac{1}{1 + \frac{1-\pi_1}{1-\pi_2}\cdot \frac{p}{1-p}}\,: \label{eq:P_NoInf_Neg_1} 
\end{eqnarray*} 
in this case, as we have learned, in order to be quite confident that the negative test implies no infection, we need $(1-\pi_1)\ll 1$,
that is, for realistic values of $\pi_2$, a value of
$\pi_1$ practically equal to 1, unless $p$ is rather small,
as we can see from Fig.~\ref{fig:TruePosTrueNegBayes}.
(In order to show the importance to reduce $\pi_2$, rather
than to increase $\pi_1$, in the case of low proportion
of infectees in the population,
we show in Fig.~\ref{fig:TruePosTrueNegBayes_1} 
the results based on some other
sets of parameters.) 
\begin{figure}[t]
  \begin{center}
  \epsfig{file=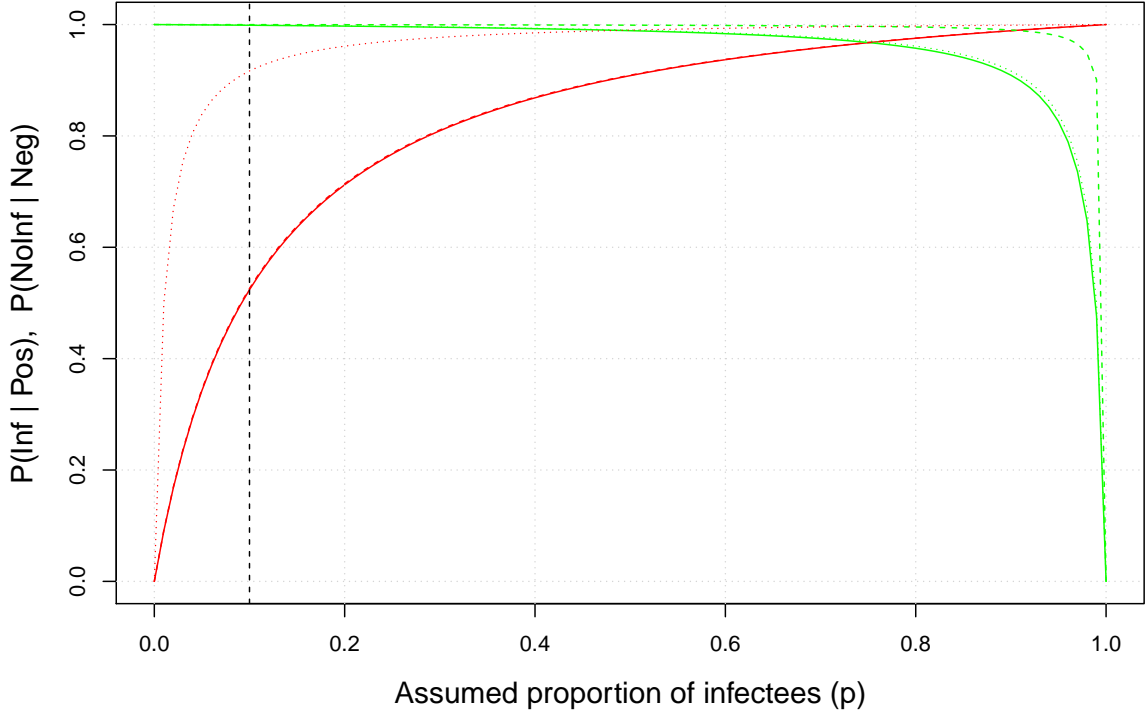,clip=,width=\linewidth}
    \\  \mbox{} \vspace{-1.0cm} \mbox{}
    \end{center}
  \caption{\small \sf Same as Fig.~\ref{fig:TruePosTrueNegBayes},
    but with different parameters. 
    Solid lines: $\pi_1=0.99$ and $\pi_2=0.10$.
    Dashed lines (the red one, describing $P(\Inf\,|\,\Pos)$
    overlaps perfectly
    with the continuous one): $\pi_1=0.999$ and $\pi_2=0.10$.
    Dotted lines (the green one, describing $P(\NoInf\,|\,\Neg)$,
    almost overlaps the solid one):   $\pi_1=0.99$ and $\pi_2=0.01$.
  }
    \label{fig:TruePosTrueNegBayes_1}
\end{figure}

\subsection{Initial odds, final odds and Bayes' factor}\label{ss:BF}
Let us go again to the above formulae, which
we rewrite in different ways in order to get some insights on what
is going on.
Before the test, if no other information is available,
the {\em initial odds} Infected vs Not Infected are
given by
\begin{eqnarray*}
 \frac{P_0(\Inf)}{P_0(\NoInf)} &=& \frac{p}{1-p}\,,
\end{eqnarray*}
equal to $1/9$ for our reference value of  $p=0.1$.
After the test has resulted in Positive
the new probability of Infected is given by Eq.~(\ref{eq:bayes0}).
The corresponding probability of Not Infected is given by
a fraction that has the same denominator but
$P(\Pos\,|\,\NoInf)\cdot P_0(\NoInf)$ as  numerator.
The {\em final odds}
are then given by
\begin{eqnarray}
  \frac{ P(\Inf\,|\,\Pos)}{ P(\NoInf\,|\,\Pos)} &=&
  \frac{P(\Pos\,|\,\Inf)}{P(\Pos\,|\,\NoInf)}
  \times  \frac{P_0(\Inf)}{P_0(\NoInf)}\,. \label{eq:BF}
\end{eqnarray}
Using our numerical values, we get
\begin{eqnarray*}
  \frac{ P(\Inf\,|\,\Pos)}{ P(\NoInf\,|\,\Pos)} &=&
  \frac{\pi_1}{\pi_2}\times \frac{p}{1-p} \\
  &\approx& 8.2 \times \frac{1}{9}\,.
\end{eqnarray*}
The effect of the test resulting in Positive
has been to modify the initial
odds by the factor
$$BF_{ \Infs \,vs\, \NoInfs}\,(\Pos)=\frac{P(\Pos\,|\,\Inf)}{P(\Pos\,|\,\NoInf)}\,,$$
known as  {\em Bayes' Factor}.\footnote{A more proper
  name could be Bayes-Turing factor, or perhaps
  even better Gauss-Turing factor\,\cite{GaussBF}, but we stick
here to the conventional name.}
In our case this factor is equal to $\pi_1/\pi_2 \approx 8.2$.
This means that after a person has been tagged as Positive,
the odds Infected vs Not Infected have increased by this factor.
But since the initial odds were $1/9$,
the final odds are just below 1, that is about 1-to-1, or 50-50. 

In the same way we can define the
Bayes Factor Not Infected vs Infected
in the case of a negative result:
\begin{eqnarray*}
  BF_{ \NoInfs \,vs\, \Infs}\,(\Neg)&=&\frac{P(\Neg\,|\,\NoInf)}{P(\Neg\,|\,\Inf)}
  = \frac{1-\pi_2}{1-\pi_1} 
  = 44 
\end{eqnarray*}  
This is the reason why, for a hypothetical proportion
of infectees in the population of $10\%$, a negative result
makes one {\em practically sure} to be not infected. The
initial odds of 9-to-1 are multiplied by a factor 44, thus
reaching 396, about 400-to-1, resulting into a probability
of not being infected of 396/397, or 99.75\%. 

\subsection{What do we learn by a second test?}\label{ss:SecondTest}
Let us imagine that the same individual undergoes  
a second test and that
the result is again Positive. How should we update
our believes that this individual is infected,
in the light of the second observation? 
The first idea would be to apply {\em Bayes' rule}
in sequence, thus getting an overall Bayes' Factor
of $(\pi_1/\pi_2)^2\approx 67$ that, multiplied by the initial
odds of $1/9$, would give posterior odds of 7.4, or a
probability of being infected of 88\%, still far from a
{\em practical certainty}. But the real question is if we can
apply twice the same kind of test to the same person.
It is easy to understand that the multiplication of the Bayes' factors
assumes (stochastic) {\em independence} among them.
In fact, according to probability theory we have to
replace now Eq.~(\ref{eq:BF}) by
\begin{eqnarray}
  \frac{ P(\Inf\,|\,\Pos^{(1)},\Pos^{(2)})}
       { P(\NoInf\,|\,\Pos^{(1)},\Pos^{(2)})} &=&
       \frac{P(\Pos^{(1)},\Pos^{(2)}\,|\,\Inf)}
            {P(\Pos^{(1)},\Pos^{(2)}\,|\,\NoInf)}
  \times  \frac{P_0(\Inf)}{P_0(\NoInf)}\,, \label{eq:BF2}
\end{eqnarray}
having indicated  by $\Pos^{(1)}$ and  $\Pos^{(2)}$
the two outcomes.
Numerator and denominator of the Bayes' Factor
are then
\begin{eqnarray*}
  P(\Pos^{(1)},\Pos^{(2)}\,|\,\Inf) &=&
  P(\Pos^{(2)}\,|\,\Pos^{(1)},\Inf)\cdot P(\Pos^{(1)}\,|\,\Inf) \\
   P(\Pos^{(1)},\Pos^{(2)}\,|\,\NoInf) &=&
  P(\Pos^{(2)}\,|\,\Pos^{(1)},\NoInf)\cdot P(\Pos^{(1)}\,|\,\NoInf) \,,
\end{eqnarray*}  
which can be rewritten as
\begin{eqnarray*}
  P(\Pos^{(1)},\Pos^{(2)}\,|\,\Inf) &=&
  P(\Pos^{(2)}\,|\,\Inf)\cdot P(\Pos^{(1)}\,|\,\Inf) \\
   P(\Pos^{(1)},\Pos^{(2)}\,|\,\NoInf) &=&
  P(\Pos^{(2)}\,|\,\NoInf)\cdot P(\Pos^{(1)}\,|\,\NoInf) \,,
\end{eqnarray*}
and therefore we can factorize the two Bayes' factors,
{\em only if the two test results are independent}.
But this is far from being obvious. If the test response
depends on something one has in the blood, different from
the virus one is searching for,
a second test of the same kind will most likely give the
same result.

\section{Uncertainty about $\pi_1$ and $\pi_2$} \label{sec:uncertainty}
Until now we have used the nominal values of 
$\pi_1$ and $\pi_2$ of Ref.~\cite{FattoQuotidiano},
and have already seen how our
probabilistic conclusions change if other
sets of values are employed. But
these two model parameters come
from tests performed on selected people,
known with certainty\footnote{This is what we assume, although we
  are not in the position to enter into the details.
}
 to be infected or not.
More precisely $\pi_1=0.98$ results from 400 {\em surely
infected}, 392 of which resulted positive; 
$\pi_2=0.12$ from 200 {\em surely not infected},
176 of which resulted negative~\cite{FattoQuotidiano}.

\subsection[From $ P(n_{P_I}\,|\,n_I,\pi_1)$ to
  $f(\pi_1\,|\,n_{P_I},n_I)$: Bayes' rule applied to `numbers']{From
  $ P(n_{P_I}\,|\,n_I,\pi_1)$ to
  $f(\pi_1\,|\,n_{P_I},n_I)$: Bayes' rule applied to `numbers'}
It is rather obvious to think that, repeating the same test with
samples of exactly the {\em same size}, but involving
{\em different individuals}, no one would be surprised
to count different numbers of positives and negatives in the two samples.
In fact, sticking for a while only to infectees and
assuming an {\em exact value} of $\pi_1$, the number $n_{P_I}$ of positives
is given by the {\em binomial distribution}, 
\begin{eqnarray}
  f(n_{P_I}\,|\,n_I,\pi_1) \equiv P(n_{P_I}\,|\,n_I,\pi_1) & = &
  \frac{n_I!}{n_{P_I}!\cdot (n_I-n_{P_I})!}\cdot
  \pi_1^{n_{P_I}}\cdot (1-\pi_1)^{n_I-n_{P_I}}\,,\ \ \ \ 
  \label{eq:f(n_P)}
\end{eqnarray}
that is, in short (with `$\sim$' to be read as `follows\ldots'), 
\begin{eqnarray*}
n_{P_I} & \sim & \mbox{Binom}(n_I, \pi_1).  
\end{eqnarray*}
The {\em probability distribution} (\ref{eq:f(n_P)}) describes
how much we have to rationally believe to observe 
the possible  values of $n_{P_I}$ (integers between 0 and $n_I$),
given $n_I$ and $\pi_1$.

An {\em inverse problem} is to {\em infer}
$\pi_1$, given  $n_I$ and the {\em observed} number $n_{P_I}$
(indeed, there is also a second inverse problem, that is inferring
$n_I$  from $n_{P_I}$ and $\pi_1$ -- the three problems
are represented graphically by the {\em networks}
of Fig.~\ref{fig:binom_and_inv}).
\begin{figure}[t]
  \begin{center}
    \epsfig{file=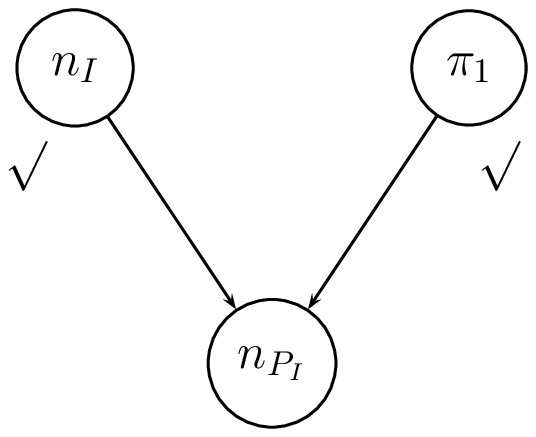,clip=,width=0.28\linewidth}
    \hspace{0.8cm}
    \epsfig{file=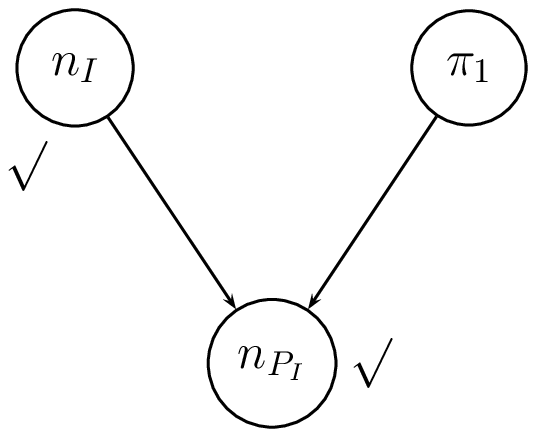,clip=,width=0.28\linewidth}
    \hspace{0.8cm}
    \epsfig{file=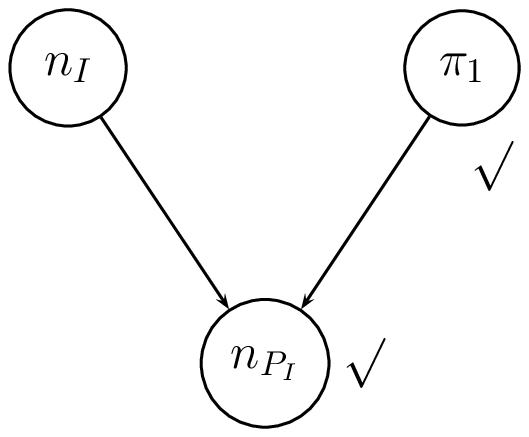,clip=,width=0.28\linewidth}
       \\  \mbox{} \vspace{-1.0cm} \mbox{}
  \end{center}  
  \caption{\small \sf Graphical models of the binomial
    distribution (left) and its `inverse problems'. The symbol
    `$\surd$' indicates the `observed' {\em nodes}
    of the {\em network}, that is the value of the quantity
    associated to it
    is ({\em assumed to be}) certain. The other node
    (only one in this simple case)
    is  `unobserved' and it is associated to
    a quantity whose value is uncertain.
  }
   \label{fig:binom_and_inv} 
\end{figure}
This is the kind of
{\em Problem in the Doctrine of Chances} first solved by
Bayes\,\cite{Bayes},
and, independently and in a more refined way, by
Laplace\,\cite{Laplace_PC} about 250 years ago.
Applying the result of probability theory that
nowadays goes under the name of
{\em Bayes' theorem} (or {\em Bayes' rule}) that we
have introduced in the previous section, we get, apart from the
normalization factor
$[$hereafter the same generic
symbol is
used for both {\em probability functions} and
{\em probability density functions} (pdf), being the meaning clear from the
context$]$:\footnote{Some clarifications are provided
  in Appendix A. With reference to Eq.~(A.8) there,
  Eq.~(\ref{eq:Inference_pi1_1a}) derives from
\begin{eqnarray*}
  f(\pi_1\,|\,n_{P_I},n_I) & \propto &   f(\pi_1,n_{P_I},n_I) \\
    &\propto& f(n_{P_I}\,|\,\pi_1,n_I)\cdot
  f(\pi_1\,|\,n_I)\cdot f(n_I) \\
    &\propto& f(n_{P_I}\,|\,\pi_1,n_I)\cdot  f(\pi_1)\,, 
\end{eqnarray*}
in which we have used a pedantic {\em chain rule} derived
from a bottom-up analysis of the second graphical model of
Fig.~\ref{fig:binom_and_inv} (the one in which $\pi_1$ is unknown)
and taking into account, in the final step, that
$\pi_1$ does not depend on $n_I$,
which has a precise, well known value in this problem.
We can note also that
$f(\pi_1,n_{P_I},n_I)$ involves the continuous variable $\pi_1$
and the discrete values $n_{P_I}$ and $n_I$, being then strictly speaking
neither a probability function nor a probability density function,
while the meaning of each term of the chain rule
is clear from the nature (continuous or discrete) of each variable
(see Appendix A for details).
\label{fn:Inference_pi1_1a}
}
\begin{eqnarray}
  f(\pi_1\,|\,n_{P_I},n_I) & \propto &
  f(n_{P_I}\,|\,\pi_1,n_I) \cdot f_0(\pi_1)
  \label{eq:Inference_pi1_1a} \\
  && \nonumber \\
  & \propto & \pi_1^{n_{P_I}}\cdot (1-\pi_1)^{n_I-n_{P_I}} \cdot f_0(\pi_1)\,,
  \label{eq:Inference_pi1_1}
\end{eqnarray}
where $f_0(\pi_1)$ is the prior pdf, that describes how we believe
in the possible values of $\pi_1$ `before' (see
footnote \ref{fn:notary} and Sec.~\ref{ss:priors_priors_priors})
we get the knowledge
of the {\em experiment}
resulting into $n_{P_I}$ {\em successes} in $n_I$ {\em trials}.
Naively one could say that all possible values of $\pi_1$ 
are equally possible, thus resulting in $f_0(\pi_1)=1$.
But this is absolutely unreasonable,\footnote{Nevertheless,
  we shall comment in  Sec.~\ref{ss:priors_priors_priors}
  about the practical importance of using a {\em flat prior},
  because it is possible to modify the result
  {\em in a second step}, `reshaping' the posterior
  by personal, informative priors based on the best
  knowledge of the problem, which might be different
  for different experts (remember that the `prior'
  does not imply time order, as remarked in
  footnote \ref{fn:notary}).
}
in the case of instrumentation
and procedures  devised by experts in order
to hopefully tag infected people as positive.
Therefore the value of $\pi_1$ should be most likely
in the region above $\approx 90\%$, though without sharp cut below it.
Similarly, reasonable values of $\pi_2$ are expected
to be in the region below $\approx 10\%$.

\subsection{Conjugate priors}\label{ss:conjugate_priors}
At this point, remembering Laplace's dictum that
 {\em ``probability is good sense reduced to a calculus''},
we need to model the prior in a reasonable but mathematically
convenient way.\footnote{See Sec.~\ref{ss:more_on_priors}
  for advice about the usage of mathematically convenient models.}
A good compromise for this kind of problem
is the {\em Beta} probability
function, which we remind here, written  
for the generic variable $x$  and neglecting
multiplicative factors in order to focus, at this point,
on its structure:\footnote{Our preferred {\em vademecum} of Probability
  Distributions is the
  homonymous {\em app}~\cite{ProbabilityDistributions}.
More details are given in Sec.~\ref{sec:direct}.}
\begin{eqnarray}
f(x\,|\,r,s) &\propto& x^{r-1}\cdot (1-x)^{s-1}
\hspace{0.6cm}\left\{\!\begin{array}{l}  r,\,s > 0 \\
   0\le x\le 1 \,.  \end{array}\right.
\label{eq:distr_beta}
\end{eqnarray}
We see that for $r=s=1$ a uniform distribution is recovered. 
An important remark is that for $r>1$ the pdf vanishes
at $x=0$; for $s>1$ it vanishes at $x=1$. It follows
that,
if $r$ and $s$ are both above 1, we can see at a glance
that the function has a single maximum. 
It is easy to calculate that it occurs at (`{\em modal value}')
\begin{eqnarray}
  x_m &=& \frac{r-1}{r + s -2}\,.
  \end{eqnarray}
Expected value and variance
($\sigma^2$) are
\begin{eqnarray}
\mu = \mbox{E}(X)&=&\frac{r}{r+s} \label{eq:Ebeta}\\
\sigma^2 = \mbox{Var}(X)&=&\frac{r\cdot s}{(r+s+1)\cdot(r+s)^2}\,.
\label{eq:Varbeta}
\end{eqnarray}
In the case of uniform distribution, recovered by $r=s=1$, we obtain the well known
$\mbox{E}(X)=1/2$ and $\sigma(X)=1/\sqrt{12}$ (and, obviously, there is no
single modal value).
For large $r=s$, we get $\sigma(X)\approx 1/\sqrt{8\,r}$: as the values
of  $r$ and $s$ increases, the
distribution becomes very narrow around $1/2$.
\begin{figure}[t]
  \begin{center}
    \epsfig{file=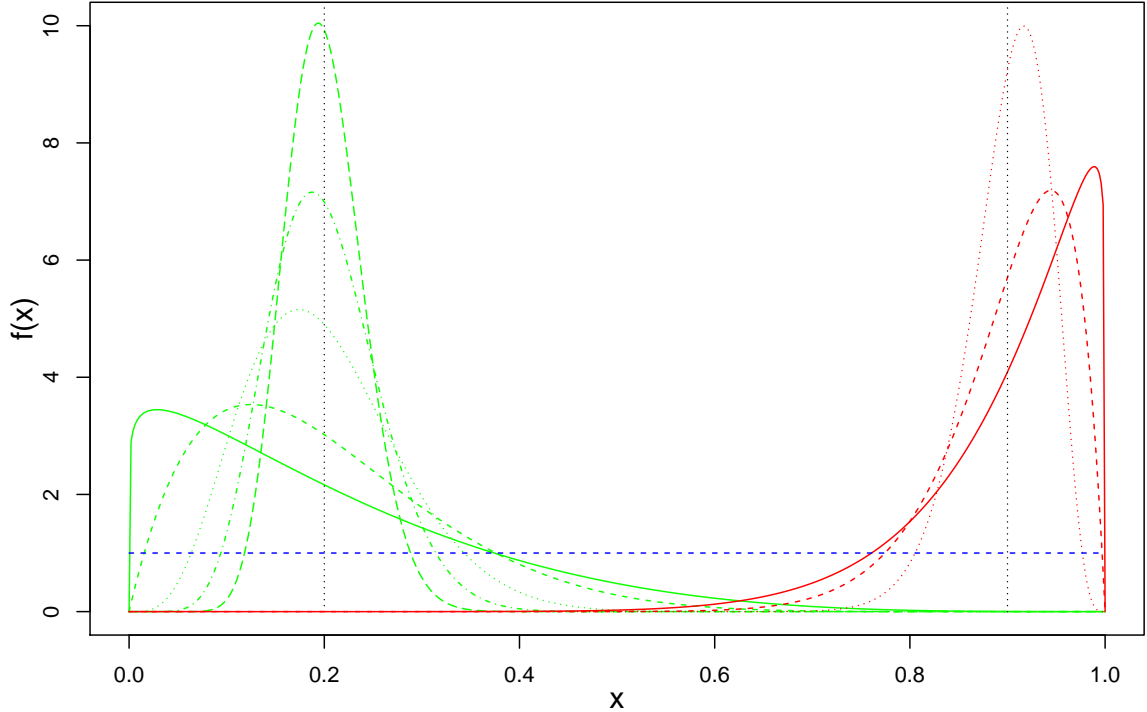,clip=,width=\linewidth}
    \\  \mbox{} \vspace{-1.0cm} \mbox{}
    \end{center}
  \caption{\small \sf Examples of Beta distributions.
    The curves preferring small values of the generic variable $x$,
    all having $\mbox{E}(X)=0.2$
    are obtained with (widest to narrowest)
    $r=1.1,\,2,\,5,\,10$ and $s=4\,r$
    ($\sigma$: 0.16, 0.12, 0.078, 0.056).
    Those preferring larger
    values of $x$,
    all having $\mbox{E}(X)=0.9$
    are obtained with (again widest to narrowest)
    $s=1.1,\,2,\,5$ and $r=9\,s$
    ($\sigma$: 0.087, 0.065, 0.042).
  }
    \label{fig:esempi_beta}
\end{figure}
Examples, with values of $r$ and $s$ to possibly model
the priors we are interested in,
are shown in Fig.\,\ref{fig:esempi_beta}. 

Using the Beta distribution for $f_0(\pi_1)$, our inferential problem
is promptly solved, since 
Eq.~(\ref{eq:Inference_pi1_1})
becomes, besides a normalization factor and with parameters indicated
as $r_0$ and $s_0$ in order to remind their role of prior parameters,
\begin{eqnarray}
  f(\pi_1\,|\,n_I,n_{P_I},r_0,s_0) & \propto &
  \pi_1^{n_{P_I}}\cdot (1-\pi_1)^{n_I-n_{P_I}} \cdot \pi_1^{r_0-1}\,(1-\pi_1)^{s_0-1} \\
   & \propto &
  \pi_1^{n_{P_I}+r_0-1}\cdot (1-\pi_1)^{(n_I-n_{P_I})+s_0-1}
  \label{eq:Inference_pi1_2}
\end{eqnarray}
So, the posterior is still a Beta distribution, with parameters
updated according to the simple rules
\begin{eqnarray}
  r_f &=& r_0 + n_{P_I}  \label{eq:Beta_update_r}\\
  s_f &=& s_0 + (n_I-n_{P_I})\,.  \label{eq:Beta_update_s}
\end{eqnarray}  
For this  
reason the Beta is known to be the {\em prior conjugate}
of the binomial distribution. In terms of our variables, 
\begin{eqnarray}
  n_{P_I} \sim \mbox{Binom}(n_I,\pi_1)
  &\Longrightarrow& \pi_1 \sim \mbox{Beta}(r_0+n_{P_I},s_0+n_I-n_{P_I})\,.  
\end{eqnarray}
The advantage of using the Beta prior conjugate
is self-evident, if we can choose values of $r_0$ and $s_0$
that reasonably
model our prior belief about $\pi_1$. For this 
reason it might be useful to invert Eq.~(\ref{eq:Ebeta})
and (\ref{eq:Varbeta}), thus getting
\begin{eqnarray}
  r_0 &=& \frac{(1-\mu_0)\cdot \mu_0^2}{\sigma_0^2} - \mu_0
  \label{eq:r_from_E-sigma}\\
  s_0  &=&  \frac{1-\mu_0}{\mu_0}\cdot r_0\,.
    \label{eq:s_from_E-sigma}
\end{eqnarray}
So, for example, if we think that $\pi_1$  should be
around 0.95 with a standard uncertainty of about $0.05$,
we get then $r_0 = 17.1$ and $s_0=0.9$, the latter
slightly increased `by hand' to
 $s_0=1.1$ because our rational
prior has to assign zero probability to $\pi_1=1$,
that would imply the possibility of
a perfect test.\footnote{To be fastidious, $s_0<1$
  is not acceptable, because we do not believe
  a priori that a test could be perfect, and therefore
  $f_0(\pi_1)$ has to vanish at $\pi_1=1$. This implies that 
  $s_0$ must be slightly above 1, for example 1.1. But in our case
  the observation of at least one Negative would automatically
  rule out  $\pi_1=1$. Anyway, although this little numerical
  difference is irrelevant in our case, we use $s_0=1.1$
  only because, since we plot priors and posteriors in
  Fig.~\ref{fig:prior_posterior_pi1_pi2}
  {\em we do not like to show a prior not vanishing at 1}.
  $[$We are admittedly a bit pedantic here for
  didactic purposes, but we shall be more pragmatic
  later (see Sec.~\ref{ss:priors_priors_priors})
  and even critical about the literal use of mathematical
  expressions that should instead only be employed for convenience and
  {\em cum grano salis} (see Sec.~\ref{ss:more_on_priors}).$]$
}
The experimental data update then
$r$ and $s$
to $\bm{r = 409.1}$ and $\bm{s=9.1}$.
For $\pi_2$ we model a symmetric prior, with expected value 0.05
and $\sigma= 0.05$. We just need to swap $r$ and $s$, thus getting 
$r_0=1.1$ and $s_0=17.1$, updated by the data
to $\bm{r=25.1}$ and $\bm{s=193.1}$. The results
are shown in Fig.~\ref{fig:prior_posterior_pi1_pi2}. 
\begin{figure}[t]
  \begin{center}
    \epsfig{file=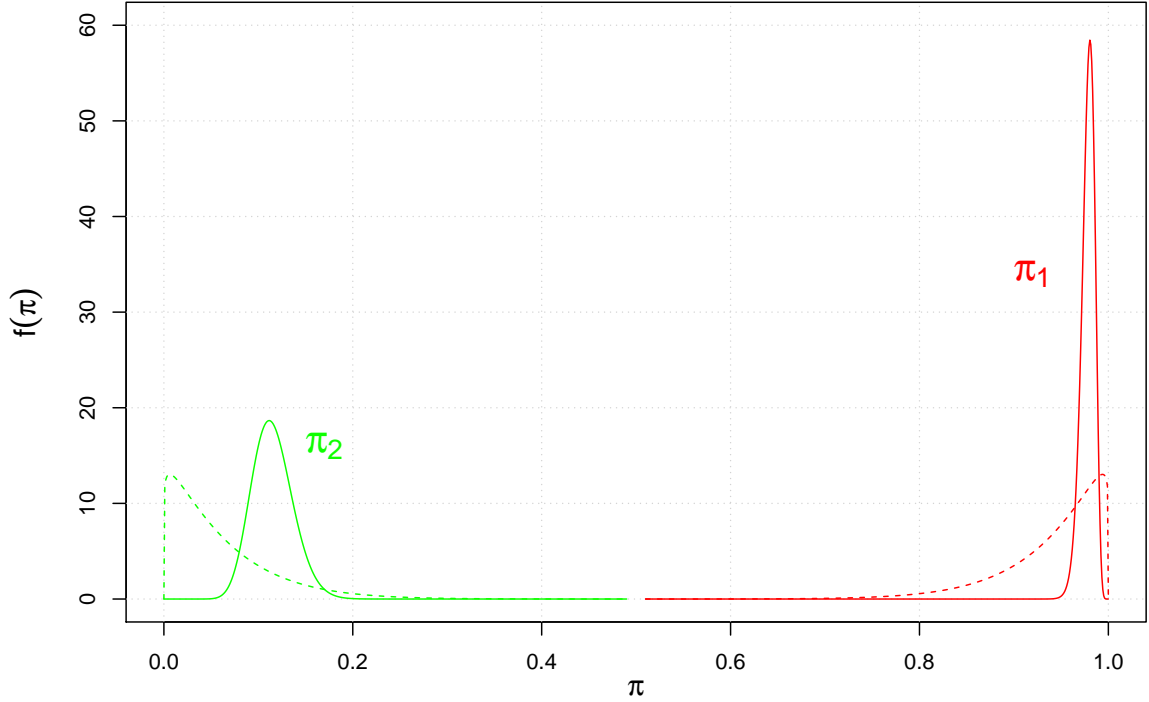,clip=,width=\linewidth}
    \\  \mbox{} \vspace{-1.0cm} \mbox{}
    \end{center}
  \caption{\small \sf Priors (dashed) and posterior (solid)
    probability density functions
    of $\pi_1$ and $\pi_2$.
  }
    \label{fig:prior_posterior_pi1_pi2} 
\end{figure}
 Expressed in terms of
expected value $\pm$ standard deviation they are
\begin{eqnarray}
  \pi_1 &=& 0.978 \pm 0.007\\
  \pi_2 &=& 0.115 \pm 0.022\,.
\end{eqnarray}
As we can easily guess, using simply 0.98 and 0.12, as we have
done in the previous sections, will give essentially the same results,
in terms of expectations. Anyway, in order to be internally consistent
hereafter {\bf our reference values
  will be  $\bm{\pi_1=0.978}$ and
  $\bm{\pi_2=0.115}$}.\footnote{If, instead, we had used
  flat prior over the two parameters, we would get, by
  the Laplace' rule of succession that we shall see in a while,
  0.978 and 0.124. The result is identical (within rounding)
  for $\pi_1$ and practically the same for  $\pi_2$, because 
  with hundreds of trials the inference is dominated by the data.
  (We insist in being fastidiously pedantic because of the didactic
  aim of this paper. For more
  on priors, and for the practical importance of routinely
  using a flat one, see Sec.~\ref{ss:priors_priors_priors}.)
  \label{fn:flat_prior}
}

\subsection{Expected value or most probable value of $\pi_1$ and $\pi_2$?}
At this point someone would object that one should use
the most probable values of $\pi_1$ and $\pi_2$, rather than their
expected values. The answer is rather simple.
Let us consider again Eq.\,(\ref{eq:P_given_pi1}).
Assuming a well precise value of $\pi_1$, the probability of Positive
if Infected is exactly equal to $\pi_1$.
However, if we want to evaluate
$P(\Pos\,|\,\Inf)$, taking into account all possible
values of $\pi_1$ and how much we believe each of them,
that is $f(\pi_1)$, 
we just to need to use a well known result of
probability theory:
\begin{eqnarray}
  P(\Pos\,|\,\Inf) &=& \int_0^1\!P(\Pos\,|\,\Inf,\pi_1)\cdot
  f(\pi_1)\,\mbox{d}\pi_1\,.
\end{eqnarray}
But, being $P(\Pos\,|\,\Inf,\pi_1) = \pi_1$, we get 
\begin{eqnarray}
   P(\Pos\,|\,\Inf)   &=& \int_0^1\!\pi_1\cdot
  f(\pi_1)\,\mbox{d}\pi_1\,,
\end{eqnarray}
in which we recognize the expected value of 
$\pi_1$.\footnote{In
  the case of a uniform prior, i.e. $r_0=s_0=1$, we get 
\begin{eqnarray*}
  P(\Pos\,|\,\Inf) &=& \frac{r_f}{r_f+s_f} = \frac{n_{P_I}+1}{n_I+2}\,, 
\end{eqnarray*}
known as {\em Laplace's rule of succession}. In particular,
for large values of $n_{P_I}$ and $n_I$,  $ P(\Pos\,|\,\Inf)\approx n_{P_I}/n_I$:
{\em more frequently} past tests applied to surely
infected individuals 
resulted in Positive,
{\em more probably} we have to expect a positive outcome of
a new test of the same kind applied
to an infected individual.
}

\subsection{Effect of 
  the
  uncertainties on $\pi_1$ and $\pi_2$ on the probabilities
of interest}\label{sec:dependence_pi1_pi2} 
The immediate question that follows is how the uncertainties
concerning these two parameters change the probabilities
of interest. We start reporting in Tab.~\ref{tab:prob_vs_parametri}
\begin{table}
  {\footnotesize
  \begin{center}
  \begin{tabular}{r|ccccccc}
    & \multicolumn{6}{|c}{$p$} \\
 \multicolumn{1}{c}{Probabilities}  & \multicolumn{6}{|c}{\mbox{}} \\   
  & $0.01$& $0.05$ & $0.10$ & $0.15$ & $0.20$ & $0.50$\\
  \hline
  $P(\Inf\,|\,\Pos,\pi_1\!=\!0.98,\pi_2\!=\!0.12,p)$
  & 0.0762 & 0.301  & 0.476  & 0.590  & 0.671  & 0.891\\ 
  $P(\NoInf\,|\,\Neg,\pi_1\!=\!0.98,\pi_2\!=\!0.12,p)$
  & 0.9998 & 0.999  & 0.997  & 0.996  & 0.994   & 0.978\\
  &&&&&&& \\
  $P(\Inf\,|\,\Pos,\bm{\pi_1=0.978,\pi_2=0.115},p)$
  & \bm{0.0791}  & \bm{0.309}  & \bm{0.486}  & \bm{0.600}  & \bm{0.680}
          & \bm{0.895}\\
 $P(\NoInf\,|\,\Neg,\bm{\pi_1=0.978,\pi_2=0.115},p)$
          & \bm{0.9997}  & \bm{0.999}  & \bm{0.997}  & \bm{0.996}
          & \bm{0.994}  & \bm{0.976} \\
  &&&&&&& \\ 
  \hline
  $P(\Inf\,|\,\Pos,\pi_1\!=\!0.985,\pi_2\!=\!0.115,p)$
 & 0.0796  & 0.311  & 0.488  & 0.602  & 0.682  & 0.895 \\
  $P(\NoInf\,|\,\Neg,\pi_1\!=\!0.985,\pi_2\!=\!0.115,p)$
& 0.9998  & 0.999  & 0.998  & 0.997  & 0.996  & 0.983 \\
  &&&&&&& \\
  $P(\Inf\,|\,\Pos,\pi_1=0.971,\pi_2=0.115,p)$
 & 0.0786  & 0.308  & 0.484  & 0.598  & 0.679  & 0.894\\
 $P(\NoInf\,|\,\Neg,\pi_1=0.971,\pi_2=0.115,p)$
  & 0.9998  & 0.998  & 0.996  & 0.994  & 0.992  & 0.968\\
   &&&&&&& \\
   \hline 
  $P(\Inf\,|\,\Pos,\pi_1\!=\!0.978,\pi_2\!=\!0.137,p)$
& 0.0673  & 0.273  & 0.442  & 0.557  & 0.641  & 0.877\\
  $P(\NoInf\,|\,\Neg,\pi_1\!=\!0.978,\pi_2\!=\!0.137,p)$
 & 0.9997  & 0.999  & 0.997  & 0.996  & 0.994  & 0.975\\
  &&&&&&& \\
  $P(\Inf\,|\,\Pos,\pi_1=0.978,\pi_2=0.093,p)$
& 0.0960  & 0.356  & 0.539  & 0.650  & 0.724  & 0.913\\
 $P(\NoInf\,|\,\Neg,\pi_1=0.978,\pi_2=0.093,p)$
& 0.9998  & 0.999  & 0.997  & 0.996  & 0.994  & 0.976\\
  &&&&&&& \\  
  \hline
  $P(\Inf\,|\,\Pos,\pi_1=0.985,\pi_2=0.093,p)$
& 0.0966  & 0.358  & 0.541  & 0.651  & 0.726  & 0.914 \\
 $P(\NoInf\,|\,\Neg,\pi_1=0.985,\pi_2=0.093,p)$
& 0.9998  & 0.999  & 0.998  & 0.997  & 0.996  & 0.984 \\
  &&&&&&& \\
  $P(\Inf\,|\,\Pos,\pi_1=0.971,\pi_2=0.137,p)$
& 0.0668  & 0.272  & 0.441  & 0.556  & 0.639  & 0.876 \\
 $P(\NoInf\,|\,\Neg,\pi_1=0.971,\pi_2=0.137,p)$
  & 0.9997  & 0.998  & 0.996  & 0.994  & 0.992  & 0.967\\
  &&&&&&& \\  
  \hline
  $P(\Inf\,|\,\Pos,\pi_1=0.985,\pi_2=0.137,p)$
& 0.0677  & 0.275  & 0.444  & 0.559  & 0.643  & 0.878\\
 $P(\NoInf\,|\,\Neg,\pi_1=0.985,\pi_2=0.137,p)$
 & 0.9998  & 0.999  & 0.998  & 0.997  & 0.996  & 0.983\\
  &&&&&&& \\
  $P(\Inf\,|\,\Pos,\pi_1=0.971,\pi_2=0.093,p)$
& 0.0954  & 0.355  & 0.537  & 0.648  & 0.723  & 0.913\\
  $P(\NoInf\,|\,\Neg,\pi_1=0.971,\pi_2=0.093,p)$
  & 0.9997  & 0.998  & 0.996  & 0.994  & 0.992  & 0.969  \\
  &&&&&&& \\ 
  \hline
  \hline
  &&&&&&& \\   
  \multicolumn{1}{c|}{ $P(\Inf\,|\,\Pos,\bm{p})$} &
  0.0815 &  0.314 & 0.490 & 0.603 & 0.682 & 0.895 \\
  \multicolumn{1}{c|}{$P(\NoInf\,|\,\Neg,\bm{p})$ } &
  0.9998 & 0.999 & 0.997 & 0.996 & 0.994 & 0.976 \\
  &&&&&&& \\
    \hline
  \hline
\end{tabular}
\end{center}
}
\caption{\small \sl Probability of Infected and Not Infected, given the
  test result, as a function of the model parameters.
  The third and fourth rows, in boldface,
  are for our reference values of $\pi_1$ and $\pi_2$. The last
  two rows are the results `integrating over' all the possibilities
  of $\pi_1$ and $\pi_2$, according to
  Eq.\,(\ref{eq:P.Inf.Pos.IntegraleDoppio}) and
  (\ref{eq:P.NoInf.Neg.IntegraleDoppio}),
  with the integrals done in practice by Monte Carlo sampling.
}
\label{tab:prob_vs_parametri}  
\end{table}
the dependence of  $P(\Inf\,|\,\Pos,\pi_1,\pi_2,p)$ and 
$P(\NoInf\,|\,\Neg,\pi_1\pi_2,p)$, on which we particularly
focused in the previous sections, on the three parameters.
The dependence on $p$ is shown in the different columns, while
the sets of $\pi_1$ and $\pi_2$ are written explicitly in the
conditionands of the different probabilities. 
We start from the nominal values of 0.98 and 0.12 taken from
Ref.~\cite{FattoQuotidiano} (first two rows of the table). 
Then we use the expected values calculated
in the previous section (third and fourth rows, in boldface), followed
by variations of $\pi_1$ and $\pi_2$ based on
$\pm$ one standard deviation
from their expected values.

We see that the probabilities
of interest do not change significantly, the main effect 
being due to the assumed proportion of infectees in the
population. One could argue that the dependence on
 $\pi_1$ and $\pi_2$  could be larger, if larger deviations 
of the parameters were considered.
Obviously this is true, but one has to
take also into account the (small)
probabilities of large deviations
from the mean values, especially if we allow
simultaneous deviations of both parameters.

A more relevant question is, instead, how do
 $P(\Inf\,|\,\Pos)$ and 
$P(\NoInf\,|\,\Neg$) change, if we take into
account {\em all}  possible variations of the two parameters
(weighed by their probabilities!).
This is easily done, applying the result of
probability theory that we have already used above.
We get, for the probabilities we are mostly interested in,  
\begin{eqnarray}
  P(\Inf\,|\,\Pos,p) &=&
  \int_0^1\!\!\int_0^1\!P(\Inf\,|\,\Pos,\pi_1,\pi_2,p)\cdot
  f(\pi_1,\pi_2)\,\mbox{d}\pi_1\mbox{d}\pi_2\,.
  \label{eq:P.Inf.Pos.IntegraleDoppio} \\
  && \nonumber \\
 P(\NoInf\,|\,\Neg,p) &=&  \int_0^1\!\!\int_0^1\!P(\NoInf\,|\,\Neg,\pi_1,\pi_2,p)\cdot
 f(\pi_1,\pi_2)\,\mbox{d}\pi_1\mbox{d}\pi_2\,,
  \label{eq:P.NoInf.Neg.IntegraleDoppio}
\end{eqnarray}
where $f(\pi_1,\pi_2)$ can be factorized into
$f(\pi_1)\cdot f(\pi_2)$.\footnote{In principle
  $\pi_1$ and $\pi_2$ are not really independent, because
  they might depend on how the test `technology' has been optimized,
  and it could be easily that aiming to
  reach high `sensitivity' affects `specificity'. But
  with the information available to us we can only take them
  independent, each one obtained by the number of positives
  and negatives observed in, hopefully, well controlled samples
  of infected and not infected individuals.
\label{fn:dipendenza_pi1_pi2}}
The integral can be easily done by Monte Carlo,\footnote{The rational
  is quite easy to understand, starting e.g. 
  from Eq.~(\ref{eq:P.Inf.Pos.IntegraleDoppio}) and remembering that
  $ f(\pi_1,\pi_2)\,\mbox{d}\pi_1\mbox{d}\pi_2$ represents the
  infinitesimal probability $\mbox{d}P$ that $\pi_1$ and $\pi_2$
  occur in the infinitesimal cell $\mbox{d}\pi_1\mbox{d}\pi_2$.
  We can discretize the plane $(\pi_1,\pi_2)$
  in $N$ cells and indicate by $P_i$ the probability that 
  a point of $\pi_1$ and $\pi_2$ falls inside it.
  Equation (\ref{eq:P.Inf.Pos.IntegraleDoppio}) can be approximated
  as
  \begin{eqnarray*}
    P(\Inf\,|\,\Pos,p) &\approx & \sum_{i=1}^N P(\Inf\,|\,\Pos,\pi_{1_i},\pi_{2_i},p)
    \cdot P_i \\
   &\approx & \sum_{i=1}^N P(\Inf\,|\,\Pos,\pi_{1_i},\pi_{2_i},p)
    \cdot f_i = \sum_{i=1}^N P(\Inf\,|\,\Pos,\pi_{1_i},\pi_{2_i},p)
    \cdot \frac{n_i}{n_{tot}}\,,
  \end{eqnarray*}
  in which we have approximated each $P_i$ by its
  expected relative frequency of occurrence $f_i=n_i/n_{tot}$
  (Bernoulli's theorem). As one can see, we have approximated
  the integral by a weighted average, in which the
  cells in the plane that are expected to be more probable count
  more. In reality we do not even need to subdivide the plane into cells.
  We just extract at random $\pi_1$ and $\pi_2$ in the plane, according
  to their probability distributions, calculate
  $P(\Inf\,|\,\Pos,\pi_{1_i},\pi_{2_i},p)$ at each point and
  calculate the average. When we consider a very large $n_{tot}$, then 
  {\em we expect that the average will not differ much from the integral}.  
}
whose implementation in the R language\,\cite{R}, both for
$P(\Inf\,|\,\Pos)$ and $P(\NoInf\,|\,\Neg)$, is
given in Appendix B.1.

We get, for our arbitrary reference value of $p=0.1$,
 $P(\Inf\,|\,\Pos,p=0.1) = 0.49038$ and 
$P(\NoInf\,|\,\Neg,p=0.1)=0.99727$, to be compared
to 0.4858 and 0.9973, respectively, if the expected values were used.
The results, shown with an exaggerated number of digits
just to appreciate tiny differences, are practically the same. 
This result could sound counter-intuitive, especially if
one thinks that $\pi_2$ has an almost 20\% intrinsic
standard uncertainty.
The reason is due to the fact that the dependence of the
probabilities of interest on $\pi_1$ and $\pi_2$
is rather linear in the region where their probability
mass is concentrated, as shown in Fig.~\ref{fig:variations_pi1_pi2}.
\begin{figure}[t]
  \begin{center}
    \epsfig{file=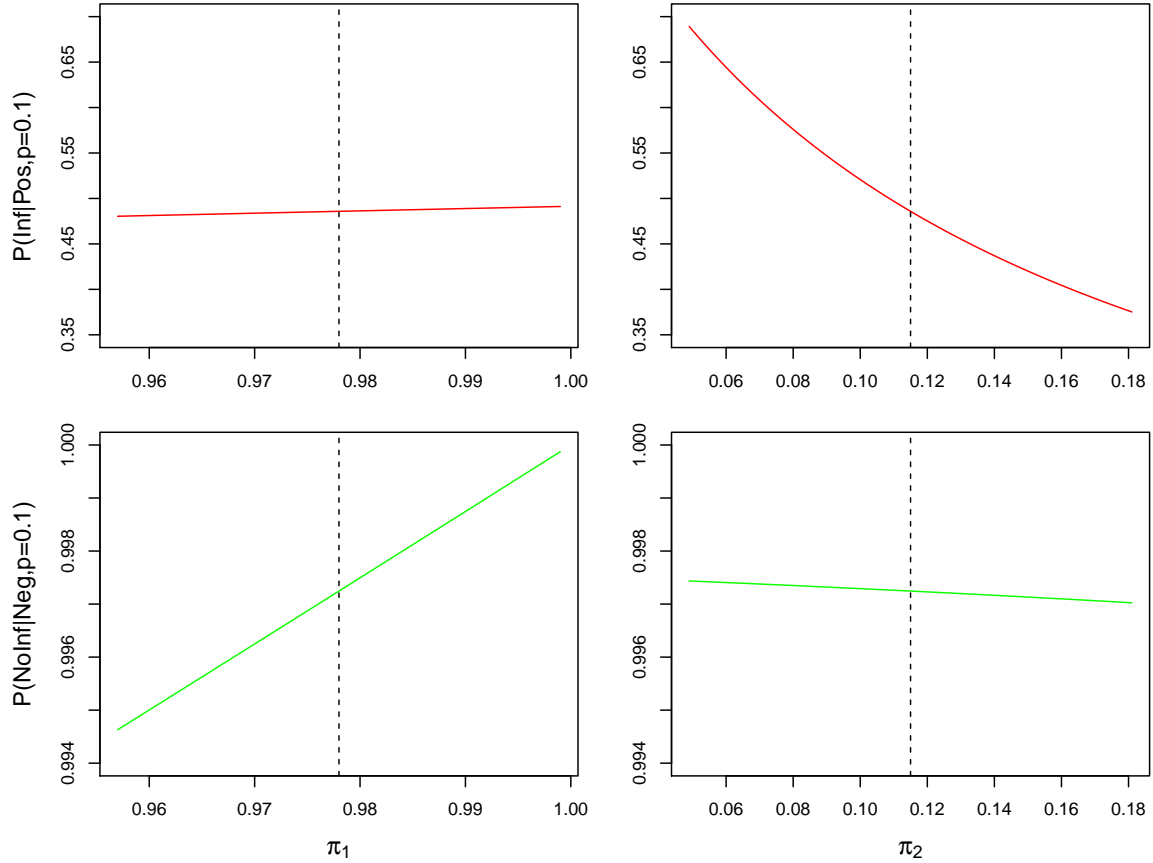,clip=,width=\linewidth}
   \\  \mbox{} \vspace{-1.2cm} \mbox{}  
    \end{center}
  \caption{\small \sf Dependence of $P(\Inf\,|\,\Pos)$ (upper plots) and
    $P(\NoInf\,|\,\Neg)$ (lower plots)
    on $\pi_1$ (left hand plots, for $\pi_2=0.115$
    and $p=0.1$) and on  $\pi_2$ (right hand plots, for $\pi_1=0.978$
    and $p=0.1$). The parameters $\pi_1$ and $\pi_2$ are
    allowed to change withing a range of $\pm 3\,\sigma$'s around
    their expected values.
  }
    \label{fig:variations_pi1_pi2} 
\end{figure}
This rather good linearity causes a high degree of cancellations
in the integral.\footnote{A similar effect happens in evaluating
the contribution of systematics on measured physical quantity.
If the dependence of the `influence factor'\,\cite{ISO}
is almost linear, then the `central value' is practically not affected,
and only its `standard uncertainty' increases. $[$But in our case
we are only interested on its `central value', that is e.g.
the result of the integrals of
Eqs.~(\ref{eq:P.Inf.Pos.IntegraleDoppio})-%
(\ref{eq:P.NoInf.Neg.IntegraleDoppio}).$]$
}
This explains why the only perceptible effect appears
in $P(\Inf\,|\,\Pos,p=0.1)$, {\em slightly} 
larger than the number calculated at the expected values
(49.04\% vs 48.58\%), caused by the small non-linearity
of that probability as a function of $\pi_2$,
as shown in the upper, right hand plot of 
Fig.~\ref{fig:variations_pi1_pi2}: symmetric
variations of $\pi_2$ cause  {\em slightly}
asymmetric variations of  $P(\Inf\,|\,\Pos,p=0.1)$,
thus {\em slightly} favoring higher
values of that probability.

\subsection{Adding also the uncertainty about $p$}\label{sec:uncertain_p} 
Now that we have learned the game, we can use it to
include also the uncertainty concerning $p$. At a given
stage of the pandemic we could have good reasons to
guess a  proportion of infected around 10\%, as we
have been done till now, with a sizable uncertainty,
for example 5\% (i.e. $p=0.10\pm 0.05$). We model, also in this case, 
$f(p)$ with a Beta distribution, getting $r=3.5$ and $s=31.5$.
Equation~(\ref{eq:P.Inf.Pos.IntegraleDoppio}) becomes then
\begin{eqnarray}
  P(\Inf\,|\,\Pos)\! &=&\!
  \int_0^1\!\!\int_0^1\!\!\int_0^1\!P(\Inf\,|\,\Pos,\pi_1,\pi_2,p)\cdot
  f(\pi_1,\pi_2,p)\,\mbox{d}\pi_1\mbox{d}\pi_2\mbox{d}p 
  \label{eq:P.Inf.Pos.IntegraleDoppio_p}  \\
  && \nonumber \\
 \!&=&\!
  \int_0^1\!\!\int_0^1\!\!\int_0^1\!P(\Inf\,|\,\Pos,\pi_1,\pi_2,p)\cdot
  f(\pi_1)\cdot f(\pi_2)\cdot f(p)\,\mbox{d}\pi_1\mbox{d}\pi_2\mbox{d}p\,,
  \ \ \  \ \  \label{eq:P.NoInf.Neg.IntegraleDoppio_p}  
\end{eqnarray}
in which we have made explicit that the joint pdf factorizes,
considering $\pi_1$, $\pi_2$ and $p$
independent.\footnote{The question could be
  a bit more sophisticated, and we have already commented
  in footnote \ref{fn:dipendenza_pi1_pi2}
  on the possible dependency of $\pi_1$ and $\pi_2$.
  But, given the information at hand and the purpose of this
  paper, this is a more than reasonable assumption.
}
With a minor modification
to the script provided in Appendix B.1\footnote{
  One just needs to replace `{\tt p = 0.1}' by
  `{\tt p = rbeta(n, 3.5, 31.5)}', to be placed
  after {\tt n} has been defined.
}
we get
$P(\Inf\,|\,\Pos) = 0.4626$ and 
$P(\NoInf\,|\,\Neg) = 0.9972$, reported again with an exaggerated number
of digits. We only note a small effect in $P(\Inf\,|\,\Pos)$.
As a further exercise, let also take into account $p=0.20\pm 0.10$,
modeled by a $\mbox{Beta}(r\!=\!3,s\!=\!12)$. In this case the Monte Carlo
integration yields
$P(\Inf\,|\,\Pos) = 0.641$ and 
$P(\NoInf\,|\,\Neg) = 0.993$, to be compared with 0.682 and 0.994
of Tab.~\ref{tab:prob_vs_parametri}.\footnote{The reason why
  the integral over all possible values of $p$ gives
  $P(\Inf\,|\,\Pos)$ smaller than that obtained
  at a fixed value of $p$ can be understood looking
  at the solid red curve of Fig.~\ref{fig:TruePosTrueNegBayes}
  showing $P(\Inf\,|\,\Pos)$ as a function of $p$ around $p=0.1$,
  indicated by the vertical dashed line. If $p$ has a symmetric variation
  around 0.1 of $\pm 0.1$ (just to make things more evident),
  than $P(\Inf\,|\,\Pos)$ has an asymmetric variation
  of $^{+0.20}_{-0.47}$ around 0.476 and therefore the Monte Carlo average 
  will be quite below 0.476 (but the Beta distribution
  used for $p$ is skewed on the right side and therefore
  there is a little compensation). For the same reason
  $P(\NoInf\,|\,\Neg)$, practically flat in that region of $p$,
  is instead rather insensitive on the exact value of $p$
  (unless we take unrealistic values around 0.9).
}

\subsection{Uncertainty about  $P(\Inf\,|\,\Pos)$ and
  $P(\NoInf\,|\,\Neg)$?}
As we have seen, the probabilities of interest, taking into account
all the possibilities of $\pi_1$, $\pi_2$ and $p$ are obtained
as weighted averages, with weights equal to $f(\pi_1,\pi_2,p)$.
One could then be tempted to evaluate the standard deviation too,
attributing to it the meaning of `standard uncertainty' about
$P(\Inf\,|\,\Pos)$ and $P(\NoInf\,|\,\Neg)$. But some care is needed.
In fact, although is quite 
obvious that, sticking again to $P(\Inf\,|\,\Pos)$, 
we can {\em form an idea about
the variability} of $P(\Inf\,|\,\Pos,\pi_1,\pi_2,p)$
varying $\pi_1$, $\pi_2$ and $p$ according to $f(\pi_1,\pi_2,p)$
(something like we have done in Tab.~\ref{tab:prob_vs_parametri},
although we have not associated probabilities to the different
entries of the table),
one has to be careful in making a further step.
The fact that the weighted average \underline{is}
$P(\Inf\,|\,\Pos)$ comes from the rules of probability theory,
namely from Eq.~(\ref{eq:P.Inf.Pos.IntegraleDoppio_p}),
but there is not an equivalent rule to evaluate
the uncertainty of $P(\Inf\,|\,\Pos)$.

In order to simplify the notation, let us indicate in the following lines
$P(\Inf\,|\,\Pos)$ by $\cal P$. In order to speak about
standard uncertainty of $\cal P$, we first need to
define the pdf  $f(\cal P)$, and then evaluate average
and standard deviation. But Eq.~(\ref{eq:P.Inf.Pos.IntegraleDoppio_p})
does not provide that, but only a single number, that is
$\cal P$ itself. 

Let us reword what we just stated
using a simple example. Given the `random variable'
$X$ and the pdf associated to it $f(x)$, mean and standard deviation
of $f(x)$ provide expected value (`$\mu_X$')  and standard
deviation \underline{of $X$}, and not of $\mu$.

\section{Predicting the number of positives resulting from
  testing a sample}\label{sec:fpN_from_ps}
The previous sections have been dedicated to the evaluation of 
the probability that a particular individual, tagged as positive
in a test, is really infected. In those sections we have understood how, 
in absence of any other hints, it is important to know 
the percentage $p$ of infectees in the population. Knowing this 
parameter is paramount also for better designing a containing strategy 
in addressing the pandemic. Therefore we move now  to the related, but quite
different problem: `counting', although not in an
exact way, the number of infected
individuals in a population. Given the didactic spirit of this paper,
we keep proceeding  step-by-step. First we focus on the number
of positives that we expect to observe if we check
a {\em sample} using the quite imperfect test
we are considering. Then we also take into account
the effect of {\em sampling a population},
since, as it is rather obvious, the
proportion of infected in a sample of size $n_s$ will not be exactly
equal to that in the whole population of  $N$ individuals.
For this reason we distinguish, hereafter, $p_s$ of the sample
from $p$ of the population.

\subsection{Expected number of positives and its standard uncertainty}
In Sec.~\ref{ss:NumberPositiveServa} we have considered
the numbers of positives and negatives that we
expect to observe, analyzing 10000 individuals,
using our initial parameters
($p_s=0.10$, $\pi_1=0.98$, $\pi_2=0.12$) but
without taking into account the unavoidable 
`statistical fluctuation'. We do it now, using the
probabilistic graphical  model shown 
in Fig.~\ref{fig:two_binom},   
\begin{figure}[t]
  \begin{center}
    \epsfig{file=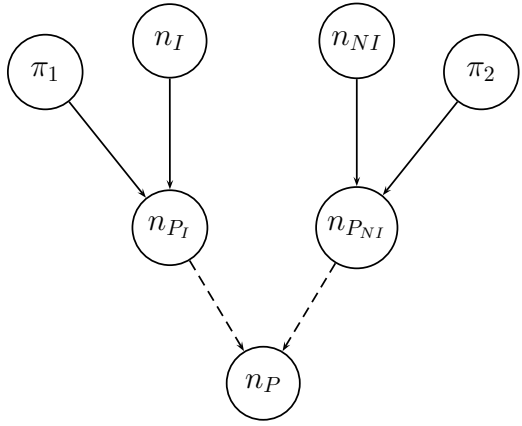,clip=,width=0.45\linewidth}
    \\  \mbox{} \vspace{-1.0cm} \mbox{}
    \end{center}
  \caption{\small \sf Graphical model in which the number of positives
    could come from infected or not infected individuals. Arrows with
    dashed lines stand for a deterministic link, being
    $n_{P}$ simply equal to the sum of  $n_{P_I}$ and  $n_{P_{NI}}$.
  }
    \label{fig:two_binom} 
\end{figure}
obtained by doubling the basic one
of Fig.~\ref{fig:binom_and_inv}, one branch
for the infectees and a second for the others. Then the numbers
of positives resulting from the two contributions are added up.
Note in Fig.~\ref{fig:two_binom}
the dashed arrows from the {\em nodes} $n_{P_I}$
and  $n_{P_{NI}}$ to the {\em node} $n_{P}$:
they indicate a deterministic link,\footnote{This convention is standard
  in the literature, although one might object -- and we agree --
  that the opposite
  one would have been a better choice, a solid line better representing
a deterministic link than a dashed one, but we stick to the convention.}
being $n_{P} = n_{P_I} + n_{P_{NI}}$.

The probability distribution of $n_P$ is with good approximation
Gaussian, due to the well known large numbers behavior of the
binomial distribution (and, moreover,
to the properties of the sum of `random variables').
On the other hand, the expected value and the standard deviation of $n_P$ can  been
calculated exactly, using the properties
of expected values and variances, thus getting
(summarizing for sake of space with the 
 symbol $I$, staying for all available {\em information},
the conditions on which the
various quantities depend):
\begin{eqnarray}
   \mbox{E}(n_P\,|\,I) &=& \mbox{E}(n_{P_I}\,|\,I)
   + \mbox{E}(n_{P_{NI}}\,|\,I) \nonumber \\
   &=&  \pi_1\cdot n_I + \pi_2\cdot n_{NI}  \nonumber \\
  &=& \pi_1\cdot p_s\cdot n_s + \pi_2\cdot (1-p_s)\cdot n_s \\  
  \sigma^2(n_P\,|\,I) &=& \sigma^2(n_{P_I}) + \sigma^2(n_{P_I})  \nonumber \\
   &=&
  \pi_1\cdot (1-\pi_1)\cdot p_s\cdot n_s +
  \pi_2\cdot (1-\pi_2)\cdot (1-p_s)\cdot n_s  \label{eq:var_nP_binom}\\
  \sigma(n_P\,|\,I) &=&
  \sqrt{ \pi_1\cdot (1-\pi_1)\cdot p_s\cdot n_s +
    \pi_2\cdot (1-\pi_2)\cdot (1-p_s)\cdot n_s}\,,   
\end{eqnarray}
with $n_s$ the sample size. 
Expected value and standard deviation
of the fraction of the number of individuals
tagged as positive ($f_P=n_P/n_s$) are then
\begin{eqnarray}
  \mbox{E}(f_P\,|\,I) = \frac{\mbox{E}(n_P\,|\,I)}{n_s} &=&
  \pi_1\cdot p_s + \pi_2\cdot (1-p_s) \\
  \sigma(f_P\,|\,I) = \frac{ \sigma(n_P\,|\,I)}{n_s} 
  &=& \sqrt{\frac{\,\pi_1\cdot (1-\pi_1)\cdot p_s +
      \pi_2\cdot (1-\pi_2)\cdot (1-p_s)\,}{\,n_s}}.
  \ \ \ \ \ 
\end{eqnarray}
For example, making use of our reference numbers
($n_s=10000$, $\pi_1= 0.978$ and  $\pi_2= 0.115$)
we get for some
values of $p_s$ (expected value $\pm$ standard uncertainty):
\begin{eqnarray*}
   \left.n_P\right|_{(n_s=10000,\,\pi_1= 0.978,\,\pi_2= 0.115,\,{\footnotesize\bm{ p_s=0.0}})}
  &=&  1150\pm 32 \ \ \ \longrightarrow \, f_P=0.1150\pm0.0032\\ 
  \left.n_P\right|_{(n_s=10000,\,\pi_1= 0.978,\,\pi_2= 0.115,\,{\footnotesize\bm{ p_s=0.1}})}
  &=&  2013\pm 31  \ \ \ \longrightarrow \, f_P=0.2013\pm0.0031\\
  \left.n_P\right|_{(n_s=10000,\,\pi_1= 0.978,\,\pi_2= 0.115,\,{\footnotesize\bm{ p_s=0.2}})}
  &=&  2876\pm 29  \ \ \ \longrightarrow \, f_P=0.2876\pm0.0029 \\
  \left.n_P\right|_{(n_s=10000,\,\pi_1= 0.978,\,\pi_2= 0.115,\,{\footnotesize\bm{ p_s=0.5}})}
  &=&  5465\pm 25  \ \ \ \longrightarrow \, f_P=0.5465\pm0.0025\,.
\end{eqnarray*}
From this numbers we can get an idea about the precision
we could get on $p_s$, {\em if $\pi_1$ and $\pi_2$ were
perfectly known}, although their values are rather far from
what one would ideally desire. For example, since under the hypotheses
$p_s=0.1$ and $p_s=0$ (and similar numbers are obtained
varying $p_s$ from $0.1$ to $0.2$) the expected difference of
positives is $\Delta_{n_P}=863\pm 45$, it follows that,
varying $p_s$ by $\pm 0.01$ the expected number of positives would vary
by $\approx \pm\, (86\pm 4.5)$. This means that, {\em roughly speaking},
it {\em could} be possible to estimate $p_s$ with an uncertainty
of $\pm 0.01$ or better.

Before taking into account the effects due to the uncertainties
of $\pi_1$ and $\pi_2$, let us
also see how the quality of the measurement depends on the sample size.
In order to do this, we fix this time $p_s$
to our arbitrary value of $0.1$
and vary the sample size by about half order of
magnitude (that is $\approx 10^{k/2}$, with $k=6, 7, \ldots, 10$),
reporting in this case directly the expected
fraction of positives:
\begin{eqnarray*}
   \left.f_P\right|_{({\footnotesize\bm{n_s=1000}},\,\pi_1= 0.978,\,\pi_2= 0.115,\,p_s=0.1)}
   &=&  0.2013\pm 0.0097 \\
  \left.f_P\right|_{({\footnotesize\bm{n_s=3000}},\,\pi_1= 0.978,\,\pi_2= 0.115,\,p_s=0.1)}
   &=&  0.2013\pm 0.0056 \\  
  \left.f_P\right|_{({\footnotesize\bm{n_s=10000}},\,\pi_1= 0.978,\,\pi_2= 0.115,\,p_s=0.1)}
  &=&  0.2013\pm 0.0031 \\
   \left.f_P\right|_{({\footnotesize\bm{n_s=30000}},\,\pi_1= 0.978,\,\pi_2= 0.115,\,p_s=0.1)}
   &=&  0.2013\pm 0.0018 \\
    \left.f_P\right|_{({\footnotesize\bm{n_s=100000}},\,\pi_1= 0.978,\,\pi_2= 0.115,\, p_s=0.1)}
   &=&  0.2013\pm 0.0010\,.  
\end{eqnarray*}
As we can see, if we knew perfectly $\pi_1$ and $\pi_2$,
already a sample of a few thousands individuals would allow us to
predict the fraction of tagged positives with a relative
uncertainty of a few percent.
However there are other effects to be taken into account:
\begin{itemize}
\item there is uncertainty about $\pi_1$ and $\pi_2$;
\item the proportion of infectees in the
  sample is different from that in the population
  (that is, in general $p_s$ differs from $p$);
\item the inference from the observed numbers of positives to
  $p_s$, and then to $p$, has to be done using
  sound probabilistic inferential methods.
\end{itemize}  

\subsection{Taking into account the uncertainty on $\pi_1$ and $\pi_2$}
\label{ss:model_uncertainties_pi1_pi2}
As we have seen in
Sec.~\ref{sec:dependence_pi1_pi2},
the way to take into account all possible values
of $\pi_1$ and $\pi_2$, using the rules of probability theory,
consists in evaluating the following integral
\begin{eqnarray}
  f(n_P\,|\,n_s,p_s) &=& \int_0^1\!\!\int_0^1\!f(n_P\,|\,n_s,p_s,\pi_1,\pi_2)\cdot
  f(\pi_1)\cdot f(\pi_2)\,\mbox{d}\pi_1\mbox{d}\pi_2\,.
  \label{eq:f_nP_int_pi1_pi2}
\end{eqnarray}
Before tacking the problem of how to evaluate this integral,
a very important remark on how we are going to {\em model
  the uncertainty about $\pi_1$ and $\pi_2$} is in order.
\begin{itemize}
\item When we write $f(n_P\,|\,n_s,p_s,\pi_1,\pi_2)$, we are assuming,
  trivially, the 
  same exact values of $\pi_1$ and $\pi_2$ for all the tests
  performed on the $n_s$ individuals of the sample.
\item If, instead, their value is uncertain, and we describe
  their uncertainty by  $f(\pi_1)$ and  $f(\pi_2)$, again
  it means that {\em the same two numerical values} influence the
  results of the $n_s$ tests. But {\em we just do not know
    with certainty which
  are these values}. 
\item
  In particular, associating to these two parameters
  the pdf's $f(\pi_1)$ and $f(\pi_2)$ \underline{does not}
  mean that $\pi_1$ and $\pi_2$ fluctuate from one test to one other.
  The two pdf's only describe the uncertainty on their numerical values.
\item
  It is however reasonable to think that,
  from how the `test devices' are built up, each
  item could perform slightly differently than the other, but we
  \underline{shall ignore}
  these possible test-to-test fluctuations, although
  they could be taken into account just extending the model.
\end{itemize}
Going back to the practical issue of evaluating the integral,
we use again Monte Carlo methods,
employing e.g. the R script provided
in Appendix B.2, for the case of  $n_s=10000$ and $p_s=0.1$.
The result, shown in the bottom plot of  Fig.~\ref{fig:PredictionPositive},
\begin{figure}
  \begin{center}
  \centering{\epsfig{file=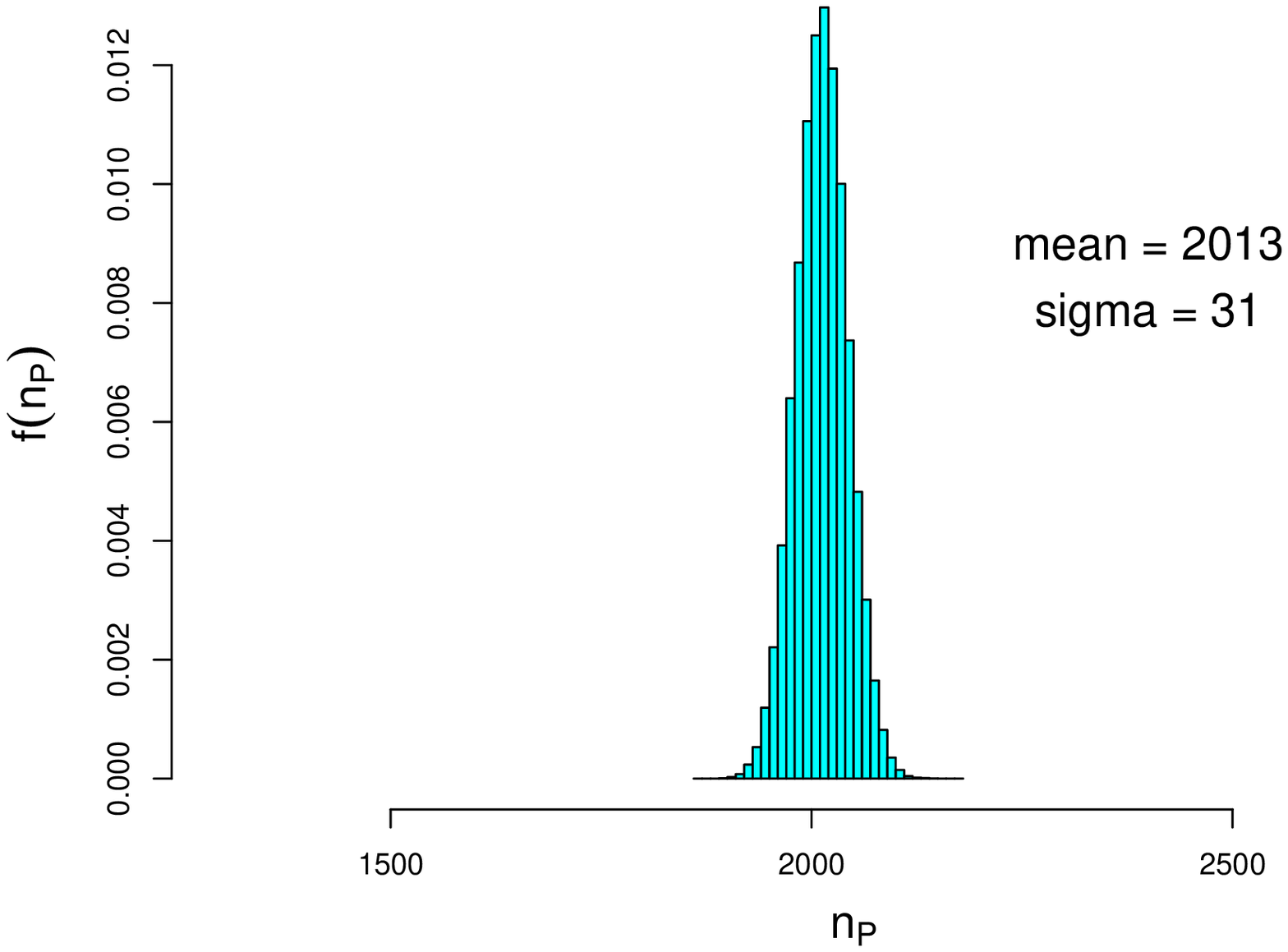,clip=,width=0.89\linewidth}}\\
  \centering{\epsfig{file=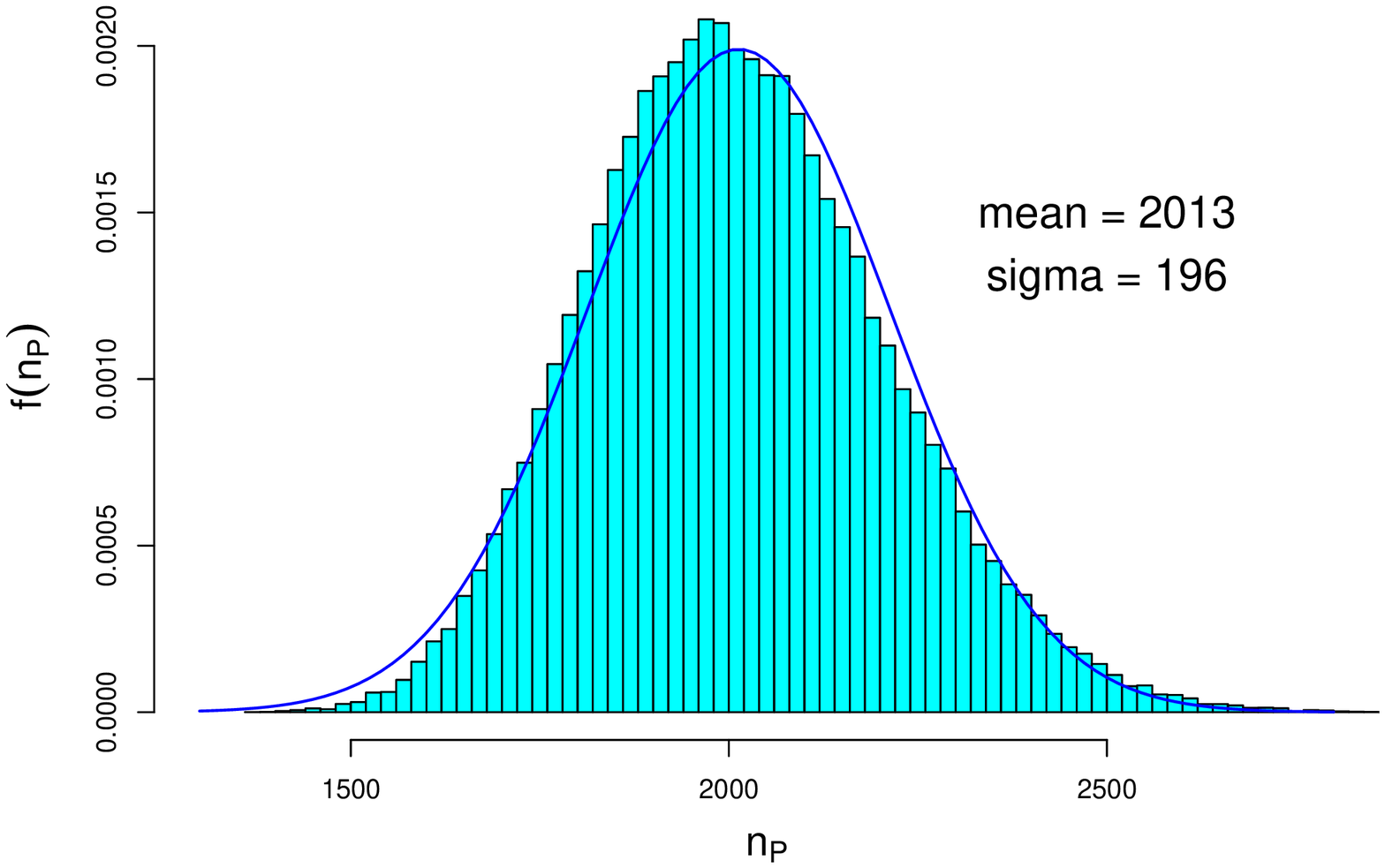,clip=,width=0.9\linewidth}}
  \mbox{}\vspace{-0.4cm}\mbox{}
  \end{center}
  \caption{\small \sf Probabilistic prediction of
    the numbers of positives, based on a hypothetical
    test on 10000 individuals, exactly 1000 of them being infected.
    In the upper plot we use $\pi_1=0.978$ and $\pi_2=0.115$.
    In the lower plot we take into account their possible variability
    (see text). The over-imposed curve shows a Gaussian with average
    2013 and standard deviation 200, values obtained by the
    approximated Eqs.~(\ref{eq:approx_E.nP}) and (\ref{eq:approx_Var.nP}).
     (The top histogram is repeated, with enlarged horizontal scale,
    in  Fig.~\ref{fig:PredictionPositive_sampling}.) 
  } 
    \label{fig:PredictionPositive} 
\end{figure}
is quite impressive, compared to the top one, in which
the precise values $\pi_1=0.978$ and $\pi_2=0.115$ were used.
The mean of the distribution is unchanged,
as more or less expected (see Sec.~\ref{sec:dependence_pi1_pi2}),  
but its standard deviation, which quantifies the uncertainty of the prediction,
increases by more than a factor six. 
We have then good reasons to expect
a similar effect when we will be interested in the `reverse' problem, 
that is inferring the number of infectees in
the sample from the resulting number of positives.
Going into details, we see that the expected number of positives
is essentially the same of Sec.~\ref{ss:NumberPositiveServa}
(the reduction from 2060 to 2013 is simply due to
the new reference values for $\pi_1$ and $\pi_2$ we
are using starting from Sec.~\ref{sec:uncertainty}).
But this number is now accounted by an uncertainty, which
rises to about 10\% of its value, when the uncertainties
about $\pi_1$ and $\pi_2$ are also taken into account.

\subsubsection{Approximated formulae}
Although Monte Carlo integration is a powerful tool to solve
at best non trivial problems of this kind, it is 
very useful to get, whenever it is possible, approximate solutions
in order to have an idea, analyzing the resulting formulae,
of how the result depends on the assumptions.
First at all, in analogy to what we have seen in
Sec.~\ref{sec:dependence_pi1_pi2}, we can be rather confident
that the expected value of $n_P$ is not
significantly affected, as also
confirmed by the Monte Carlo results shown in Fig.~\ref{fig:PredictionPositive}.
The variance, given by Eq.~(\ref{eq:var_nP_binom}) 
is, instead, increased by terms
whose approximated values can be obtained by
linearization.\footnote{See Sec.~6.4 of Ref.~\cite{RPP}
  and Sec.~8.6 of Ref.~\cite{BR}.}
These are the resulting approximated expressions:\footnote{The first two terms
  of the r.h.s. of Eq.~(\ref{eq:approx_Var.nP}) come from
  Eq.~(\ref{eq:var_nP_binom}), in which the precise values
  $\pi_1$ and $\pi_2$ have been replaced by their expected value.
  The other two terms are obtained by linearization, yielding
  e.g. for the contribution due to $\pi_1$
  (remember that $p_s$ is, so far, a precise parameter)
\begin{eqnarray*}
 \left. \left(\frac{\partial}{\partial \pi_1} \big(\pi_1\cdot p_s\cdot n_s +
 \pi_2\cdot(1-p_s)\cdot n_s\big)\right)^2\right|_{\mbox{E}(\pi_1,\pi_2)}
 \! \cdot \sigma^2(\pi_1) & = & \big(p_s\cdot n_s\big)^2 \cdot \sigma^2(\pi_1)\,.
\end{eqnarray*}
\mbox{}\vspace{-0.5cm}\mbox{}
}
\begin{eqnarray}
  \mbox{E}(n_P) &\approx&  \mbox{E}(\pi_1)\cdot p_s\cdot n_s
  +  \mbox{E}(\pi_2)\cdot (1-p_s)\cdot n_s  \label{eq:approx_E.nP}\\
  \sigma^2(n_P) &\approx & \mbox{E}(\pi_1)\cdot (1-\mbox{E}(\pi_1))
  \cdot p_s\cdot n_s
  + \mbox{E}(\pi_2)\cdot (1-\mbox{E}(\pi_2))\cdot (1-p_s)\cdot n_s
  \nonumber \\
  && + \,\sigma^2(\pi_1)\cdot p_s^2\cdot n_s^2 +
  \sigma^2(\pi_2)\cdot (1-p_s)^2\cdot n_s^2\,.   \label{eq:approx_Var.nP}
\end{eqnarray}
Applying them to the case shown in Fig.~\ref{fig:PredictionPositive}
we obtain an expected value of 2013 
and a standard deviation of 200, in excellent agreement
with the Monte Carlo result.
In order to have an idea of the deviation from `normality'
we also over-impose,
to the bottom histogram of the figure, the Gaussian
having average and standard deviation
calculated by Eq.~(\ref{eq:approx_E.nP}) and (\ref{eq:approx_Var.nP})
-- we remind that the top histogram has instead strong
theoretical reasons to be, with
very good approximation, normally distributed
(a zoomed version of the same histogram is reported
in Fig.~\ref{fig:PredictionPositive_sampling}).

As a further check, let us see what happens
in the case of no infected individuals in the sample, that is $p_s=0$.
The Monte Carlo results
are shown in Fig.~\ref{fig:PredictionPositive_p0},
\begin{figure}
  \begin{center}
  \centering{\epsfig{file=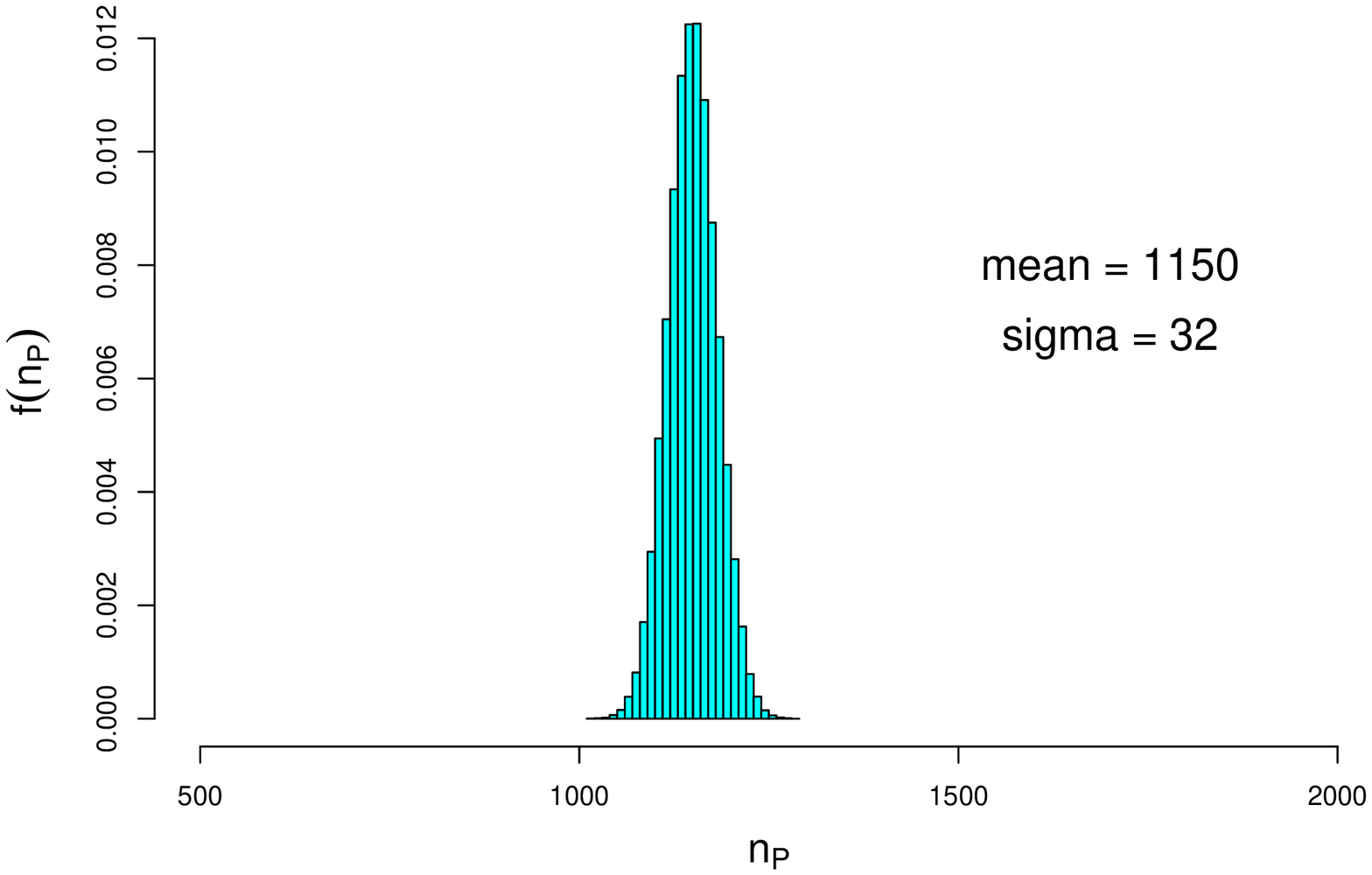,clip=,width=0.89\linewidth}}\\
  \centering{\epsfig{file=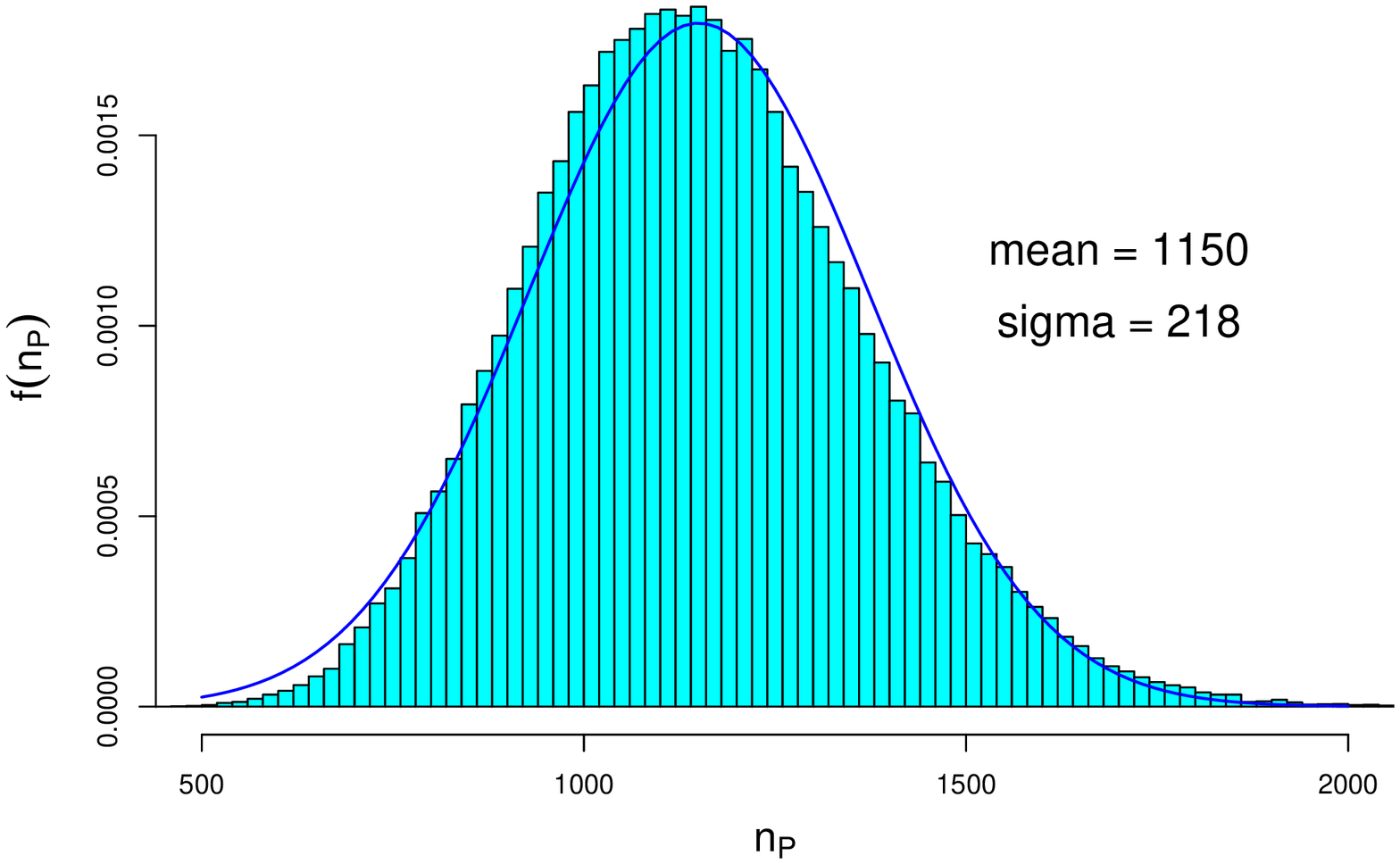,clip=,width=0.89\linewidth}}  
  \end{center}
  \caption{\small \sf Same as Fig.~\ref{fig:PredictionPositive}, but in the
    case of no infected individual in the sample ($p_s=0$).
    The over-imposed Gaussian has average
    1150 and standard deviation 222.
  }
    \label{fig:PredictionPositive_p0} 
\end{figure}
We see that, as already stated qualitatively in Sec.~\ref{sec:rough},
number of positives can occur well below the value one would compute
only reasoning on rough estimates (1150 in this case).
Therefore, since the formulae derived in that way were unreliable,
a probabilistic treatment
of the problem is needed in order to take into account the fact
that {\em fluctuations around expected values do usually occur}.
Also in this case the approximated results obtained by
Eqs.~(\ref{eq:approx_E.nP}) and (\ref{eq:approx_Var.nP})
are in excellent agreement with the Monte Carlo estimates,
yielding $1150 \pm 222$ (and, again, the Gaussian approximation
is not too bad, at least within
a couple of standard deviations from the mean value).
The approximation remains good also for high values
of $p_s$. For example, for the quite high value of $p_s=0.5$, the Monte Carlo
integration gives $5465 \pm 116$ versus an approximated result
of $5465 \pm 118$.

A natural question is how the results change not only with
the proportion of infectees in the sample,
but also with the size of the sample. The answer
is given in Tab.~\ref{tab:predicting_positives}, with $n_s$ varying,
in steps of roughly half order of magnitude, 
from the ridiculous value of 100 up to 100000
(that is $\approx 10^{k/2}$, with $k=4,5,\ldots,10$).
\begin{table}[!t]
{\small
  \begin{center}
    \begin{tabular}{r|cccccc}
\hline      
$\bm{p_s} \rightarrow $  & {\bf 0.01} &  {\bf 0.05}  &  {\bf 0.10}  & {\bf 0.15} & {\bf 0.20} & {\bf 0.50}\\
\hline
$\bm{\mbox{\bf E}(f_P)} \rightarrow $ &  {\bf 0.124} & {\bf 0.158} & {\bf 0.201} & {\bf 0.244} & {\bf 0.287} & {\bf 0.546} \\
\hline
\multicolumn{1}{c|}{{\bf $\bm{n_s}$:}}  %
 & \multicolumn{6}{|c}{{\em standard uncertainties}} \\
 {\bf 100}  
       & (0.032) & (0.031) & (0.031) & (0.030) & (0.029) & (0.025) \\
       & [0.038] & [0.037] & [0.036] & [0.035] & [0.034] & [0.027] \\
 {\bf 300}  
       & (0.018) & (0.018) & (0.018) & (0.017) & (0.017) & (0.014) \\
       & [0.028] & [0.027] & [0.026] & [0.025] & [0.024] & [0.018] \\
 {\bf 1000}
       & (0.010) & (0.010) & (0.010) & (0.009) & (0.009) & (0.008) \\
       & [0.024] & [0.023] & [0.022] & [0.021] & [0.020] & [0.014] \\
 {\bf 3000}
       & (0.006) & (0.006) & (0.006) & (0.005) & (0.005) & (0.005) \\
       & [0.022] & [0.021] & [0.020] & [0.019] & [0.018] & [0.012] \\
  {\bf 10000} 
     & (0.003) & (0.003) & (0.003)& (0.003)& (0.003)&  (0.002) \\
     & [0.022] & [0.021] & [0.020]& [0.019]& [0.018]&  [0.012] \\
 {\bf  30000}
       & (0.002) & (0.002) & (0.002) & (0.002) & (0.002) & (0.001) \\
       & [0.021] & [0.021] & [0.020] & [0.018] & [0.017] & [0.011] \\
 {\bf  100000} 
       & (0.001) & (0.001) & (0.001) & (0.001) & (0.001) & (0.001) \\
       & [0.021] & [0.021] & [0.019] & [0.018] & [0.017] & [0.011] \\
   \hline
\end{tabular}
\end{center}
}
  \caption{\small \sl Predicted fraction of tagged positives
    in a sample ($f_P$) as a function of the assumed proportion
    of infected individuals {\em in the} sample ($p_s$), also 
    taking into account the uncertainty on the test
    parameters $\pi_1$ and $\pi_2$ (numbers in squared brackets --
    those in round brackets are evaluated at the expected values
    of  $\pi_1$ and $\pi_2$). 
}
\label{tab:predicting_positives}  
\end{table}
The chosen values of $p_s$ are the same of $p$ of
Tab.~\ref{tab:prob_vs_parametri}. For an easier comparison,
the fraction $f_p$ of positively tagged individual
is provided.
The expected value of $f_P$, depending essentially
only on $p_s$, is reported in the second row of the table.
Two standard uncertainties are reported for each combination
of $p_s$ and $n_s$: the first, in round brackets, only takes
into account the two binomial distributions (`statistical errors',
in old style\footnote{For this question see the ISO's GUM\,\cite{ISO}.}
physicist's jargon); the second, in square brackets takes 
into account {\em also} the possible variability of $\pi_1$ and $\pi_2$
(`systematic error',  in the same jargon). They have all been evaluated
by Monte Carlo, but the agreement with the approximated formula
(\ref{eq:approx_Var.nP}) has been
checked.

\subsection[General considerations on the approximated
evaluation of  $\sigma(n_P)$]{General considerations on the approximated
evaluation of  $\sigma(n_P)$ by Eq.~(\ref{eq:approx_Var.nP})}
At this point some further remarks on the utility
of  Eq.~(\ref{eq:approx_Var.nP}) is in order.
Its  advantage,
within its limits of validity (checked in our case),
is that it allows to disentangle the contributions to the overall uncertainty.
In particular we can rewrite it as
\begin{eqnarray}
 \sigma(n_P) &\approx&   \sigma_R(n_P) \oplus  
 \sigma_{\pi_1}(n_P) \oplus  \sigma_{\pi_2}(n_P)\,,
\end{eqnarray}
that is a 
`quadratic sum' (or `quadratic combination', indicated by the symbol `$\oplus$')
of three contributions,
\begin{eqnarray*}
  \sigma_R(n_P) &=&  \sqrt{\mbox{E}(\pi_1)\cdot (1-\mbox{E}(\pi_1))\cdot p_s\cdot n_s
      + \mbox{E}(\pi_2)\cdot (1-\mbox{E}(\pi_2))\cdot (1-p_s)\cdot n_s} \\
   \sigma_{\pi_1}(n_P) &=&  \sigma(\pi_1)\cdot p_s\cdot n_s \\ 
  \sigma_{\pi_2}(n_P)  &=&   \sigma(\pi_2)\cdot (1-p_s)\cdot n_s\,,
\end{eqnarray*}
due, as indicated by the suffixes, to the binomials (`$R$'
standing for `random'), to the uncertainty on $\pi_1$
and to that on $\pi_2$.

This quadratic combination of the contributions
can be easily extended, just dividing by $n_s$,
to the uncertainty on the fraction of positives,
thus getting
\begin{eqnarray}
 \sigma(f_P) &\approx&   \sigma_R(f_P) \oplus  
 \sigma_{\pi_1}(f_P) \oplus  \sigma_{\pi_2}(f_P)\,,
\end{eqnarray} 
quadratic sum of
\begin{eqnarray}
  \sigma_R(f_P) &=& 
  \sqrt{\mbox{E}(\pi_1)\cdot (1-\mbox{E}(\pi_1))\cdot p_s
    + \mbox{E}(\pi_2)\cdot (1-\mbox{E}(\pi_2))
    \cdot (1-p_s)}/\sqrt{n_s} \ \ \label{eq:sigma_R}\\
   \sigma_{\pi_1}(f_P) &=&  \sigma(\pi_1)\cdot p_s  \label{eq:sigma_pi1}\\ 
  \sigma_{\pi_2}(f_P)  &=&   \sigma(\pi_2)\cdot (1-p_s)\,.  \label{eq:sigma_pi2}
\end{eqnarray} 
We see immediately, for example,
that for $p_s$ around 0.1 the contribution due to
$\pi_2$ dominates over that due to $\pi_1$
by a factor $0.022/0.007\times 0.9/0.1 \approx 30$. 
This allows us 
to evaluate, on the basis of the Monte Carlo
results shown in  Tab.~\ref{tab:predicting_positives}, the
contribution due the systematic effects alone.
For example  we get, for our customary
values of $p_s=0.1$ and $n_s=10000$,
$\sigma(f_P)$ equal to
0.003 and 0.020, respectively.
Assuming a quadratic combination,
the contribution due to systematics is then
$\sqrt{0.020^2-0.003^2} = 0.0198$. Besides questions of
rounding,\footnote{Using the values 0.0196 and 0.0031 of
Fig.\ref{fig:PredictionPositive} we would get 0.194.}
it is clear that {\em the uncertainty is largely dominated
by the uncertainty on $\pi_1$ and $\pi_2$.}
We can check this result by a direct, although
approximated, calculation using
Eq.~(\ref{eq:sigma_pi1}) and (\ref{eq:sigma_pi2}):
\begin{eqnarray*}
  \sigma_{\pi_1}(f_P) &=& 0.007\times 0.1 = 0.0007 \\
  \sigma_{\pi_2}(f_P) &=& 0.022\times 0.9 = 0.0198 \\
  \sigma_{\pi_1}(f_P) \oplus   \sigma_{\pi_2}(f_P) &\approx &
  \sigma_{\pi_2}(f_P) =  0.0198\,,  
\end{eqnarray*}
getting the same result.

Looking at the numbers of Tab.~\ref{tab:predicting_positives},
we see that this effect starts already at $n_s=1000$.
For example, for $p_s=0.1$ we get $\sqrt{0.022^2-0.010^2}=0.0196$,
twice the standard uncertainty of 0.010 due to the binomials alone.
The sample size at which the two contributions have the
same weight in the global uncertainty is around 300
(for example, for $p_s=0.1$ we get $\sqrt{0.026^2-0.018^2} = 0.019$).
The take-home message is, at this point,
rather clear (and well known to physicists and other scientists):
{\em unless we are able to make our knowledge about $\pi_1$ and $\pi_2$ more
accurate, using sample sizes much larger than 1000 is
only a waste of time}.

However, there is still another important effect  we need
to consider, due to the fact that we
are indeed sampling {\em a population}. 
This effect leads  unavoidably to extra variability
and therefore to a new contribution to the uncertainty in prediction
(which will be somehow reflected into uncertainty in the inferential process).

Before moving to this other important effect, let us
exploit a bit more the approximated evaluation of $ \sigma(f_P)$.
For example, 
solving  with respect to $n_s$ the condition
\begin{eqnarray*}
  \sigma_R(f_P) &=& \sigma_{\pi_1}(f_P) \oplus \sigma_{\pi_2}(f_P)
  \label{eq:condizione_n*_parziale} 
\end{eqnarray*}  
we get from Eqs.\,(\ref{eq:sigma_R})-(\ref{eq:sigma_pi2})
\begin{eqnarray}
  n_s^* & \approx & 
  \frac{\mbox{E}(\pi_1)\cdot (1-\mbox{E}(\pi_1))\cdot p_s
    + \mbox{E}(\pi_2)\cdot (1-\mbox{E}(\pi_2))
    \cdot (1-p_s)}{ \sigma^2(\pi_1)\cdot p_s^2 + \sigma^2(\pi_2)\cdot (1-p_s)^2}\,, 
\end{eqnarray}
which gives a {\em rough idea} of the sample size above which
the uncertainty due to systematics starts to dominate.
For example,  for $p_s = 0.1$ we get $n_s=240$
of the order of magnitude ($\approx 300$)
got from the Monte Carlo study.
If we require, to be safe,
$\sigma_{\pi_1}(f_P) \oplus \sigma_{\pi_2}(f_P) = (\mbox{2-3})
\times  \sigma_R(f_P)$
we get $n_s\approx 1000$ and  $n_s\approx 2200$, 
again in reasonable 
agreement with the results of  Tab.~\ref{tab:predicting_positives}.
We shall go through a more complete analysis of $n_s^*$
in Sec.~\ref{ss:Balance_Stat_Syst}, in which 
a further contribution to the uncertainty will be also taken
into account.

\subsubsection[Contribution of the uncertainty on $p_s$
  due to sampling]{Including in the approximated formulae
  the contribution of the uncertainty on $p_s$ due to sampling}
Next section will be dedicated to the effect of sampling $n_s$
individuals from a population. However,  
having taken some confidence with the approximated formulae,
we can already extend them in order to see how the uncertain
$p_s$, characterized by its expected value
$\mbox{E}(p_s)$ and standard uncertainty $\sigma(p_s)$,
whose evaluation will be the subject of Sec.~\ref{sec:fpN_from_p},
affects our prediction about the number of individuals
resulting positive in the test. In the
approximated expression for the expected value of $n_P$ 
(Eq. \ref{eq:approx_E.nP}) we have
to replace $p_s$ by its expected value $\mbox{E}(p_s)$, while
in the variance we have to add a term again obtained by
linearization,\footnote{The contribution
  to $\sigma^2(n_P)$ due to  $\sigma(p_s)$,
evaluated by linearization, is given by
\begin{eqnarray*}
 \left. \left(\frac{\partial}{\partial p_s} \big(\pi_1\cdot p_s\cdot n_s +
 \pi_2\cdot(1-p_s)\cdot n_s\big)\right)^2\right|_{\mbox{E}(\pi_1,\pi_2,p_s)}
\! \cdot \sigma^2(p_s) & = & \big(\mbox{E}(\pi_1)\cdot n_s -
                          \mbox{E}(\pi_2)\cdot n_s\big)^2 \cdot \sigma^2(p_s)\,.
\end{eqnarray*}
\mbox{}\vspace{-0.5cm}\mbox{}
} thus getting
\begin{eqnarray}
  \mbox{E}(n_P) &\approx&  \mbox{E}(\pi_1)\cdot \mbox{E}(p_s)\cdot n_s
  +  \mbox{E}(\pi_2)\cdot (1-\mbox{E}(p_s))\cdot n_s  \label{eq:approx_E.nP_s}\\
  \sigma^2(n_P) &\approx & \mbox{E}(\pi_1)\cdot (1-\mbox{E}(\pi_1))
  \cdot \mbox{E}(p_s)\cdot n_s
  + \mbox{E}(\pi_2)\cdot (1-\mbox{E}(\pi_2))\cdot (1-\mbox{E}(p_s))\cdot n_s
  \nonumber \\
  && + \,\sigma^2(\pi_1)\cdot \mbox{E}^2(p_s)\cdot n_s^2 +
  \sigma^2(\pi_2)\cdot (1-\mbox{E}(p_s))^2\cdot n_s^2 \nonumber \\
  && + \,\sigma^2(p_s)\cdot (\mbox{E}(\pi_1) - \mbox{E}(\pi_2))^2\cdot
  n_s^2\,.
  \label{eq:approx_Var.nP_s}
\end{eqnarray}
As far as  the fraction of positives is concerned, we have
the following four contributions to the global uncertainty,
\begin{eqnarray}
 \sigma(f_P) &\approx&   \sigma_R(f_P) \oplus  
 \sigma_{\pi_1}(f_P) \oplus  \sigma_{\pi_2}(f_P)
 \oplus  \sigma_{p_s}(f_P) \,, \label{eq:sigma_fP_approx_4}
\end{eqnarray}
the first three given by Eqs.~(\ref{eq:sigma_R}-\ref{eq:sigma_pi2}),
in which  $p_s$ has to be replaced by its expected value $\mbox{E}(p_s)$, 
and the fourth term being
\begin{eqnarray}
  \sigma_{p_s}(f_P)  &=&
  \sigma(p_s)\cdot |\mbox{E}(\pi_1) - \mbox{E}(\pi_2)|\,.
  \label{eq:sigma_fP_sigma_ps}  
\end{eqnarray}
(Note that also the fourth term is of `random nature', although,
from the `perspective' we are now seeing the problem
it could be considered as a third contribution
to systematics.\footnote{Note that this terminology is a matter
  of convention and habits. From a probabilistic point of view
  we just apply probability theory to all quantities with respect to
  which we are in condition of uncertainty, considering
  the `fixed ones' as conditionands.
})

\section{Sampling a population}\label{sec:fpN_from_p}
In Sec.~\ref{sec:fpN_from_ps}  we went through the question
of predicting the number of positives when we plan to test an entire
sample of  $n_s$ individuals, a fraction $p_s$ of which
is assumed to be infected. 
At this point we have to take into account the last source of
uncertainty we have to deal with. If we sample
at random $n_s$ individuals out of the $N$ of the entire 
population, the sample will contain a fraction
of infected $p_s$ usually different from the (`{\em true}')
fraction $p$
of the population and described by 
$f(p_s\,|\,n_s,N,p)$. Once the pdf of $p_s$ has been
somehow evaluated,
we can get the pdf 
of interest, that is $f(n_P\,|\,n_s,N,p)$, extending
Eq.~(\ref{eq:f_nP_int_pi1_pi2}) to \\
\mbox{}\vspace{-1.1cm}\mbox{}\\
\begin{eqnarray}
  f(n_P\,|\,n_s,N,p)\!\!\! &=&\!\!\! \int_0^1\!\!\!\int_0^1\!\!\!\!\int_0^1
  \!\!f(n_P\,|\,n_s,p_s,\pi_1,\pi_2)
  \!\cdot\! f(p_s\,|\,p,n_s,N)
  \!\cdot\! f(\pi_1) \!\cdot\!  f(\pi_2)\,
  \mbox{d}p_s\mbox{d}\pi_1\mbox{d}\pi_2\,. \nonumber \\
&&  \label{eq:f_nP_int_pi1_pi2_ps}
\end{eqnarray}

\subsection[Proportion of infected individuals
  in the random sample]{Proportion of infected individuals
  in the random sample
  -- Binomial and hypergeometric distributions}
\label{ss:binom_hg}
We have already reminded and made use of the binomial
distribution,  assumed well known to the reader.
A {\em related} problem in probability theory
is that of {\em extraction without replacement},
which we introduce here for two reasons.
The first is that it is little known even by many practitioners
(we think e.g. to ourselves and to our colleagues physicists).
The second is that some care is needed
with the parameters used in literature and in
scientific/statistical libraries 
of computer languages.

Let us imagine an urn containing  $m$ white and 
$n$ black balls. Let us imagine then that we are going to take out of it,
at random, $k$ balls
and that we are interested in the number $X$ of
white balls that we shall get
(for convenience of the reader, and also for us who never 
worked before with such a distribution,
we use the same idealized objects and symbols
of the R {\em help page} -- obtained e.g. by `{\tt ?dhyper}').  
The probability distribution of $X$ is known
as {\em hypergeometric}.\footnote{Some care is needed
  with this distribution because, as it is easy
  to understand, different sets of parameters can be used.
  For example, the app already
  suggested\,\cite{ProbabilityDistributions} uses
  \\ \mbox{}\vspace{-0.6cm}\mbox{}
  \begin{eqnarray*}
    X & \sim & \mbox{HG}(n, N, M)\,, 
  \end{eqnarray*}
 \mbox{}\vspace{-0.6cm}\mbox{} \\
  with $n$ the sample size, $N$ the population size and
  $M$ the number of {\em white balls}, thus leading to the
  following correspondence with respect to the parameters
  of the probability functions of the R language, to which
  we are going to adhere in the text\\
    \mbox{}\vspace{-0.8cm}\mbox{}
 \begin{eqnarray*}
     \mbox{app} & \longleftrightarrow & \mbox{R} \\
     n   & \longleftrightarrow & k \\
     N     & \longleftrightarrow & m+n \\
     M   & \longleftrightarrow & m\,.      
 \end{eqnarray*}
 Expected value and variance are, using the app convention,
  \begin{eqnarray*}
    \mbox{E}(X) &=& n\,\frac{M}{N} \\
    \sigma^2(X) &=&  n\,\frac{M}{N}\cdot\left(1 - \frac{M}{N}\right)
    \cdot \left(\frac{N-n}{N-1}\right)\,.
  \end{eqnarray*}
  (In Wikipedia\,\cite{WikiHG} there is a similar convention,
  apart from the names, being the `random variable'
  indicated by $k$ and the number of `white balls in the urn'
  by $K$.) \label{fn:hypergeometric}
}
In short, referring to the parameters of the probability functions
of the R language (see footnote \ref{fn:hypergeometric}),
\begin{eqnarray*}
X & \sim & \mbox{HG}(m, n, k) 
\end{eqnarray*}
with expected value and variance
 \begin{eqnarray*}
    \mbox{E}(X) &=& k\cdot \frac{m}{m+n} \\
    \sigma^2(X) &=&  k\cdot \frac{m}{m+n}\cdot\left(\frac{n}{m+n}\right)
    \cdot \left(\frac{m+n-k}{m+n-1}\right)\,.
 \end{eqnarray*}
 In terms of the proportion of `objects' having the characteristic
 of interest (`white'), their fraction in the urn is then assumed to be
 $p=m/(m+n)$, corresponding,
 in our problem, to the proportion of infectees.
 Using the symbol $n_s$ for the sample size $k$,
 as we have done so far, and $N$ for the total
 number of individuals in the population, 
 the above equations can be conveniently rewritten as
 \begin{eqnarray}
    \mbox{E}(X) &=& p\cdot n_s \\
    \sigma^2(X) &=&  n_s\cdot p\cdot (1-p)
    \cdot \left(\frac{N-n_s}{N-1}\right)\,.
 \end{eqnarray} 
The expression of the expected value is identical to that
of a binomial distribution, while that of the variance differs from it by a
factor depending on the difference between the population size and
the sample size,  vanishing when $n_s$ is equal to $N$. 
That is simply because in that case
we are going to empty the `urn' and therefore we
shall count exactly the number of `white balls'.
When, instead,
$n_s$ is much smaller than $N$ (and then $N\gg 1$), we recover the variance
of the binomial. In practice it means that the effect of
replacement, related to the chance to extract more than once
the same object, becomes negligible.

Moving to our problem, the role of the generic
variable $X$ is played by the number of infectees
in the sample, indicated by $n_I$ in the previous sections.   
In terms of their proportion, being  $p_s=X/k = n_I/n_s$,
we get
\begin{eqnarray}
  \mbox{E}(p_s) = \mbox{E}\left(\frac{n_I}{n_s}\right) &=&
  \frac{m}{m+n} = p\,,
\end{eqnarray}
as intuitively expected. As far as the variance is concerned,
being simply $\sigma(p_s) = \sigma(n_I)/n_s$, we get
\begin{eqnarray}
  \sigma^2(p_s) = \frac{\sigma^2(n_I)}{n_s^2}
  &=&  \frac{1}{n_s}\cdot p\cdot\left(1-p\right)
  \cdot   \left(\frac{N-n_s}{N-1}\right)\\
  &\approx&  \frac{1}{n_s}\cdot p\cdot\left(1-p\right)\cdot
  \left(1-\frac{n_s}{N}\right) \label{eq:var_ps_no-1}
 \end{eqnarray} 
 being $N\gg 1$ in all practical cases of (our) interest.
 
Finally, if the sample size is much smaller
than the population size, then the last
factor can be neglected and the variance
can be approximated by
$p\cdot(1-p)/n_s$, thus yielding
\begin{eqnarray}
  \left.\sigma(p_s)\right|_{n_s\ll N} &\approx&
  \sqrt{\frac{p\cdot (1-p)}{n_s}}\,,  \label{eq:var_ps_approx}
\end{eqnarray}
the well known standard deviation
of the fraction of successes in a binomial distribution
with $n_s$ trials, each with probability $p$. The reason is that
-- it is worth repeating it -- 
 when the sample size is much smaller than the population size, 
then we can neglect the effects of no-replacement
and consider the trials as ({\em conditionally}) independent
Bernoulli processes, each with probability of success $p$.

\subsection[Expected number of positives
assuming exact values of $\pi_1$ and $\pi_2$]{Expected number of positives 
  sampling of a population (assuming exact values of $\pi_1$ and $\pi_2$)}
At this point we can convolute the uncertainty on the
number of positives in a sample,
analyzed in Sec.~\ref{sec:fpN_from_ps},
with the uncertain value of $p_s$ due to sampling:
\begin{eqnarray}
  f(n_P\,|\,n_s,N,p,\pi_1,\pi_2)  &=& \int_0^1
  \!\!f(n_P\,|\,n_s,p_s,\pi_1,\pi_2)
  \!\cdot\! f(p_s\,|\,p,n_s,N)\,
  \mbox{d}p_s\,.   \label{eq:f_nP_int_ps}
\end{eqnarray}
We start, as usual, with our exact reference values
of test sensitivity and specificity of 97.8\% and
88.5\% ($\pi_1=0.978$ and $\pi_2=0.115$), respectively, and perform the
integration by Monte Carlo.\footnote{The R code
for $N=10^5$, $n_s=10^4$ and $p=0.1$ is provided in
Appendix B.3.}
\begin{figure}
  \begin{center}
  \centering{\epsfig{file=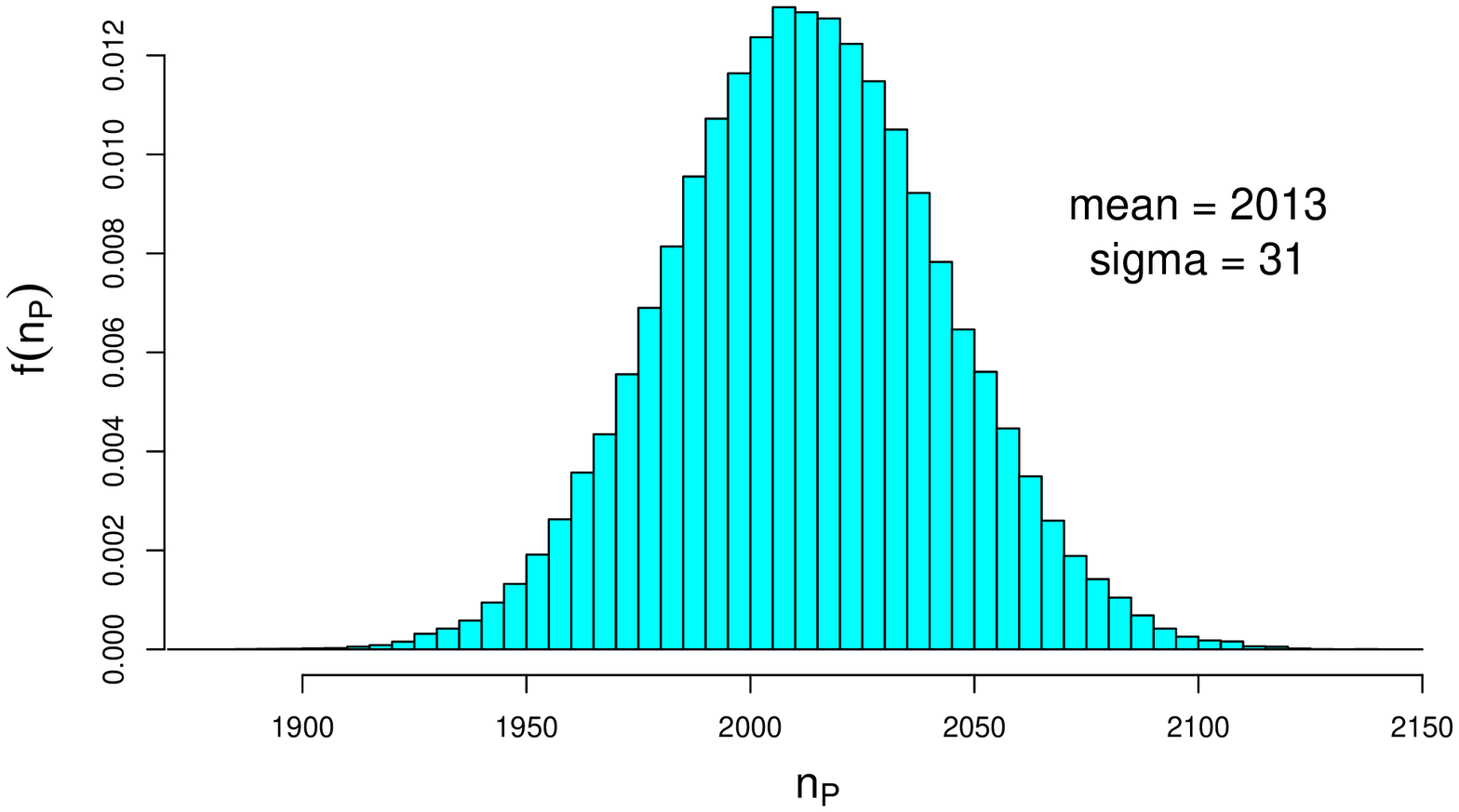,clip=,width=0.70\linewidth}}\\
  \centering{\epsfig{file=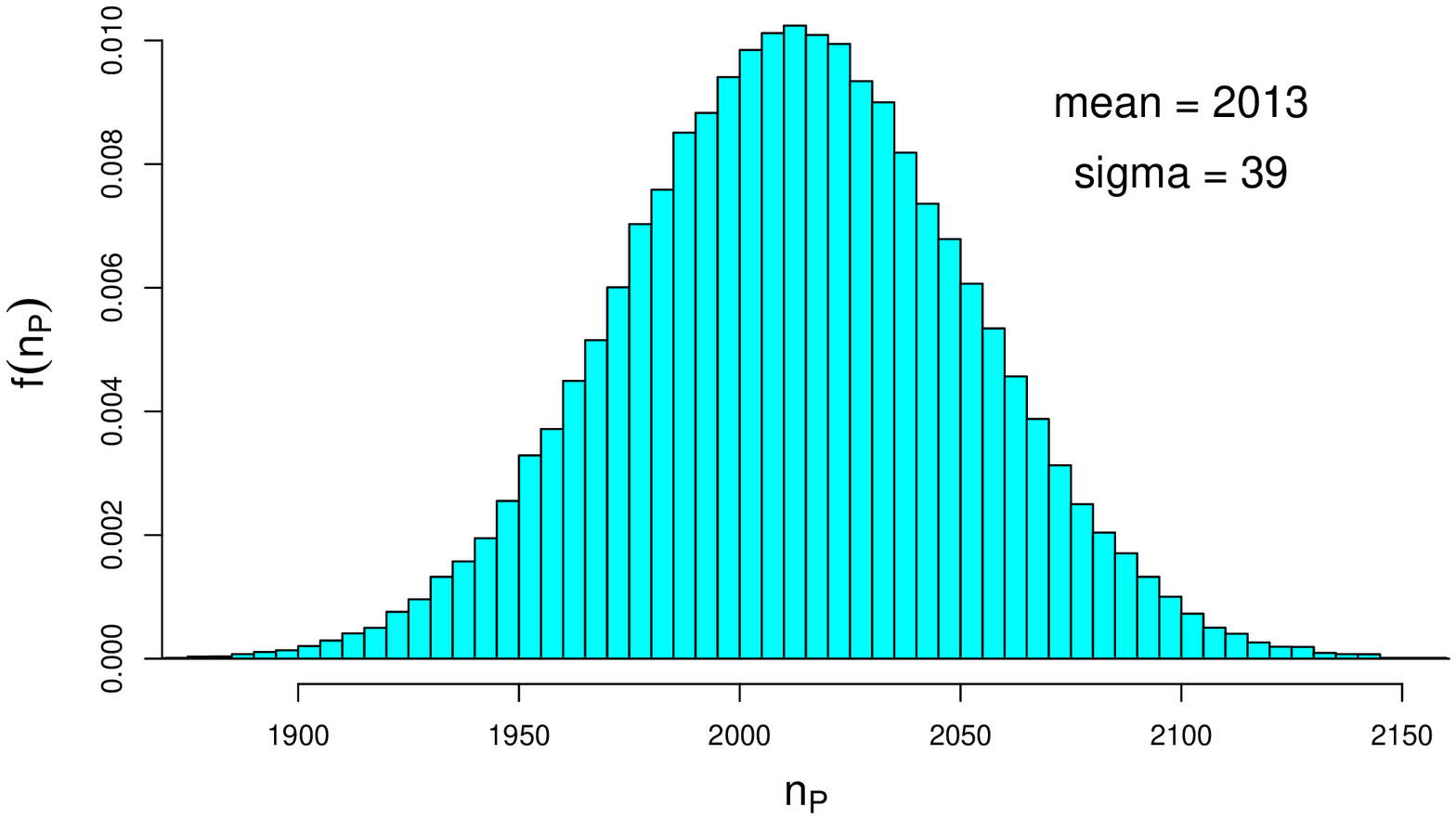,clip=,width=0.70\linewidth}}  \\
  \centering{\epsfig{file=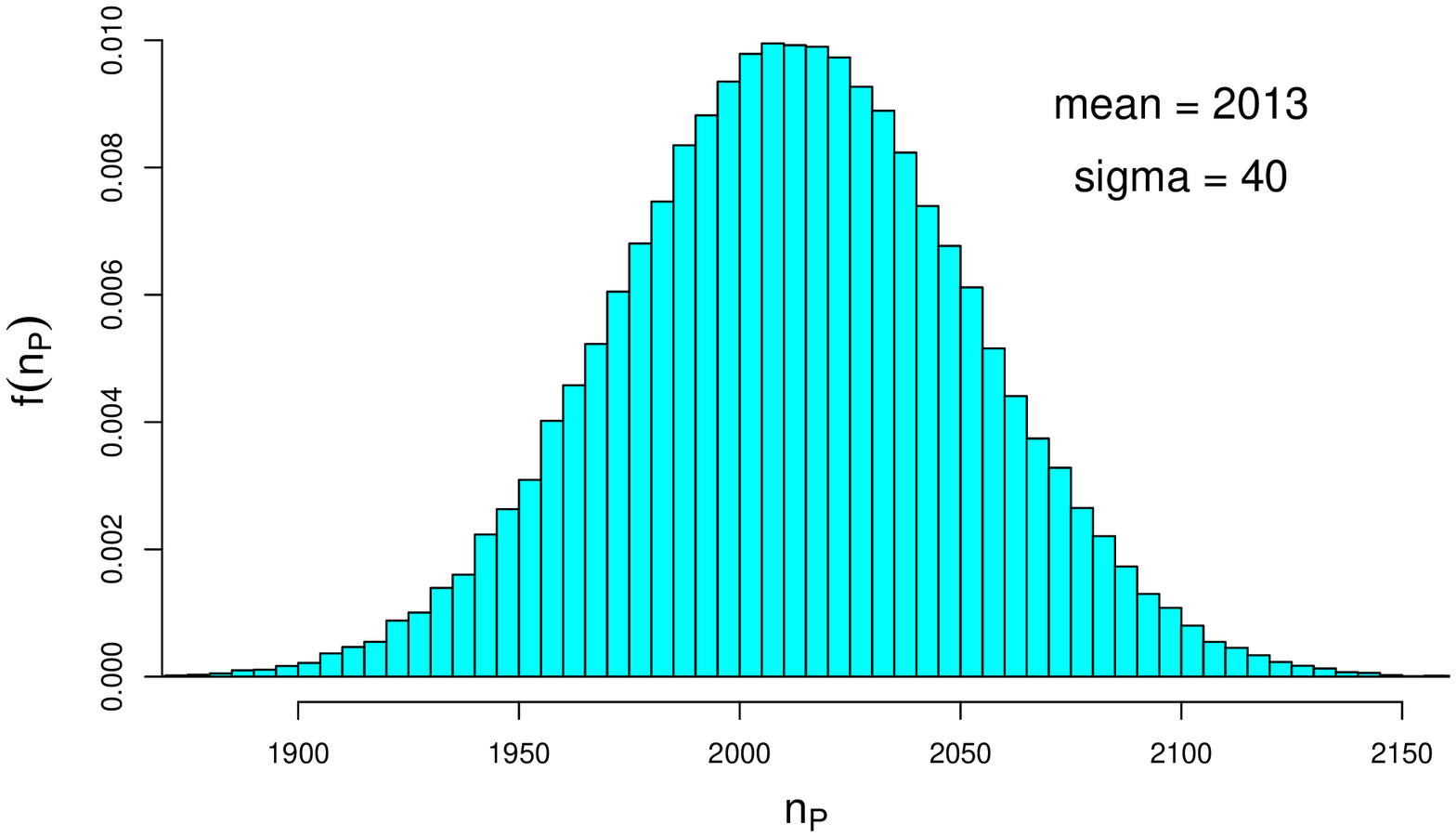,clip=,width=0.70\linewidth}}\\
  \mbox{}\vspace{-0.9cm}\mbox{}
  \end{center}
  \caption{\small \sf Probabilistic prediction of
    the numbers of positives in a sample of 10000 individuals
    taken from a population of 10000, 100000 and 1000000 individuals
    (in order, from top to bottom), 10\% of which
    are infected ($p=0.1$),
    assuming $\pi_1=0.978$ and $\pi_2=0.115$. 
  }
    \label{fig:PredictionPositive_sampling} 
\end{figure}
Some results are shown in Fig.~\ref{fig:PredictionPositive_sampling},
where, for comparison with what we have seen in the previous sections,
a sample size of 10000 individuals is used, 
taken from a population of 10000 (top histogram),
100000 (middle) and 1000000 (bottom), and assuming $p=0.1$. 
$[$Note that first case
corresponds exactly to the assumed value of
$p_s=0.1$ shown in the top plot of
Fig.~\ref{fig:PredictionPositive}, since, being $n_s=N$,
the standard uncertainty on $p_s$ vanishes.$]$
Increasing the population size
the standard deviation increases, as an effect of
$\sigma(p_s)$, although this growth saturates
for $N$ a bit higher than  $\approx 10\times n_s$, above which
the size dependent factor of Eq.~(\ref{eq:var_ps_no-1})
becomes negligible. In fact, the asymptotic value,
given by  Eq.~(\ref{eq:var_ps_approx}) is in this case
$\left.\sigma(p_s)\right|_{N\rightarrow\infty} = 0.0030$.
For $N/n_s=10$ the standard uncertainty on $p_s$
becomes 0.00285, vanishing for $n_s=N$
(the value of 0.0015, half of the asymptotic one, is
reached for $N=4/3\times n_s$).

\subsubsection{Approximated results}
It is interesting to compare the Monte Carlo results
of  Fig.~\ref{fig:PredictionPositive_sampling}
to those obtained by the approximated
values of expected value and standard deviation 
given by
Eqs.~(\ref{eq:approx_E.nP_s})-(\ref{eq:approx_Var.nP_s}) just putting
$\sigma(\pi_1)=\sigma(\pi_2)=0$. 
The contribution to the uncertainty
due to the two binomials of Fig.~\ref{fig:two_binom} 
is $\sigma_R(n_P)= 30.6$ (rounded to 31
in Fig.~\ref{fig:PredictionPositive_sampling}), while 
those due to $\sigma(p_s)$ are 
equal to 0, 24.6 and 25.9, for the three population sizes. 
The combined standard uncertainties
 are then 30.6, 39.3 and 40.1, in perfect agreement with
the results shown in Fig.~\ref{fig:PredictionPositive_sampling}.

\subsection{Detailed study of the four contributions to $\sigma(f_P)$}
\label{sec:detailed_contributions}
At this point it is time to release the limiting assumption
of exact values of sensitivity and specificity, i.e. 
$\sigma(\pi_1)=\sigma(\pi_2)=0$. 
Moreover, having checked that the approximated formulae
can take into account with great accuracy also the
contribution due to the uncertain value of $p_s$,
we find it interesting and useful to study
the individual contributions to the uncertainty
with which we can forecast the fraction $f_P$ of tested individuals
resulting positive. For the reader's
convenience, we summarize here the relevant,
approximated expressions, making also use,
in order to simplify them, of the equality $\mbox{E}(p_s) = p$\,:
\begin{eqnarray}
  \mbox{E}(f_P) &\approx&  \mbox{E}(\pi_1)\cdot p
  +  \mbox{E}(\pi_2)\cdot (1-p)  \label{eq:approx_E.fP_s} \\
 \sigma(f_P) &\approx&   \sigma_R(f_P) \oplus \sigma_{p_s}(f_P) \oplus
 \sigma_{\pi_1}(f_P) \oplus  \sigma_{\pi_2}(f_P) \\
  \sigma_R(f_P) &=& 
  \sqrt{\mbox{E}(\pi_1)\cdot (1-\mbox{E}(\pi_1))\cdot p
    + \mbox{E}(\pi_2)\cdot (1-\mbox{E}(\pi_2))
    \cdot (1-p)}/\sqrt{n_s} \ \ \label{eq:sigma_R_bis}\\
   \sigma_{\pi_1}(f_P) &=&  \sigma(\pi_1)\cdot p  \label{eq:sigma_pi1_bis}\\ 
  \sigma_{\pi_2}(f_P)  &=&   \sigma(\pi_2)\cdot (1-p)\,.  \label{eq:sigma_pi2_bis}\\
  \sigma_{p_s}(f_P)  &=&
  \sigma(p_s)\cdot |\mbox{E}(\pi_1) - \mbox{E}(\pi_2)| \nonumber  \\
     & \approx &
  |\mbox{E}(\pi_1) - \mbox{E}(\pi_2)| \cdot
  \sqrt{p\cdot (1-p)\cdot (1-n_s/N)}/\sqrt{n_s} 
  \label{eq:sigma_fP_sigma_ps_bis} 
\end{eqnarray}
We can note that  $\sigma_{\pi_2}(f_P)$ and $\sigma_{\pi_1}(f_P)$
are independent of the sample size $n_s$,
while  $\sigma_R(f_P)$ and  $\sigma_{p_s}(f_P)$
exhibit the typical `statistical dependence' $\propto 1/\sqrt{n_s}$.
Therefore we shall refer hereafter to
$\sigma_R(f_P)$ and  $\sigma_{p_s}(f_P)$
as {\em random} (or {\em statistical}) {\em contributions};
to the others
as {\em contributions due to systematics},
which cannot be improved increasing the sample size.

The upper plot of Fig.~\ref{fig:Contributions_uncertanties_p0.1}
\begin{figure}
  \begin{center}
  \epsfig{file=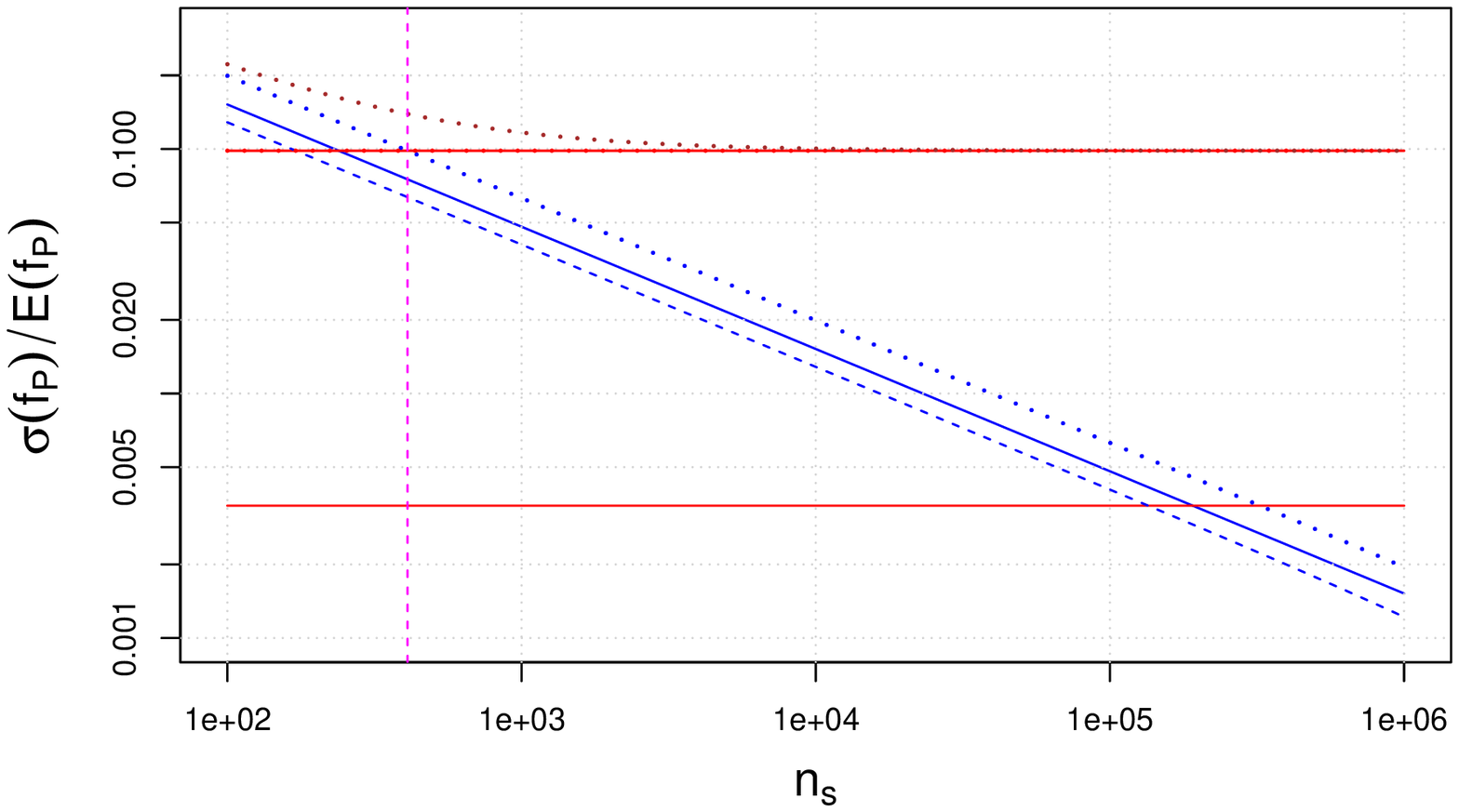,clip=,width=0.92\linewidth}
  \epsfig{file=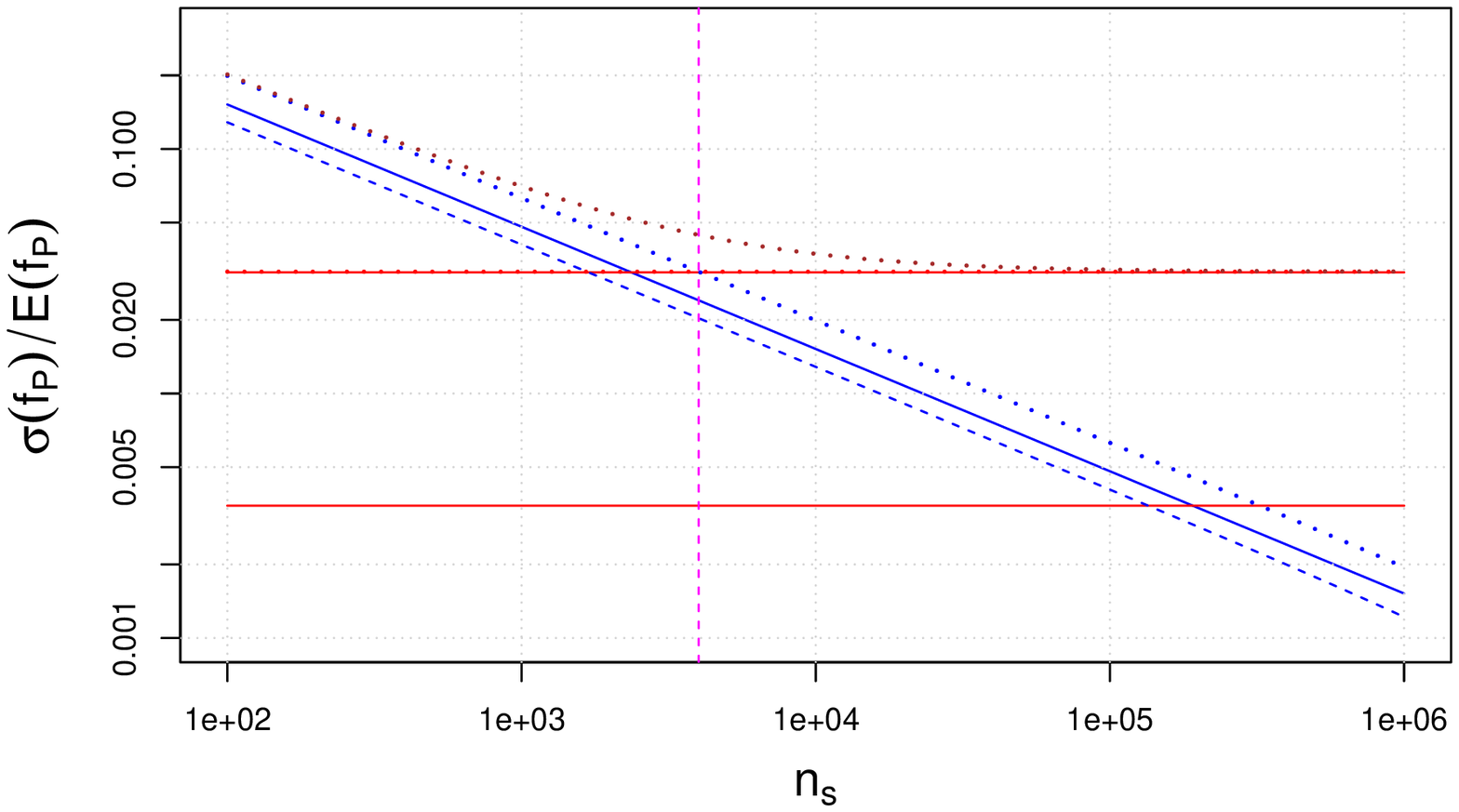,clip=,width=0.92\linewidth}
   \\ \mbox{} \vspace{-1.0cm} \mbox{}
  \end{center}  
  \caption{\small \sf Contributions to the relative uncertainty
    on the fraction of positives as a function of the sample size $n_s$,
    assuming it much smaller than the population size $N$, for a proportion
    of infected individuals $\bm{p=0.1}$. 
    The solid blue line with negative slope is the contribution
    from $\sigma_R(f_P)$, the dashed blue one is the contribution
    from $\sigma_{p_s}(f_P)$, the dotted line is the
    `quadratic sum' of the two; the lower horizontal red one
    is the contribution from  $\sigma_{\pi_1}(f_P)$ and the 
    upper horizontal one is the contribution from
    $\sigma_{\pi_2}(f_P)$ (a dotted red line,
    showing their `quadratic sum' is
    indeed overlapping the $\pi_2$ contribution).
    The overall uncertainty is shown by the
    uppest curve (dotted brown). The upper plot is for
    a standard uncertainty on $\pi_2$  $\sigma(\pi_2)=0.022$.
    The lower plot is for the case of uncertainty reduced to    
    $\sigma(\pi_2)=\sigma(\pi_1)=0.007$.
  }
    \label{fig:Contributions_uncertanties_p0.1} 
\end{figure}
shows,
for our reference value of $p=0.1$ and for uncertain  $\pi_1$ and $\pi_2$ 
(summarized as $\pi_1=0.978\pm 0.007$ and $\pi_2=0.115\pm 0.022$),
the relative uncertainty on $f_{P}$,
that is $\sigma(f_{P})/\mbox{E}(f_{P})$,  as a function of $n_s$,
highlighting the different contributions to the total uncertainty.
The horizontal lines represent the two systematic contributions,
independent from $n_s$, while their quadratic sum does not
appears in the plot, because it overlaps practically exactly
with the dominant systematic contribution, due to the uncertain $\pi_2$.
The `straight lines with negative slopes'
(in log-log plot, which notoriously
linearizes power laws) are the individual
statistical contributions (solid and dashed, respectively --
see the figure caption for details) and their quadratic sum
(dotted). The uppest (dotted brown) curve is the overall uncertainty,
dominated at small $n_s$ by the statistical contributions
and at high $n_s$ by the systematic ones,
namely by  $\sigma_{\pi_2}(f_P)$. (We shall come in a while
into the meaning and the importance
of the vertical line.)

Since the dominant contribution due to  $\sigma(\pi_2)$
limits the relative uncertainty on $f_P$ to about $10\%$, reached
for $n_s$ above a few thousands, 
it is  interesting to see what we would gain reducing $\sigma(\pi_2)$
to the value of  $\sigma(\pi_1)$. This is done in the bottom plot
of Fig.~\ref{fig:Contributions_uncertanties_p0.1}, which shows a 
 clear improvement, although the contribution
due to  $\sigma(\pi_2)$ still dominates with respect to that due
to  $\sigma(\pi_1)$, 
because the former enters, for $p=0.1$, with a weight 9 times higher
than the latter,
as it results from Eqs.~(\ref{eq:sigma_pi1_bis}) and (\ref{eq:sigma_pi2_bis}).
Moreover, since all contributions to the uncertainty on $f_P$ depend also
on  $p$,  
we report in Fig. \ref{fig:Contributions_uncertanties_p0.5}
the case of a supposed proportion of
infectees\footnote{We remind once more that this paper is rather
  general, although motivated by Covid-19 related issues, and therefore
we also analyze the possibility of very large $p$.}
as high as $50\%$ (i.e.  $p=0.5$).
\begin{figure}
  \begin{center}
  \centering{\epsfig{file=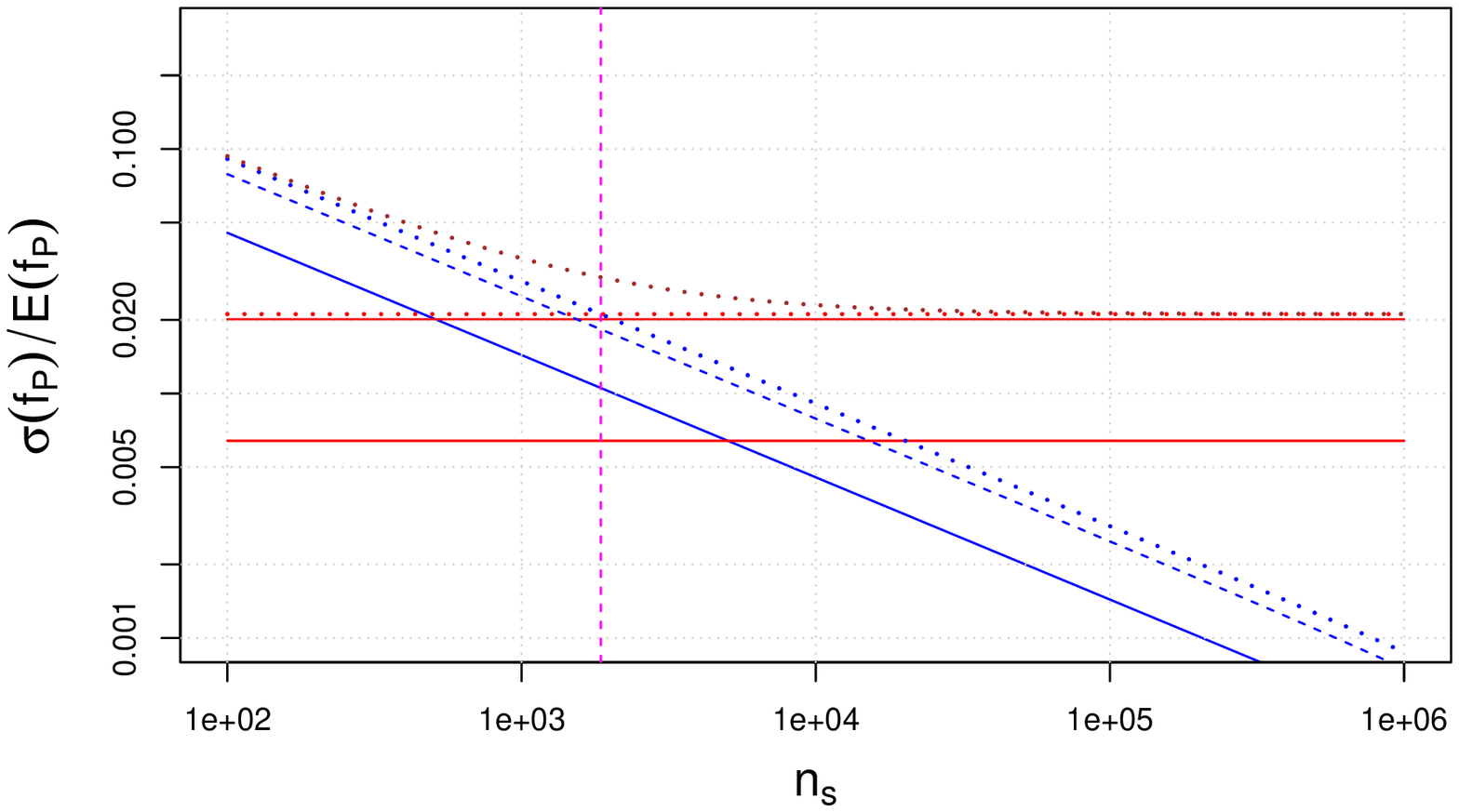,clip=,width=0.92\linewidth}}\\
  \centering{\epsfig{file=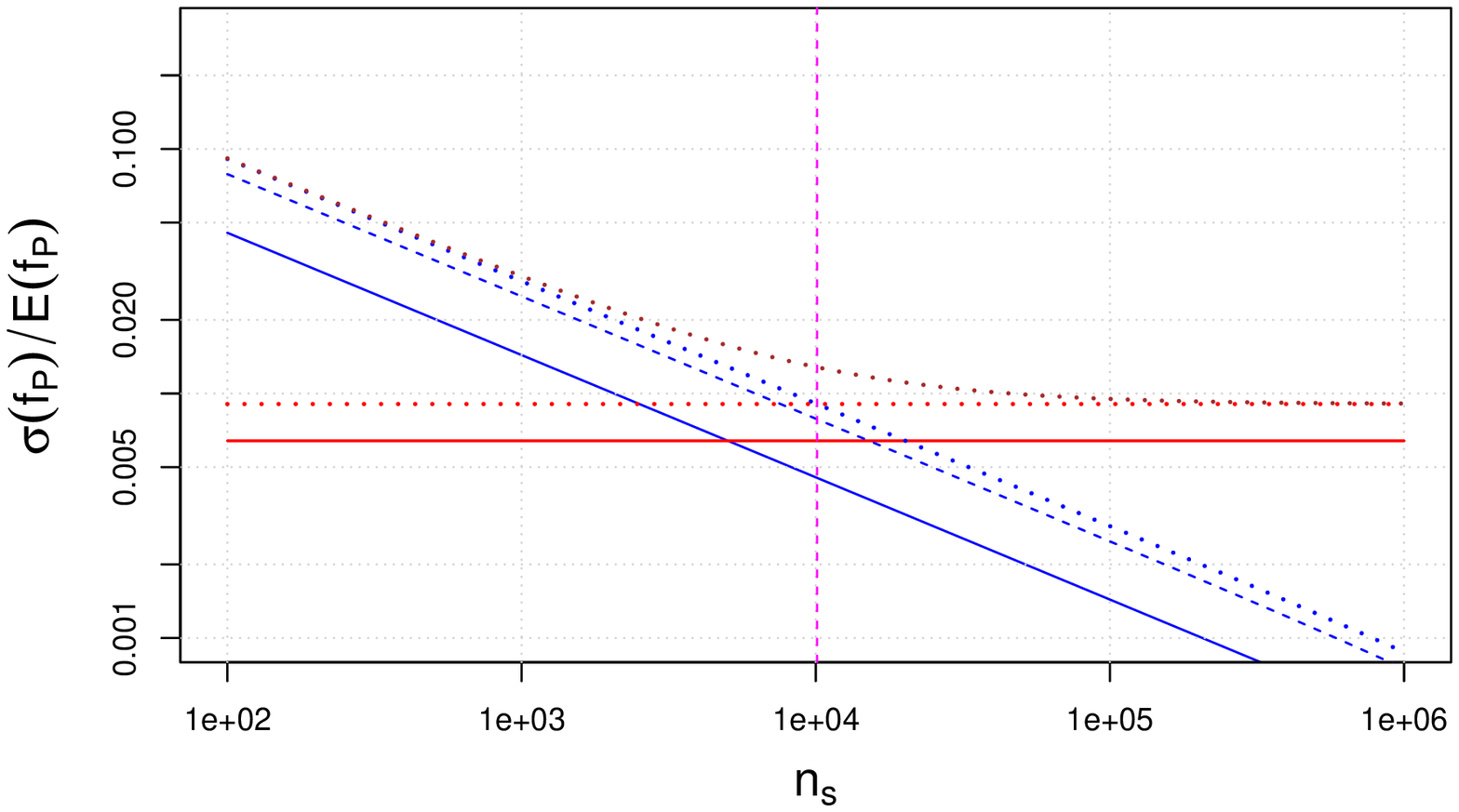,clip=,width=0.92\linewidth}}
  \end{center}
  \caption{\small \sf Same as Fig.~\ref{fig:Contributions_uncertanties_p0.1}
    for a proportion of infected individuals of $50\%$ ($\bm{p=0.5}$).
    In this case the contribution from sampling the population
    $\sigma_{p_s}(f_P)$ is larger than that from $\sigma_R(f_P)$. Note that
    in the lower plot the two solid horizontal lines collapse into a single one,
    being the contribution from  $\sigma_{\pi_1}(f_P)$ and  $\sigma_{\pi_2}(f_P)$
    equal. It is, instead, visible, with respect to the plots of 
    Fig.~\ref{fig:Contributions_uncertanties_p0.1} the horizontal
    dotted line showing the quadratic combination of the systematic
    contributions, reached asymptotically by the
    top dotted curve representing the global relative uncertainty on $f_P$.
  }
  \label{fig:Contributions_uncertanties_p0.5} 
\end{figure}
One of the remarkable difference with respect to
Fig.~\ref{fig:Contributions_uncertanties_p0.1} is that
the contribution from  $\sigma_{p_s}(f_P)$ becomes
larger than that from  $\sigma_R(f_P)$ (remaining
always `parallel' as a function of $n_s$ in `log-log' plots,
since they depend on the same power of the sample size).
Indeed, $\sigma_{p_s}(f_P)$ starts dominating from $p\approx 0.15$
up to  $p\approx 0.95$, as shown in  Fig.~\ref{fig:Rapporto_incertezze_sampling},
\begin{figure}
  \begin{center}
    \epsfig{file=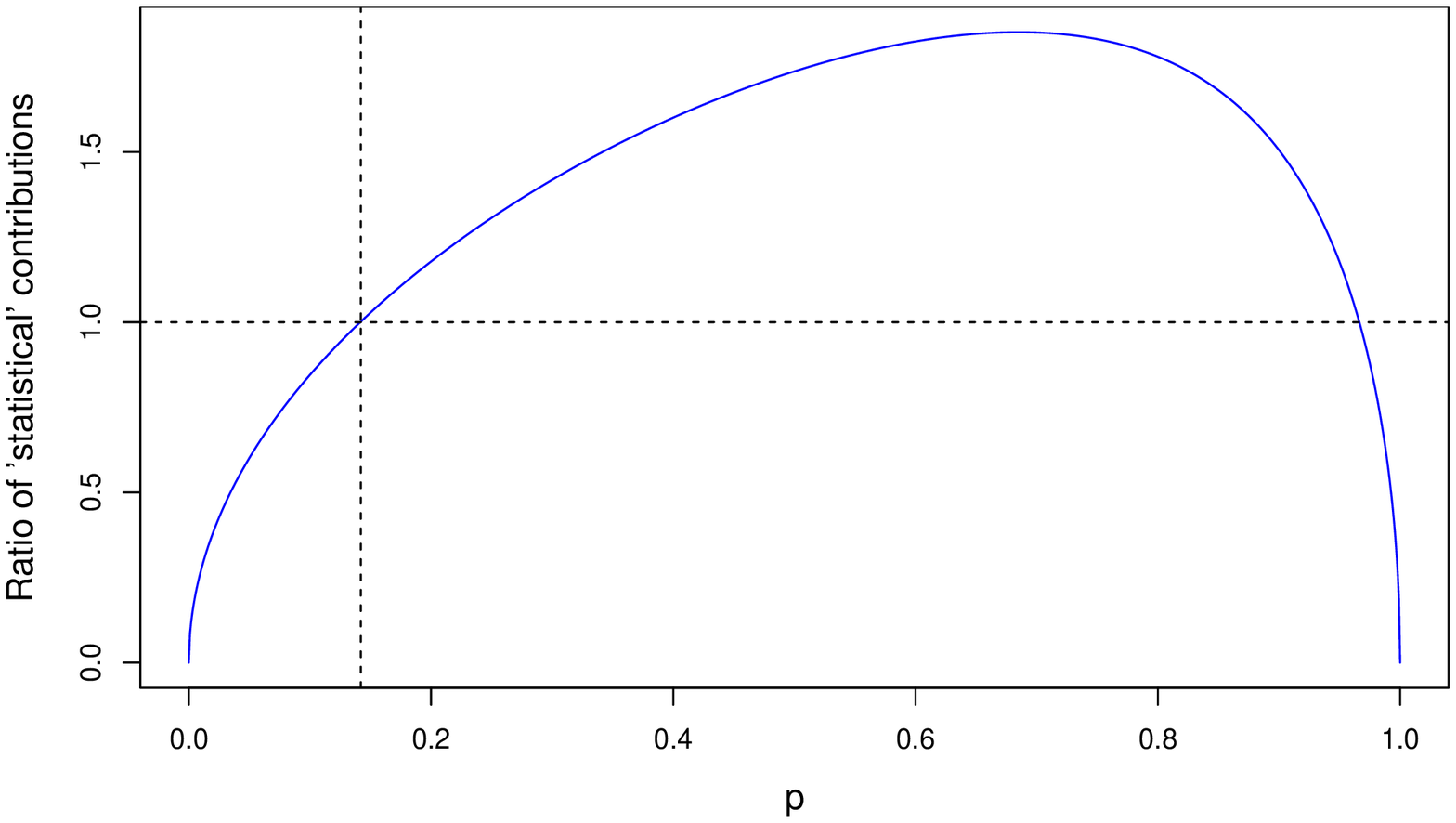,clip=,width=\linewidth}
    \\  \mbox{} \vspace{-1.3cm} \mbox{}
  \end{center}
  \caption{\small \sf
  Ratio of $\sigma_{p_s}(f_P)$ to
  $\sigma_R(f_P)$ as a function of the population
  fraction of infected $p$.
  }
    \label{fig:Rapporto_incertezze_sampling} 
\end{figure}
in which the
ratio $\sigma_{p_s}(f_P)/\sigma_R(f_P)$ as a function of $p$, is reported, 
exhibiting a {\em whale}-like shape.

As a further example we show in
Fig.~\ref{fig:Contributions_uncertanties_sym} the contributions
\begin{figure}
  \begin{center}
  \centering{\epsfig{file=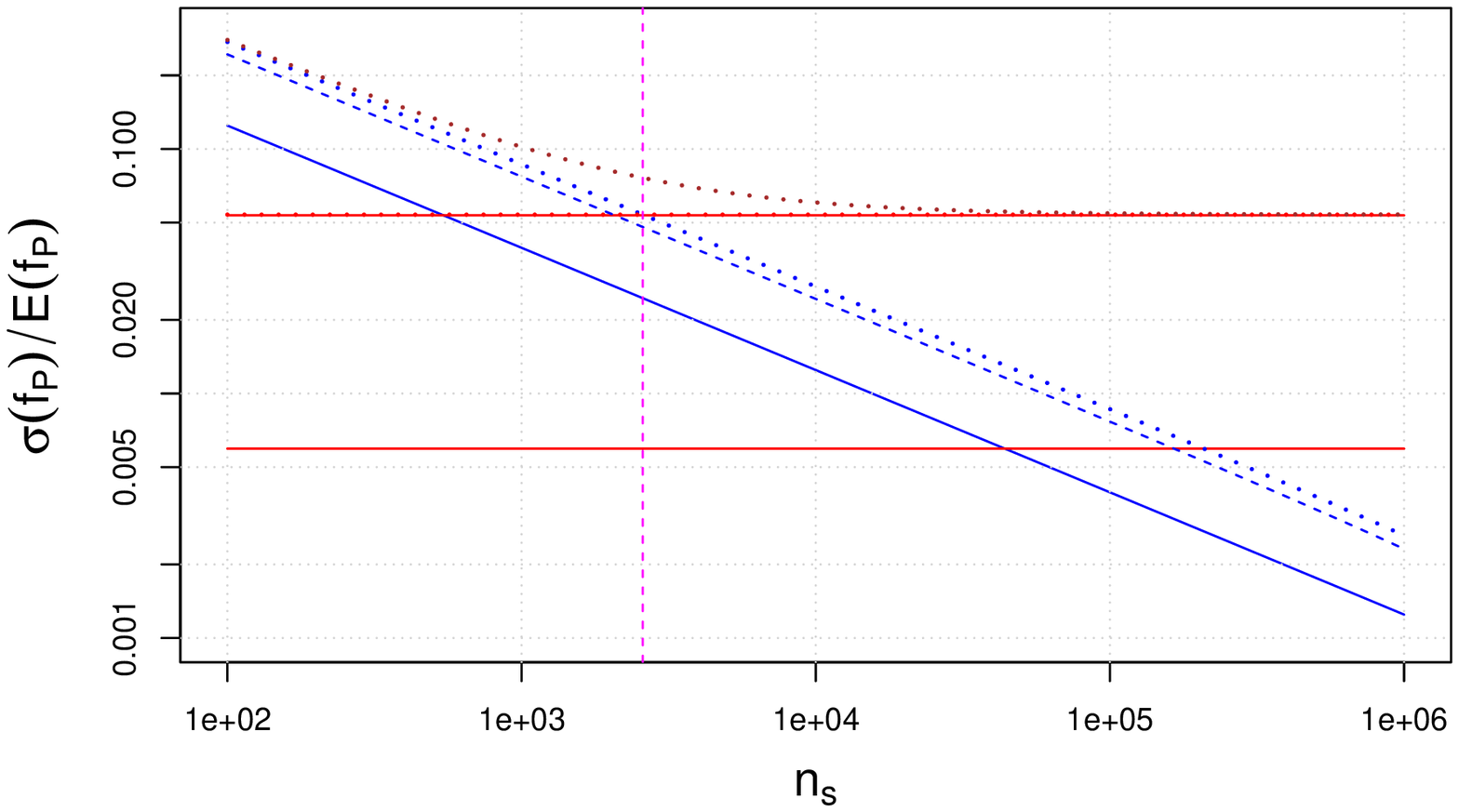,clip=,width=0.92\linewidth}}\\
  \centering{\epsfig{file=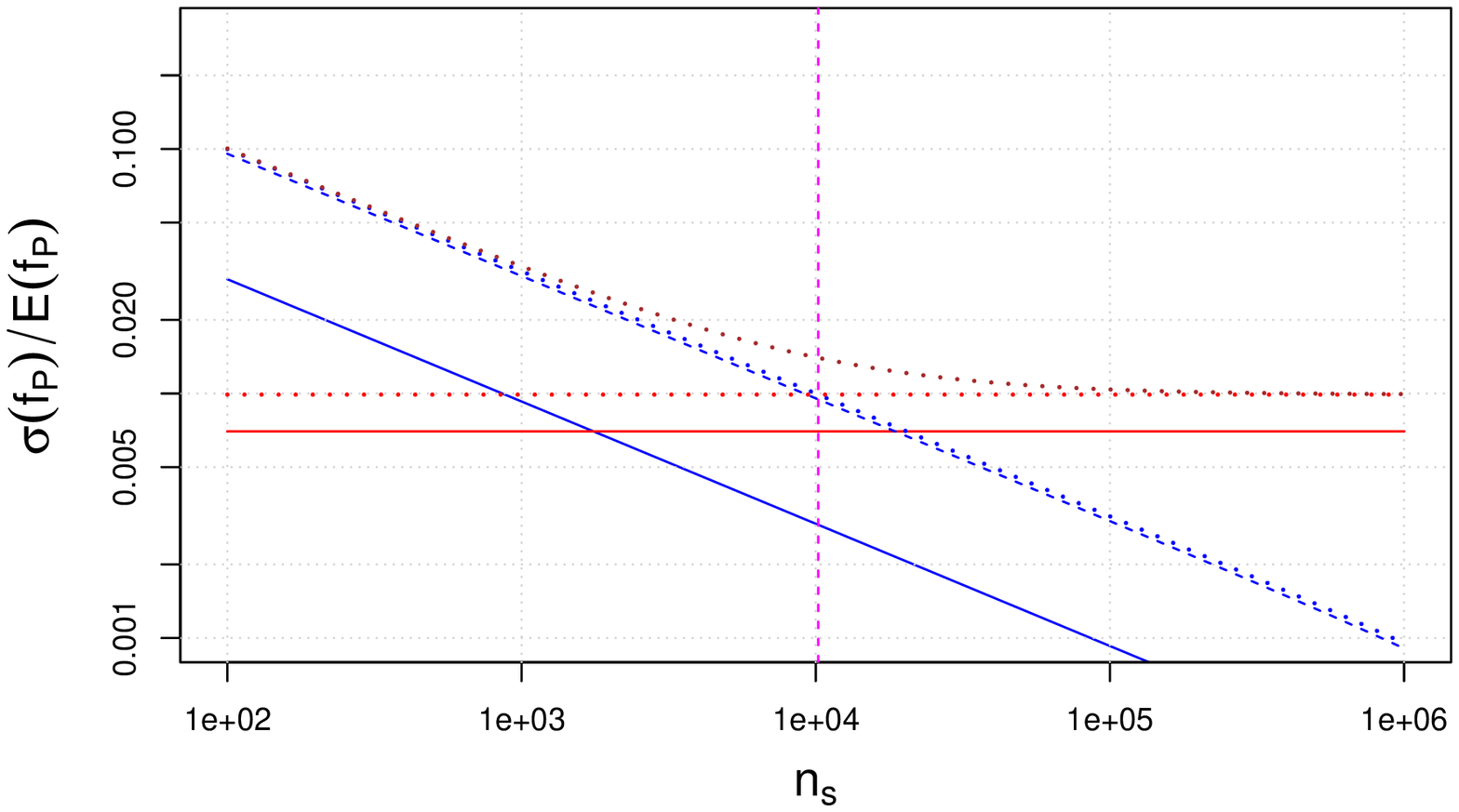,clip=,width=0.92\linewidth}}
  \end{center}
  \caption{\small \sf Same quantities of
    Figs.~\ref{fig:Contributions_uncertanties_p0.1}
    and \ref{fig:Contributions_uncertanties_p0.5},
    but in the {\em symmetric case of specificity equal to sensitivity},
    i.e. $\mbox{E}(\pi_2)=1-\mbox{E}(\pi_1)=0.022$,
    again with equal uncertainties,
    i.e. $\sigma(\pi_2)=\sigma(\pi_1)=0.007$.
    The upper plot, for $\bm{p=0.1}$, has to be compared to the lower plot
    of Fig.~\ref{fig:Contributions_uncertanties_p0.1}; the lower plot,
    for $\bm{p=0.5}$, has to be compared to the lower plot of 
    Fig.~\ref{fig:Contributions_uncertanties_p0.5}.
  }
  \label{fig:Contributions_uncertanties_sym} 
\end{figure}
to the relative uncertainty of $f_P$ for the case of
improved specificity of the test, i.e. reducing the expected value
of $\pi_2$ from 0.115 to 0.022, keeping its uncertainty
equal to that of $\pi_1$, that is 0.007. This means that we consider
specificity equal to sensitivity, both in expected value and in uncertainty.
In practice this is done swapping the parameters of the
related Beta distributions, that is
$r_2=s_1$ and $s_2=r_1$ (see Sec. \ref{ss:conjugate_priors}).

In order to make evident the differences with what has been shown
 in the previous cases,
we plot  $\sigma_{p_s}(f_P)/\mbox{E}(f_P)$ for both $p=0.1$ (upper plot)
and $p=0.5$ (lower plot). In particular, in order to see
the effect of this last improvement of the specificity
(i.e. increasing its expected value from 0.885 to 0.978,
keeping the same standard uncertainty)
we need to compare the upper plot of 
Fig.~\ref{fig:Contributions_uncertanties_sym}
with the lower plot
of  Fig.~\ref{fig:Contributions_uncertanties_p0.1};
the lower plot of Fig.~\ref{fig:Contributions_uncertanties_sym}
with the lower plot of
Fig.~\ref{fig:Contributions_uncertanties_p0.5}.
The result is, at least at a first sight, quite counter-intuitive,
since to a sizable improvement in specificity
there is a reduction in the relative accuracy with which
the fraction of positives is expected (effect particularly
important for $p=0.1$). We shall comment about it in the
next sub-section, in which we start
describing the vertical
lines in the plots of Figs.~\ref{fig:Contributions_uncertanties_p0.1},
\ref{fig:Contributions_uncertanties_p0.5} and
\ref{fig:Contributions_uncertanties_sym}, commenting on their importance.

\subsection[Statistical and systematic contributions
  to $\sigma(f_P)$]{Balance between statistical
  and systematic contributions
to the uncertainty on $f_P$}\label{ss:Balance_Stat_Syst}
The vertical dashed line in the plots of
Figs.~\ref{fig:Contributions_uncertanties_p0.1},
\ref{fig:Contributions_uncertanties_p0.5}
and \ref{fig:Contributions_uncertanties_sym} 
indicates the critical value
$n_s^*$ at which the contribution to total uncertainty
due to $\sigma_{\pi_2}(f_P)$ and $\sigma_{\pi_1}(f_P)$ is equal to
that due to  $\sigma_R(f_P)$ and  $\sigma_{p_s}(f_P)$, that is
{\em for $n_s=n_s^*$ 
 statistical and systematic contributions are equal}.
It follows that, due to the quadratic combination rule,
the global uncertainty at that critical value of the sample size
will be larger than each of them by a factor $\sqrt{2}$.

Being $n_s^*$ an important parameter in order to plan
a test campaign, it is worth getting its closed, although approximated
expression, obtained extending the
condition  (\ref{eq:condizione_n*_parziale}) 
to
\begin{eqnarray}
  \sigma_R^2(f_P) +  \sigma_{p_s}^2(f_P)
  &=& \sigma_{\pi_1}^2(f_P) +  \sigma_{\pi_2}^2(f_P)\,,
  \label{eq:condizione_critica}
\end{eqnarray}  
The result,
under the minimal assumption $N\gg 1$, is
{\small
\begin{equation}
\begin{split}
  n_s^*=\frac{\big[\big(\mbox{E}(\pi_1)-\mbox{E}(\pi_2)\big)^2\!\cdot\! p\!\cdot\!(1-p)\big]
    +\big[\mbox{E}(\pi_1)\!\cdot\!(1-\mbox{E}(\pi_1))\!\cdot\!p +
      \mbox{E}(\pi_2)\!\cdot\! (1-\mbox{E}(\pi_2))\!\cdot\!(1-p)\big]}
  {\big[\sigma^2(\pi_1) \cdot p^2+\sigma^2(\pi_2)  \cdot (1-p)^2\big]
    + \big[\left(\mbox{E}(\pi_1)-\mbox{E}(\pi_2)\right)^2 \cdot
      p \cdot (1-p)\big]/N}\,.
  \label{eq:condizione_critica_esatta}
\end{split}
\end{equation}
}
(Note how in the limit $N\gg n_s$,
  i.e. $N\rightarrow\infty$,
  the second term at the denominator
  of Eq.~(\ref{eq:condizione_critica_esatta})
  can be neglected.\footnote{Indeed, in such a limit
  the  condition (\ref{eq:condizione_critica}) becomes
\begin{equation}
\begin{split}    
  \big[\mbox{E}(\pi_1)\cdot (1-\mbox{E}(\pi_1))\cdot p
    + \mbox{E}(\pi_2)\cdot (1-\mbox{E}(\pi_2))\cdot (1-p)\big]
 +  \big[\left(\mbox{E}(\pi_1) - \mbox{E}(\pi_2)\right)^2 \cdot
  p\cdot (1-p)\big]      
\,  \approx \\
n_s\cdot \big[\sigma^2(\pi_1) \cdot p^2+\sigma^2(\pi_2)
  \cdot (1-p)^2\big]\,,
  \nonumber
\end{split}  
\end{equation}
whose solution is trivial, differing from 
Eq.~(\ref{eq:condizione_critica_esatta}) just
for the term at the denominator containing the factor $N^{-1}$.
})
The top plot of Fig~\ref{fig:ns_vs_p} shows the dependence
of $n_s^*$ on $p$, 
\begin{figure}
  \begin{center}
    \epsfig{file=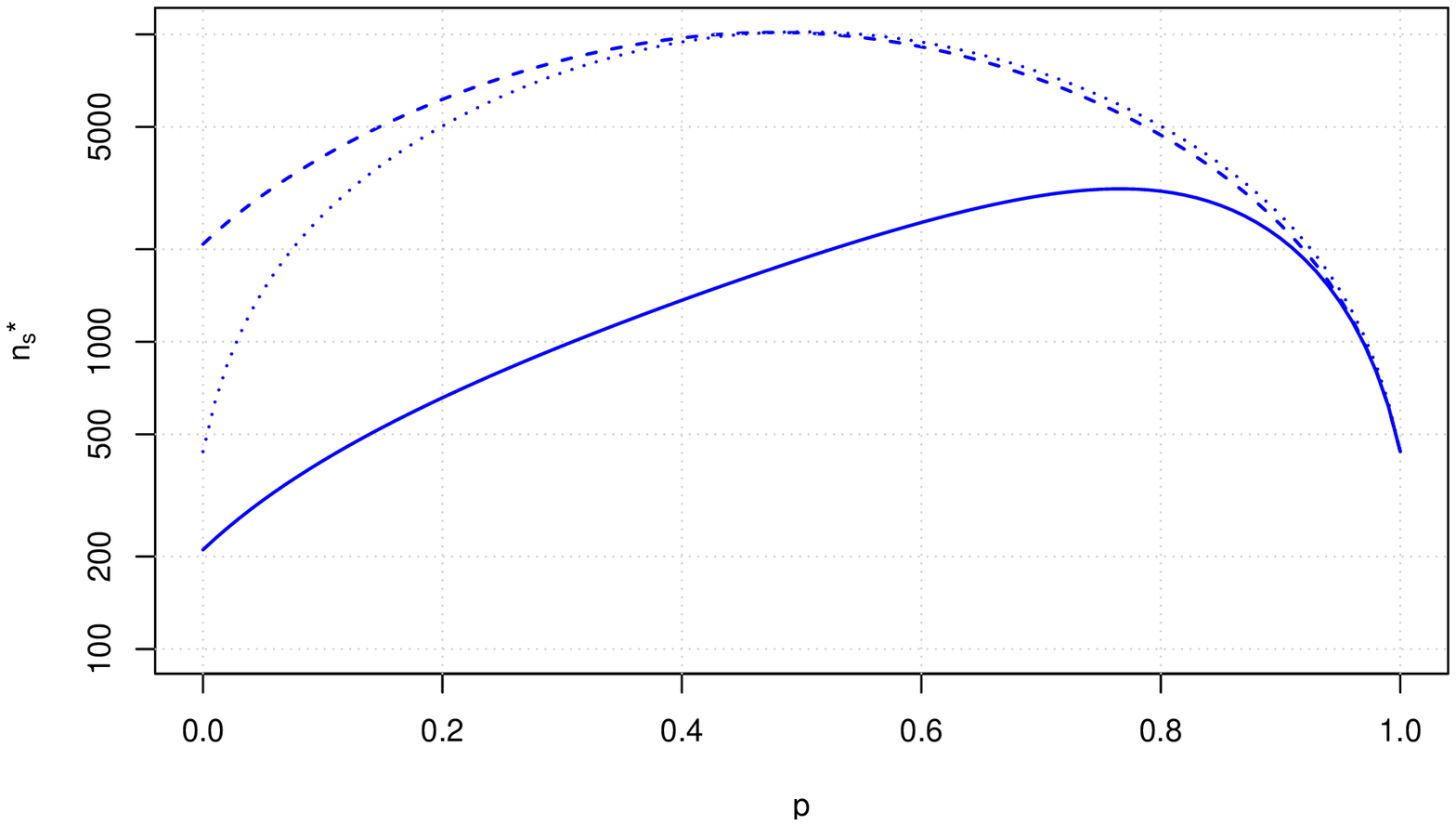,clip=,width=0.95\linewidth}\\
    \mbox{} \\ \mbox{} \\
    \epsfig{file=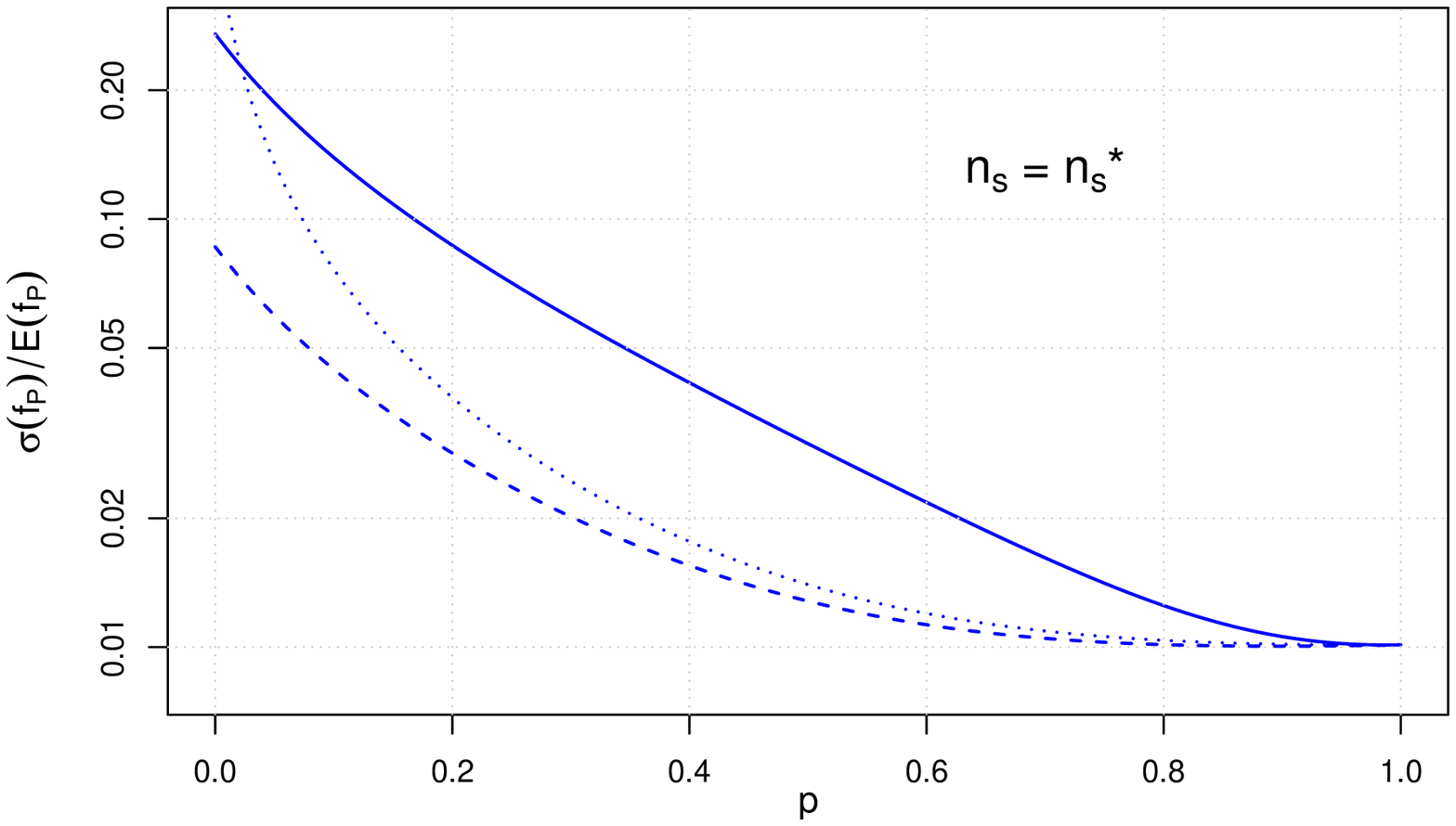,clip=,width=0.95\linewidth}
  \end{center}
  \caption{\small \sf Top plot: dependence of $n_s^*$ on $p$ for the standard
    values of $\sigma_{\pi_1}$ and $\sigma_{\pi_2}$ (solid line),
    for $\sigma_{\pi_1} = \sigma_{\pi_2} = 0.007$ (dashed line) and for
    specificity	equal to sensitivity, i.e. 
    $\mbox{E}(\pi_2) = 1 - \mbox{E}(\pi_1) = 0.022$ (dotted line).
    Bottom plot:  relative uncertainty on $f_P$ at $n_s=n_s^*$
    for the same cases. 
  }
    \label{fig:ns_vs_p} 
\end{figure}
for: our reference values of $\sigma(\pi_1)$ and $\sigma(\pi_2)$
(solid line -- see also top plots of Figs.~\ref{fig:Contributions_uncertanties_p0.1}
and  \ref{fig:Contributions_uncertanties_p0.5});
the improved case of  $\sigma(\pi_2) = \sigma(\pi_1) = 0.007$
(dashed line -- see also bottom plots of Figs.~\ref{fig:Contributions_uncertanties_p0.1}
and  \ref{fig:Contributions_uncertanties_p0.5});
the mirror-symmetric case in which 
$\mbox{E}(\pi_2) = 1 - \mbox{E}(\pi_1) = 0.022$
and  $\sigma(\pi_2) = \sigma(\pi_1) = 0.007$
(dotted line -- see also Fig.~\ref{fig:Contributions_uncertanties_sym}).
Once we know the dependence of $n_s^*$ on $p$,
since the uncertainty
on $f_P$ depends on  $n_s$ and $p$,
we can evaluate
the relative uncertainty on the predicted fraction of positives
that will result from the test campaign, as a function of $p$ 
under the condition $n_s=n_s^*$,
that is
$\left.\sigma(f_P)/\mbox{E}(f_P)\right|_{n_s=n_s^*}$.
The result is shown
in the bottom plot of Fig~\ref{fig:ns_vs_p}
for the three cases of the upper plot of the same figure.

When we reduce the uncertainty about $\sigma(\pi_2)$, keeping constant
its expected value, the systematic contribution to the uncertainty
is reduced and then, as we have already learned from
Figs.~\ref{fig:Contributions_uncertanties_p0.1},
\ref{fig:Contributions_uncertanties_p0.5}
and \ref{fig:Contributions_uncertanties_sym},
it becomes meaningful to analyze larger samples.
We can then predict the fraction
of individuals tagged as positive with improved
accuracy, i.e. $\sigma(f_P)/\mbox{E}(f_P)$ decreases.
This intuitive reasoning is confirmed by
the plots of Fig~\ref{fig:ns_vs_p}, moving from the
solid curves to the dashed ones.
Instead, improving the specificity to 0.885
to 0.978, i.e. reducing $\mbox{E}(\pi_2)$ from 0.115 to 0.022,
keeping the same uncertainty of 0.007,
leads to surprising results at low values of $p$, at least at a first
sight (dashed curves $\rightarrow$ dotted curves).
In fact, one would expect that from this further improvement
in the quality of the test (which definitively makes
a difference
when testing a single individual,
as discussed in Sec.~\ref{sec:uncertainty})
should follow a general improvement
in the prediction of the fraction of positives.

The reason of this counter-intuitive outcome
is due to the combination of two effects.
The first is the dependence on $\mbox{E}(\pi_1)$ and
$\mbox{E}(\pi_2)$ of the statistical contributions to the
uncertainty, as we can see from Eqs.~(\ref{eq:sigma_R_bis}) and
(\ref{eq:sigma_fP_sigma_ps_bis}). The second is that,
decreasing $\mbox{E}(\pi_2)$, the expected value of $f_P$
decreases too (less `false positives') and therefore
the relative uncertainty on  $f_P$, i.e.
 $\sigma(f_P)/\mbox{E}(f_P)$, increases. 
While the second effect is rather obvious and there is
little to comment, we show the first one graphically,
for $p=0.1$ at which the effects becomes sizable,
in the three plots of
Fig.~\ref{fig:Contributions_uncertanties_abs_p0.1}:
\begin{figure}
  \begin{center}
  \centering{\epsfig{file=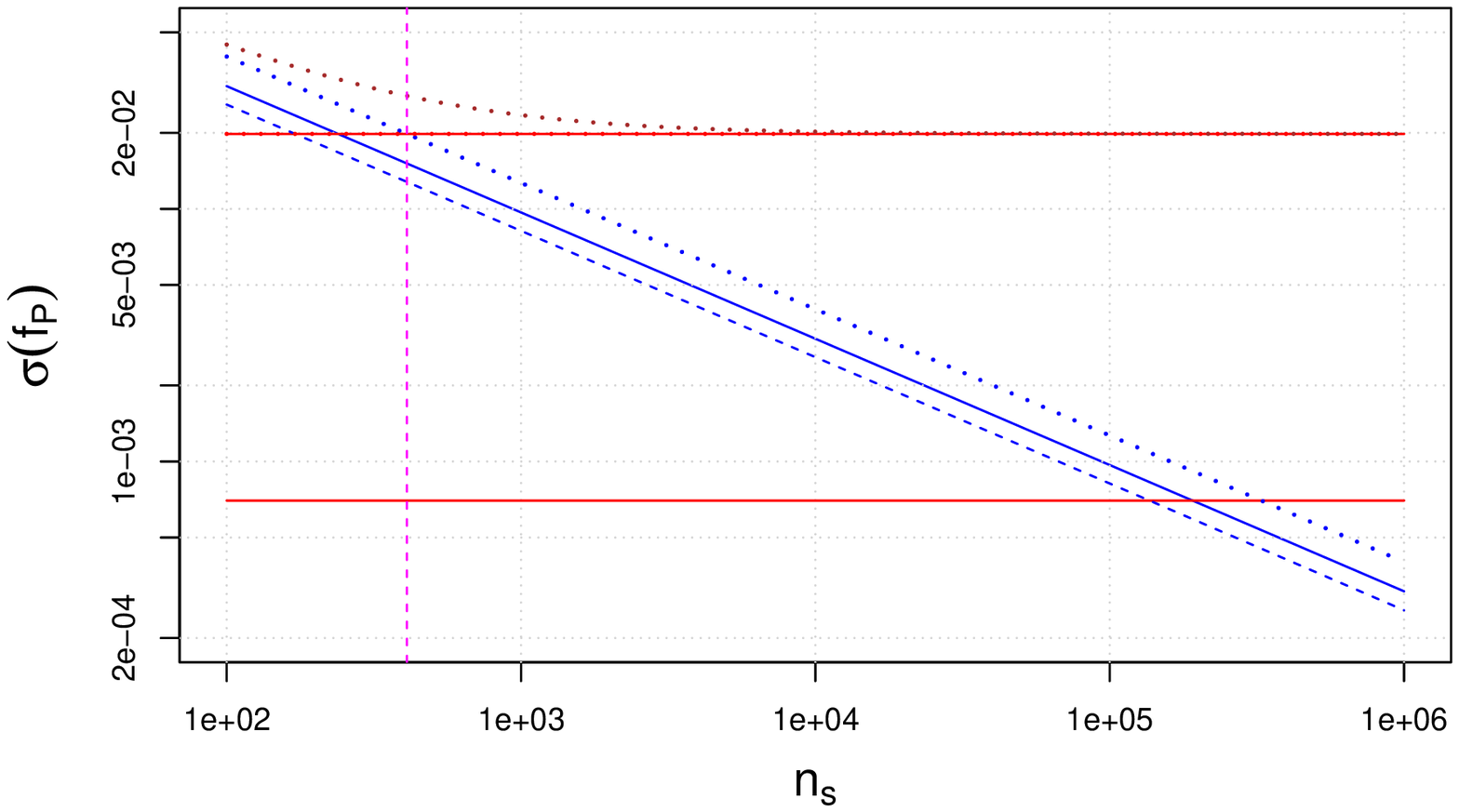,clip=,width=0.81\linewidth}}\\
  \centering{\epsfig{file=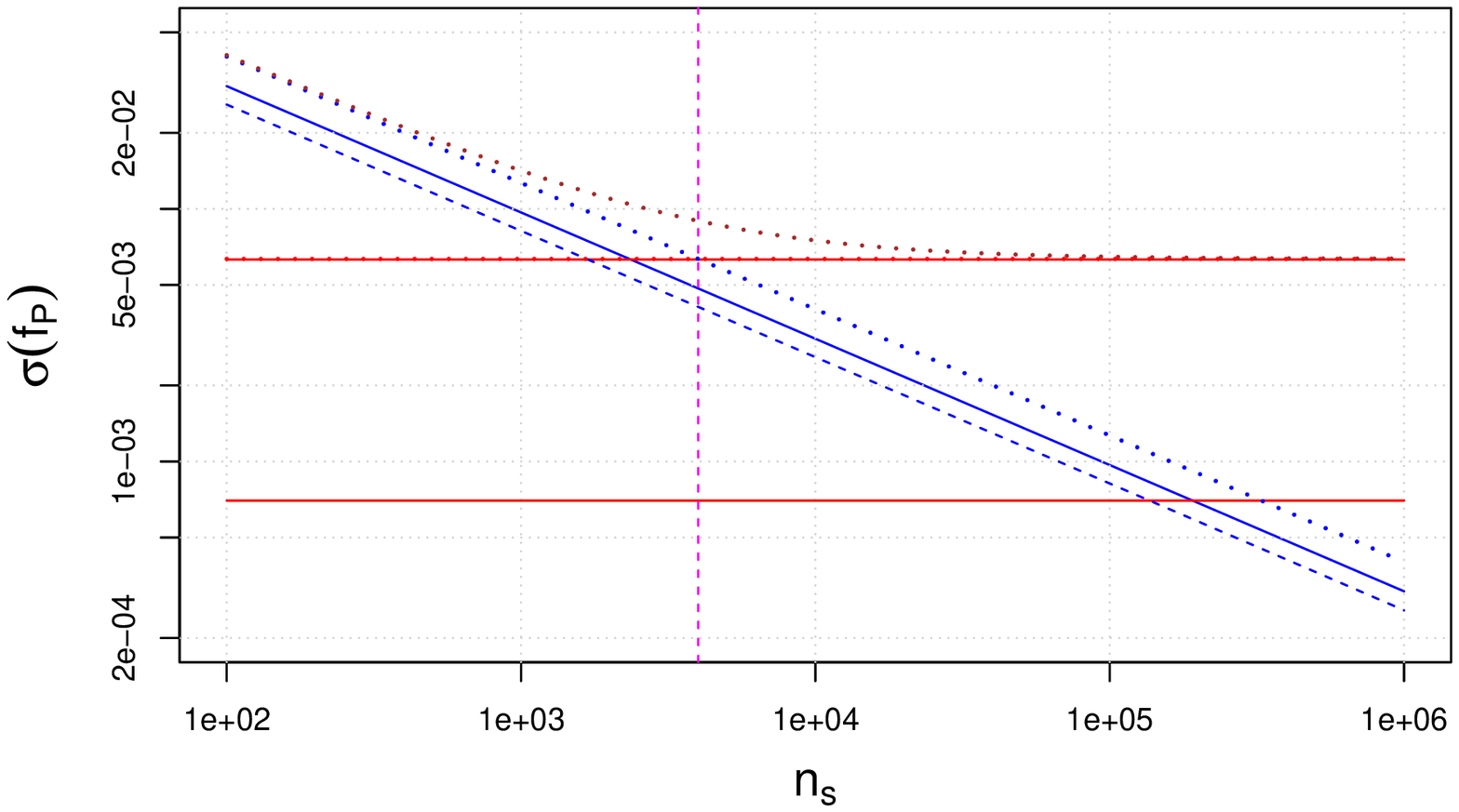,clip=,width=0.81\linewidth}}
  \\
  \centering{\epsfig{file=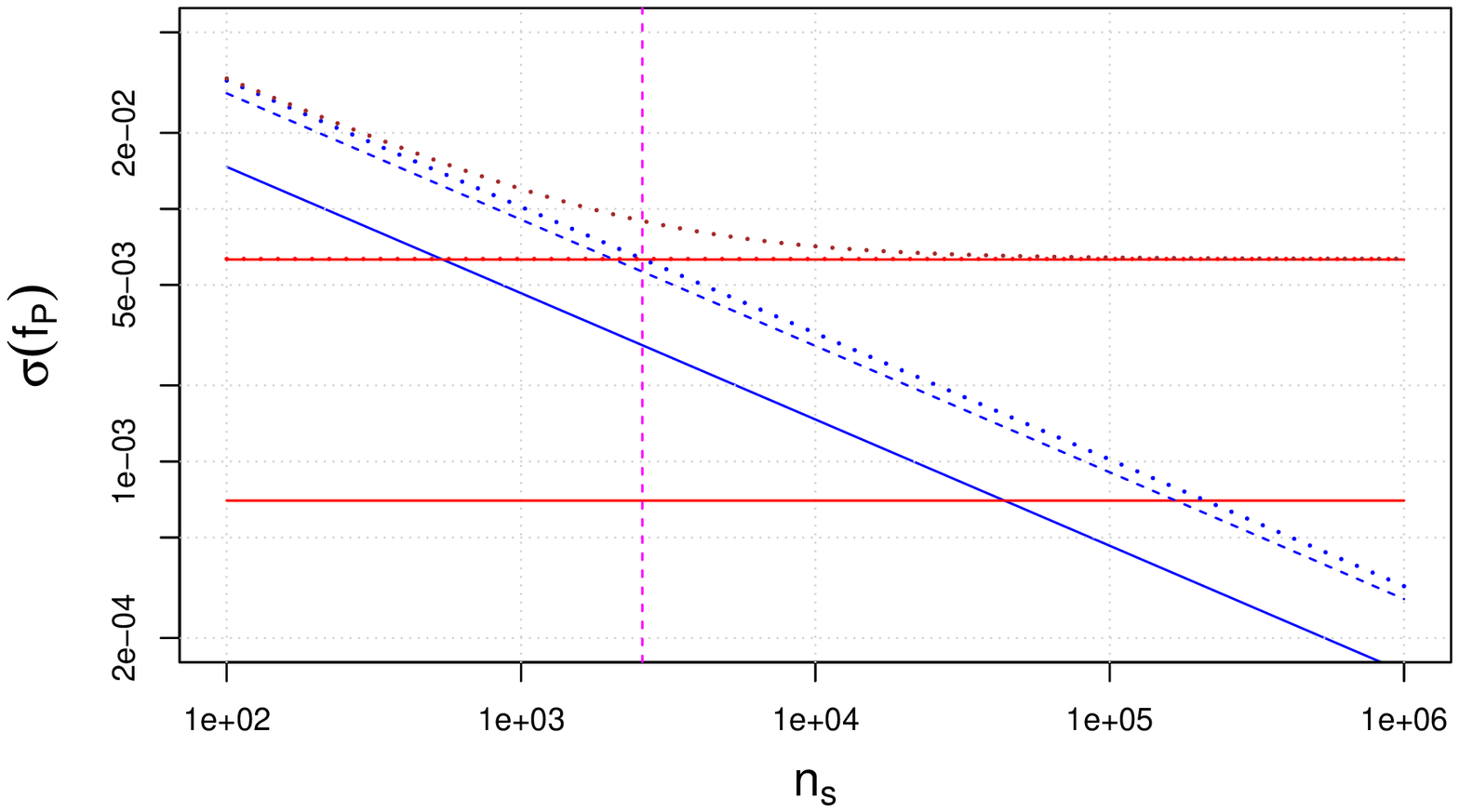,clip=,width=0.81\linewidth}}
  \end{center}
  \mbox{} \vspace{-1.1cm} \mbox{}
  \caption{\small \sf Contributions to $\sigma(f_P)$ varying $\sigma(\pi_2)$
    and  $\mbox{E}(\pi_2)$ for $\bm{p=0.1}$ (see text).
  }
  \label{fig:Contributions_uncertanties_abs_p0.1} 
\end{figure}
the upper plot for our reference values of $\pi_1$ and $\pi_2$,
the middle one improving $\sigma(\pi_2)$ to 0.007,
and the bottom one also reducing the expected value of $\pi_2$ to 0.022. 
But differently from
Figs.~\ref{fig:Contributions_uncertanties_p0.1},
\ref{fig:Contributions_uncertanties_p0.5}
and \ref{fig:Contributions_uncertanties_sym}, these plots show $\sigma(f_P)$
instead of $\sigma(f_P)/\mbox{E}(f_P)$, so that we can 
focus only
on the contributions to the uncertainty, not `distracted' by the
variation of the expected value of $f_P$.
Moving from the top plot to the middle one,
only the contribution due to $\pi_2$ is reduced, 
all the others remaining exactly the same.
Then, when we increase the specificity, i.e. we reduce
$\mbox{E}(\pi_2)$ from 0.115 to 0.022, keeping unaltered
its uncertainty, its contribution to  $\sigma(f_P)$
is unaffected, while the statistical contributions do change.
In particular  $\sigma_R(f_P)$ is strongly reduced, while
$\sigma_{p_s}(f_P)$ increases a little bit.
The combined effect is a decrease of the overall
statistical contribution, thus lowering $n_s^*$.

Summing up,  the combination of the two plots of
Fig.~\ref{fig:ns_vs_p}
gives at a glance,
for an assumed proportion of infectees $p$, an idea
of the `optimal' relative uncertainty we can get on $f_P$ (bottom plot)
and the sample size needed to achieve it (upper plot).
We remind that the lowest relative uncertainty, equal to $1/\sqrt{2}$
of the value shown in the plot, is reached
when the sample size $n_s$ is about one order of magnitude larger
than $n_s^*$, i.e. when the random contribution to the uncertainty
is absolutely negligible and any further increase of $n_s$
not justifiable. But, anyway -- think about it --
being $1/\sqrt{2}\approx 0.7$, is it worth increasing
so much ($\approx 10$ times) the sample size in order
to reduce $\sigma(f_P)$ by only 30\%?

\begin{figure}[t]
\begin{center}
  \epsfig{file=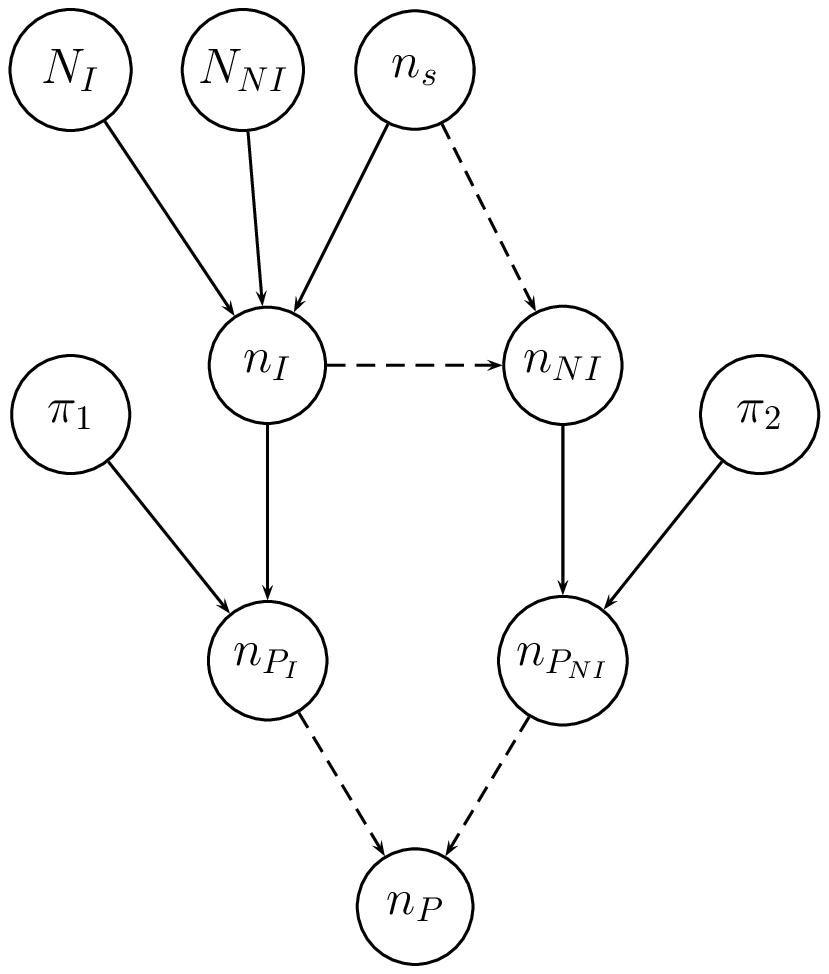,clip=,width=0.45\linewidth}
  \mbox{} \hspace{1.0cm}
  \epsfig{file=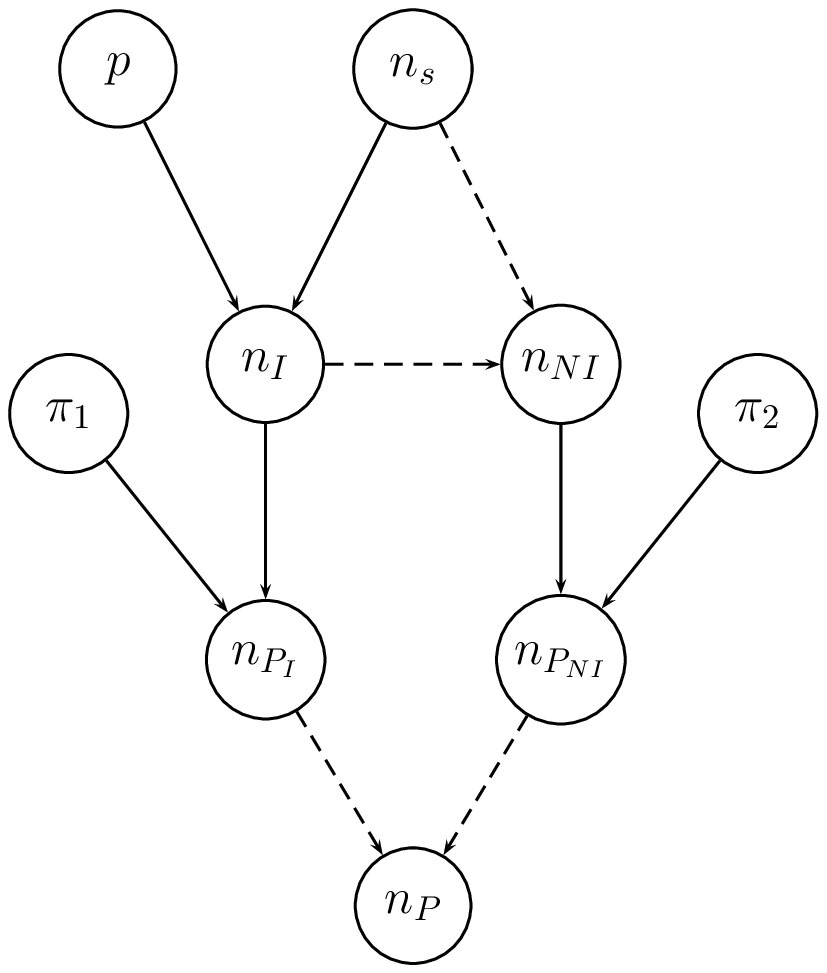,clip=,width=0.45\linewidth}
  \\  \mbox{} \vspace{-0.5cm} \mbox{} 
\end{center}
\caption{\small \sf Graphical network of Fig.~\ref{fig:two_binom},
  augmented by the sampling process, modeled by a
  hypergeometric distribution (left) or by
  a binomial distribution (right) with $p=N_I/(N_I+N_{NI})$.
}
\label{fig:sampling_binom-hg} 
\end{figure}
\section{Measurability of $p$}\label{sec:measurability_p}
It is now time to put together all the items discussed so far 
and to move to the question of how well we can
{\em measure} the proportion $p$ of infectees 
in a population of $N$ individuals, out of which
$n_s$ have been sampled at random and tested, resulting in
a total of $n_P$ positives. Again we proceed
by steps. In this section we tackle the question of how
the expected number of positives and its uncertainty
depend on the proportion $p$ of infectees in the population.
The real inferential problem, consisting in stating
which values of $p$ are more or less believable,  
will be analyzed in Secs.~\ref{sec:interring_p}
and \ref{sec:direct}.

\subsection{Probabilistic model}
The graphical model describing the quantities of interest
is shown in the left hand network of  Fig.~\ref{fig:sampling_binom-hg},
based on that of Fig.~\ref{fig:two_binom},
to which we have added {\em parents} to the nodes $n_I$
and $n_{NI}$, the number
of Infected and Not Infected in the sample, respectively.
More precisely, the number of infectees 
$n_I$ in the sample is described by a hypergeometric distribution,
that is
\begin{eqnarray*}
  n_I &\sim&  \mbox{HG}(N_I,N_{NI},n_s)\,,
\end{eqnarray*}
with $N_I$ and $N_{NI}$ the numbers of infected and not infected individuals
in the population. Then, the number  $n_{NI}$ of not
infected people in the sample is
deterministically related to $n_{I}$, being $n_{NI} = n_s - n_I$.

However, since  in this paper
we are interested in sample sizes
much smaller than those of the populations, we can remodel
the problem according to the right hand network
of  Fig.~\ref{fig:sampling_binom-hg}, in which $n_I$
is described by a binomial distribution, that is
\begin{eqnarray*}
  n_I &\sim&  \mbox{Binom}(n_s, p)\,,
\end{eqnarray*}
with  $p=N_I/(N_I+N_{NI})$.
This simplified model has been re-drawn in the network
shown in the left hand side
of Fig.~\ref{fig:sampling_binom_pred}, 
\begin{figure}[t]
\begin{center}
  \epsfig{file=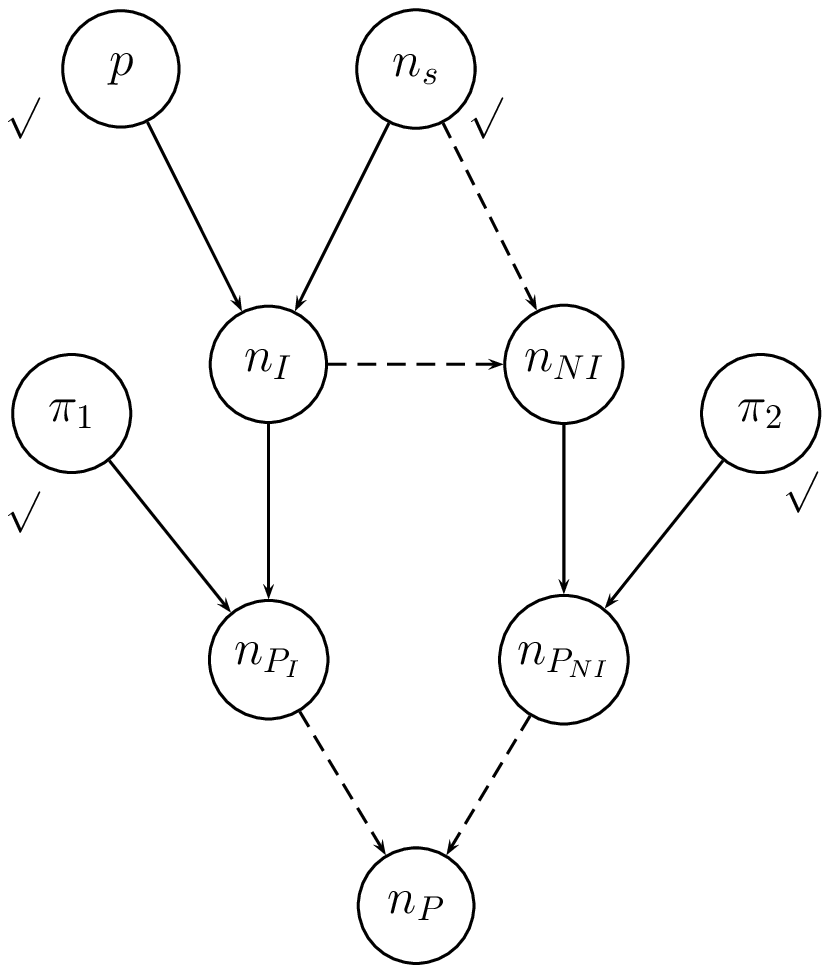,clip=,width=0.43\linewidth}
  \mbox{} \hspace{0.3cm}
  \epsfig{file=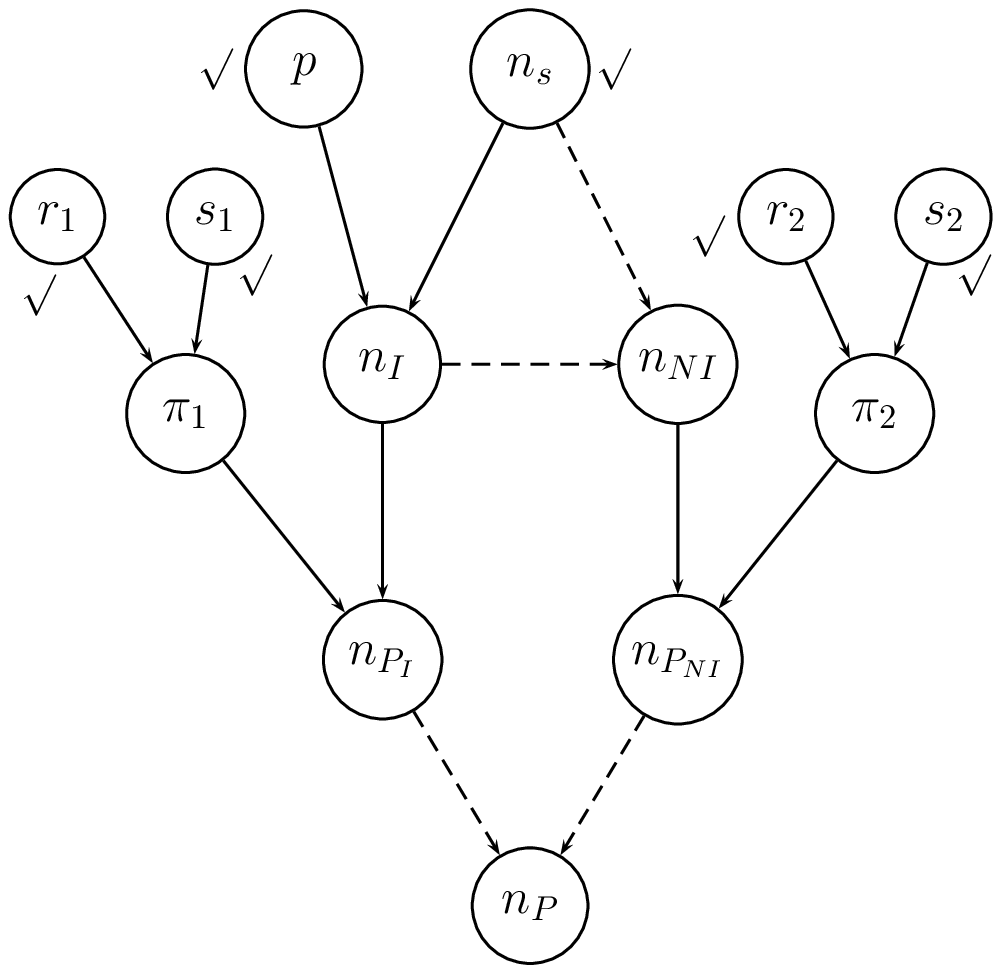,clip=,width=0.52\linewidth}
   \\  \mbox{} \vspace{-0.5cm} \mbox{}
\end{center}
\caption{\small \sf Simplified graphical model of
  Fig.~\ref{fig:sampling_binom-hg} rewritten in order to
  make explicit `known'/`assumed' quantities, tagged by the symbol
  '$\surd\,$', and the uncertain ones. In particular, in the left
  hand diagram precise values of $\pi_1$ and $\pi_2$ are assumed,
  while in the the right hand one the uncertainty on their values
  is modeled with Beta pdf's with parameters $(r_1,s_1)$ and $(r_2,s_2)$.
}
\label{fig:sampling_binom_pred} 
\end{figure}
indicating by the symbol `$\surd\,$' the {\em certain} variables
in the game (indeed those which are for some
reason \underline{\em assumed}),
in contrast to the others, which are uncertain 
and whose values will be ranked in degree of belief
following the rules of probability theory.
Note that in this diagram $\pi_1$ and $\pi_2$ are assumed to be
exactly known. Instead, as we have already seen in Sec.~\ref{sec:uncertainty},
their values are uncertain and their probability distribution can
be conveniently modeled by Beta probability functions characterized 
by parameters $r$'s and $s$'s. The graphical
model which takes into account also
the uncertainty about $\pi_1$ and $\pi_2$
is drawn in the same Fig.~\ref{fig:sampling_binom_pred}
(right side).

We have already discussed extensively, in Sec.~\ref{sec:fpN_from_p},
how the expectation of $n_P$, and therefore of the fraction
on positives in the sample, $f_P$, depends on the model parameters.
Now we go a bit deeper into the question of the dependence
of $f_P$ on the fraction of infectees in the population and,
more precisely, which are the `closest' (to be defined somehow)
two values of $p$, such that the resulting  $f_P$'s
are `reasonably separated' (again to be defined somehow)
from each other. Moreover, instead of simply relying
on the approximated
formulae developed in Sec.~\ref{sec:fpN_from_p}, we are going
to use Monte Carlo methods in different ways: initially
just based on R random number generators; then using
(well below its potentials!) the program JAGS, which will then
be used in Sec.~\ref{sec:interring_p} for inferences.
However we shall keep using 
the approximated formulae for cross check and 
to derive some useful, although approximated, results in closed form.

\subsection{Monte Carlo estimates of $f(n_P)$ and $f(f_P)$}
Analyzing the graphical model in the right hand side
of Fig.~\ref{fig:sampling_binom_pred}, 
the steps we have to go through become self-evident:
\begin{enumerate}
\item generate a value of $n_I$ according to a binomial distribution,
  then calculate $n_{NI}$;
\item generate a value of $\pi_1$ and of $\pi_2$
  according to Beta distributions;
\item generate a value of $n_{P_I}$ and a value of
  $n_{P_{NI}}$ according to binomial distributions;
\item sum up  $n_{P_I}$ and $n_{P_{NI}}$ in order to get
  $n_P$.
\end{enumerate}

\subsubsection{Using the R random number generators}\label{sss:MC_R}
The implementation in R is rather simple, thanks
also to the capability of the language to handle `vectors', meant
as one dimensional arrays, in a compact way. The steps described above
can then be implemented into the following
lines of code:
\begin{verbatim}
n.I   <- rbinom(nr, ns, p)       # 1. 
n.NI  <- ns - n.I
pi1   <- rbeta(nr, r1, s1)       # 2.
pi2   <- rbeta(nr, r2, s2)
nP.I  <- rbinom(nr, n.I, pi1)    # 3.
nP.NI <- rbinom(nr, n.NI, pi2)
nP    <- nP.I + nP.NI            # 4.
\end{verbatim}
We just need to define the parameters of interest,
including {\tt nr}, number of extractions, run
the code and add other instructions in order
to print and plot the results (a minimalist script performing all tasks
is provided in Appendix B.5). The only instructions
worth some further comment are the two related to the step nr. 3.
A curious feature of the `statistical functions' of R
is that they can accept vectors for the number of trials and for
probability at each trial, as it is done here. It means that
internally the random generator is called {\tt nr} times,
the first time e.g. with {\tt n.I$[$1$]$}
and  {\tt pi1$[$1$]$}, the second time with
{\tt n.I$[$2$]$} and  {\tt pi1$[$2$]$}, and so on, thus
avoiding us to use explicit loops.
Note that, if precise values of $\pi_1$ and $\pi_2$ were assumed,
then we just need to replace the two lines of step nr. 2
with the assignment of their numeric values. 

Figure \ref{fig:PredictionPositive_by01} shows the results
\begin{figure}
  \begin{center}
  \epsfig{file=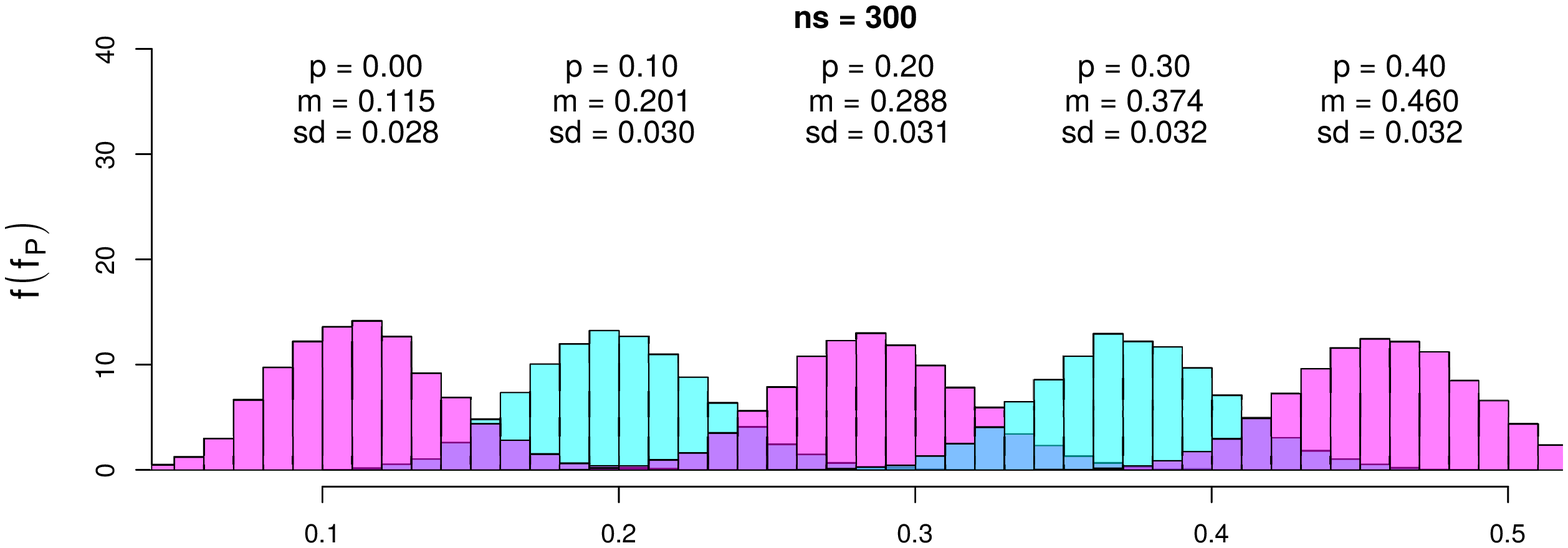,clip=,width=0.9\linewidth} \\
  \epsfig{file=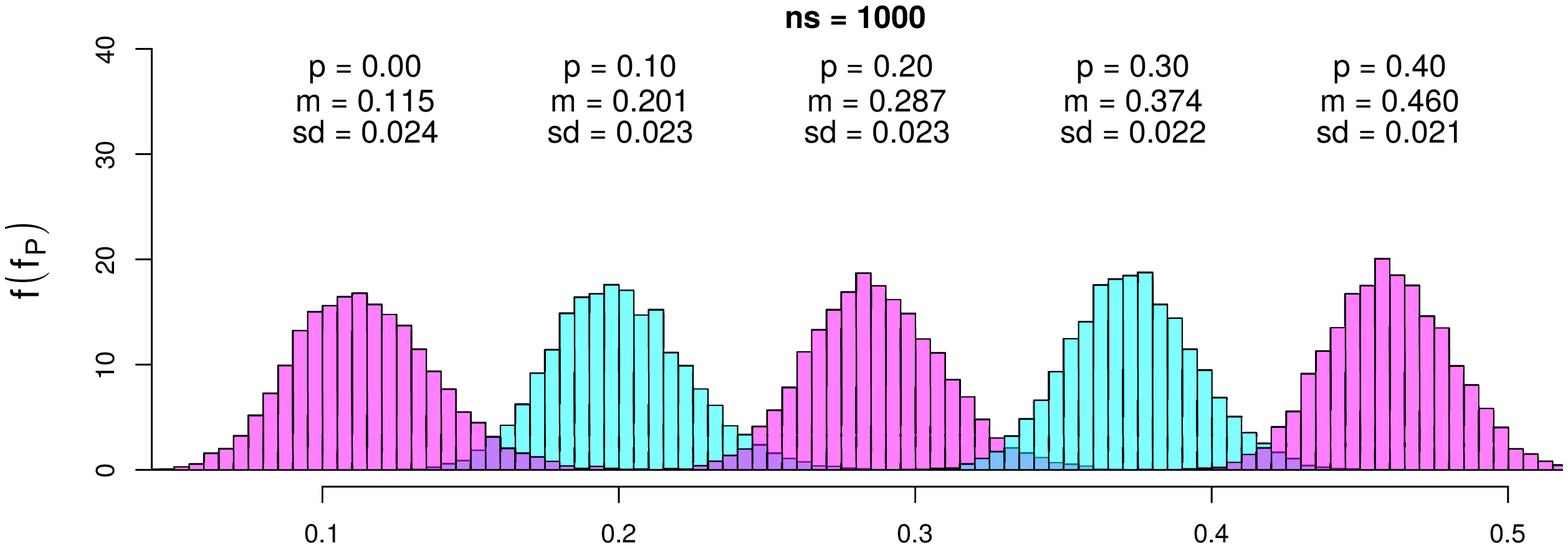,clip=,width=0.9\linewidth} \\
  \epsfig{file=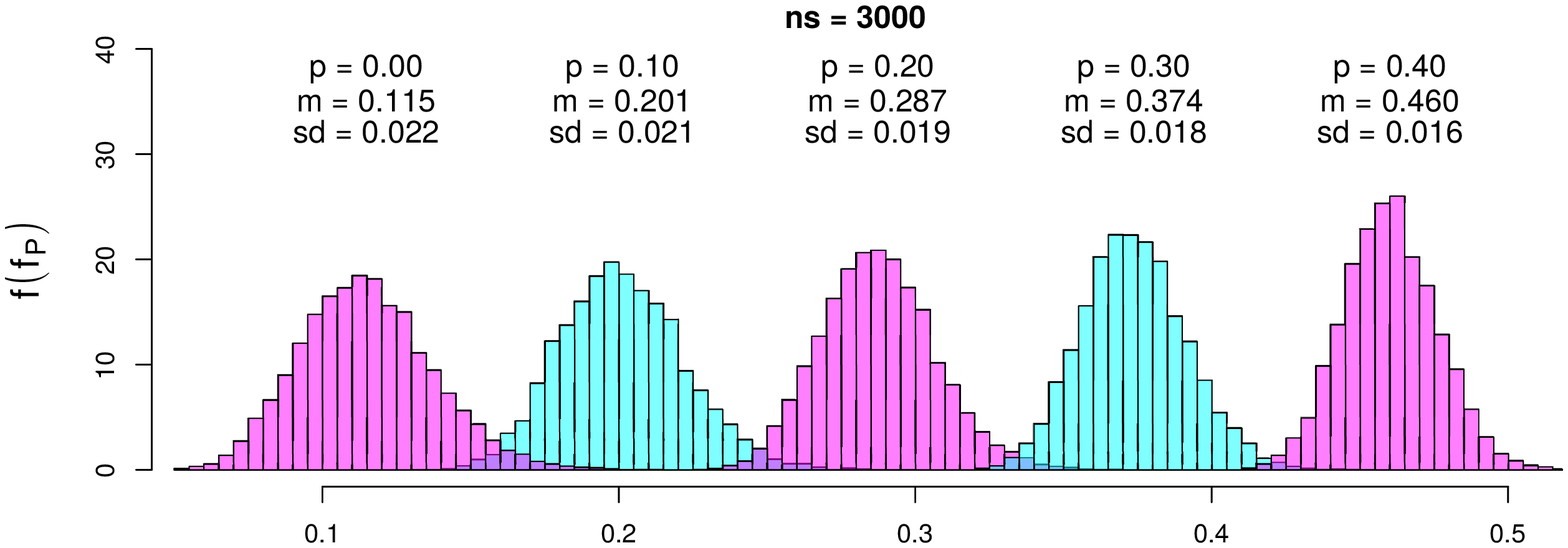,clip=,width=0.9\linewidth} \\
  \epsfig{file=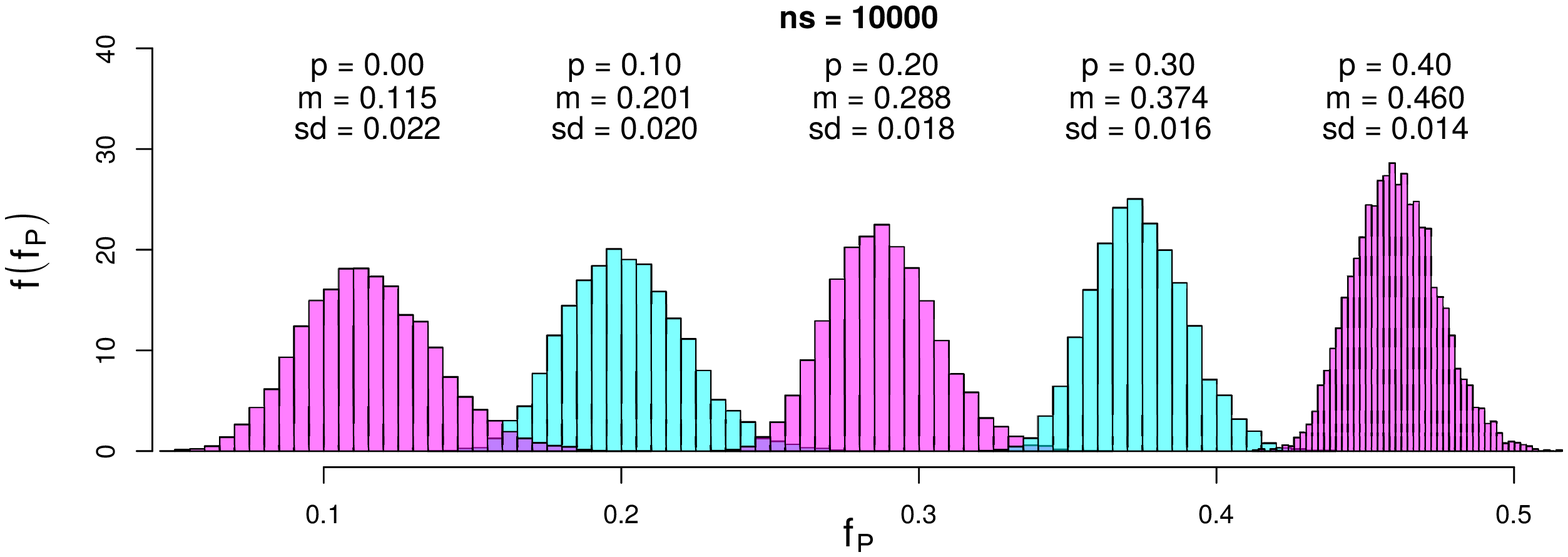,clip=,width=0.9\linewidth}
 \\  \mbox{} \vspace{-1.0cm} \mbox{}
  \end{center}
  \caption{\small \sf Predictive distributions of $f_P$
    as a function of $p$ and $n_s$ for our default
    uncertainty on $\pi_2$, summarized as $\pi_2=0.115\pm 0.022$.    
  }
    \label{fig:PredictionPositive_by01} 
\end{figure}
obtained for some values of $p$ and $n_s$, and modeling the
uncertainty of $\pi_1$ and $\pi_2$ in our default way,
summarized by $\pi_1=0.978\pm 0.007$ and  $\pi_2=0.115\pm 0.022$. 
The values of $n_s$ have been chosen in steps of roughly
half order of magnitude
in the region of $n_s^*$ of interest,
as we have learned in Sec.~\ref{sec:fpN_from_p}.
We see that
for the smallest $n_s$ shown in the figure, equal to 300,
varying $p$ by 0.1 produces distributions of $f_P$ with
quite some overlap. Therefore with samples of such a small
size we can say, {\em very qualitatively} that we
can resolve different values of $p$ if they do not
differ less than $\approx {\cal O}(0.1)$. The situations improves
(the separation roughly doubles)
when we increase  $n_s$ to 1000 or even to 3000, while there is no further
gain reaching $n_s=10000$. This is in agreement with what
we have learned in the previous section.

Since, as we have already seen, the limiting effect is due to
 systematics, and in particular, in our case, to the uncertainty
about $\pi_2$, we show in Fig.~\ref{fig:PredictionPositive_by01_equal-sigmas}
\begin{figure}
  \begin{center}
  \epsfig{file=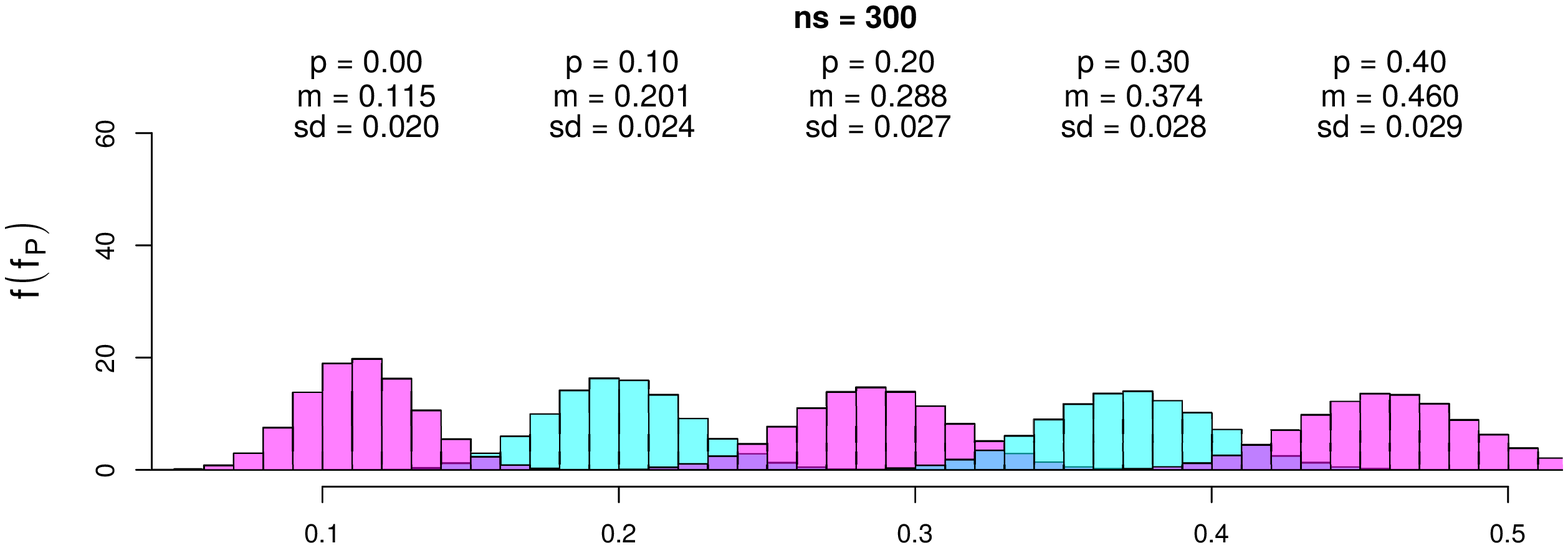,clip=,width=0.9\linewidth} \\
  \epsfig{file=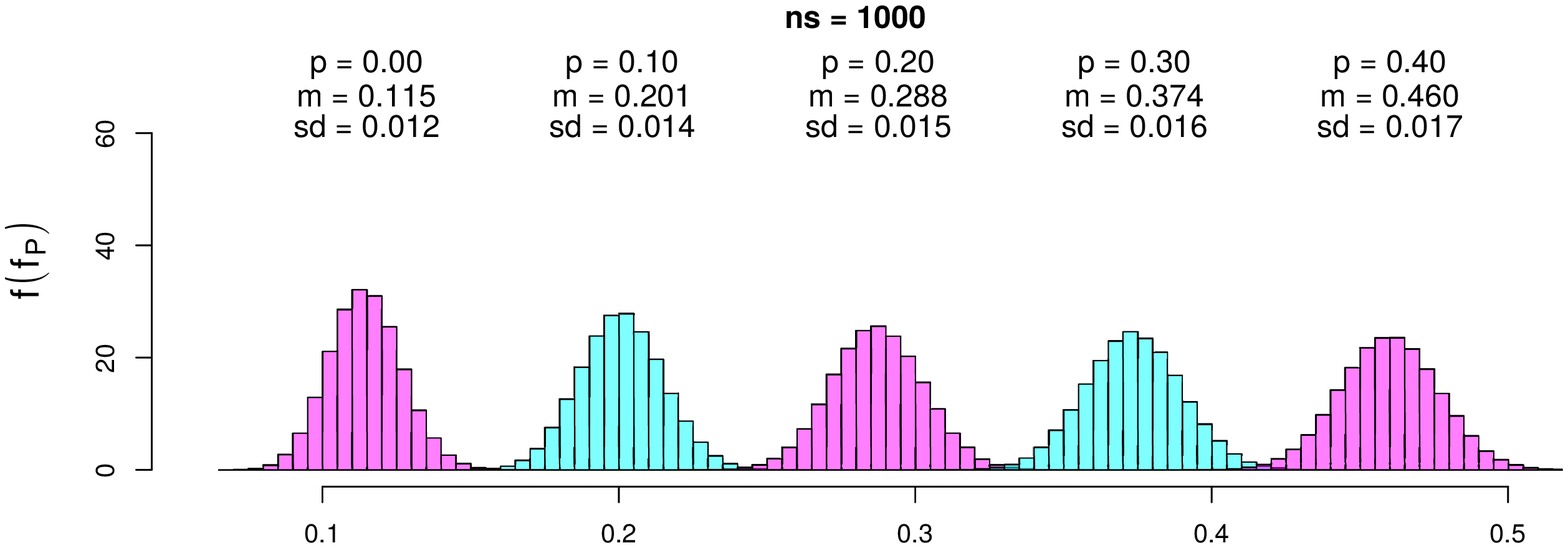,clip=,width=0.9\linewidth} \\
  \epsfig{file=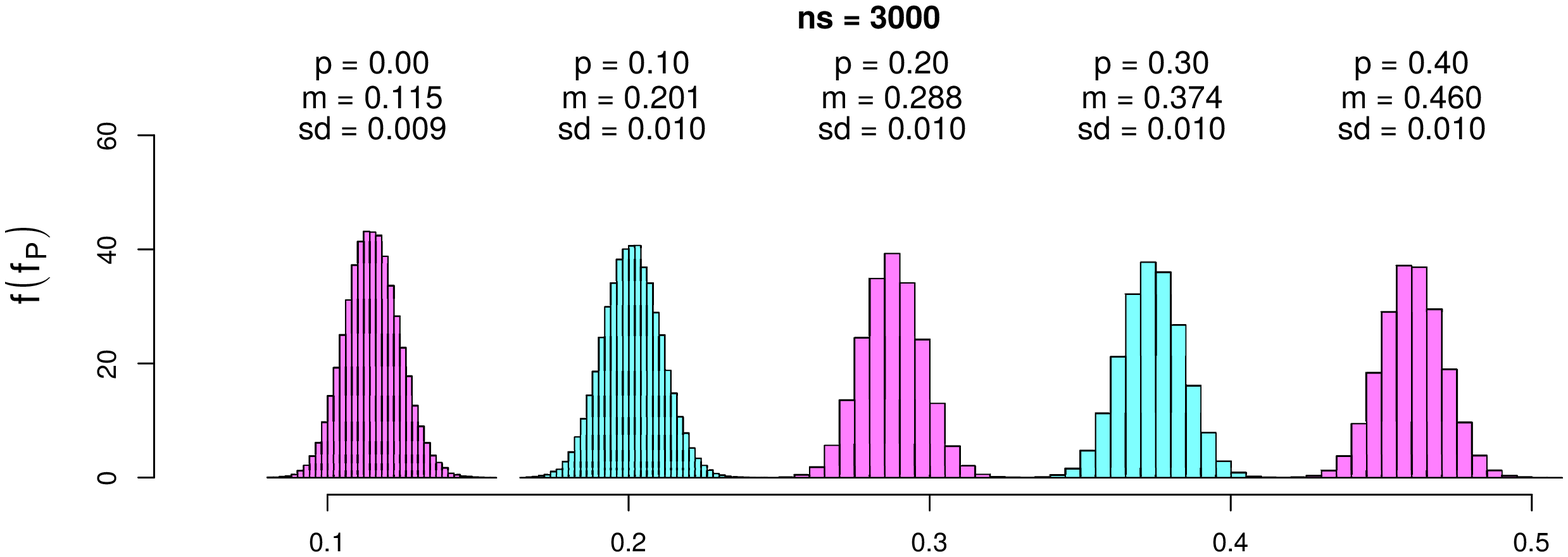,clip=,width=0.9\linewidth} \\
  \epsfig{file=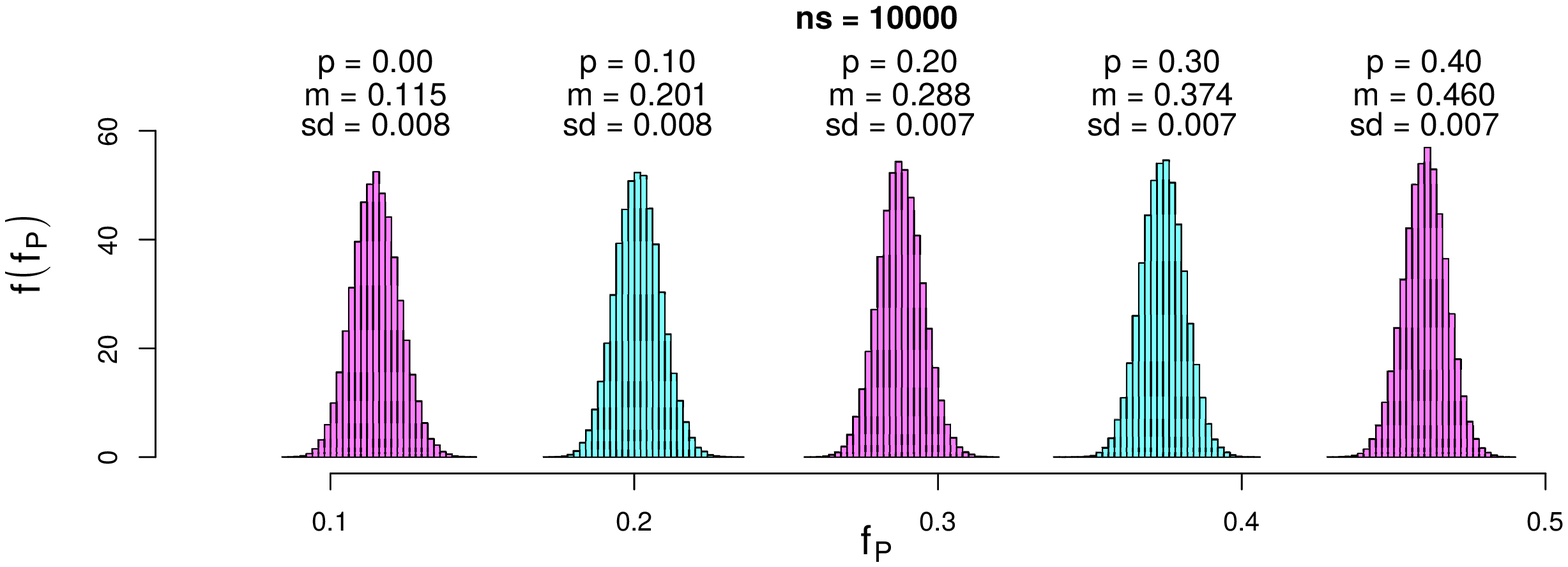,clip=,width=0.9\linewidth}
   \\  \mbox{} \vspace{-1.0cm} \mbox{}
  \end{center}  
  \caption{\small \sf Same as Fig.~\ref{fig:PredictionPositive_by01}, but for
    an improved knowledge of $\pi_2$, summarized as $\pi_2=0.115\pm 0.007$. 
  }
    \label{fig:PredictionPositive_by01_equal-sigmas} 
\end{figure}
how the result changes if we reduce $\sigma(\pi_2)$ to the level
of $\sigma(\pi_1)$.\footnote{This can be done evaluating
  $r_2$ and $s_2$
  from
  Eqs.~(\ref{eq:r_from_E-sigma}) and (\ref{eq:s_from_E-sigma})
with $\mu=0.978$ and $\sigma=0.007$.}
As we can see (a result largely expected), there is quite
a sizable improvement
in {\em separability} of values of $p$ for large values of $n_s$.
Again {\em qualitatively}, we can see that values of $p$
which differ by $\approx {\cal O}(0.01)$ can be resolved.

Finally we show in Fig.~\ref{fig:PredictionPositive_by01_sym}
\begin{figure}
  \begin{center}
  \epsfig{file=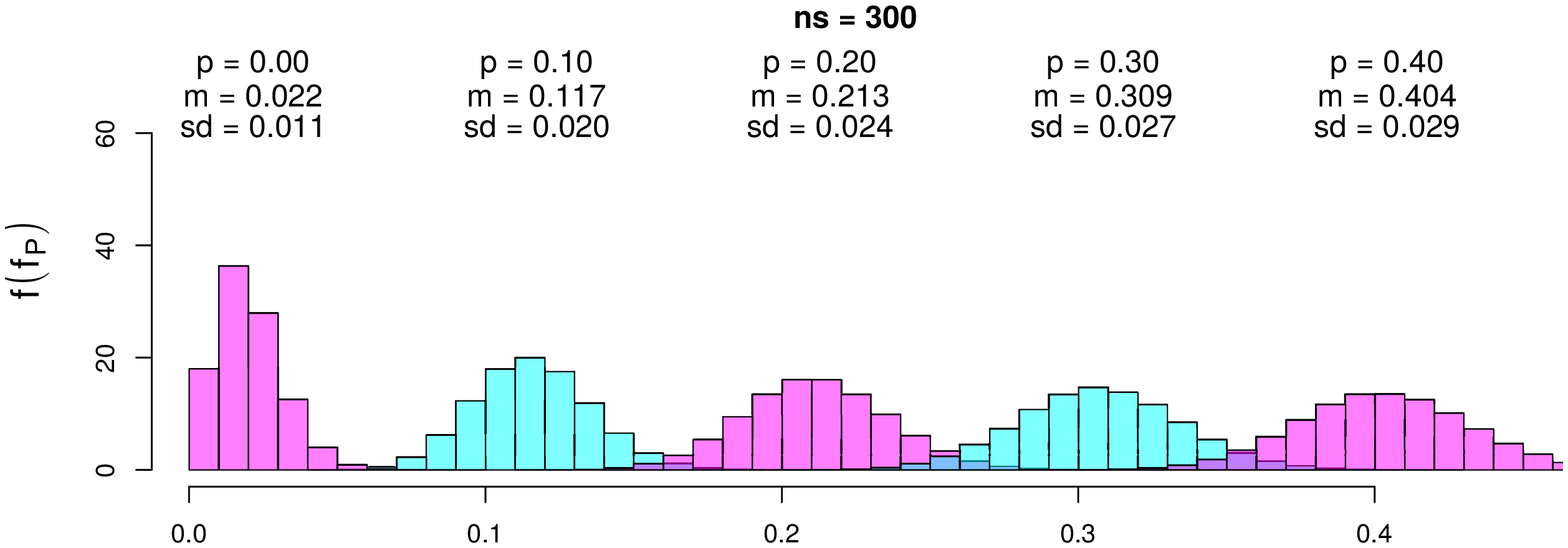,clip=,width=0.9\linewidth} \\
  \epsfig{file=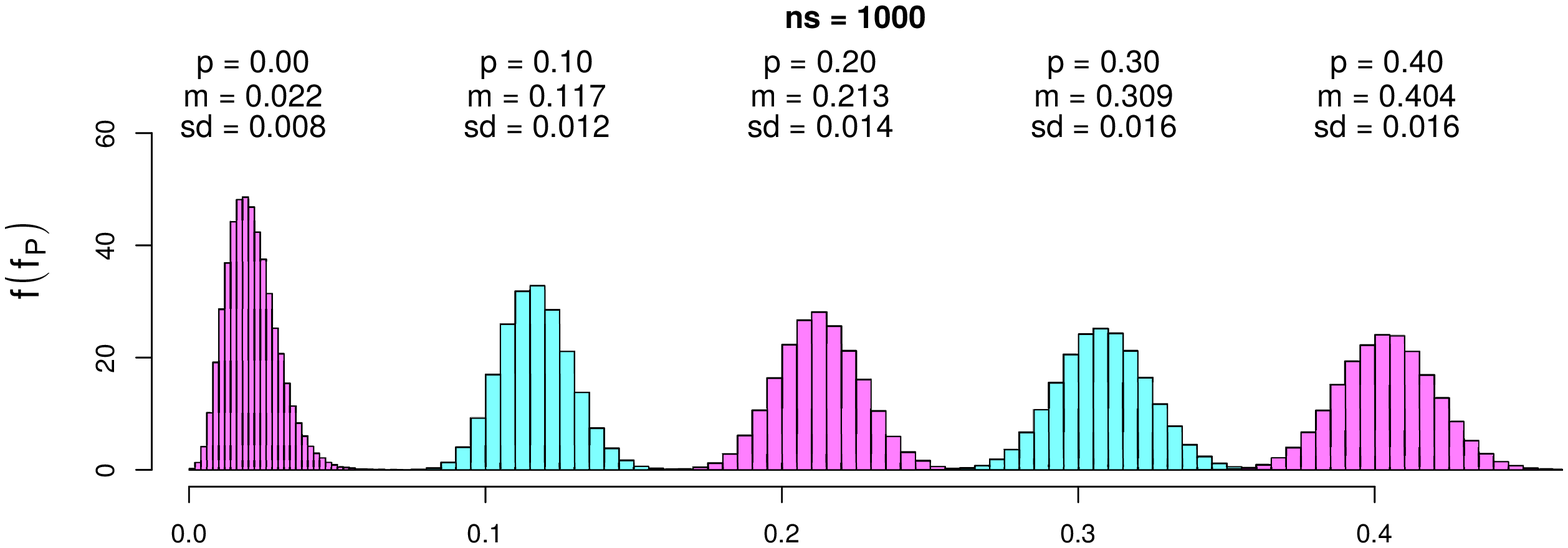,clip=,width=0.9\linewidth} \\
  \epsfig{file=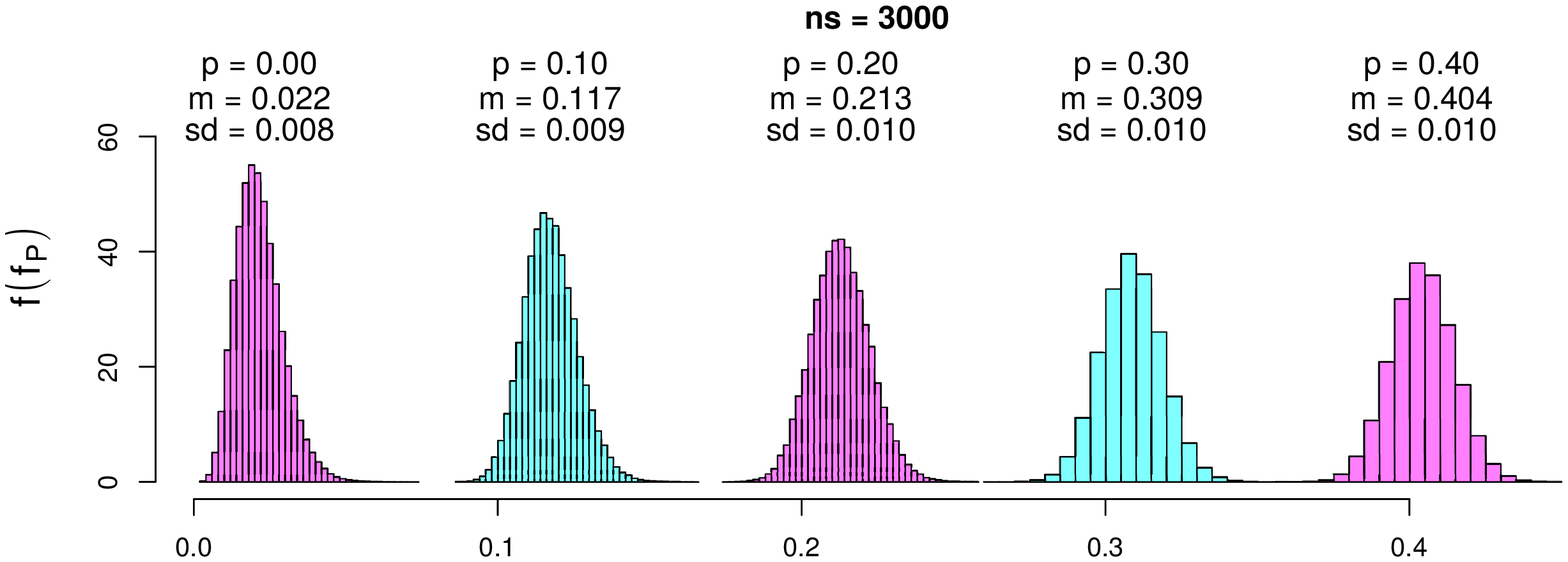,clip=,width=0.9\linewidth} \\
  \epsfig{file=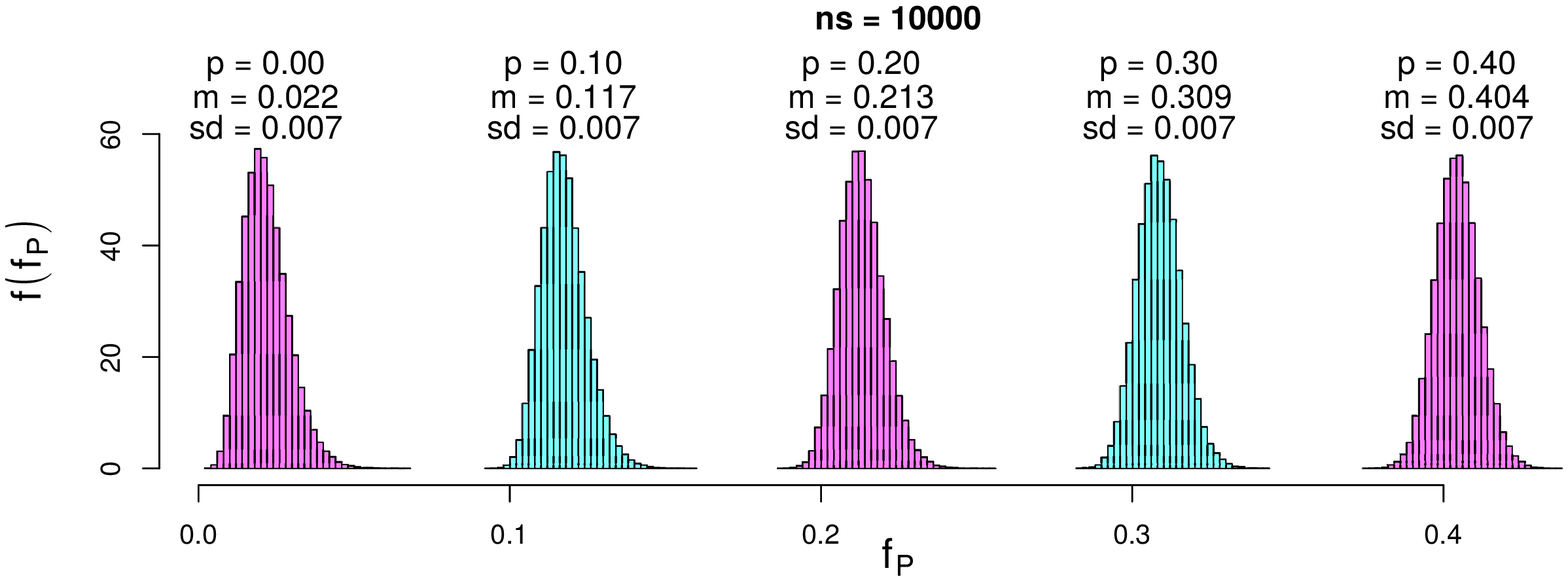,clip=,width=0.9\linewidth}
   \\  \mbox{} \vspace{-1.0cm} \mbox{}
  \end{center}
  \caption{\small \sf Same as
    Fig.~\ref{fig:PredictionPositive_by01_equal-sigmas}, but for
    an improved specificity, summarized as $\pi_2=0.022\pm 0.007$. 
  }
    \label{fig:PredictionPositive_by01_sym} 
\end{figure}
the case in which sensitivity and specificity are equal, both
as expected value and standard uncertainty. The first thing we note in these new
histograms is that for $p=0$ they are no longer symmetric and
Gaussian-like. This is due to the fact that no negative values of $n_P$
are possible, and then there is a kind of `accumulation'
for low values of $n_P$, and therefore of $f_P$
(this kind of {\em skewness} is typical of all probability distributions
of variables defined to be positive and whose standard deviation is not
much smaller than the expected value -- think e.g.
at a Poissonian with $\lambda=2$).

\subsubsection{Using JAGS}\label{sss:usingJAGS}
Let us repeat the Monte Carlo simulation
{\em improperly} using the program JAGS~\cite{JAGS},  
interfaced to R via the package {\tt rjags}~\cite{rjags}.
JAGS is a powerful tool developed,
as open source, multi-platform clone of {\tt BUGS},\footnote{
  Introducing  MCMC and related algorithms
  goes well beyond the purpose of this paper
  and we recommend Ref.~\cite{Andrieu}.  
  Moreover, mentioning  the Gibbs Sampler algorithm applied to
  probabilistic inference (and forecasting) it is impossible 
  not to refer to the {\em BUGS project} ~\cite{BUGSpaper},
  whose acronym stands
for \underline{B}ayesian inference \underline{u}sing
\underline{G}ibbs \underline{S}ampler, that has
been a kind of revolution in Bayesian analysis,
decades ago limited to simple cases because of computational problems
(see also Sec.~1 of Ref.\cite{JAGS}).
In the project web site~\cite{BUGSsite}
  it is possible to find packages with excellent Graphical User Interface,
  tutorials and many examples \cite{BUGSexamples}.
  \label{fn:MCMC_BUGS}
} 
to perform Bayesian inference
by {\em Markov Chain Monte Carlo} (MCMC)
using the {\em Gibbs Sampler} algorithm, as its name
reminds, acronym of {\em Just Another Gibbs Sampler}. 
For the moment we just get familiar with JAGS using it as
kind of `curious' random generator.

The first thing to do is to write down
the probabilistic model that relates the different variables
that enter the game. For example, the
left hand graphical model of Fig.~\ref{fig:sampling_binom_pred}
is implemented in JAGS by the
following self-explaining piece of code
\begin{verbatim}
model {
  n.I ~ dbin(p, ns)
  n.NI <- ns - n.I 
  nP.I ~ dbin(pi1, n.I)
  nP.NI ~ dbin(pi2, n.NI)
  nP ~ sum(nP.I, nP.NI)
  fP <- nP / ns 
}
\end{verbatim}
in which we have added the last instruction
to model the trivial node (not shown in Fig.~\ref{fig:sampling_binom_pred})
relating in a deterministic way {\tt fP} to {\tt nP} and {\tt ns}.
In the model the symbol `$\sim$' indicates that, e.g. {\tt n.I} is described
by a binomial distribution defined by {\tt p} and {\tt ns}
(be aware of the different order of the parameters with respect to the
R function!), while   `{\small $<\!\!\!-$}' 
stands for a deterministic relation
(indeed the symbols of assignment in R).
A nice thing of such a model is that the order of the instructions
is not relevant.
In fact it is only needed -- let us put it so --
to describe the related graphical model. All the rest will
be done internally by JAGS at the compilation step.

Then, obviously, we have to
\begin{itemize}
\item pass to the program
the model parameters ({\em observed nodes}), which are
{\tt p}, {\tt ns}, {\tt pi1} and {\tt pi2};
\item instruct it on how many `iterations' to do;
\item analyze,  among all {\em unobserved nodes}
({\tt n.I}, {\tt n.NI}, {\tt nP.I}, {\tt nP.NI}, {\tt nP}
and {\tt fP}), the ones of interest, the most important one
being, for us, {\tt fP}.
\end{itemize}
Moving to the second model of Fig.~\ref{fig:sampling_binom_pred},
in which we also take into account the uncertainty about
$\pi_1$ and $\pi_2$, is straightforward: we just need to add
two instructions to tell JAGS that
{\tt pi1} and {\tt pi2} are indeed {\em unobserved}
and that they depend on {\tt r1}, {\tt s1}, {\tt r2} and {\tt s2}:
\begin{verbatim}
model {
  n.I ~ dbin(p, ns)
  n.NI <- ns - n.I 
  nP.I ~ dbin(pi1, n.I)
  nP.NI ~ dbin(pi2, n.NI)
  pi1 ~ dbeta(r1, s1)
  pi2 ~ dbeta(r2, s2)
  nP ~ sum(nP.I, nP.NI)
  fP <- nP / ns 
}
\end{verbatim}
Once the model is defined, it has to be saved into a file,
whose location is then 
passed to JAGS
(we shall regularly use the temporary file {\tt tmp\_model.bug},
whose extension `.bug' is the BUGS/JAGS default). 

At this point, moving to the R code to interact
with JAGS, we need to
\begin{itemize}
\item load the interfacing package {\tt rjags};
\item  prepare a R `list' containing the data (in particular, the values
  of the `observed' nodes);
\item call {\tt jags.model()} to `setup' the model;
\item call {\tt coda.samples()} to ask JAGS to perform the sampling,
  also specifying  the variables we want to monitor,
  whose `histories' will be returned in a single
  `object'.\footnote{To the returned object is assigned the name
    `{\tt chain}' in the script of Appendix B.6.
    In order to get information about the kind of object,
    just issue the command `{\tt str(chain)}'.
  }
\end{itemize}
Finally we have to show the result. All this is done, for example, 
in the R script provided
in Appendix B.6 (note that the temporary model file is written
directly from R, a convenient solution for small models).
Needless to say, we get, apart from statistical fluctuations
inherent to
Monte Carlo methods, `exactly' the same results obtained with the
script of Appendix B.5, which only uses R statistical functions.

\subsubsection{Further check of the approximated formulae}
\label{sss:check_approximations}
Finally, we have checked the validity of the approximated formulae
(\ref{eq:approx_E.fP_s})-(\ref{eq:sigma_fP_sigma_ps_bis})
to evaluate the expected value and the standard uncertainty of
$f_P$ in all cases considered in
Figs.~ \ref{fig:PredictionPositive_by01}-\ref{fig:PredictionPositive_by01_sym}.
The agreement is indeed excellent, even in the cases
of $p=0$ of Fig.~\ref{fig:PredictionPositive_by01_sym},
characterized by skewed distributions.
The R script to reproduce all numbers of all three figures
is provided in Appendix B.7.

\subsection{Resolution power}
Having to turn the qualitative judgment regarding the `separation'
of the distributions of $f_P$ for different $p$, as it results
from 
Figs.~\ref{fig:PredictionPositive_by01}-\ref{fig:PredictionPositive_by01_sym},
into a {\em resolution power}, one needs some convention.
First, we remind that,  unless $p$ is very small, 
we have good theoretical reasons, confirmed by Monte Carlo
simulations, that $f_p$ is about Gaussian, at least in the
range of a few standard deviations around its mean value. 
But Gaussian curves are, strictly speaking, never separated from
each other, because they have as domain the entire
real axis for all  $\mu$'s and $\sigma$'s.
In fact this is a {\sl ``defect''} of such
distribution,
as Gauss himself called it~\cite{Gauss} and some {\em grain of salt}
is required using it. 
In order to form an idea of how one could define conventionally
`resolution', the upper plot of
Fig.~\ref{fig:resolution_power} shows
\begin{figure}
\begin{center}
  \epsfig{file=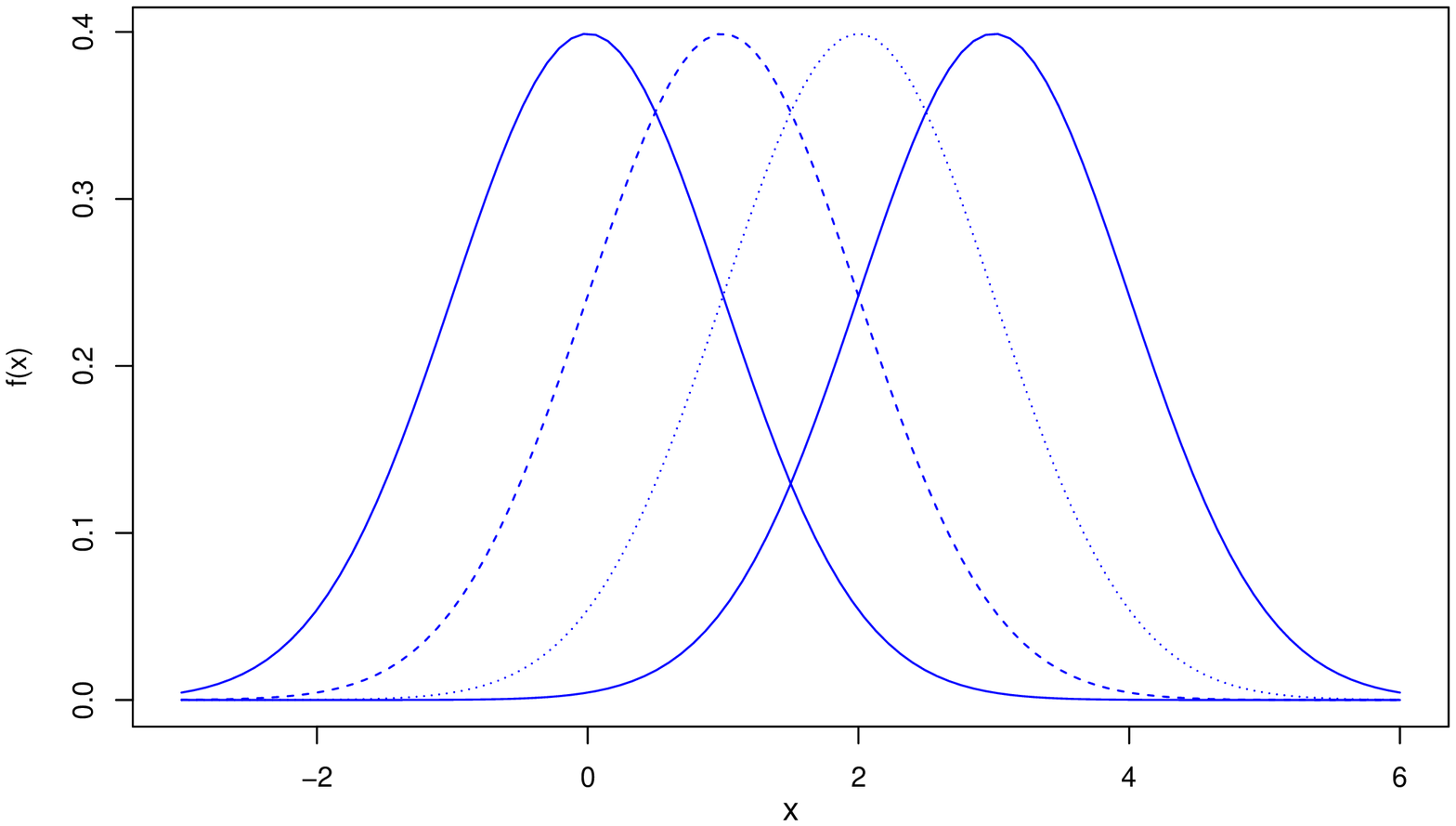,clip=,width=\linewidth}\\
  \mbox{} \\
  \epsfig{file=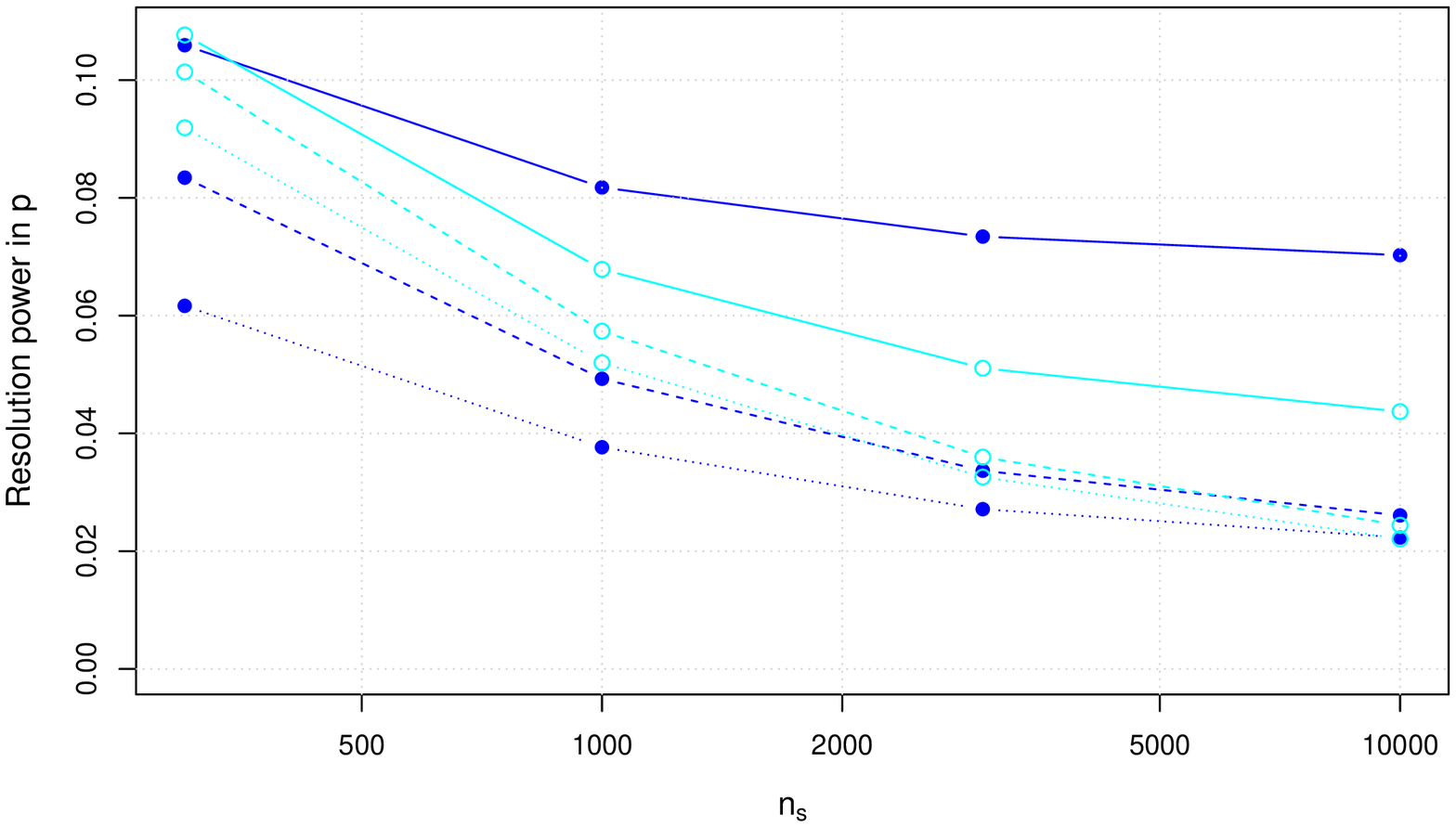,clip=,width=\linewidth}
    \\  \mbox{} \vspace{-1.0cm} \mbox{}
\end{center} 
\caption{\small \sf Upper plot: Examples of Gaussians whose $\mu$ parameters
  are separated by 1 $\sigma$. Bottom plot:
  resolution power in $p$, defined by Eq.~(\ref{eq:resolution_power}),
  for $\kappa=3$
  (lines between points just to guide the eye). 
  Filled (blue) circles for $p=0.1$ and open (cyan) circles for $p=0.5$.
  Solid lines for
  $\pi_2=0.115\pm 0.022$, dashed lines for $\pi_2=0.115\pm 0.007$
  and dotted lines for  $\pi_2=0.022\pm 0.007$
  ($\pi_1 = 0.978\pm 0.007$ in all cases).
}
\label{fig:resolution_power} 
\end{figure}
some Gaussians having unitary $\sigma$, with $\mu$'s differing
by  one $1\sigma$. 

We see that a `reasonable separation' is achieved
when they differ by a few $\sigma$'s -- let us say, generally
speaking, $\kappa\, \sigma$, although absolute separation
can never occur, for the already quoted intrinsic {\em ``defect''} of the distribution.
Having to choose a value, we just
opt {\em arbitrarily} for $\kappa=3$, corresponding to the
two solid lines of the figure, although the
conclusions that follow from this choice can be easily
rescaled at wish.
Moreover, as we can see from
Figs.~\ref{fig:PredictionPositive_by01}-\ref{fig:PredictionPositive_by01_sym}
(and as it results from the approximated formulae)
\begin{itemize}
\item the standard deviation of the distributions varies
  smoothly with $p$;
\item the mean value depends linearly on $p$
for obvious reasons.\footnote{Deviations from linearity
  are expected for $p\approx 0$
  and rather small $n_s$, but, as we have checked with approximated formulae,
  the effect is negligible for the values of interest.}
\end{itemize}
Therefore the resolution power in the interval
$[p,p+\Delta p]$ can be
evaluated by a simple proportion
\begin{eqnarray}
  {\cal R}(p,p+\Delta p) &\approx&  \frac{\Delta p}
  {\left.\mbox{E}(f_P)\right|_{p+\Delta p}-\left.\mbox{E}(f_P)\right|_p}
  \cdot \kappa \cdot \left.\sigma(f_P)\right|_{p+\Delta p/2}\,. 
\end{eqnarray}
For example, using the numbers of the Monte Carlo evaluations
shown in Fig.~\ref{fig:PredictionPositive_by01}, 
for $p=0.1$ and $n_s=300$ we get
$ 0.1 / (0.288-0.201) \times 3 \times 0.0305 = 0.105$,
reaching at best $\approx 0.022$ in the case of $n_s=10000$ shown in
Fig.~\ref{fig:PredictionPositive_by01_sym}.
The resolution power at a given value of $p$, is obtained
in the limit `$\Delta p\rightarrow 0$':
\begin{eqnarray}
  {\cal R}(p) &\approx&  \frac{\Delta p}
  {\left.\mbox{E}(f_P)\right|_{p+\Delta p}-\left.\mbox{E}(f_P)\right|_p}
  \cdot \kappa \cdot  \left.\sigma(f_P)\right|_p\,
\hspace{0.8cm}(\Delta p \rightarrow 0).
  \label{eq:resolution_power}
\end{eqnarray}
The bottom plot of Fig.~\ref{fig:resolution_power}
shows the variation of the resolution power in $p$
for the same
values of $n_s$ of
Figs.~\ref{fig:PredictionPositive_by01}-\ref{fig:PredictionPositive_by01_sym}
and for the usual cases of $\pi_{1}$ and $\pi_{2}$ of those
figures (in the order: solid, dashed and dotted line --
the lines are drawn just
to guide the eye and to easily identify the conditions).
The resolution power has been evaluated using the approximated
formulae, for $\kappa=3$, around
$p=0.1$ (blue filled circles) and around $p=0.5$
(cyan open circles),
using for the gradient $\Delta p=0.01$ (the exact value
is irrelevant for the numerical evaluation, provided
it is {\em small enough}).
$[$Obviously, if one prefers a different value
of $\kappa$ (in particular one might like $\kappa=1$), then
one just needs to rescale the results.$]$

\subsection[Predicting fractions of positives
  sampling two populations]{Predicting the fractions
  of positives obtained
  sampling two different
  populations}\label{ss:predict_Delta_fP}
An interesting question then arises: what happens if we measure,
using tests having the same uncertainties on sensitivity and
specificity, two different populations, having proportions of infectees
$p^{(1)}$ and $p^{(2)}$, respectively? For example, in order to make use
of results we have got above, let us take the results shown in
Fig.~\ref{fig:PredictionPositive_by01}
for $n_s=10000$, $p^{(1)}=0.1$ and $p^{(2)}=0.2$.
For this value of the sample size and for our standard hypotheses 
for sensitivity and
specificity, summarized as $\pi_1=0.978\pm0.007$ and $\pi_2=0.115\pm 0.022$,
the uncertainties are dominated by the systematic contributions. 
Our expectations are then $f_P^{(1)} = 0.201\pm 0.020$
and $f_P^{(2)} = 0.288\pm 0.018$. The difference of expectations
is therefore $\Delta f_P = f_P^{(2)} - f_P^{(1)} = 0.087$.

Now it is interesting to know how much uncertain this number is. 
One could {\em improperly} use a quadratic
combination of the two standard uncertainties, thus getting
$\Delta f_P = 0.087 \pm 0.027$. But this evaluation of the uncertainty
on the difference 
is incorrect because  $f_P^{(1)}$ and  $f_P^{(2)}$
are obtained from the same knowledge of $\pi_1$ and $\pi_2$, and 
are therefore {\em correlated}. Indeed, in the limit of
negligible uncertainties on these two parameters,
the expectations would be much more precise, as we can see from the
upper plot of Fig.~\ref{fig:Contributions_uncertanties_p0.1},
with a consequent reduction of $\sigma(\Delta f_P)$. 
These are the results, obtained by Monte Carlo evaluation using only 
R commands
(see script in Appendix B.8),\footnote{As alternative, one
  could use JAGS, of which we provide the model
  in Appendix B.9, leaving the R steering commands as exercise.
  JAGS will be instead used in Sec.~\ref{ss:inferring_Delta_p}
  to infer $p^{(1)}$,
  $p^{(2)}$ and $\Delta p = p^{(2)}-p^{(1)}$.
}
with one extra digit
with respect to Fig.~\ref{fig:PredictionPositive_by01}
and adding
also the correlation coefficient:
\begin{eqnarray*}
  f_P^{(1)} & = & 0.2013  \pm 0.0199  \\
  f_P^{(2)} & = & 0.2876  \pm 0.0179  \\
  \Delta f_P &= & 0.0863 \pm 0.0064 \\
   \rho\left(f_P^{(1)},f_P^{(2)}\right) &=& 0.9470
\end{eqnarray*}
The uncertainty on $\Delta f_P$ is about {\em one fourth of
what naively evaluated} above and about {\em one third
of the individual predictions}, due to the well known effect of
(at least partial) {\em cancellations of uncertainties in differences,
  due to common systematic
  contributions}.
In this case, in fact, the standard deviation of $\Delta f_P$,
calculated from standard deviations and correlation coefficient,
is given by\footnote{It might be useful to remind
  that, given a linear combination $Y=c_1\cdot X_1+c_2\cdot X_2$,
  the variance of $Y$ is given by
\begin{eqnarray*}
  \sigma^2(Y) &=& c_1^2 \cdot \sigma^2(X_1) +
  c_2^2\cdot  \sigma^2(X_2) + 2\,c_1\cdot c_2\cdot \mbox{Cov}(X_1,X_2)\\
  &=&
  c_1^2 \cdot \sigma^2(X_1) +
  c_2^2 \cdot \sigma^2(X_2) + 2\,c_1\cdot c_2\cdot \rho(X_1,X_2)
  \cdot \sigma(X_1)\cdot  \sigma(X_2)\,.
\end{eqnarray*}
\mbox{}\vspace{-0.5cm}\mbox{}
}   
\begin{eqnarray*}
  \sigma(\Delta f_P)  & = &
  \sqrt{\sigma^2(f_P^{(1)}) +  \sigma^2(f_P^{(2)})
    - 2\,\rho\left(f_P^{(1)},f_P^{(2)}\right)\cdot
  \sigma(f_P^{(1)})\cdot \sigma(f_P^{(2)}) }\,\, = \,\,0.0064\,,
\end{eqnarray*}  
in perfect agreement with what we get from
Monte Carlo sampling.

An important consequence of the correlation among the predictions of
the numbers of positives in different populations
is that we have to expect a similar {\em correlation in the
  inference of  the proportion of infectees in different populations}.
This implies that we can measure their difference
much better than how we can measure a single proportion.
And, if one of the two proportions is precisely known
using a different kind of test, we can take its value
as kind of {\em calibration point}, which will allow
a better determination also of the other proportion.
We shall return to this interesting point in Sec.~\ref{ss:inferring_Delta_p}.

\section{Inferring $p$ from the observed number of positives in the sample}
\label{sec:interring_p}
Let us finally move to the probabilistic inference of the proportion
of infected individuals, $p$, based on the number of positives $n_P$ 
in a sample of size $n_s$ and given our best knowledge of the
performance of the test, all summarized in the graphical
model of Fig.~\ref{fig:sampling_binom_inf},
\begin{figure}[t]
\begin{center}
  \epsfig{file=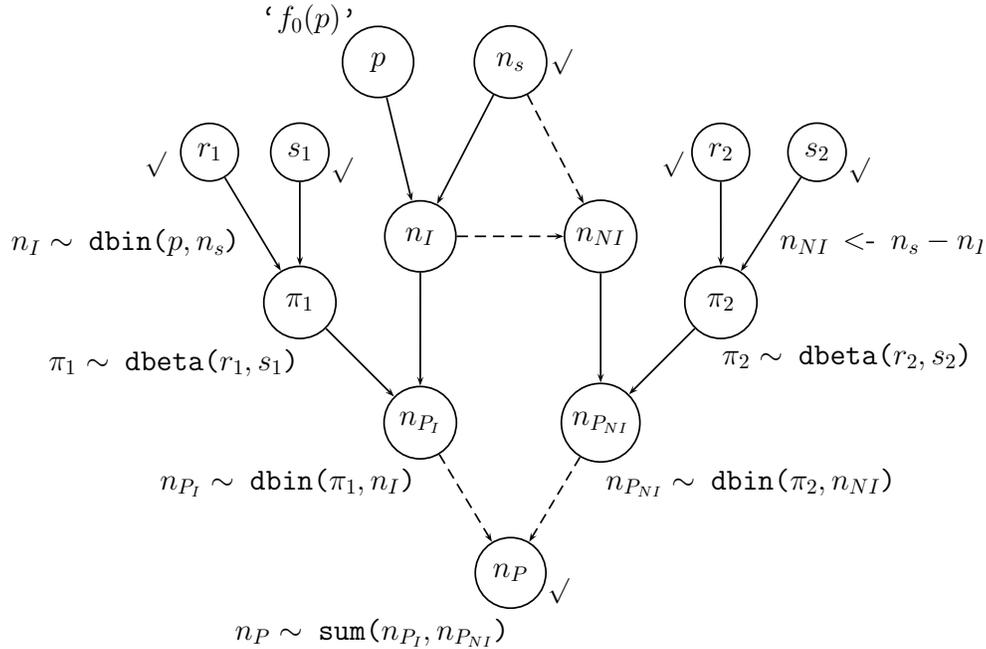,clip=,width=0.85\linewidth}
  \\  \mbox{} \vspace{-0.9cm} \mbox{}
\end{center}
\caption{\small \sf Graphical model of Fig.~\ref{fig:sampling_binom_pred},
  re-drawn in order to emphasize its inferential use and
  including the commands to build up the JAGS model.  `$f_0(p)$',
  left open in this diagram, stands for the prior distribution of $p$.
}
\label{fig:sampling_binom_inf} 
\end{figure}
which differs from that of  Fig.~\ref{fig:sampling_binom_pred}
only for the symbol `$\surd$\,' moved
from node $p$ (now `unobserved') to node $n_P$ (now `observed').
The diagram contains also the probabilistic and deterministic
relations among the nodes, written directly using the
JAGS language.\footnote{The relation,
  `$n_P \sim \mbox{\tt sum}(n_{P_I},n_{P_{NI}})$' is
  logically equivalent to
  `$n_P <\mbox{- } n_{P_I}+n_{P_{NI}}$', but
  the latter instruction would not work
  because JAGS prohibits `observed nodes'
  to be defined by a deterministic assignment,
  as, instead, it has been done in the case of $n_{NI}$, defined
  as  `$n_{NI} <\!\!\mbox{-- } n_s - n_I$'.
}
\subsection{From the general problem to its implementation in JAGS}
The most general problem would
be to evaluate the joint conditional probability
of the uncertain (`unobserved') quantities,
conditioned by the `observed' (`known'/ `assumed'/`postulated') ones,
that is, in this case\footnote{Note that we can in principle
  learn something
  also about $\pi_1$ and $\pi_2$, because we can properly marginalize
  Eq.~(\ref{eq:inferenza_tutte}) in order to get
  $f(\pi_1,\pi_2\,|\,n_P,n_s,r_1,s_1,r_2,s_2)$.
  In the limit that they are very well known (condition
  reflected
  into  very large $r_i$ and $s_i$) we expect that their joint probability
  distribution is not updated much by the new pieces of information.
  But, if instead they are poorly known, we get some information
  on them, at the expense of the quality of information
  we can get on the main quantity of interest, that is $p$
  (although we are not going into the
  details, see Sec.~\ref{ss:updated_pi2} for a case in which
  $\pi_2$ is updated by the data).
} (see Appendix A),
\begin{eqnarray}
  f(p,n_I,n_{NI},n_{P_I},n_{P_{NI}},\pi_1,\pi_2\,|\,n_P,n_s,r_1,s_1,r_2,s_2)\,,
  \label{eq:inferenza_tutte}
\end{eqnarray}
although in practice we are indeed interested in
$f(p\,|\,n_P,n_s,r_1,s_1,r_2,s_2)$, and perhaps
in $f(n_I\,|\,n_P,n_s,r_1,s_1,r_2,s_2)$ and 
$f(n_{P_I}\,|\,n_P,n_s,r_1,s_1,r_2,s_2)$.
This is done marginalizing Eq.~(\ref{eq:inferenza_tutte}) 
i.e. summing (or integrating, depending on their nature)
over the variables on which we are not interested
(see Appendix A).
As commented in the same appendix, Eq.~(\ref{eq:inferenza_tutte}) 
is obtained, apart from a normalization factor,
from
\begin{eqnarray}
  f(p,n_I,n_{NI},n_{P_I},n_{P_{NI}},\pi_1,\pi_2,n_P,n_s,r_1,s_1,r_2,s_2),
  \label{eq:joint_tutte}
\end{eqnarray}
and the latter from a {\em properly chosen chain rule}.
The steps used to build up Eq.~(\ref{eq:joint_tutte}) by 
 the proper chain rule
are exactly the instructions given to JAGS to set up
the model, if we start from the bottom of
the diagram of Fig.~\ref{fig:sampling_binom_inf}
and ascend through the {\em parents} (see Sec.~\ref{sss:usingJAGS}):
\begin{verbatim}
model {
  nP ~ sum(nP.I, nP.NI)
  nP.I ~ dbin(pi1, n.I)
  nP.NI ~ dbin(pi2, n.NI)
  pi1 ~ dbeta(r1, s1)
  pi2 ~ dbeta(r2, s2)
  n.I ~ dbin(p, ns)
  n.NI <- ns - n.I
  p ~ dbeta(r0,s0)       
}
\end{verbatim}
The differences with respect to the JAGS model of
Sec.~\ref{sss:usingJAGS} are
\begin{itemize}
\item the last instruction there, `{\tt fP $<$- nP/ns}', is here
  irrelevant;
\item we have to add a prior to {\tt p}, because
  {\em all unobserved nodes having no parents need a prior}
  (for practical convenience,
   as we have seen in Sec.~\ref{ss:conjugate_priors},
  we shall use a Beta distribution, as indicated in the code);
\item the sequence of the statements has been changed, but this has
  been done only in order to stress the analogy with
  the chain rule constructed ascending the graphical model
  of Fig.~\ref{fig:sampling_binom_inf} (let us remind that the order is irrelevant
  for JAGS, which organizes all statements at the stage
  of compilation).
\end{itemize}
Hereafter we proceed using, very conveniently,
JAGS,
showing in Sec.~\ref{sec:direct} the steps needed from
writing down the chain rule till the exact evaluation of $f(p)$
after marginalization.

\subsection{Inferring $p$ and $n_I$ with our `standard parameters'}
\label{ss:infer_p_nI} 
Let us start using as $n_P$ the expected value of positives of
$\approx 2010$,
obtained from what has been our starting set of parameters through the paper,
that is $p=0.1$ with $n_s=10000$, with the uncertain parameters $\pi_1$
and $\pi_2$ modeled by Beta distributions with $(r_1=409.1,\,s_1=9.1)$
and $(r_2=25.2,\,s_2=193.1)$, respectively. Also for the prior of
$p$ we use a Beta, starting with $r_0=s_0=1$, that models a flat
prior, although we obviously do not believe that $p=0$ or $p=1$ are
possible. We shall discuss in Sec.~\ref{ss:priors_priors_priors}
the role of such
at a first glance an {\em insane prior} (see also Sec.~\ref{sec:direct}).

These are the R command to set the parameters of the game,
call JAGS and show some results (for the complete script
see Appendix B.10). 
\begin{verbatim}
#---- data and parameters
nr = 1000000
ns = 10000
nP = 2010
r0 = s0 = 1
r1 = 409.1; s1 = 9.1
r2 = 25.2;  s2 = 193.1 

# define the model and load rjags (omitted)
# ......................................... 

#---- call JAGS ---------
data <- list(ns=ns, nP=nP, r0=s0, s0=s0, r1=r1, s1=s1, r2=r2, s2=s2)  
jm <- jags.model(model, data)
update(jm, 10000)
to.monitor <-  c('p', 'n.I')
chain <- coda.samples(jm, to.monitor, n.iter=nr)

#---- show results
print(summary(chain))
plot(chain, col='blue')
\end{verbatim}
\mbox{}\\
Here are the results shown by `{\tt summary(chain)}'
\begin{verbatim}
1. Empirical mean and standard deviation for each variable,
   plus standard error of the mean:

         Mean        SD  Naive SE Time-series SE
n.I 991.12477 225.85901 2.259e-01      16.079460
p     0.09919   0.02278 2.278e-05       0.001601

2. Quantiles for each variable:

         2.5%       25%      50%       75%     97.5%
n.I 506.00000 838.00000 1012.000 1153.0000 1389.0000
p     0.05046   0.08372    0.101    0.1155    0.1396
\end{verbatim}
\mbox{} \\
So, for {\em this run} we get $p=0.0992 \pm 0.023$
and a number of infectees in the sample equal
to $991\pm 226$, in agreement with our expectations.
The results of the Monte Carlo sampling are
shown in the `densities' 
of Fig.~\ref{fig:JAGS_plots_standard_conf}, 
\begin{figure}[t]
\begin{center}
  \epsfig{file=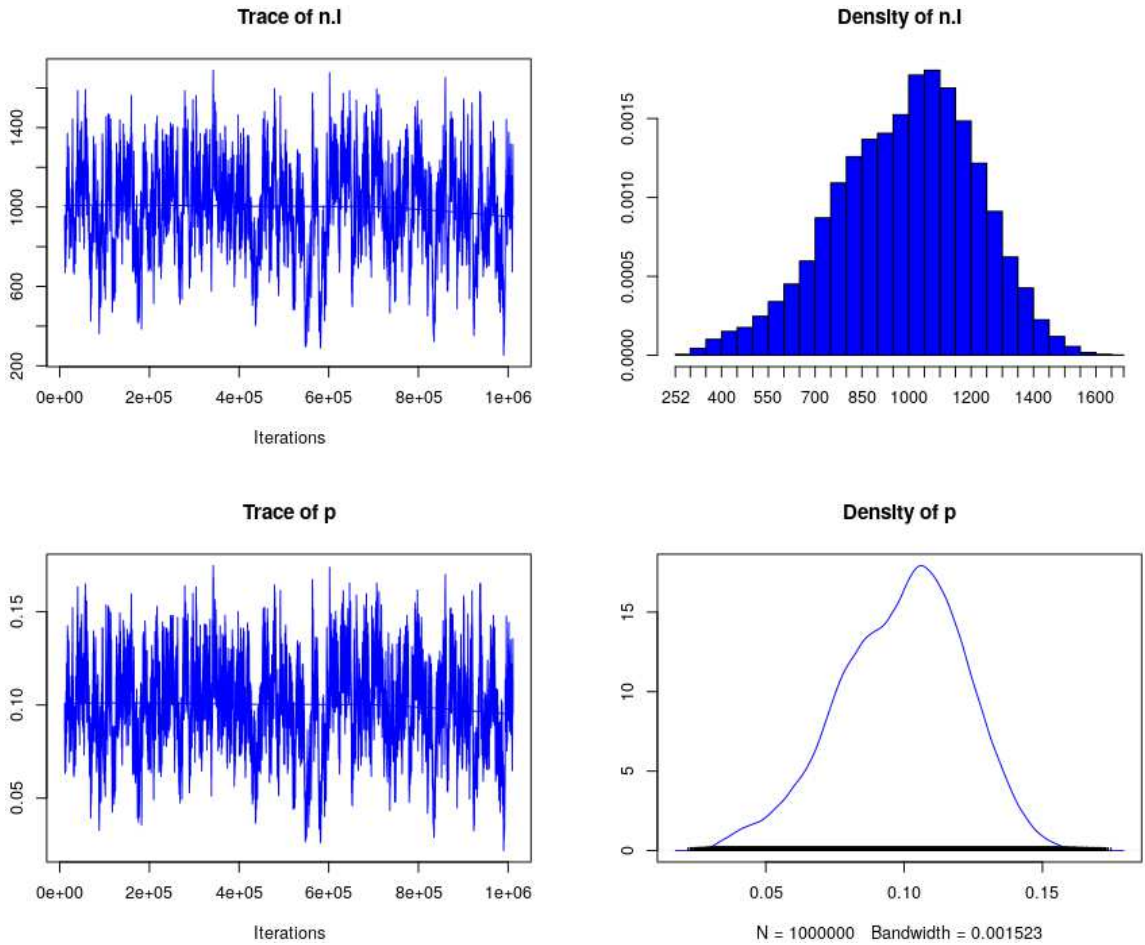,clip=,width=\linewidth}
   \\  \mbox{} \vspace{-1.0cm} \mbox{}
\end{center}
\caption{\small \sf Plots showing some JAGS results (see text).
}
\label{fig:JAGS_plots_standard_conf} 
\end{figure}
together with the `traces', i.e.
the values of the sampled variables during the
$10^6$ iterations.\footnote{Indeed
  the traces show that the sampling is, so to say,
  not optimal, and more iterations would be needed.
  But for our needs here and for reminding the care
  needed in applying this powerful tool,
  we prefer to show this not ideal case of
  sampling  with a quite larger but not large enough
  number of iterations. (Later on, when critical, we shall increase
  {\tt nr} up to $10^7$.)
}
As it is easy to guess and as it appears from the two traces
of the figure, there is some degree of correlation
between the two variables, because
they are obtained in a joint inference.
The correlation is made evident in the scatter plot
of Fig.~\ref{fig:JAGS_plots_standard_conf_correlation}
\begin{figure}[t]
\begin{center}
  \epsfig{file=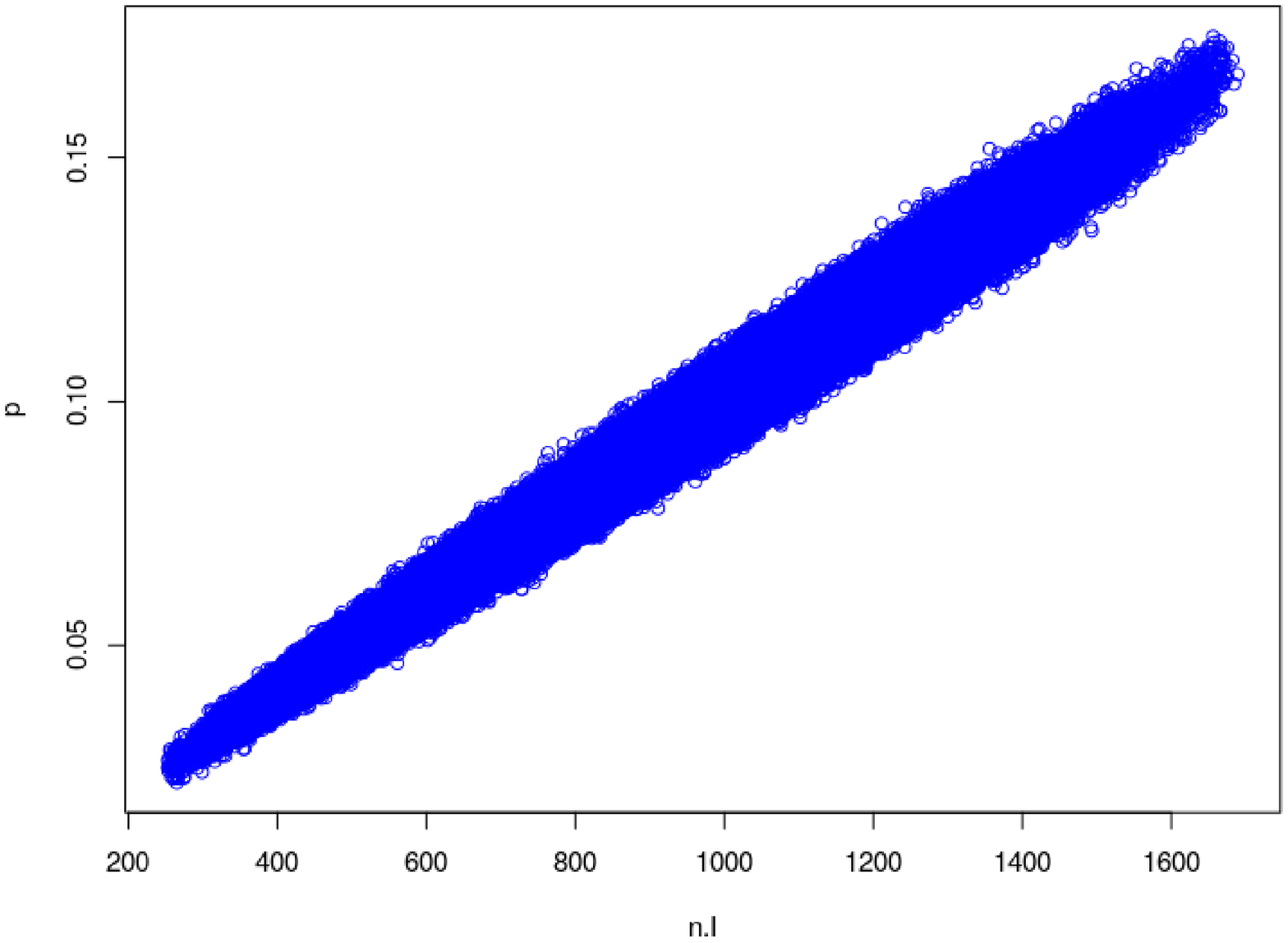,clip=,width=0.8\linewidth}
   \\  \mbox{} \vspace{-1.0cm} \mbox{}
\end{center}
\caption{\small \sf Scatter plot of $p$ vs $n_I$,
  showing the very high correlation
  between the two variables.
}
\label{fig:JAGS_plots_standard_conf_correlation} 
\end{figure}
and quantified
by $\rho(p,n_I) = 0.9914.\,$\footnote{Plot and correlation coefficient
are obtained by the following R commands
   \\  \mbox{} \vspace{-.7cm} \mbox{}  
\begin{verbatim}
chain.df <- as.data.frame( as.mcmc(chain) )
plot(chain.df, col='blue')
cor(chain.df)
\end{verbatim}
}
\subsection{Dependence on our knowledge concerning $\pi_1$ and $\pi_2$}
As we have already well understood, the uncertainty
on the result is highly dependent of the
uncertainty concerning $\pi_1$ and $\pi_2$.
Therefore, as we have done in the previous sections,
let us also change here our
assumptions and see how the main result changes
accordingly ($n_I$ is of little interest, at this point,
also because of its very high correlation with $p$,
and hence we shall not monitor it any longer in further examples).
\begin{enumerate}
\item First we start assuming negligible uncertainty
  on sensitivity and specificity.\footnote{In order
  to avoid to modify the JAGS model, we simply multiply
  all relevant Beta parameters by the large factor
  $a=10^6$, thus reducing all uncertainties by a factor thousand
  (see Eq.~(\ref{eq:Varbeta})).
  This is done by adding the following command
   \\  \mbox{} \vspace{-.7cm} \mbox{}  
\begin{verbatim}
a=1e6; r1=r1*a; s1=s1*a; r2=r2*a; s2=s2*a
\end{verbatim}
  \mbox{} \vspace{-0.6cm} \mbox{}  
}
 As a result, the standard
 uncertainty $\sigma(p)$ becomes $0.0046$,
 that is about a factor $\approx 5$ smaller.
\item Then, as we have done in the above sections,
  we keep $\pi_1$ to its default value ($r_1= 409.1, s_1 = 9.1$),
  only reducing the uncertainty of $\pi_2$ to $0.007$.\footnote{We exploit
    the same trick of the previous item redefining the Beta parameters
    as follows
   \\  \mbox{} \vspace{-.7cm} \mbox{}  
\begin{verbatim}
a=(22/7)^2;  r2=r2*a; s2=s2*a
\end{verbatim}
  \mbox{} \vspace{-0.6cm} \mbox{}  
  }
  Also in this case the uncertainty decreases, 
  getting $\sigma(p)=0.0085$.
\item Finally, we make the pdf of $\pi_2$ mirror symmetric
  with respect to that of $\pi_1$, that is $r_2= 9.1, s_2 = 409.1$.
  But, obviously we need to change the number of the
  observed positives, choosing this time 1170,
  as suggested by our expectations
  (see Fig.~\ref{fig:PredictionPositive_by01_sym}).
  As a result we get  $p=0.0995\pm 0.0076$, with an uncertainty
  not differing so much with respect to the previous case.
  Indeed, as we have
  have already noted in the previous sections,
  {\em improving the specificity} ($\pi_2$ reduced by a factor five)
  {\em has only a little effect on the quality of the measurement,
  being more important the uncertainty with which
  that test parameter is known}. (And we expect that something like that
  is also true for the sensitivity.)
\end{enumerate}  

\subsection[Quality of the inference as a function of $n_s$
and $f_P$]{Quality of the inference as a function of the sample size
  and of the fraction of positives in sample}
A more systematic study of the quality of the inference is shown
in Tab.~\ref{tab:tabella_inferenza_p},
\begin{table}
{\footnotesize
\begin{center}
\begin{tabular}{r|cccccc}
 \multicolumn{1}{c}{{\normalsize $n_s$}}   & \multicolumn{6}{|c}{{\normalsize $[n_P]$}} \\
 \multicolumn{1}{c}{} &   \multicolumn{6}{|c}{{\normalsize $\mbox{E}(p)\pm \sigma(p)$}} \\
   &  &  &  &  &  &  \\
 \hline
  &  &  &  &  &  &  \\ 
  300   & [34]  &   [60]  & [86] & [112] & [138] & [164] \\
        &  $0.026\pm 0.019$ & $0.100\pm 0.034$ & $0.200\pm 0.036$  &
  $0.299\pm 0.037$ & $0.399\pm 0.037$ & $0.495\pm 0.036$ \\
   &  &  &  &  &  &  \\
  1000  &  [115]      &  [201]      &  [288]      &  [374]      &  [460]      &  [546] \\
        &  $0.021\pm 0.015$ & $0.099\pm 0.028$  &  $0.198\pm 0.026$
        & $0.298\pm 0.025$ & $0.399\pm 0.024$ & $0.498\pm 0.023$ \\
        &  &  &  &  &  &  \\ 
  3000  &  [345]      &  [604]      &  [863]      &  [1122]      &  [1381]      &  [1640]\\
        & $0.018\pm 0.014$ & $0.099\pm 0.024$ & $0.198\pm 0.023$
        & $0.299\pm 0.020$ & $0.399\pm 0.019$ & $0.499\pm 0.017$ \\
        &  &  &  &  &  &  \\
  10000  &  [1150]      &  [2013]      &  [2876]      &  [3739]      &  [4602]   &  [5465]\\
         & $0.018\pm 0.013$  & $0.099\pm 0.022$ & $0.198\pm 0.020$
         & $0.299\pm 0.019$  & $0.399\pm 0.016$ &  $0.499\pm 0.015$ \\
         &  &  &  &  &  &  \\
  \hline
    &  &  &  &  &  &  \\ 
  300   & [34]  &   [60]  & [86] & [112] & [138] & [164] \\
        & $0.019\pm 0.015$ & $0.101\pm 0.028$ & $0.201\pm 0.031$ 
        & $0.300\pm 0.033$ & $0.400\pm 0.034$ &  $0.496\pm 0.034$ \\
    &  &  &  &  &  &  \\
  1000  &  [115]      &  [201]      &  [288]      &  [374]      &  [460]      &  [546] \\
  & $0.011\pm 0.009$  & $0.100\pm 0.016$ & $0.200\pm 0.018$
  & $0.299\pm 0.019$ & $0.400\pm 0.019$ & $0.499\pm 0.019$ \\
    &  &  &  &  &  &  \\
  3000  &  [345]      &  [604]      &  [863]      &  [1122]      &  [1381]      &  [1640]\\
  & $0.009\pm 0.006$  & $0.100\pm 0.011$  & $0.200\pm 0.012$
  & $0.300\pm 0.012$ & $0.400\pm 0.012$ & $0.500\pm 0.012$ \\
    &  &  &  &  &  &  \\
  10000  &  [1150]      &  [2013]      &  [2876]      &  [3739]      &  [4602]   &  [5465]\\
  & $0.007\pm 0.005$ & $0.100\pm 0.009$  & $0.200\pm 0.008$
  & $0.300\pm 0.008$ & $0.400\pm 0.008$ &  $0.500\pm 0.008$\\
    &  &  &  &  &  &  \\
  \hline
    &  &  &  &  &  &  \\ 
  300  &  [7]      &  [35]      &  [64]      &  [93]      &  [121]      &  [150]  \\
  & $0.010\pm 0.008$ & $0.102\pm 0.021$  & $0.199\pm 0.025$
  & $0.299\pm 0.028$ & $0.400\pm 0.030$ & $0.500\pm 0.031$ \\
    &  &  &  &  &  &  \\
  1000  &  [22]      &  [118]      &  [213]      &  [309]      &  [404]      &  [500] \\
  & $0.007\pm 0.005$ & $0.100\pm 0.013$ & $0.200\pm 0.015$
  & $0.300\pm 0.016$ & $0.400\pm 0.017$ & $0.500\pm 0.017$ \\
    &  &  &  &  &  &  \\
  3000  &  [66]      &  [353]      &  [640]      &  [926]      &  [1213]      &  [1500] \\
  & $0.006\pm 0.004$ & $0.100\pm 0.009$ & $0.200\pm 0.010$
  & $0.300\pm 0.011$ & $0.400\pm 0.011$ & $0.500\pm 0.011$ \\
    &  &  &  &  &  &  \\
  10000  &  [220]      &  [1176]      &  [2132]      &  [3088]      &  [4044]      &  [5000]\\
  & $0.006\pm 0.004$ & $0.100\pm 0.008$ & $0.200\pm 0.008$
  & $0.300\pm 0.007$ & $0.400\pm 0.007$  & $0.500\pm 0.007$  \\
    &  &  &  &  &  &  \\
\end{tabular}
\end{center}
}
\caption{\small \sl Proportion $p$ of infected in a population, inferred from
  the number $n_P$ of positives in a sample of $n_S$ individuals. The three blocks
  of the table corresponds to the assumptions summarized by $\pi_1 = 0.978\pm 0.007$ and $\pi_2 =(0.115\pm 0.022$, $0.115\pm 0.007$, $0.022\pm 0.007)$.}
\label{tab:tabella_inferenza_p}  
\end{table}
 which reports the 
inferred value of $p$, {\em summarized by the expected value and its standard
deviation} evaluated by sampling, as a function
of the sample size and the number of positives in the sample.
The three blocks of the
table correspond to our typical hypotheses on the knowledge of
sensitivity and specificity, and summarized, from top to bottom,
by $(\pi_1=0.978\pm 0.007, \pi_2=0.115\pm 0.022)$,
$(\pi_1=0.978\pm 0.007, \pi_2=0.115\pm 0.007)$
and $(\pi_1=0.978\pm 0.007, \pi_2=0.022\pm 0.007)$,
corresponding then to the cases shown, in the same order, in
Figs.~\ref{fig:PredictionPositive_by01}-\ref{fig:PredictionPositive_by01_sym}
(we have added an extra column with the numbers of
positives yielding $p\approx 0.5$).
We see that, from columns 2 to 6, we get
$p$ ranging from $0.1$ to $0.5$ at steps of $0.1$, with
standard uncertainty varying with $n_s$ and $n_P$ (and therefore with
the fraction of positives $f_P$) in agreement with what
we have learned in Sec.~\ref{sec:measurability_p}, studying the
predictive distributions (note the difference between {\em resolution
power}, used there, and {\em standard uncertainty}, used here).

We note that, instead, the results of the first column is
{\em ``not around zero, as expected}'' ({\em naively}).
The reason is very simple and it is illustrated in
Fig.~\ref{fig:JAGS_p0_ns10000_standard} for the case of $n_s=10000$.
\begin{figure}[t]
\begin{center}
  \epsfig{file=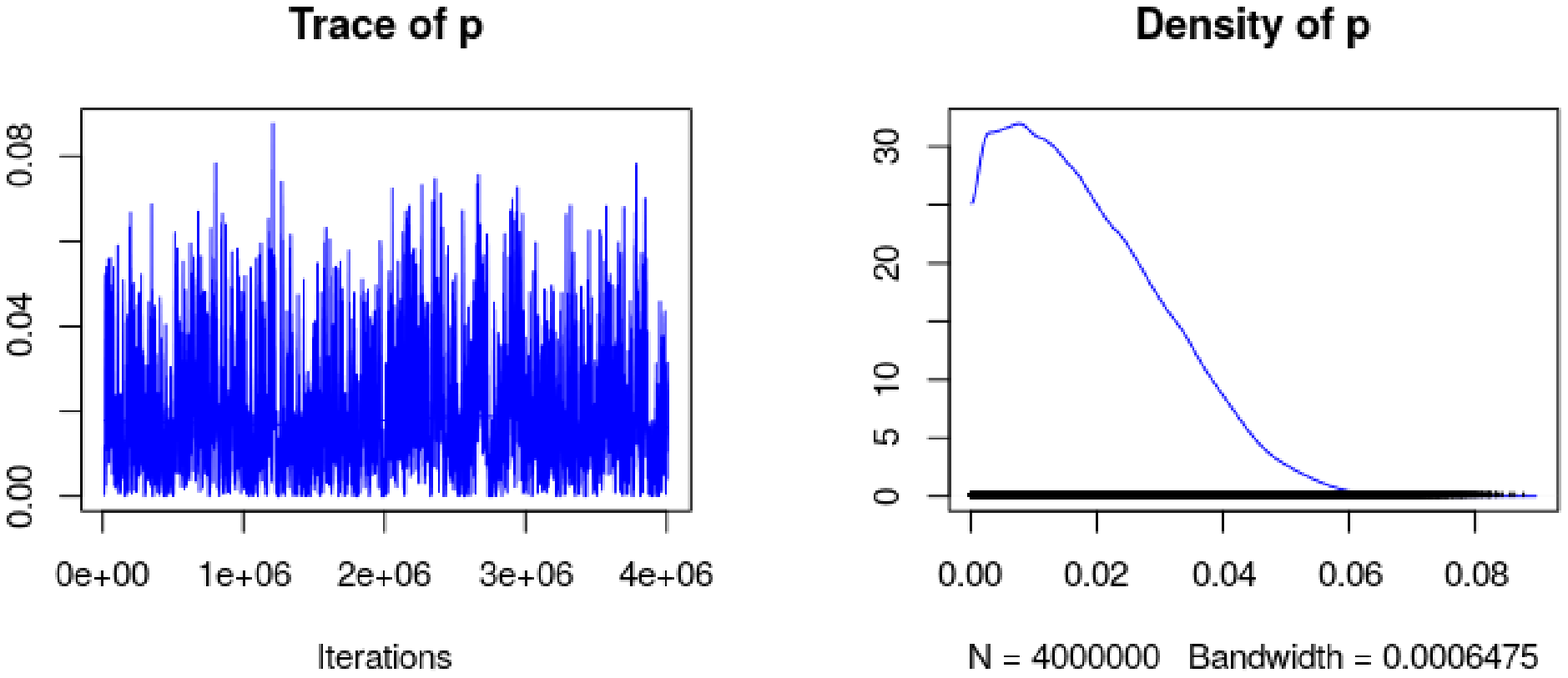,clip=,width=\linewidth}
   \\  \mbox{} \vspace{-1.0cm} \mbox{}
\end{center}
\caption{\small \sf Inference of $p$ from $ns=10000$ and $n_P=1150$.
}
\label{fig:JAGS_p0_ns10000_standard} 
\end{figure}
It is true that, if there were no infected in the population,
then we would expect $n_P \approx 1150$ (with a standard uncertainty of 220),
but the distribution of $p$ provided by the inference {\em cannot have}
a mean value zero, simply because negative values of $p$ are
impossible.\footnote{In particular, we would like to point
  out that this question has nothing to do with the story
  of the `biased estimators' of frequentists. In probabilistic
  inference the result is not just
  a single number (the famous `estimator'), but rather
  the distribution of the quantity of interest, of which
  mean and standard deviation are only some of the possible
  summaries, certainly the most convenient for several purposes.}
Obviously the smaller is the number of positives in the sample
and more peaked is the distribution of $p$ close to 0. But what happens
if, for $n_s=10000$, $n_P$ is much smaller of 1150? 
This interesting case will be the subject of the next subsection.

\subsection[Updated $f(\pi_1)$ and $f(\pi_2)$ in the case of
  `anomalous' number of positives]{Updated knowledge of
  $\pi_1$ and $\pi_2$ in the case of
  `anomalous' number of positives}\label{ss:updated_pi2}
Let us imagine that, instead of 1150 positives, we `had observed'
a much smaller number (in terms of standard deviation of prediction,
that, we remind, is about 220). For example, an under-fluctuation
of 3 $\sigma$'s would yield 490 positives. But let us exaggerate
and take as few as 50 positives, corresponding to $-5\,\sigma$'s.
The JAGS result (this time
monitoring also $\pi_1$ and $\pi_2$),
obtained using our usual uncertainties concerning
$\pi_1$ and $\pi_2$ ($0.978\pm 0.007$ and $0.115 \pm 0.022$, respectively),
is showed in Fig.~\ref{fig:jags_inf_p_pi1_pi2}
\begin{figure}
\begin{center}
  \epsfig{file=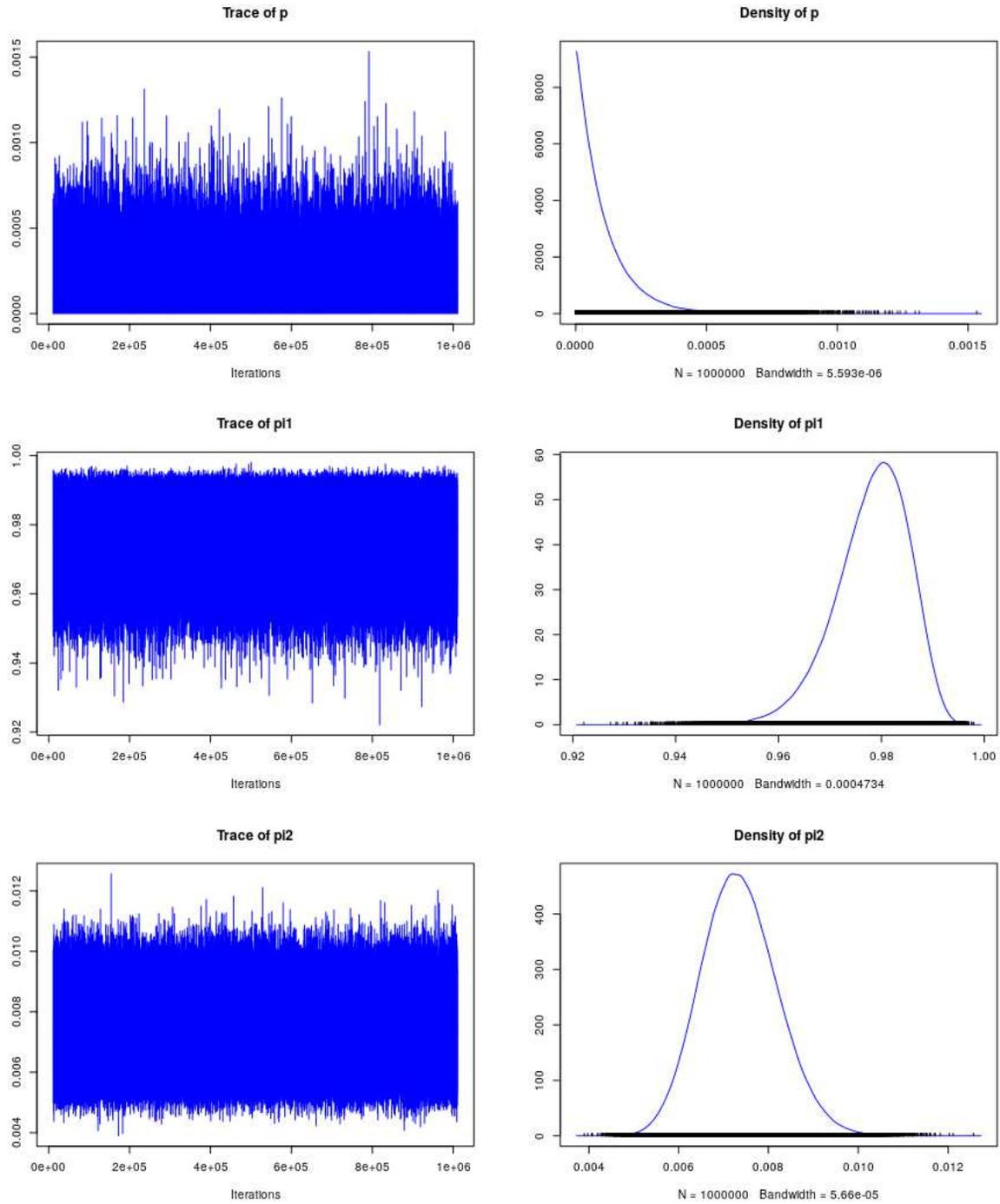,clip=,width=\linewidth} 
   \\  \mbox{} \vspace{-0.5cm} \mbox{}
\end{center}
\caption{\small \sf JAGS inference of $p$, $\pi_1$ and $\pi_2$
  from $ns=10000$ and $n_P=50$ (see text).
}
\label{fig:jags_inf_p_pi1_pi2} 
\end{figure}
and summarized as
\begin{verbatim}
1. Empirical mean and standard deviation for each variable,
   plus standard error of the mean:

         Mean        SD  Naive SE Time-series SE
p   0.0001022 0.0001023 1.023e-07      1.998e-07
pi1 0.9781819 0.0071445 7.145e-06      7.161e-06
pi2 0.0073581 0.0008463 8.463e-07      8.463e-07

2. Quantiles for each variable:

         2.5%       25%       50%       75%     97.5%
p   2.586e-06 2.949e-05 7.091e-05 0.0001415 0.0003771
pi1 9.622e-01 9.738e-01 9.789e-01 0.9833334 0.9899013
pi2 5.790e-03 6.771e-03 7.327e-03 0.0079113 0.0091035
\end{verbatim}
As we see, the distribution of $p$ looks exponential,
with mean and standard deviation practically identical
and equal to $1.0\times 10^{-4}$ (we remind that it is
a property of the  exponential distribution to have
 expected value and standard deviation equal).
In this case the {\em quantiles} produced by $R$ are particularly
interesting, providing e.g. $P(p\le 3.77\times 10^{-4}) = 97.5\%$.

The fact that a small number of infectees squeezes the distribution of $p$
towards zero follows the expectations. More surprising, at first sight,
is the fact that also the value of  $\pi_2$ does change:
\begin{eqnarray*}
  \pi_2: & 0.115\pm 0.022 & \ \ \longrightarrow \ \ 0.0074\pm 0.0008\,.
\end{eqnarray*}  
The reason why $\pi_2$ can change
(also $\pi_1$ could, although JAGS `thinks' this is not the case)
is due to the fact that it is now a unobserved node,
and the $\mbox{Beta}(r_2,s_2)$ with
which we model it is just the {\em prior distribution} we assign to it.
In other words,
the very small number of positives could be not only due to a very small value
of $p$, but also to the possibility that $\pi_2$ is indeed
substantially smaller than what we
initially thought. This sounds absolutely reasonable,
but telling exactly what the result will
be can only be done using strictly the rules of
probability theory,
although with the help of MCMC, because
in multivariate problems of this kind intuition can easily
fail.\footnote{Although we cannot go through the details
  in this paper, it would be interesting to 
  use `wider priors' about $\pi_1$ and $\pi_2$
  in order to see how they get updated by JAGS, and then
  try to understand
  what is going on making pairwise scatter plots of
  the resulting $p$, $\pi_1$ and $\pi_2$.
}

\subsection{Inferring the proportions of infectees in two different
  populations}\label{ss:inferring_Delta_p}
Let us now go through what
has been anticipated in Sec.~\ref{ss:predict_Delta_fP},
talking about predictions. We have seen that, since
(at least in our model)
an important contribution to the uncertainty is due to systematics,
related to the uncertain knowledge of $\pi_1$ and $\pi_2$,
we cannot increase at will the sample size
with the hope to reduce the uncertainty on $p$.
Nevertheless, as a consequence of what we have seen in
Sec.~\ref{ss:predict_Delta_fP}, we expect to be able to measure
the difference of proportions of infectees in
two populations much better
than how we can measure a single proportion.

Let us use again sample sizes of 10000 (they could be different
for the different populations)
and imagine that we get numbers of positives rather `close', as we
know from the predictive distribution: $n_P^{(1)}=2000$
and  $n_P^{(2)}=2200$. As far as sensitivity and specificity
are concerned, since we have learned their effect, let us stick,
for this exercise, to our default case,
summarized by $\pi_1=0.978\pm 0.007$ and $\pi_2=0.115\pm 0.022$.
The R script is given in Appendix B.11.
Here is the result of the joint inference and of the difference
of the proportions:
\begin{eqnarray*}
  p^{(1)} &=& 0.097 \pm 0.023   \\
  p^{(2)} &=& 0.120 \pm 0.022   \\
  \Delta p = p^{(2)} - p^{(1)} &=&  0.023 \pm 0.007 \\
  \rho(p^{(1)},p^{(2)}) &=&   0.955\,.
\end{eqnarray*}  
As we see,  $p^{(1)}$ and  $p^{(2)}$ are, as
we use to say, `equal within the uncertainties',
but nevertheless their difference is rather `significative'.
This is due to the fact that the common systematics
induce a quite strong positive correlation
among the determination of the two proportions,
quantified by the correlation coefficient.
The relevance of measuring differences has been already commented
in Sec.~\ref{ss:predict_Delta_fP}, in which we also provided
some details on how to evaluate the uncertainty
of the difference from the other pieces of information.
We would just like to stress its practical/economical
importance. For example, dozens of regions of a state could be
sampled and tested with `rather cheap' kits, with performances
of the kind we have seen here (but it is important that they
are the same!), and only one region (or a couple of them,
just for cross-checks) \underline{also}
with a more expensive (and hopefully more accurate) one.
The region(s) tested with the high quality kit could then be
used as calibration point(s) for the
others and the practical impact in planning a test campaign
is rather evident.

\subsection{Which priors?}\label{ss:priors_priors_priors}
After having read in the first part of the paper
the dramatic role of the prior, when we
had to evaluate the probability of individual
being infected, given the test result,
one might be surprised by the regular use of
a flat prior of $p$ throughout the present section.
First at all, we would like to point out that we are
doing so, in this case, \underline{not}
{``in order to leave the data to `speak' by themselves''},
as someone says. It is, instead, the other way
around: the values of $p$ preferred by the data,
starting from a uniform prior, are characterized
by a distribution much narrower than what we 
could reasonably judge, based on previous rational knowledge.
In other words,
they are not at odds with what we could believe
independently of the data. But this is not always the case,
and experts could have more precise expectation,
grounded on their knowledge.

Anyway, a prior distribution is something
that we have to plug in the
model, if we want to perform a probabilistic inference.
In practice -- and let us remind again that
{\em ``probability is good sense reduced to a calculus''} --
we  model  the prior in a reasonable and mathematically
convenient way, and the Beta distribution is well suited
for this case, also due to the flexibility of the shapes
that it can assume, as seen in Sec.~\ref{ss:conjugate_priors}.
Once we have opted for a Beta, a uniform prior is recovered
for $r=1$ and $s=1$, although  we are far from
thinking that
$p=0$ or  $p=1$ are possible, as well as that 
$p$ could be above 0.9  with 10\% chance, and so on.

\subsubsection{Symmetric role of prior and `integrated likelihood'}
Since we cannot go into indefinite and sterile discussions
on all the possible priors that we might use
(remember that if we collect and analyze data is to improve our knowledge,
often used to make practical decision in a finite time scale!)
it is important to understand a bit deeper their role in the inference.
This can be done 
factorizing  Eq.~(\ref{eq:joint_tutte}), written here
in compact notation as
\begin{eqnarray}
f(\ldots)& =& f(p,n_I,n_{NI},n_{P_I},n_{P_{NI}},\pi_1,\pi_2,n_P,n_s,r_1,s_1,r_2,s_2)\,,  
\end{eqnarray}
into two parts:
one that only contains
$f_0(p)$ and the other containing the remaining factors of the `chain',
indicated here as $f_{\emptyset}(\ldots)$:
\begin{eqnarray}
f(\ldots) &=&  f_{\emptyset}(\ldots)\cdot f_0(p)\,. 
\end{eqnarray}
The unnormalized pdf of $p$, conditioned by data and parameters,
can be then rewritten (see Appendix A) as
\begin{eqnarray}
 f(p\,|\,n_P,n_s,r_1,s_1,r_2,s_2) &\propto&
 \left[ \sum_{n_I}\sum_{n_{NI}}\sum_{n_{P_I}}\sum_{n_{P_{NI}}}
   \int\!\!\int\! f_{\emptyset}(\ldots)\,\mbox{d}\pi_1
  \mbox{d}\pi_2\right]\! \cdot f_0(p) \ \ \ \mbox{} \\
 &\propto& {\cal L}(p\,;\,n_P,n_s,r_1,s_1,r_2,s_2)\cdot  f_0(p)\,,
\end{eqnarray}
in which we have indicated with the usual symbol used
for the (`integrated') {\em likelihood}
(in which constant factors are irrelevant)
the part which multiplies $f_0(p)$. 
It is then rather evident the role of ${\cal L}$ in
`reshaping' $f_0(p)$.\footnote{It is worth pointing out
  the cases, occurring especially in frontier science,
  in which the likelihood is constant in some regions, and
  therefore it does not update/reshape $f_0(v)$, where
  `$v$' stands for the generic variable of interest
  (see chapter 13 of Ref.~\cite{BR}).
  An interesting instance,
  in which $v$ has the role of {\em rate of gravitational
  waves} $r$, is discussed in Ref.~\cite{conPia}, where
  the concept of {\em relative belief updating ratio} was first
  introduced. Another frontier physics case, applied to the
  {\em Higgs boson mass} $m_H$,
  on the basis of the experimental and theoretical information available
  before year 1999, is reported in Ref.~\cite{conPeppe}.
  The two cases are complementary because in the first one
  {\em sensitivity is lost} for $r\rightarrow 0$ (`likelihood {\em open}
  on the left side'), while in the
  second for $m_H\rightarrow \infty$
  (`likelihood {\em open} on the right side').
  (For recent developments and applications, see
  Ref.~\cite{Gariazzo}.)
}
In the particular case in which  $f_0(p)=1$ the inference
is simply given by
\begin{eqnarray}
 f(p\,|\,n_P,n_s,r_1,s_1,r_2,s_2,f_0(p)=1)
&\propto& {\cal L}(p\,;\,n_P,n_s,r_1,s_1,r_2,s_2) 
\end{eqnarray}
(``the inference is determined by the likelihood'').

If, instead, the prior is not flat, then {\em it does reshape}
the posterior obtained by ${\cal L}$ alone.
Therefore there are two alternative
ways to see the contributions
of ${\cal L}$ and $f_0(p)$:  {\bf each one reshapes the other}.
In particular
\begin{itemize}
\item in the regions of $p$ set to zero by either function
  the posterior vanishes;
\item the function which is more narrow around its maximum
  `wins' against the smoother one.
\end{itemize}
Therefore, for the case shown in Fig.~\ref{fig:JAGS_plots_standard_conf},
obtained by a flat prior, the `density of $p$'
is nothing but the shape of ${\cal L}(p;n_P,n_s,r_1,s_1,r_2,s_2)$.
If $f_0(p)$ is constant, or varies slowly, in the range $[0.02,0.17]$
it provides null or little effect. If, instead, it is very peaked
around 0.15 (e.g. with a standard deviation of $\approx 0.01$)
it dominates the inference.

But what is more interesting is that {\em the reshape
  by $f_0(p)$ can be done in a second step}.\footnote{As
  already remarked in footnote \ref{fn:notary}, `prior' does not
  mean that you have to declare `before' you sit down to
  make the inference! It just means that it is based on other
  pieces of information (`knowledge')
  on the quantity under study.
}
This is the importance of choosing a flat prior
(and not just a question of laziness): the data analysis expert
could then present a result of the kind
of Fig.~\ref{fig:JAGS_plots_standard_conf}
to an epidemiologist who could then 
reshape her priors (or, equivalently, reshape the curve provided
by the data analyst with her priors).
But she could also have such a strong prior on the variable under study,
that she could reject {\em tout court} the result, blaming the
data analysis expert that there must be something wrong in 
the analysis or in the data -- see Sec.~\ref{ss:more_on_priors}.

\subsubsection{Some examples}
Let us illustrate these ideas with a simple case on which
exact calculations can be also done:
the inference of $p$ of a binomial distribution, based on
$n$ successes got in $N$ trials.
We went through it in Sec.~\ref{sec:uncertainty},
but we do it solve it now with JAGS in order to provide
some details on `reshaping'.
The model is really trivial
\begin{verbatim}
model {
  n ~ dbin(p, N)
  p ~ dbeta(r0,s0)   
}
\end{verbatim}
and the full script is provided in Appendix B.12.
For $N=10$ and $n=3$ and a flat prior the JAGS result
is shown by the histogram of Fig.~\ref{fig:test_reshape}.
\begin{figure}[t]
\begin{center}
  \epsfig{file=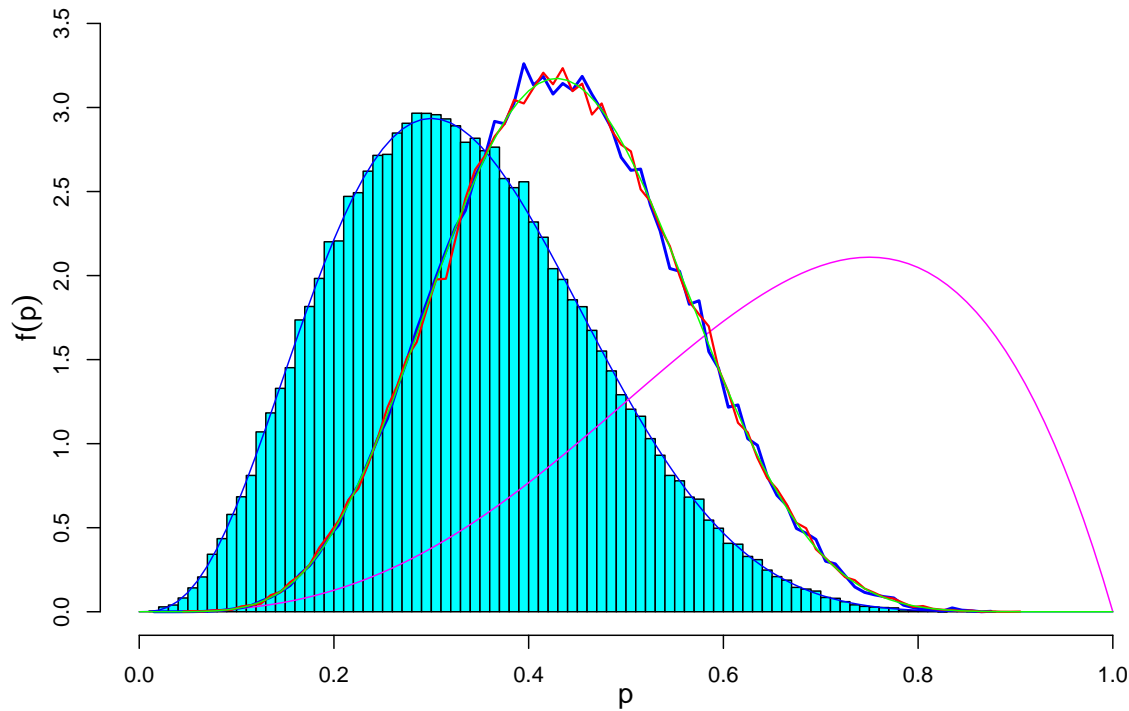,clip=,width=\linewidth}
   \\  \mbox{} \vspace{-1.1cm} \mbox{}
\end{center}
\caption{\small \sf Inference of $p$ with binomial
  distributions obtained with different priors and different ways
  to make use of an `informative prior' (see text).
}
\label{fig:test_reshape} 
\end{figure}
The blue line along the profile of the histogram
is the analytic result obtained starting from
of a Beta prior with $r_0=s_0=1$, that is
$\mbox{Beta}(1\!+\!n,\, 1\!+\!N\!-\!n)$. Then the
`informative prior'
(rather vague indeed), modeled by a $\mbox{Beta}(4, 2)$
and therefore having a mean value of $4/(4+2)=2/3$,
is shown by the magenta curve having the maximum
value at $3/4\,=[(4-1)/(4+2-2)]$.
The distribution obtained reweighing the
posterior got from a flat prior
(histogram) by this new prior is shown by the blue broken curve,
while the red broken curve shows the JAGS result obtained
using  the new prior (the latter curves overlap so much
that they can only be identified by color code).
Finally, the green continuous curve is the analytic
posterior obtained updating the Beta parameters, that is
$\mbox{Beta}(4\!+\!3,\, 2\!+\!7)$.
The agreement of the three results is `perfect'
(taking into account that two of them are got by sampling).

The second example is our familiar case of 2010 positives
in a sample of 10000 individuals shown in detail in
Sec.~\ref{ss:infer_p_nI} and of which a different Monte Carlo
run, with $n_r=4\times 10^6$ in order to get
a smoother histogram, is shown in Fig.~\ref{fig:inf_p_reshape_pub}. 
\begin{figure}[t]
\begin{center}
  \epsfig{file=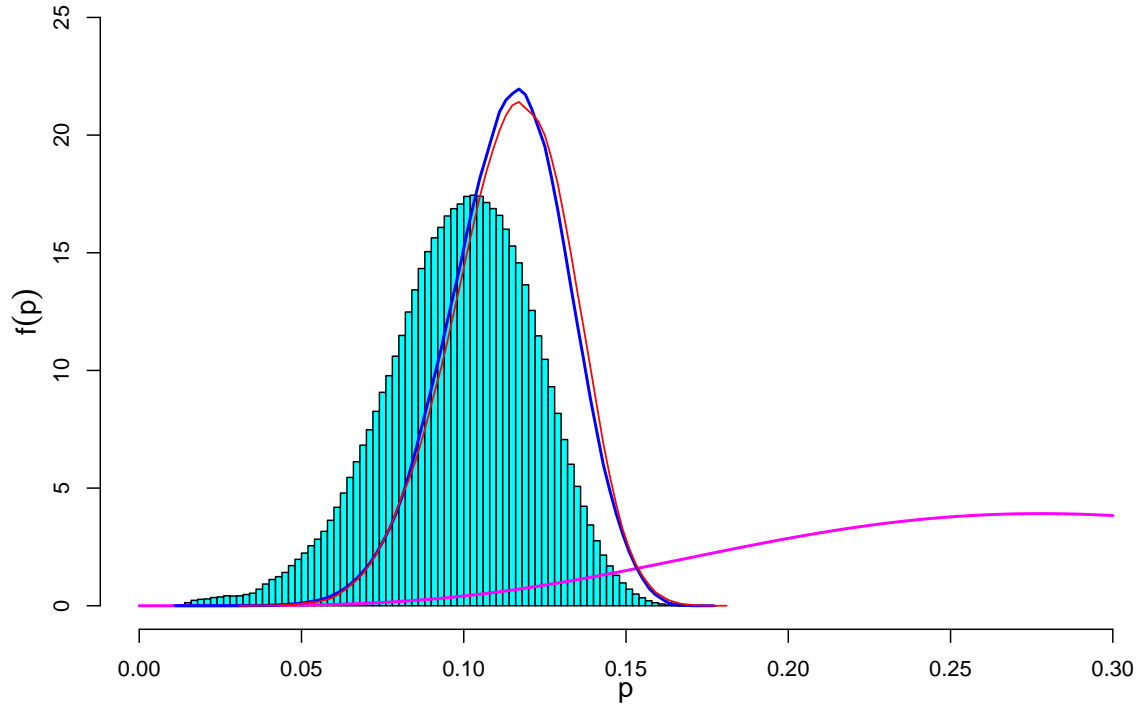,clip=,width=\linewidth}
   \\  \mbox{} \vspace{-1.1cm} \mbox{}
\end{center}
\caption{\small \sf Inference of the proportion $p$
  of infected in a population, having measured 2010 positives
  in a sample of 10000 individuals: JAGS result based on a
  flat prior (histograms) and effect of `reshaping' based
  on an informative prior.
  (see text).
}
\label{fig:inf_p_reshape_pub} 
\end{figure}
The new prior is indicated by the magenta curve, modeled
by a $\mbox{Beta}(6,14)$, having its mode at $5/18\approx 0.28$.
The reshaped posterior is indicated by the blue curve,
having mean  $0.1134$ and standard deviation $0.0182$.
The result of JAGS using as prior the  $\mbox{Beta}(6,14)$
is shown by the red curve, characterized by a
mean of 0.1145 and a standard deviation 0.0184
(we are using an exaggerated number of digits just for checking --
using one digit for the uncertainty both results
become `$0.11\pm 0.02$').
The degree of agreement is excellent, also taking into account that
they have intrinsic Monte Carlo fluctuations.
It is interesting to note that, besides increasing slightly the mean
values (but one could object
that ``they are equal within the uncertainties''),
the main effect of the new prior is to practically rule out
values of $p$ below 0.05.

\subsubsection{Some approximated rules}
Having seen the utility of reshaping the posterior
got from a flat prior, once a different prior is assumed,
we also try to find some practical rules based on the
mean and the standard deviations of the distributions involved.
\begin{enumerate}
\item The first is based on Gaussian approximation, and it
  holds if both the prior and the posterior got by JAGS
  assuming a uniform prior appear somehow `bell-shaped',
  although we cannot expect that they are perfectly symmetric,
  especially if small or large values of $p$ are preferred.
  In this case the following  (very rough) approximation
  is obtained for the mean and the
  standard deviation\footnote{See e.g. Sec. 2 of Ref.~\cite{skept}.}
\begin{eqnarray}
  \mu_p &=& \frac{\mu_{\cal L}/\sigma^2_{\cal L} + \mu_0/\sigma_0^2}
     {1/\sigma^2_{\cal L} + 1/\sigma_0^2} \\
     \frac{1}{\sigma^2_p} &=& \frac{1}{\sigma^2_{\cal L}}
     + \frac{1}{\sigma_0^2}\,,
\end{eqnarray}
where $\mu_{\cal L}$ and $\sigma_{\cal L}$ are the mean and the standard deviation
got from JAGS with a flat prior;  $\mu_0$ and $\sigma_0$
are those summarizing the priors; $\mu_p$ and $\sigma_p$ should be
(approximately) equal  to the JAGS results we had got using the prior
summarized by  $\mu_0$ and $\sigma_0$.
Applying this rule to the case in Fig.~\ref{fig:inf_p_reshape_pub},
for which $\mu_{\cal L}=0.0987$, $\sigma_{\cal L}=0.0229$,
$\mu_0=0.30$ and $\sigma_0=0.10$, we get $p=0.1087\pm 0.022$,
that, rounding the uncertainty to one digit becomes
`$0.11\pm 0.02$', equal to the one obtained above
by reshaping or re-running JAGS with the new prior.
\item
The second rule makes use of the Beta and its usage as prior
conjugate when inferring $p$ of a binomial, as we have seen in 
Sec.~\ref{ss:conjugate_priors}. The idea is to see the pdf
estimated by JAGS with flat prior as a `rough Beta'
whose parameters can be estimated from the mean and the standard deviation
using Eqs.~(\ref{eq:r_from_E-sigma})-(\ref{eq:s_from_E-sigma}).
We can then imagine that the pdf of $p$ could have
been estimated by a `virtual' Poisson processes whose outcomes
update the parameters of the Beta according to
Eqs.~(\ref{eq:Beta_update_r})-(\ref{eq:Beta_update_s}).
The trick consists then in modifying the Beta parameters
according to the simple rules:
\begin{eqnarray*}
  r_p &=& r_{\cal L} + r_0 - 1 \\
  s_p &=& s_{\cal L} + s_0 - 1\,, 
\end{eqnarray*}
where $r_{\cal L}$ and $s_{\cal L}$ are evaluated
from  $\mu_{\cal L}$ and $\sigma_{\cal L}$ making use of
Eqs.~(\ref{eq:r_from_E-sigma}) and (\ref{eq:s_from_E-sigma}).
Then the new mean and standard deviation are evaluated from $r_p$ and $s_p$
(see Sec.~\ref{ss:conjugate_priors}).

For example, in the case of  Fig.~\ref{fig:test_reshape}
we have (with an exaggerated number of digits)
$p=0.0987\pm 0.0229$, which could derive from a Beta
having  $r_{\cal L} = 16.7$ and  $s_{\cal L} = 152.4$.
If we have a prior somehow peaked around 0.3, e.g.
$p_0=0.3\pm 0.1$, it can be parameterized by a Beta with $r_0=6$
and $s_0=14$. Applying the above rule we get
\begin{eqnarray*}
  r_p &=& 16.7 + 6 - 1 = 21.7  \\
  s_p &=& 152.4 + 14 - 1 = 165.5  \,, 
\end{eqnarray*}
which yield then $p = 0.1159 \pm 0.0223$, very similar to
what was obtained by reshaping or re-running JAGS
($0.12\pm 0.02$ at two decimal digits).
\end{enumerate}
As we see, these approximated rules are rather rough, but
they have the advantage of being fast to apply, if one
wants to arrive quickly to some reasonable conclusions, based
on her personal priors.\footnote{The main reason of
  the not excellent level of agreement
is due to the quite pronounced tail on the left
side of the distribution. The rule could work better
for other values of $n_P$, given $n_s$, but we have no interest
in showing the best case and try to sell it as `typical'.
We just stuck to the numeric case we have used mostly throughout
the paper.}

\section{Exact evaluation of $f(p)$}\label{sec:direct}
After having solved the inferential task
by MCMC making use of JAGS,
let us now attempt to solve our problem exactly, although
limiting ourselves to the inference of $p$. 
\subsection{Setting up the problem}
As we have seen in Sec.~\ref{sec:interring_p}, the inference
of the `unobserved' variables, based on the `observed' one,
for the problem represented graphically in the `Bayesian'
network of Fig.~\ref{fig:sampling_binom_inf}, consists
in evaluating `somehow'
\begin{eqnarray}
  \hspace{1cm}&&f(p,n_I,n_{NI},n_{P_I},n_{P_{NI}},\pi_1,\pi_2\,|\,n_P,n_s,r_1,s_1,r_2,s_2)\,,
 \label{eq:appc_1}
\end{eqnarray}
from which the most interesting probability distribution, at least
for the purpose of this paper,
\begin{eqnarray*}
\hspace{1cm}&&f(p\,|\,n_P,n_s,r_1,s_1,r_2,s_2)
\end{eqnarray*}
 can be obtained by marginalization
(see also Appendix A). Besides a normalization factor,
Eq.~(\ref{eq:appc_1}) is proportional to Eq.~(\ref{eq:joint_tutte}),
hereafter indicated by `$f(\ldots)$' for compactness, which can be
written making use of the chain rule obtained
following the bottom-up analysis of the graphical model of
Fig.~\ref{fig:sampling_binom_inf}:
\begin{eqnarray}
  f(\ldots) &=& f(n_P\,|\,n_{P_I},n_{P_{NI}}) \cdot
  f(n_{P_I}\,|\,\pi_1,n_I) \cdot 
  f(n_{P_{NI}}\,|\,\pi_2,n_{NI})  \cdot 
f(\pi_1\,|\,r_1,s_1)\cdot   \nonumber \\
&&   f(\pi_2\,|\,r_2,s_2) \cdot  f(n_{NI}\,|\,n_s,n_I)  \cdot f(n_I\,|\,p,n_s)
\cdot f_0(p) \nonumber
\end{eqnarray}
in which
\begin{eqnarray}
 f(n_P\,|\,n_{P_I},n_{P_{NI}}) &=& \delta_{n_p,\,n_{P_I}+n_{P_{NI}}} \label{eq:kron1}
  \\
 f(n_{P_I}\,|\,\pi_1,n_I) &=& \binom{n_I} {n_{P_I}} \cdot
 \pi_1^{n_I}\cdot (1-\pi_1)^{n_I-n_{P_I}} \\
 f(n_{P_{NI}}\,|\,\pi_2,n_{NI}) &=&  \binom{n_{NI}} {n_{P_{NI}}}\cdot 
 \pi_2^{n_{P_{NI}} } \cdot (1-\pi_2)^{n_{NI}-n_{P_{NI}`}}
\end{eqnarray}
\begin{eqnarray}
 f(\pi_1\,|\,r_1,s_1) &=&  \frac{\pi_1^{r_1-1}\cdot(1-\pi_1)^{s_1-1}}{\beta(r_1,s_1)} \\
 f(\pi_2\,|\,r_2,s_2) &=&  \frac{\pi_2^{r_2-1}\cdot(1-\pi_2)^{s_2-1}}{\beta(r_2,s_2)} \\
 f(n_{NI}\,|\,n_s,n_I) &=& \delta_{n_{NI},\,n_s-n_{I}} \label{eq:kron2}\\
 f(n_I\,|\,p,n_s) &=& \binom{n_s}{n_I}\cdot p^{n_I}\cdot(1-p)^{n_s-n_I}  \,, 
\end{eqnarray}
where $\delta_{m,k}$ is the Kroneker delta (all other symbols belong to the
definitions of the binomial and the Beta distributions) and we have left
to define the prior distribution $f_0(p)$. 
The distribution of interest is then obtained by summing up/integrating 
\begin{eqnarray*}
 f(p\,|\,n_P,n_s,r_1,s_1,r_2,s_2) &\propto&
 \sum_{n_I}\sum_{n_{NI}}\sum_{n_{P_I}}\sum_{n_{P_{NI}}}\int\!\!\int\! f(\ldots)\,\mbox{d}\pi_1
 \mbox{d}\pi_2 \,,
\end{eqnarray*}
where the limits of sums and integration will be written in
detail in the sequel.

As a first step we simplify the equation by summing
over $n_{P_{NI}}$ and $n_{NI}$ and exploiting the Kroneker delta
terms (\ref{eq:kron1}) and (\ref{eq:kron2}).
We can then replace $n_{P_{NI}}$ with $n_P - n_{P_{I}}$ and $n_{NI}$ with $n_s-n_I$ 
\begin{eqnarray}
f(n_P\!-\!n_{P_I}| n_s\!-\!n_{I},\pi_2)\! &=& \!{\binom{n_s\!-\!n_I} {n_P\!-\!n_{P_I}}}
\cdot  {\pi_2}^{(n_P-n_{P_I})}  \cdot  (1-\pi_2)^{(n_s-n_I)-(n_P-n_{P_I})}\ \  \ \ 
\end{eqnarray}
with the obvious constraints $ n_s-n_I > n_P-n_{P_{I}}$
(i.e. $n_{NI} > n_{P_{NI}}$) and $n_{P_I} < n_I$.

The inferential distribution of interest
$ f(p\,|\,n_P,n_s,r_1,s_1,r_2,s_2)$,
becomes then, besides constant factors and indicating
all the status of information on which the inference is based
as `$I$', that is $I \equiv \{ n_P,n_s,r_1,s_1,r_2,s_2\}$,
\begin{eqnarray}
f(p\,|\,I) &\propto& f_0(p) \cdot \sum_{n_{P_I}=0}^{n_P}\sum_{n_I=0}^{n_s}
                 \int_{0}^{1} \int_{0}^{1}\!f({n_{P_I}}\,|\, n_I,\pi_1)
                 \cdot f(n_P-n_{P_I}| n_s-n_I,\pi_2) \cdot \nonumber \\
        && f(\pi_1\,|\,r_1,s_1)  \cdot  f(\pi_2\,|\,r_2,s_2)
           \cdot  f(n_I\,|\,p,n_s)\, \mbox{d}\pi_1\, \mbox{d}\pi_2 \nonumber \\
 & \propto& f_0(p) \cdot \sum_{n_{P_I}=0}^{n_P}\sum_{n_I=0}^{n_s}\,
   \int_{0}^{1}\! \int_{0}^{1} {\binom{n_I} {n_{P_I}}}  \cdot {\pi_1}^{n_{P_I}}
   \cdot  (1-\pi_1)^{n_I-n_{P_I}} \cdot \nonumber \\
& & {\binom{n_s-n_I} {n_P-n_{P_I}}} \cdot  {\pi_2}^{(n_P-n_{P_I})}  \cdot
    (1-\pi_2)^{(n_s-n_I)-(n_P-n_{P_I})} \cdot \nonumber \\
& & \pi_1^{r_1-1} \cdot (1-\pi_1)^{s_1-1}  \cdot \nonumber
    \pi_2^{r_2-1} \cdot (1-\pi_2)^{s_2-1} \cdot  \nonumber
    \binom{n_s}{n_I}  \cdot  p^{n_I}  \cdot  (1-p)^{n_s-n_I}  \mbox{d}\pi_1  \mbox{d}\pi_2 \nonumber
\end{eqnarray}
\begin{eqnarray}
&\propto& f_0(p) \cdot \sum_{n_{P_I}=0}^{n_P}\sum_{n_I=0}^{n_s}
 \binom{n_I} {n_{P_I}}  \cdot  \binom{n_s-n_I} {n_P-n_{P_I}}  \cdot
 \binom{n_s}{n_I} \! \cdot \! p^{n_I} \! \cdot \! (1-p)^{n_s-n_I}\cdot\! \nonumber\\
&& \int_{0}^{1}\! {\pi_1}^{n_{P_I}+r_1-1}\cdot  (1-\pi_1)^{n_I-n_{P_I}+s_1-1}\,\mbox{d}\pi_1 \label{eq:C.4}\\
&& \int_{0}^{1}\! {\pi_2}^{(n_P-n_{P_I}+r_2-1)}  \cdot
(1-\pi_2)^{(n_s-n_I)-(n_P-n_{P_I})+s_2-1}\,\mbox{d}\pi_2 \nonumber 
\end{eqnarray}
where we have dropped all the terms not depending on the variables
summed up/integrated.

The two integrals appearing
in Eq.~(\ref{eq:C.4}) are, in terms of the generic variable $x$,
of the form
$\int_0^1x^{\alpha-1}\cdot (1-x)^{\beta-1}\mbox{d}x$,
which defines the special function {\em beta}
$\mbox{\tt \large B}(\alpha,\beta)$,
whose value
can be expressed
in terms of Gamma function as
$\mbox{\tt \large B}(\alpha,\beta) = \Gamma(\alpha)\cdot \Gamma(\beta)/
\Gamma(\alpha+\beta)$.
We get then
\begin{eqnarray}
f(p\,|\,I) &\propto&   f_0(p) \cdot \sum_{n_{P_I}=0}^{n_P}\sum_{n_I=0}^{n_s}
            \left[\binom{n_I} {n_{P_I}}  \cdot  \binom{n_s-n_I} {n_P-n_{P_I}} \cdot
            \binom{n_s}{n_I}  \cdot  p^{n_I}   \cdot  (1-p)^{n_s-n_I} \cdot \right. \nonumber \\
&& \left.\frac{\Gamma (n_{P_I}+r_1)\cdot \Gamma(n_I-n_{P_I}+s_1)}
              {\Gamma(r_1 + n_I+s_1)} \cdot\right.  \label{eq:C.5}  \\
&&\left.\frac{\Gamma (n_P-n_{P_I}+r_2)\cdot \Gamma(n_s-n_I-n_P+n_{P_I}+s_2)}
        { \Gamma(n_s-n_I+s_2+r_2)} \right] \nonumber \,.
\end{eqnarray}

\subsection{Normalization factor and other moments of interest}
The normalization factor $N_f$ is given by the integral 
in d$p$ of this expression, once $f_0(p)$
has been chosen.
As we have done in the previous section, we opt for  $\mbox{Beta}(r_0, s_0)$,
taking the advantage not only of the flexibility
of the probability distribution to model our `prior
judgment' on $p$, but also of its mathematical
convenience. In fact, with this choice, the resulting term
in Eq.~(\ref{eq:C.5}) depending on $p$  is given
by $p^{r_0-1+n_I}\cdot (1-p)^{s_0-1+(n_s-n_I)}$. The integral
over $p$ from 0 to 1 yields again a Beta function, that is
$\mbox{\tt \large B}(r_0+n_I,s_0+n_s-n_I)$, thus
getting
\begin{eqnarray}
N_f\!\! & = &\!\!\sum_{n_{p_I}=0}^{n_P}\sum_{n_I=0}^{n_s}
\left[\!\binom{n_I}{n_{P_I}} \cdot  
  \binom{n_s-n_I} {n_P-n_{P_I}} \cdot  \binom{n_s}{n_I} \cdot
  \frac{\Gamma (n_{P_I}+r_1)\cdot \Gamma(n_I-n_{P_I}+s_1)}
 { \Gamma(r_1 + n_I+s_1)} \cdot
  \right. \nonumber \\
&&\left. \frac{\Gamma (n_P-n_{P_I}+r_2)\cdot \Gamma(n_s-n_I-n_P+n_{P_I}+s_2)}
             {\Gamma(n_s-n_I+s_2+r_2)} \right. \cdot  \label{eq:C.6}\\ 
&&\left.\frac{\Gamma(r_0+n_I)\cdot\Gamma(s_0+n_s-n_I)}{\Gamma(r_0+s_0+n_s)}\right] \nonumber
\end{eqnarray}
Similarly, we can evaluate
the expression of the expected values of $p$ and of $p^2$,
from which  the variance follows, being
$\sigma^2(p)=\mbox{E}(p^2)-\mbox{E}^2(p)$.  For example,
being $\mbox{E}(p)$
given by
\begin{eqnarray*}
  \mbox{E}(p) &=& \int_0^1p\cdot f(p\,|\,I)\mbox{d}p\,,
\end{eqnarray*}
in the integral the term depending on $p$
becomes   $p\cdot p^{r_0-1+n_I}\cdot (1-p)^{s_0-1+(n_s-n_I)}$,
increasing the power of $p$ by 1 and 
thus yielding 
\begin{eqnarray}  
 \mbox{E}(p) \!& = &\!\frac{1}{N_f}\cdot \sum_{n_{p_I}=0}^{n_P}\sum_{n_I=0}^{n_s}
\left[\!\binom{n_I}{n_{P_I}}\!\cdot\! \nonumber
 \binom{n_s-n_I} {n_P-n_{P_I}}\!\cdot\!  \binom{n_s}{n_I}  \cdot 
 \frac{\Gamma (n_{P_I}+r_1)\cdot \Gamma(n_I-n_{P_I}+s_1)}
 { \Gamma(r_1 + n_I+s_1)}\right.\!\cdot \nonumber \\
&&\left. \frac{\Gamma (n_P-n_{P_I}+r_2)\cdot \Gamma(n_s-n_I-n_P+n_{P_I}+s_2)}
             {\Gamma(n_s-n_I+s_2+r_2)} \right.\!\cdot\! \label{eq:calcolo_esatto} \\
&&\left.\frac{\Gamma(r_0+n_I\bm{+1})\cdot\Gamma(s_0+n_s-n_I)}
           {\Gamma(r_0+s_0+n_s\bm{+1})}\right]\,, \nonumber
\end{eqnarray}
while $\mbox{E}(p^2)$ is obtained replacing `$\bm{+1}$' by
 `$\bm{+2}$'. A script to evaluate expected value and standard deviation
of $p$ is provided in Appendix B.13.

 The expression can be extended to  `$\bm{+3}$' by `$\bm{+4}$',
 thus getting  $\mbox{E}(p^3)$ and  $\mbox{E}(p^4)$, from which
 {\em skewness} and {\em kurtosis} can be evaluated.
 Finally, making use of the so called
 Pearson Distribution System implemented in R~\cite{PearsonDS}, 
 $f(p)$ can be obtained with a quite high degree of accuracy, unless
 the distribution is squeezed towards 0 o 1, as
 e.g. in Fig.~\ref{fig:jags_inf_p_pi1_pi2}.\footnote{The R package
   PearsonDS~\cite{PearsonDS} also contains
 a random number generator, used in the script,
 very convenient if further Monte Carlo integrations/simulations
 starting from $f(p)$ are needed.
  \label{fn:calcolo_esatto_momenti}}
 A script to evaluate mean, variance, skewness and kurtosis, and
 from them $f(p)$ by the Pearson Distribution System is shown
 in Appendix B.14.
 
\subsection{Result and comparison with JAGS} 
The pdf of $p$, given the set of conditions $I$, to which we have
added $r_0$ and $s_0$ in order to remind that it also depends on the
chosen family for the prior, is finally
\begin{eqnarray}
f(p\,|\,I,r_0,s_0) &=& \frac{1}{N_f}\cdot \Large[ p^{r_0-1}\cdot (1-p)^{s_0-1}\Large]\cdot
       \!\sum_{n_{P_I}=0}^{n_P}\sum_{n_I=0}^{n_s}
            \left[\binom{n_I} {n_{P_I}}  \cdot  \binom{n_s-n_I} {n_P-n_{P_I}} \cdot
            \binom{n_s}{n_I}  \cdot \right. \nonumber \\
&&   p^{n_I}   \cdot  (1-p)^{n_s-n_I} \cdot
       \frac{\Gamma (n_{P_I}+r_1)\cdot \Gamma(n_I-n_{P_I}+s_1)}
              {\Gamma(r_1 + n_I+s_1)} \cdot  \label{eq:C.7}\\ 
&&\left.\frac{\Gamma (n_P-n_{P_I}+r_2)\cdot \Gamma(n_s-n_I-n_P+n_{P_I}+s_2)}
        { \Gamma(n_s-n_I+s_2+r_2)} \right]\,. \nonumber
       \end{eqnarray}
So, although we have not been able to get an analytic solution,
which for problems of this kind is out of hope,
we have got an expression for $f(p\,|\,I,r_0,s_0)$,
that we can compute numerically and check against the JAGS results seen in Sec.~\ref{sec:interring_p}.
For the purpose of this work, we did not put particular effort
in trying to speed up the calculation
of Eqs.~(\ref{eq:C.5})-(\ref{eq:C.6}) and therefore the comparison
concerns only the result, and not the computer time or other technical
issues. The agreement is excellent, even when we are dealing with
numbers as large as 10000 for $n_s$
(and a few thousands for $n_P$).
For example, the comparison using
the same values of $n_P=2010$ and $n_s=10000$  
of Sec.~\ref{sec:interring_p} is shown in the upper plot of Fig.~\ref{fig:Jagsvsdirect}:
\begin{figure}
\begin{center}
 \epsfig{file=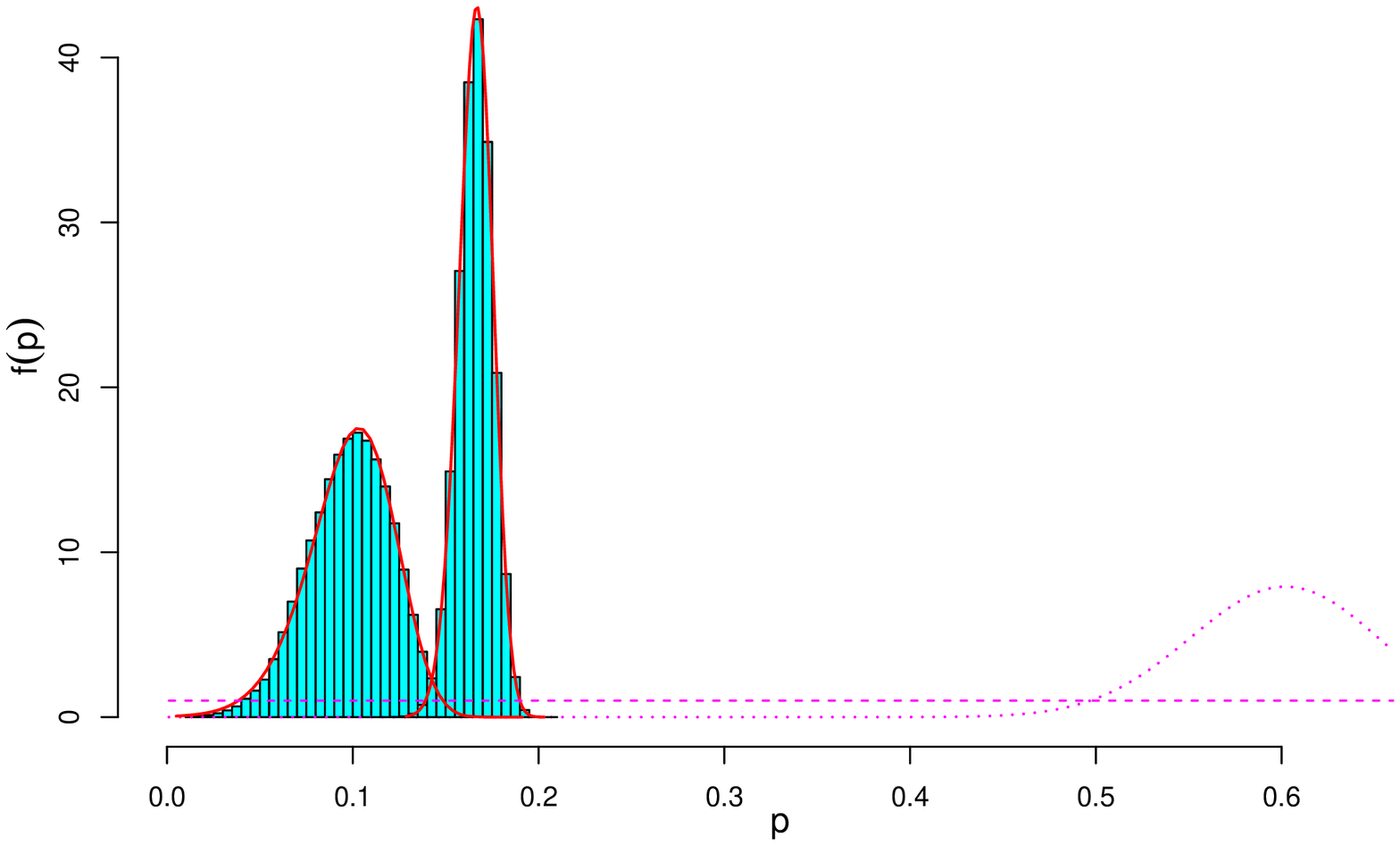,clip=,width=0.9\linewidth}\\
 \mbox{}\vspace{-1.1cm}\mbox{}\\
 \epsfig{file=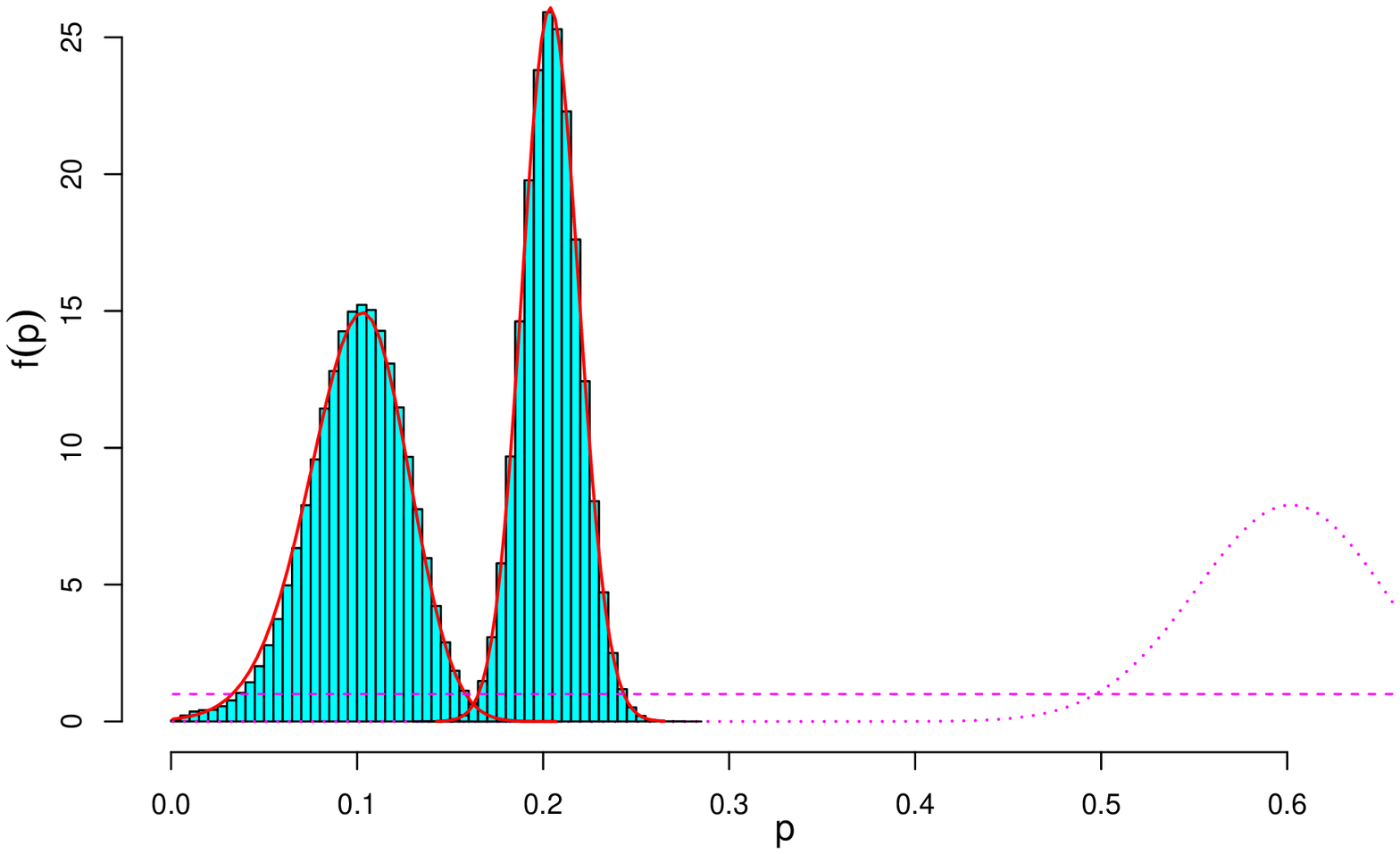,clip=,width=0.9\linewidth}
\end{center}
\mbox{}\vspace{-0.9cm}\mbox{}\\
\caption{\small \sf Direct computation of Eq.~(\ref{eq:C.7}) (solid lines)
vs JAGS results (histograms) for the flat prior
(magenta dashed line) and for a $\mbox{Beta}(57,38)$ (magenta dotted line).
Upper plot: $n_P=2010$ and $n_s=10000$. Lower plot: $n_P=201$ and $n_s=1000$.
}
\mbox{}\vspace{-0.5cm}\mbox{}
\label{fig:Jagsvsdirect}
\end{figure}
\begin{itemize}
\item The histogram peaked around $p=0.1$ is the JAGS result obtained by
$10^7$ iterations,
with over-imposed the pdf evaluated making use of Eq.~(\ref{eq:C.7}),
starting from a uniform prior (magenta dashed line).
In terms of expected value $\pm$ standard uncertainty
the direct calculation (exact -- see Eq.~(\ref{eq:calcolo_esatto}) and Appendix B.13) gives
$p=0.0981\pm 0.0233$ versus $p=0.0984\pm 0.0224$ of JAGS
(with exaggerated number of decimal digits just for detailed comparison).
\item Then we have changed the prior, choosing one strongly 
      preferring high values of $p$ (dotted magenta curve)
      with rather small uncertainty:  
      a $\mbox{Beta}(57,38)$, yielding 
      an expected value of 0.60 with standard deviation of 0.05.
      This new prior has the effect of `pulling' the
      distribution of $p$ on the right side. The agreement
      of the results obtained by the two methods is again excellent,
      resulting in $p=0.1669\pm 0.0093$ and $p=0.01658\pm0.0093$
      (direct and JAGS, respectively).
\end{itemize}
Then we repeat the game with a sample ten times smaller, that
is $n_s=1000$, and assuming $n_P=201$ (lower plot in the same figure).
Again the agreement between direct calculation and MCMC sampling
is excellent.

It is worth noting that the possibility to write down an
expression for the pdf of interest for an inferential
problem with several nodes, after marginalization
over six variables, has to be considered a lucky case,
thanks also to the approximation of modeling
the sampling by a binomial rather than a hypergeometric
and to the use of conjugate priors. The purpose of this
section is then mainly didactic, being the valuation
of the pdf's of other variables (and of several variables all together)
and of their moments prohibitive. 
It is then clear the superiority of estimates based
on MCMC methods, whose advent several decades ago has been a kind
of revolution, which have given a boost to Bayesian methods
for `serious' multidimensional applications, tasks before
not even imaginable.\footnote{It seems
  (the episode has be referred to one of us
  by a statistician present at the lectures) that in the 80's
  Dennis Lindley ended a lecture series telling something like
  {\em ``You see, I have shown you
    a wonderful, logically consistent theory.
    There is only a `little' problem.
    We are unable to do the calculations
    for the high dimensional problems that occur in real applications.''
  }
}

\subsection{More remarks on the role of priors}\label{ss:more_on_priors}
Having checked the agreement between the two methods,
let us now focus the attention on the
results themselves.
Looking at the results from the smaller sample we note:
 \begin{itemize}
 \item The width of the distribution using a flat prior
  is wider for the small sample than that obtained with larger `statistics',
  as expected, with a tiny variation
  in the mean value: $p=0.099\pm 0.027$.
 \item The prior $\mbox{Beta}(57,38)$ causes a larger shift of the
 distribution towards higher values of $p$, thus
 yielding $p=0.204 \pm 0.015$. 
 \end{itemize}
It is interesting to compare these results with what
we have seen in Sec.~\ref{ss:priors_priors_priors}
(see Fig.~\ref{fig:inf_p_reshape_pub}).
In that case the non-flat, `informative'  prior
had the role of `reshaping'  
the posterior derived by a flat prior, making
thus the result acceptable by the `expert', 
because the outcome was not in contrast with her prior belief.
Here, instead, the result provided
by a flat prior is so far from the rational belief
(most likely shared by the relevant scientific community)
of the expert, that the result would not be
accepted acritically. Most likely 
the expert would mistrust the data analysis, or the data themselves.
But she would perhaps also analyze critically her
prior beliefs in order to understand on what they were really grounded
and how solid they were.
As a matter of fact, scientists are ready to
modify their opinion, but with some care,
and, as the famous motto says,
{\em ``extraordinary claims require extraordinary evidence''}. 

Since scientific priors are usually strongly based 
on previous experimental information, the problem
of `logically merging' a prior preference summarized
by $\approx 0.60\pm 0.05$ and a new experimental
results preferring `by itself' (that is when the result
is dominated by the `likelihood'
-- see Sec.~\ref{ss:priors_priors_priors}),
summarized as $0.098\pm 0.023$ (or $\pm 0.027$, depending on $n_s$)
is similar to that of `combining apparently incompatible results.'
Also in that case, nobody would acritically
accept the `weighted average'
of the two results which appear to be in mutual disagreement.
A so called `skeptical combination' should be preferred,
which would  even yield a multi-modal distribution~\cite{skept}.
This means that in a case like those of Fig.~\ref{fig:Jagsvsdirect}
the expert could think that either
\begin{itemize}
\item she is right, with probability ${\cal P}$,
      and she would just stick to her prior $f_0(p)$;
\item she is wrong, with probability $1-{\cal P}$,
      and she would switch to the posterior
      provided by the likelihood alone, let us indicate it with
      $f_{\cal L}(p)$. 
\end{itemize}
Therefore the degrees of belief of $p$ will be
described by $f(p) = {\cal P}\cdot f_0(p) +
(1-{\cal P})\cdot f_{\cal L}(p)$. As far as we understand
from our experience she would hardly believe
the result obtained, `technically', plugging
her prior in the formulae -- and we keep repeating
once more
 Laplace's dictum that
 {\em ``probability is good sense reduced to a calculus''}.

In order to make our point more clear, let us look into the details
of the situation depicted in Fig.~\ref{fig:Jagsvsdirect} with the help of Fig.~\ref{fig:fig.C.2},
in which 
\begin{figure}[t]
\begin{center}
 \epsfig{file=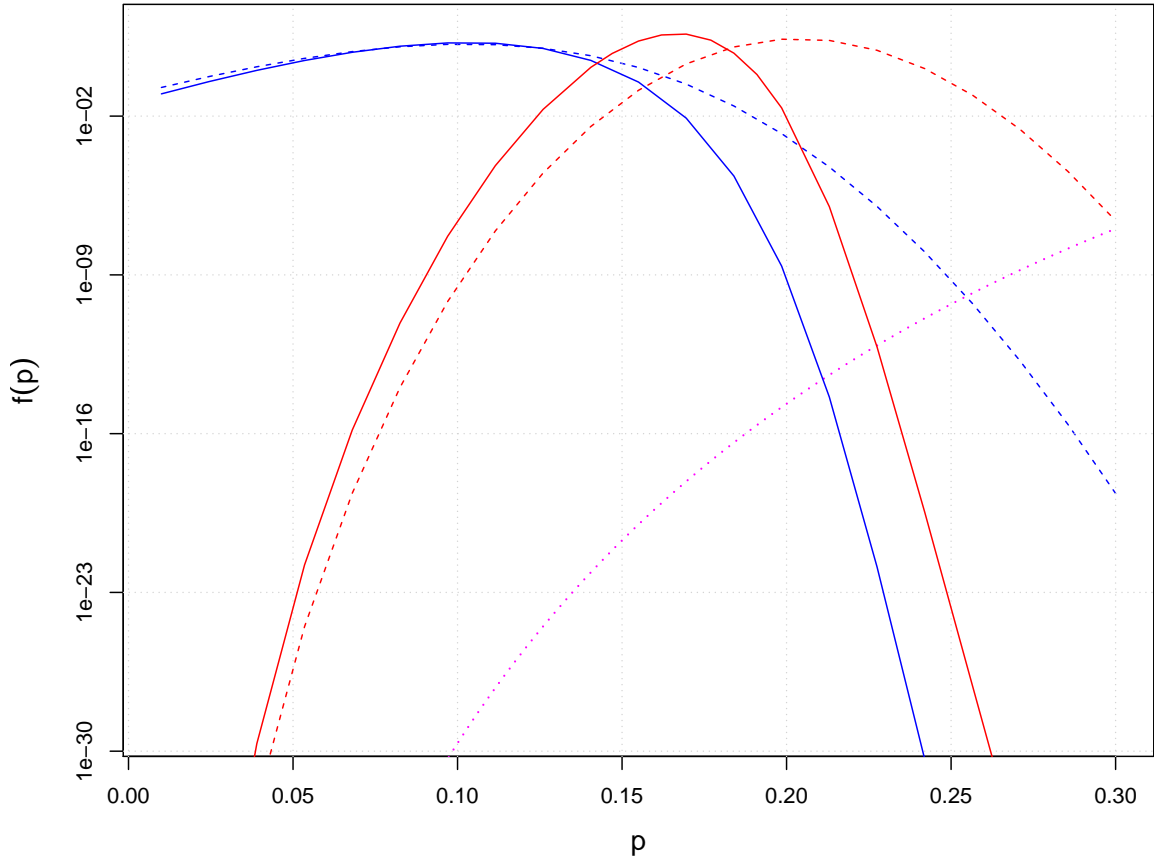,clip=,width=\linewidth}
\end{center}
\mbox{}\vspace{-0.5cm}\mbox{}\\
\caption{\small \sf Closer look at the effect of the prior
$\mbox{Beta}(57,38)$ shown in Fig.~\ref{fig:Jagsvsdirect}.
}
\mbox{}\vspace{-0.5cm}\mbox{}
\label{fig:fig.C.2}
\end{figure}
$f(p)$ is reported in log scale, and the abscissa limited to the
region of interest. The blue curves, which are dominant below
$p\approx0.10$, represent the posteriors obtained by a flat prior
(solid for $n_s=10000$ and $n_P=2010$;
dashed for $n_s=1000$ and $n_P=201$).
Then, the dotted magenta curve is the tail at small  $p$ of the 
prior $\mbox{Beta}(57,58)$, which prefers values of $p$ around $\approx 0.60\pm 0.05$.
Then the red curves (solid and dashed as previously)
show the posterior distributions obtained by this new prior.

The shift of both distributions towards the right side 
is caused by the {\em dramatic reshaping} due to 
prior in the region between $p\approx 0.1$ and $p\approx 0.3$
in which $f_0(p\,|\,\mbox{Beta}(57,38))$ varies by
{\em about 25 orders of magnitudes}\,(!).
The question is then that no expert, who believes {\em a priori} that
$p$ should be {\em most likely} in the region between
0.5 and 0.7 (and almost certainly not below 0.40-0.45),
can have a defensible, rational belief that values
of $p$ around 0.3 are $10^{\approx 25}$ times more probable
than values around 0.1. More likely, once she has to give
up her prior, she would consider small values of $p$
equally likely. For this reason -- let us put in this way
what we have said just above -- she will be in the situation
either to completely mistrust the new outcome, thus keeping her
prior, or the other way around.
The take-away message is therefore just 
the (trivial) reminder
that mathematical models are in most practical cases
just dictated
by practical convenience and should not been taken literally in their
extreme consequences, as  Gauss promptly commented on
the {\em ``defect''} of {\em his}
error function immediately after he had derived it~\cite{Gauss}.
Therefore our addendum
to Laplace's dictum reminded above is {\em don't get fooled by math}.

\section{Conclusions}\label{sec:conclusions}
In this paper we went through the issues of `stating'
if an individual belong to a particular class
and in `counting' the number of individuals in a population
belonging to that class.
Since the {\em casus belli} was the Covid-19 pandemic,
we have been constantly speaking of (currently and past)
`infectees', although our work is rather general.
A well understood complication related to the above tasks
is due to the fact that the assignment of an individual to
the class of interest is performed by 'proxies'
provided by the test result,
in this case `positive' or `negative'.
Having defined $\pi_1$ the probability that
the test result gives positive if the individual
is infected (`sensitivity') and $\pi_2$
the probability of positive if not infected
($1-\pi_2$ being the `specificity'),
we have analyzed the impact on the results
of the fact that not only these `test parameters' are
far from being ideal ($\pi_1\ne 1$ and $\pi_2\ne 0$),
but that their values are indeed uncertain.

We have started our work using parameters that can be
summarized as $\pi_1=0.978\pm 0.007$ and  $\pi_2=0.115\pm 0.022$,
based on the nominal data provided by Ref.~\cite{FattoQuotidiano} ($\pi_1=0.98$ and  $\pi_2=0.12$),
and used probability theory, and in particular
the so called Bayes' rule, in order to
\begin{itemize}
\item evaluate the probability that an individual declared
  positive is infected (and so on for the other possibilities);
\item evaluate the proportion of infectees in a population,
  based on the number of positive in a tested sample. 
\end{itemize}  
In both problems the role of `priors' is logically crucial,
although in practice it has a different impact on the numerical result:
\begin{itemize}
\item  the probability that an individual tagged
as positive is infected depends strongly on the probability
of being infected based on other pieces of information
and knowledge (in the idealistic case of `zero knowledge'
this prior probability is just the assumed proportion of infectees
in the population);
\item the probability density function of the proportion of infectees in the
  population has, instead, {\em usually} a weak dependence
  on the prior beliefs about the same proportion. 
\end{itemize} 
The dependence on the fact that the tests are `imperfect' has
a different impact on the result:
\begin{itemize}
\item the probability of infected if positive depends strongly,
  as expected, on the values
  (`expected values', in probabilistic terms)
  of $\pi_1$ and $\pi_2$, while, rather surprisingly,
  it depends very little on their
  uncertainty;
\item the inference of the proportion of infectees, instead,
  depends strongly on their uncertainty, but very little on their expected values.
\end{itemize}  
The latter outcome is important for planning test campaigns to
count and regularly monitor the number of infectees in a population,
for which tests with relatively low sensitivity and specificity
can be employed. This second task has been analyzed in detail
by exact evaluations, Monte Carlo methods and approximated formulae,
first to understand the accuracy of the predictions of the number of
positives that would result in a sample of the population,
assuming a given proportion of infectees in the population;
then to infer the proportion of infectees in the population
from the observed number of positives.

The preliminary work of predicting the number of positives has been
particularly important because it has allowed us to produce
approximated formulae with which we can disentangle the
contributions to the overall uncertainty of prediction,
which has a somehow specular relation with the uncertainty in inference.
This allows to classify then the contributions
into 'statistics' (those depending on the
sample size, due to the probabilistic effects of sampling)
and `systematics' (those not depending on the sample
size, due then to the uncertainties on $\pi_1$ and $\pi_2$).
As a consequence it is possible to evaluate the {\em critical sample size},
above which uncertainties due to systematics  
are dominant, and therefore it is not worth
increasing the sample size.

Moreover, the fact that the uncertainties about $\pi_1$ and $\pi_2$
act as systematics (within the limitation of our model,
clearly stated in Sec.~\ref{ss:model_uncertainties_pi1_pi2})
suggests that we can evaluate differences of proportions of infectees in
different populations much better than how we can measure a single proportion.
This observation has an important practical consequence, because
one could measure the proportion of infectees in a subpopulation
(think e.g. to a Region of a Country) \underline{both} 
with a test of higher quality (and presumably more expensive) \underline{and} with a cheaper, rapid and less accurate one 
and therefore use the result as calibration point for the other subpopulations.


\newpage
\section*{Appendix A -- Some remarks on `{\em Bayes' formulae}'}
Equation (\ref{eq:P_Inf_Pos_first}) is
  a straight consequence of the probability rule relating
  joint probability to conditional probability, that is,
  for the generic `events' $A$ and $B$,
\begin{eqnarray*}
  P(A\cap B) &=&  P(B\,|\,A)\cdot P(A) =
  P(A\,|\,B)\cdot P(B)\,, \hspace{2.3cm}\,\mbox{(A.1)}
\end{eqnarray*}  
  having added to  $P(\Inf)$ of Eq.~(\ref{eq:P_Inf_Pos_first}) 
  the suffix `{\small 0}' in order  to emphasize its
  role of `prior' probability.
   Equation (A.1) yields trivially 
\begin{eqnarray*}
  P(A\,|\,B) &=& \frac{P(B\,|\,A)\cdot P(A)}{P(B)}
  \ \ \longrightarrow \ \  \frac{P(B\,|\,A)\cdot P_0(A)}{P(B)}\,,
  \hspace{1.5cm}\,\mbox{(A.2)}
\end{eqnarray*}
having also emphasized that $P(A)$ in r.h.s. is the probability
of $A$ {\em before} it is updated by the {\em new condition}
$B$.\footnote{Remember that all elicitations of probabilities
\underline{always} depend on some conditions/hypotheses/assumptions.
Therefore Eq.~(A.2) should be written, more properly,
as
\begin{eqnarray*}
  P(A\,|\,B,I) &=& \frac{P(B\,|\,A,I)\cdot P_0(A\,|\,I)}{P(B\,|\,I)}
\end{eqnarray*}
with $I$ the (common!) {\em background status of information} under
which all probabilities appearing in the equation are evaluated,
although it is usually implicit in the equations to make them more
compact, as we have done in this paper.
}
But, indeed, the essence of the {\em Bayes' rule} is given by
\begin{eqnarray*}
  P(A\,|\,B) &=& \frac{P(A\cap B)}{P(B)} =  \frac{P(A, B)}{P(B)}\,,
  \hspace{4.9cm}\,\mbox{(A.3)}
\end{eqnarray*}
in which we have rewritten the `$A \cap B$' in the way it is
custom for {\em uncertain numbers} (`random variables'),
as we shall see in while. Moreover, as we can `expand' the numerator
(using the so called {\em chain rule})
to go from Eq.~(A.3) to Eq.~(A.2), and then
Eq.~(\ref{eq:P_Inf_Pos_first}), similarly we can
expand the denominator in two steps. 
We start `decomposing' $B$ into $B\cap A$ and $B\cap \overline{A}$,
from which it follows
\begin{eqnarray*}
  B &=& (B\cap A) \cup  (B\cap \overline{A})\\
  P(B) &=& P(B\cap A) + P(B\cap \overline{A}) \\
  &=&  P(B\,|\,A)\cdot P(A) + P(B\,|\,\overline{A})\cdot P(\overline{A})
\end{eqnarray*}
After the various `expansions' we can rewrite Eq.~(A.3) as
\begin{eqnarray*}
  P(A\,|\,B) &=& \frac{P(B\,|\,A)\cdot P(A)}
  { P(B\,|\,A)\cdot P(A) + P(B\,|\,\overline{A})\cdot P(\overline{A})}\,.
  \hspace{2.7cm}\,\mbox{(A.4)}
\end{eqnarray*}
Finally, if instead of only two possibilities $A$ and  
$\overline{A}$, we have a {\em complete class} of {\em hypotheses} $H_i$,
i.e. such that $\sum_iP(H_i)=1$ and $P(H_i\cap H_j)=0$ for $i\ne j$,
we get the famous
\begin{eqnarray*}
  P(H_i\,|\,E) &=& \frac{P(E\,|\,H_i)\cdot P(H_i)}
  { \sum_i P(E\,|\,H_i)\cdot P(H_i)}
\  \ \ \longleftarrow \ \ 
  \frac{P(H_i\cap E)}{P(E)}\,,
  \hspace{1.3cm}\,\mbox{(A.5)}
\end{eqnarray*}
having also replaced the symbol $B$ by $E$, given its meaning
of {\em effect}, upon which the probabilities of
the different hypotheses $H_i$ are updated.
Moreover, the sum in the denominator of the first
r.h.s. of Eq.~(A.5) makes it explicit that the
denominator is just a normalization factor, and therefore the 
essence of the reasoning can be expressed as
\begin{eqnarray*}
  P(H_i\,|\,E)  &\propto& P(E\,|\,H_i)\cdot P(H_i)
   =  P(H_i\cap E)  \hspace{3.4cm}\,\mbox{(A.6)} 
\end{eqnarray*}
The extension to discrete `random variables' is straightforward,
since the probability distribution $f(x)$ has the meaning
of $P(X=x)$, with $X$ the name of the variable and
$x$ one of the possible values that it can assume.
Similarly, $f(x,y)$ stands for
$P(X=x,Y=y)\equiv P\left((X=x)\cap (Y=y)\right)$, 
$f(x\,|\,y)$ for $P(X=x\,|\,Y=y)$, and so on.
Moreover all possible values of $X$, as well as all possible
values of $Y$, form a complete class of hypotheses
(the distributions are normalized). Equation (A.3) 
and its variations and `expansions' becomes then, for $X$ and $Y$,
\begin{eqnarray*}
f(x\,|\,y) &=& \frac{f(x,y)}{f(y)}
= \frac{f(y\,|\,x)\cdot f(x)}{f(y)} =
   \frac{f(y\,|\,x)\cdot f(x)}{\sum_x f(y,x)} = 
    \frac{f(y\,|\,x)\cdot f(x)}{\sum_x f(y\,|\,x)\cdot f(x)}
  \hspace{0.6cm}\\
  &\propto&  f(y\,|\,x)\cdot f(x) = f(x,y)\,,
   \hspace{7.3cm}\,\mbox{(A.7)} 
\end{eqnarray*}
which can be further extended to several other variables.
For example,
adding $Z$, $V$ and $W$ and being interested to the joint
probability that $X$ and $Z$ assume the values $x$ and $z$,
conditioned by $Y=y$, $V=v$ and $W=w$, we get
\begin{eqnarray*}
f(x,z\,|\,y,v,w) &=& \frac{f(x,y,v,w,z)}{f(y,v,w)}\,.
  \hspace{5.4cm}\,\mbox{(A.8)}
\end{eqnarray*}
To conclude, some remarks are important, especially for the applications:
\begin{enumerate}
\item Equations (A.7) and (A.8) are valid also for continuous variables,
in which case the various `$f()$' have the meaning of
probability density function, and the sums needed to get
the (possibly joint) marginal in the
denominator are replaced by integration.
\item The numerator of Eq.~(8)
is `expanded' using {\em a} chain rule,
choosing, among the several possibilities, that which
makes explicit the
(assumed) causal connections\footnote{Causality
is notoriously something tricky,
and conditioning does not necessarily imply causation!}
of the different variables in the game, as stressed in the proper
places through the paper (see e.g. footnote~\ref{fn:Inference_pi1_1a},
Sec.~\ref{sec:measurability_p} and Sec.~\ref{sec:direct}).
\item A related remark is that, among the variables entering the game,
as those of Eq.~(A.8), some may be continuous
and other discrete and the probabilistic meaning of `$f(\ldots)$',
taking the example of
a bivariate case $f(x,y)$ with $x$ discrete and
$y$ continuous, is given by
$P(X=x, y\le Y \le y+\mbox{d}y) = f(x,y)\,\mbox{d}y$,
with the normalization condition
given by $\sum_x\int f(x,y)\,\mbox{d}y = 1$.
\item Finally, a crucial observation is that, {\em given the model}
which connects the variables (the graphical representations
of the kinds shown in the paper are very useful to understand it) 
{\em and its parameters, the denominator} of Eq.~(A.8)
{\em is just a number}
(although often very difficult to evaluate!),
and therefore, as we have seen in Eq.~(A.7),
the last equation can be rewritten as$^{(*)}$
\begin{eqnarray*}
f(x,z\,|\,y,v,w) &\propto& f(x,y,v,w,z)\,,
  \hspace{5.4cm}\,\mbox{(A.9)}
\end{eqnarray*}
\mbox{}\vspace{-0.6cm}\mbox{}\\
or, denoting by $\tilde{f}()$ the {\em un-normalized
posterior distribution},
\mbox{}\vspace{-0.1cm}\mbox{}
\begin{eqnarray*}
\tilde f(x,z\,|\,y,v,w) &=& f(x,y,v,w,z)\,.
  \hspace{5.4cm}\,\mbox{(A.10)}
\end{eqnarray*}
The importance of this remark is that, {\em although a closed
form of posterior is often prohibitive in practical cases,
an approximation of it
can be obtained by Monte Carlo techniques},
which allow us to evaluate the quantities of interest,
like averages, probability intervals, and so on
(see references in footnote \ref{fn:MCMC_BUGS}).
\end{enumerate}
\mbox{}\vspace{-0.7cm}\mbox{}\\
--------------------------------------------------------------------------------------%
----------------\mbox{}\\
{\footnotesize
$^{(*)}$\,Perhaps
a better way to rewrite (A.9) and (A.10), in order
to avoid confusion,
could be
\begin{eqnarray*}
f(x,z\,|\,y=y_0,v=v_0,w=w_0) &\propto& f(x,y_0,v_0,w_0,z) \\
\tilde f(x,z\,|\,y=y_0,v=v_0,w=w_0) &=& f(x,y_0,v_0,w_0,z)\,,
\end{eqnarray*}
in order to emphasize the fact that $y$, $v$ and $w$ assume precise values,
under which the possible values of $x$ and $z$ are conditioned.
Anyway, it is just a question of getting used with
that notation. For example, sticking to a textbook two dimensional case,
the bivariate normal distribution is given by
\begin{eqnarray*}
f(x,y)\! &=&\!
\frac{1}{2\,\pi\,\sigma_x\,\sigma_y\,\sqrt{1-\rho^2}}\,  
  \exp \left\{
                -\frac{1}{2\,(1-\rho^2)} 
                 \left[  \frac{(x-\mu_x)^2}{\sigma_x^2} - 2\,\rho\,\frac{(x-\mu_x)(y-\mu_y)}{\sigma_x\,\sigma_y}
        + \frac{(y-\mu_y)^2}{\sigma_y^2}
                 \right]
         \right\} \,. 
\end{eqnarray*}
\mbox{}\vspace{0.8cm}\mbox{}\\
The distribution of $x$, conditioned by $y=y_0$ is then
\begin{eqnarray*}
f(x\,|\,y_0)\! &\propto& \!
\frac{1}{2\,\pi\,\sigma_x\,\sigma_y\,\sqrt{1-\rho^2}}\,  
  \exp \left\{
                -\frac{1}{2\,(1-\rho^2)} 
                 \left[  \frac{(x-\mu_x)^2}{\sigma_x^2} - 2\,\rho\,\frac{(x-\mu_x)(y_0-\mu_y)}{\sigma_x\,\sigma_y}
        + \frac{(y_0-\mu_y)^2}{\sigma_y^2}
                 \right]
         \right\} \\
   &\propto& \!  \exp \left\{
                -\frac{1}{2\,(1-\rho^2)} 
                 \left[  \frac{(x-\mu_x)^2}{\sigma_x^2}
                 - 2\,\rho\,\frac{x\,(y_0-\mu_y)}{\sigma_x\,\sigma_y}
        \right]
         \right\}   \\
  &\propto& \!  \exp \left\{
                -\frac{1}{2\,(1-\rho^2)\,\sigma_x^2} 
                 \left[x^2 -2\,x\,\left(\mu_x+\rho\frac{\sigma_x}{\sigma_y}\,
                 (y_0-\mu_y)\right)
        \right]
         \right\}     \\
   &\propto& \!  \exp \left\{-\frac{\left[x^2 -2\,x\,\left(\mu_x+\rho\frac{\sigma_x}{\sigma_y}\,
                 (y_0-\mu_y)\right)
        \right]}
   {2\,(1-\rho^2)\,\sigma_x^2}
         \right\} \\
        & \propto &
   \exp \left\{-\frac{\left[x -\,\left(\mu_x+\rho\frac{\sigma_x}{\sigma_y}\,
                 (y_0-\mu_y)\right)
        \right]^2}
  {2\,(1-\rho^2)\,\sigma_x^2}
         \right\} \,,     
\end{eqnarray*}
in which we recognize a Gaussian distribution
with
$\mu_{x|y_0} = x +\rho\frac{\sigma_x}{\sigma_y}\,(y_0-\mu_y)$
and $\sigma_{x|y_0} = \sqrt{1-\rho^2}\,\sigma_x$. \\
$[$In the various steps all factors (and hence
all addends at the exponent) not depending on $x$ have been
ignored. Finally, in the last step the `trick'
of {\em complementing the exponential} has been used, because
{\em adding}
$ \left(\mu_x+\rho\frac{\sigma_x}{\sigma_y}\,
                 (y_0-\mu_y)\right)^2\!/ \,(2\,(1-\rho^2)\,\sigma_x^2)
$
at the exponent is the same as multiplying by a constant factor.$]$ 
}

\newpage
{\small 
\section*{Appendix B -- R and JAGS code}
\subsection*{B.1 -- Monte Carlo evaluation of
  Eqs.~(\ref{eq:P.Inf.Pos.IntegraleDoppio}) and
  (\ref{eq:P.NoInf.Neg.IntegraleDoppio})}
\begin{verbatim}
r.pi1 = 409.1; s.pi1 = 9.1
r.pi2 =  25.1; s.pi2 = 193.1
p = 0.1
n = 100000
pi1 <- rbeta(n, r.pi1, s.pi1)
pi2 <- rbeta(n, r.pi2, s.pi2)
P.Inf.Pos.i   <-  pi1*p/(pi1*p + pi2*(1-p))
P.NoInf.Neg.i <- (1-pi2)*(1-p) / ((1-pi1)*p + (1-pi2)*(1-p))
P.Inf.Pos   <- mean(P.Inf.Pos.i)
P.NoInf.Neg <- mean(P.NoInf.Neg.i)
cat(sprintf("Integral (by MC): P(Inf|Pos) = %.4f; P(NoInf|Neg) = %.4f \n",
            P.Inf.Pos, P.NoInf.Neg))
E.pi1 = r.pi1 / (r.pi1 + s.pi1)
E.pi2 = r.pi2 / (r.pi2 + s.pi2)
cat(sprintf("Using E(..): P(Inf|Pos) = %.4f;  P(NoInf|Neg) = %.4f \n",
            E.pi1*p/(E.pi1*p + E.pi2*(1-p)),
            (1-E.pi2)*(1-p) / ((1-E.pi1)*p + (1-E.pi2)*(1-p))))
\end{verbatim}

\subsection*{B.2 -- Monte Carlo evaluation of
   Eq.~(\ref{eq:f_nP_int_pi1_pi2})}
\begin{verbatim}
ns = 10000
ps = 0.1
n.I = ps * ns
n.NI = (1 - ps) * ns
r1=409.1; s1=9.1
r2=25.1; s2=193.2
nr =100000
pi1 <- rbeta(nr, r1, s1)
pi2 <- rbeta(nr, r2, s2)
nP.I  <- rbinom(nr, n.I, pi1)
nP.NI <- rbinom(nr, n.NI, pi2)
nP <- nP.I + nP.NI
hist(nP, nc=100, col='cyan', freq=FALSE) 
cat(sprintf("nP: mean = %.1f, sd = %.1f\n",mean(nP),sd(nP)))
\end{verbatim}

\subsection*{B.3 -- Monte Carlo estimate of $f(n_P\,|\,N=10^5,n_s=10^4, p=0.1)$ reported in 
Fig.~\ref{fig:PredictionPositive_sampling}}
\begin{verbatim}
N = 100000; ns  = 10000; p = 0.1
N.I  <- as.integer(p*N)
N.NI <- N - N.I
E.pi1 = 0.978; E.pi2 = 0.115

nr =100000
n.I  <- rhyper(nr, m=N.I, n=N.NI, k=ns)
n.NI <- ns - n.I

nP.I  <- rbinom(nr, n.I, E.pi1)
nP.NI <- rbinom(nr, n.NI, E.pi2)
nP <- nP.I + nP.NI
cat(sprintf("mean = %.0f,  sigma = %.0f\n", mean(nP),  sd(nP) ))
hist(nP, nc=90, col='cyan', freq=FALSE)
\end{verbatim}

\subsection*{B.4 -- $\sigma(f_P)$ and $\sigma(f_P)/\mbox{\rm E}(f_P)$,
with detailed contributions to them, using the {\em approximated}
Eqs.~(\ref{eq:approx_E.fP_s})-(\ref{eq:sigma_fP_sigma_ps_bis})} 
\begin{verbatim}
p  = 0.1
N  = 10^7    # N >> ns --> binomial
ns = 1000
E.pi1 = 0.978;  sigma.pi1 = 0.007
E.pi2 = 0.115;  sigma.pi2 = 0.022
# E.pi2 = 0.115;  sigma.pi2 = sigma.pi1     # reduced sigma.pi2  
# E.pi2 = 1 - E.pi1; sigma.pi2 = sigma.pi1  # improved specificity

E.ps <- p
E.fP <- E.pi1*E.ps + E.pi2*(1-E.ps)

s.fP.R   <- sqrt( E.pi1*(1-E.pi1)*E.ps +  E.pi2*(1-E.pi2)*(1-E.ps) )/sqrt(ns)  
s.fP.pi1 <- sigma.pi1*E.ps
s.fP.pi2 <- sigma.pi2*(1-E.ps)
s.ps     <- sqrt(p*(1-p)*(1-ns/N))/sqrt(ns)  
s.fP.ps  <- s.ps*abs(E.pi1-E.pi2)           

s.fP.stat <- sqrt(s.fP.R^2+s.fP.ps^2)
s.fP.syst <- sqrt(s.fP.pi1^2+s.fP.pi2^2)
s.fP = sqrt(s.fP.stat^2 + s.fP.syst^2)

cat(sprintf("p = %.2f; ns = %d\n", p, ns))
cat(sprintf("E(fN) = %.3f; sigma(fN) = %.3f; sigma(fN)/E(fN) = %.2f\n",
            E.fP, s.fP, s.fP/E.fP))
cat("Contributions : \n")
cat(sprintf("    s.fP.R    =  %.3e;  s.fP.R/E.fP    = %.2e\n",
            s.fP.R, s.fP.R/E.fP))
cat(sprintf("    s.fP.p1   =  %.3e;  s.fP.p1/E.fP   = %.2e\n",
            s.fP.pi1, s.fP.pi1/E.fP))
cat(sprintf("    s.fP.p2   =  %.3e;  s.fP.p2/E.fP   = %.2e\n",
            s.fP.pi2, s.fP.pi2/E.fP))
cat(sprintf("    s.fP.ps   =  %.3e;  s.fP.ps/E.fP   = %.2e\n",
            s.fP.ps, s.fP.ps/E.fP))
cat(sprintf("    s.fP.stat =  %.3e;  s.fP.stat/E.fP = %.2e\n",
            s.fP.stat, s.fP.stat/E.fP))
cat(sprintf("    s.fP.syst =  %.3e;  s.fP.syst/E.fP = %.2e\n",
            s.fP.syst, s.fP.syst/E.fP))
\end{verbatim}

\subsection*{B.5 -- Monte Carlo estimate of $f_P$ using
R functions, as described
in Sec.~\ref{sss:MC_R} (see Fig.\ref{fig:PredictionPositive_by01})}
\begin{verbatim}
p  = 0.1
ns = 1000
r1 = 409.1;  s1 =  9.1
r2 =  25.1;  s2 =193.2

nr = 10000
n.I   <- rbinom(nr, ns, p)       # 1. 
n.NI  <- ns - n.I
pi1   <- rbeta(nr, r1, s1)       # 2.
pi2   <- rbeta(nr, r2, s2)
nP.I  <- rbinom(nr, n.I, pi1)    # 3.
nP.NI <- rbinom(nr, n.NI, pi2)
nP    <- nP.I + nP.NI            # 4.

fP <- nP/ns
cat(sprintf("fP: %.3f +- %.3f\n", mean(fP), sd(fP)))
hist(fP, col='cyan')
# barplot(table(fP), col='cyan')  # alternative, for small ns and p
\end{verbatim}

\subsection*{B.6 -- Monte Carlo estimate of $f_P$ JAGS from R via
rjags (Sec.~\ref{sss:usingJAGS})}
\begin{verbatim}
#--------------------------  JAGS model ------------------------------
model.name = "tmp_model.bug"  
write("
model {
  n.I ~ dbin(p, ns)
  n.NI <- ns - n.I 
  nP.I ~ dbin(pi1, n.I)
  nP.NI ~ dbin(pi2, n.NI)
  pi1 ~ dbeta(r1, s1)
  pi2 ~ dbeta(r2, s2)
  nP ~ sum(nP.I, nP.NI)
  fP <- nP / ns 
}
", model)

#--------------------------- call JAGS ---------------------------------
library(rjags)
data <- list(ns=1000, p=0.1, r1=409.1, s1=9.1, r2=25.2, s2=193.1)  
jm <- jags.model(model, data)
chain <- coda.samples(jm, c('n.I', 'fP'), n.iter=10000)

#--------------------------- Results ------------------------------------
print(summary(chain))
plot(chain, col='blue')
\end{verbatim}

\subsection*{B.7 -- Check of approximated formulae}
\begin{verbatim}
get.prediction <- function(ns, N, p, E.pi1, sigma.pi1, E.pi2, sigma.pi2) {
   E.ps <- p
   E.fP <- E.pi1*E.ps + E.pi2*(1-E.ps)

   s.fP.R   <- sqrt( E.pi1*(1-E.pi1)*E.ps +  E.pi2*(1-E.pi2)*(1-E.ps) )/sqrt(ns)  
   s.fP.pi1 <- sigma.pi1*E.ps
   s.fP.pi2 <- sigma.pi2*(1-E.ps)
   s.ps     <- sqrt(p*(1-p)*(1-ns/N))/sqrt(ns)  
   s.fP.ps  <- s.ps*abs(E.pi1-E.pi2)           

   s.fP.stat <- sqrt(s.fP.R^2+s.fP.ps^2)
   s.fP.syst <- sqrt(s.fP.pi1^2+s.fP.pi2^2)
   s.fP = sqrt(s.fP.stat^2 + s.fP.syst^2)
   return(list(E.fP=E.fP, s.fP=s.fP))
} 

N  = 10^7  
p.v  <- 0:4 / 10
ns.v <- c(300, 1000, 3000, 10000)

E.pi1 = 0.978;  sigma.pi1 = 0.007
E.pi2 = 0.115;  sigma.pi2 = 0.022
for(case in 1:3) {
  if(case == 2) sigma.pi2 = sigma.pi1
  if(case == 3) E.pi2 = 1 - E.pi1
  cat(sprintf("\n [pi1 = %.3f+-%.3f;", E.pi1, sigma.pi1))
  cat(sprintf(" pi2 = %.3f+-%.3f]\n", E.pi2, sigma.pi2))
  for(i in 1:length(ns.v)) {
    ns <- ns.v[i]
    cat(sprintf(" ns = %d\n p: ", ns))
    for(j in 1:length(p.v))  cat(sprintf("   %.2f      ", p.v[j]))
    cat("\nfP: ")
    for(j in 1:length(p.v)) {
     p <- p.v[j]   
      pred <- get.prediction(ns, N, p, E.pi1, sigma.pi1, E.pi2, sigma.pi2)
     cat(sprintf("%.3f+-%.3f ", pred$E.fP, pred$s.fP))
    }
    cat("\n")
  }
}
\end{verbatim}

\subsection*{B.8 -- Predicting the fractions of positives obtained sampling
two different  populations (Sec.~\ref{ss:predict_Delta_fP})}
\begin{verbatim}
get.fP  <- function(nr, p, ns, pi1, pi2) {
  n.I   <- rbinom(nr, ns, p)
  n.NI  <- ns - n.I
  nP.I  <- rbinom(nr, n.I, pi1)    
  nP.NI <- rbinom(nr, n.NI, pi2)
  nP    <- nP.I + nP.NI            
  fP    <- nP/ns
  return(fP)
}

p1 = 0.1
p2 = 0.2
ns = 10000
r1 = 409.1;  s1 =  9.1
r2 =  25.1;  s2 =193.2

nr = 100000
pi1 <- rbeta(nr, r1, s1)   
pi2 <- rbeta(nr, r2, s2)
fP1 <- get.fP(nr, p1, ns, pi1, pi2)
fP2 <- get.fP(nr, p2, ns, pi1, pi2)

cat(sprintf("fP1: %.4f +- %.4f\n", mean(fP1), sd(fP1)))
cat(sprintf("fP2: %.4f +- %.4f\n", mean(fP2), sd(fP2)))
cat(sprintf("fP2-fP1: %.4f +- %.4f\n", mean(fP2-fP1), sd(fP2-fP1)))
cat(sprintf("rho(fP1,fP2): %.4f\n", cor(fP1,fP2)))
cat(sprintf("Check of sigma(fP2-fP1) from correlation: %.4f\n",
            sqrt(var(fP1)+var(fP2)-2*cov(fP1,fP2))))
\end{verbatim}
(Note how the `random' sequences of values of {\tt pi1} and  {\tt pi2}
are generated \underline{before} the calls to
{\tt get.fP()}. This is crucial in order to get the correlation
among {\tt fP1} and {\tt fP2} discussed in Sec.~\ref{ss:predict_Delta_fP}.
If these two sequences are defined, each time,
inside the function, or they a generated before each call
to the function, the correlation will disappear.
This way of generating the events is consequence of our model
assumptions, as stressed in Sec.~\ref{ss:model_uncertainties_pi1_pi2}.)

\subsection*{B.9 -- JAGS model to perform the same Monte Carlo evaluation
done in Appendix B.8}
(Only the model is provided -- steering R commands
are left as exercise.) 
\begin{verbatim}
model {
  n.I.1 ~ dbin(p1, ns1)
  n.NI.1 <- ns1 - n.I.1 
  nP.I.1 ~ dbin(pi1, n.I.1)
  nP.NI.1 ~ dbin(pi2, n.NI.1)
  nP.1 ~ sum(nP.I.1, nP.NI.1)
  fP.1 <- nP.1 / ns1

  n.I.2 ~ dbin(p2, ns2)
  n.NI.2 <- ns2 - n.I.2 
  nP.I.2 ~ dbin(pi1, n.I.2)
  nP.NI.2 ~ dbin(pi2, n.NI.2)
  nP.2 ~ sum(nP.I.2, nP.NI.2)
  fP.2 <- nP.2 / ns2

  D.fP <- fP.2 - fP.1

  pi1 ~ dbeta(r1, s1)
  pi2 ~ dbeta(r2, s2) 
}
\end{verbatim}

\subsection*{B.10 -- JAGS model to infer $p$ (see Sec.~\ref{ss:infer_p_nI})}
\begin{verbatim}
#---- data and parameters
nr = 1000000
ns = 10000
nP = 2010
r0 = s0 = 1
r1 = 409.1; s1 = 9.1
r2 = 25.2;  s2 = 193.1 

#---- JAGS model ------------------------------
model = "tmp_model.bug"    # name of the model file ('temporary')
write("
model {
  nP ~ sum(nP.I, nP.NI)
  nP.I ~ dbin(pi1, n.I)
  nP.NI ~ dbin(pi2, n.NI)
  pi1 ~ dbeta(r1, s1)
  pi2 ~ dbeta(r2, s2)
  n.I ~ dbin(p, ns)
  n.NI <- ns - n.I
  p ~ dbeta(r0,s0)        
}
", model)

#---- call JAGS ---------------------------------------------
library(rjags)
data <- list(ns=ns, nP=nP, r0=r0, s0=s0, r1=r1, s1=s1, r2=r2, s2=s2)  
jm <- jags.model(model, data)
update(jm, 10000)
to.monitor <-  c('p', 'n.I')
chain <- coda.samples(jm, to.monitor, n.iter=nr)

#---- show results
print(summary(chain))
plot(chain, col='blue')
\end{verbatim}

\subsection*{B.11 -- Inferring the  proportions of infected
in two different populations (see Sec.~\ref{ss:inferring_Delta_p})}
\begin{verbatim}
model = "tmp_model.bug" 
write("
model {
  n.I.1 ~ dbin(p1, ns1)
  p1 ~ dbeta(r0, s0)     
  n.NI.1 <- ns1 - n.I.1 
  nP.I.1 ~ dbin(pi1, n.I.1)
  nP.NI.1 ~ dbin(pi2, n.NI.1)
  nP.1 ~ sum(nP.I.1, nP.NI.1)

  n.I.2 ~ dbin(p2, ns2)
  p2 ~ dbeta(r0, r0)    
  n.NI.2 <- ns2 - n.I.2 
  nP.I.2 ~ dbin(pi1, n.I.2)
  nP.NI.2 ~ dbin(pi2, n.NI.2)
  nP.2 ~ sum(nP.I.2, nP.NI.2)

  Dp <- p2 - p1

  pi1 ~ dbeta(r1, s1)
  pi2 ~ dbeta(r2, s2)
  }
", model)

library(rjags)
set.seed(193)
nr = 1000000
ns1 = 10000
ns2 = 10000   # they could be different
nP.1 = 2000
nP.2 = 2200
r0 = s0 = 1   # flat prior
r1 = 409.1; s1 = 9.1
r2 = 25.2;  s2 = 193.1

data <- list(ns1=ns1, ns2=ns2, nP.1=nP.1, nP.2=nP.2,
             r0=r0, s0=s0, r1=r1, s1=s1, r2=r2, s2=s2)  
jm <- jags.model(model, data)
update(jm, 10000)
to.monitor <-  c('p1', 'p2', 'Dp')
chain <- coda.samples(jm, to.monitor, n.iter=nr)
print(summary(chain))
\end{verbatim}

\subsection*{B.12 -- Reshaping by an informative prior
the posterior distribuition obtained starting from a flat prior
(see Sec.~\ref{ss:priors_priors_priors})}
\begin{verbatim}
pause <- function() { cat ("\n >> press Enter to continue\n"); scan() }

call.jags <- function(model, data, nr) {
  jm <- jags.model(model, data)
  update(jm, 100)
  chain <- coda.samples(jm, 'p', n.iter=nr)
  print(summary(chain))
  chain.df <- as.data.frame( as.mcmc(chain) )
  return(chain.df$p)
}

library(rjags)
model = "tmp_model.bug"    # name of the model file ('temporary')
write("
model {
  n ~ dbin(p, N)
  p ~ dbeta(r0,s0)   
}
", model)

nr = 100000
N = 10; n = 3 
r0 = s0 = 1    # flat prior

# First call to JAGS
data <- list(N=N, n=n, r0=r0, s0=s0)
p <- call.jags(model, data, nr)
pause()

h <- hist(p, freq=FALSE, nc=100, col='cyan', xlim=c(0,1), ylim=c(0,4))
p.m <- sum(h$mids*h$counts)/sum(h$counts)
p.s <- sqrt(sum(h$mids^2*h$counts)/sum(h$counts) - p.m^2)
cat(sprintf(">>> JAGS:  %.4f +- %.4f\n", p.m, p.s))
pause()

# overimpose the posterion got from a Beta conjugate
curve(dbeta(x,r0+n,s0+(N-n)), col='blue', add=TRUE)
pause()

# Not-flat prior used to reshape the JAGS result
curve(dbeta(x,r,s), col='magenta', add=TRUE)
pause()

# reweighing
w <- dbeta(h$mids, r, s)
f.p.w <- h$density*w
f.p.w <- f.p.w/sum(f.p.w)/(h$mids[2]-h$mids[1])
points(h$mids, f.p.w, col='blue', ty='l', lwd=2)     
p.w.m <- sum(h$mids*f.p.w)/sum(f.p.w)
p.w.s <- sqrt(sum(h$mids^2*f.p.w)/sum(f.p.w) - p.w.m^2)
cat(sprintf(">>> Reweighed:  %.4f +- %.4f\n\n", p.w.m, p.w.s))
pause()

# second call to JAGS, with the new prior
data <- list(N=N, n=n, r0=r, s0=s)
p1 <- call.jags(model, data, nr)
h1 <-  hist(p1, nc=100, plot=FALSE)
points(h1$mids, h1$density, col='red', ty='l', lwd=1.5)  
p1.m <- sum(h1$mids*h1$counts)/sum(h1$counts)
p1.s <- sqrt(sum(h1$mids^2*h1$counts)/sum(h1$counts) - p1.m^2)
cat(sprintf(">>> JAGS(%d,%d):  %.4f +- %.4f\n", r,s, p1.m, p1.s))
pause()

# New Beta obtained by the well known updating rule
curve(dbeta(x,r+n,s+(N-n)), col='green', add=TRUE)
\end{verbatim}

\subsection*{B.13 -- Exact calculation of $\mbox{E}(p)$
and $\sigma(p)$ using a Beta prior (see Sec.~\ref{sec:direct}, footnote
\ref{fn:calcolo_esatto_momenti})}
\begin{verbatim}
nP = 201; ns = 1000
r0 = 1; s0 = 1
r1 = 409.1; s1 = 9.1
r2 =  25.1; s2 = 193.1

Nf   <- 0
sum.p  <- 0
sum.p2 <- 0
for (nPI in 0:nP) {
    for (nI in 0:ns) {
     l0 <-  ( lchoose(nI, nPI) + lchoose(ns-nI,nP-nPI) + lchoose(ns,nI)
          + lgamma(nPI+r1) + lgamma(nI-nPI+s1) - lgamma(r1+nI+s1)
          + lgamma (nP-nPI+r2) + lgamma(ns-nI-nP+nPI+s2) - lgamma(ns-nI+s2+r2) 
          + lgamma(s0+ns-nI) )  
     Nf     <- Nf     + exp( l0 +  lgamma(r0+nI)   - lgamma(r0+s0+ns) )
     sum.p  <- sum.p  + exp( l0 +  lgamma(r0+nI+1) - lgamma(r0+s0+ns+1) )
     sum.p2 <- sum.p2 + exp( l0 +  lgamma(r0+nI+2) - lgamma(r0+s0+ns+2) )
  }
}
E.p  <- sum.p/Nf
E.p2 <- sum.p2/Nf
cat(sprintf("nP=%d, ns=%d\n", nP, ns))
cat(sprintf("E(p)    : %.4f\n",E.p)) 
cat(sprintf("sigma(p): %.4f\n",sqrt(E.p2-E.p^2))) 
\end{verbatim}

\subsection*{B.14 -- Approximated  $f(p)$ from
the first four moments of the distribution (see Sec.~\ref{sec:direct}, Eq.~(\ref{eq:calcolo_esatto}))}
\begin{verbatim}
library("PearsonDS") # (package needs to be installed)
pause <- function() { cat ("\n >> press Enter to continue\n"); scan() }
nP = 201; ns = 1000
r0 = 1; s0 = 1
r1 = 409.1; s1 = 9.1
r2 =  25.1; s2 = 193.1
Nf = sum.p = sum.p2 = sum.p3 = sum.p4 = 0
for (nPI in 0:nP) {
    for (nI in 0:ns) {
     l0 <-  ( lchoose(nI, nPI) + lchoose(ns-nI,nP-nPI) + lchoose(ns,nI)
          + lgamma(nPI+r1) + lgamma(nI-nPI+s1) - lgamma(r1+nI+s1)
          + lgamma (nP-nPI+r2) + lgamma(ns-nI-nP+nPI+s2) - lgamma(ns-nI+s2+r2) 
          + lgamma(s0+ns-nI) )    
     Nf     <- Nf     + exp( l0 +  lgamma(r0+nI)   - lgamma(r0+s0+ns) )
     sum.p  <- sum.p  + exp( l0 +  lgamma(r0+nI+1) - lgamma(r0+s0+ns+1) )
     sum.p2 <- sum.p2 + exp( l0 +  lgamma(r0+nI+2) - lgamma(r0+s0+ns+2) )
     sum.p3 <- sum.p3 + exp( l0 +  lgamma(r0+nI+3) - lgamma(r0+s0+ns+3) )
     sum.p4 <- sum.p4 + exp( l0 +  lgamma(r0+nI+4) - lgamma(r0+s0+ns+4) )
  }
}
E.p  <- sum.p/Nf
E.p2 <- sum.p2/Nf
s2.p <- E.p2-E.p^2
s.p  <- sqrt(s2.p)
cat(sprintf("nP=%d, ns=%d\n", nP, ns))
cat(sprintf("E(p)    : %.4f\n",E.p)) 
cat(sprintf("sigma(p): %.4f\n",sqrt(E.p2-E.p^2))) 
E.p3 <- sum.p3/Nf
Skew <- ( E.p3 - 3*E.p2*E.p + 2*E.p^3  )  / s.p^3   
cat(sprintf("E(p^3) = %.3e; Skewness = %.3f \n",E.p3, Skew))
E.p4 <- sum.p4/Nf
Kurt <-  ( E.p4 - 4*E.p3*E.p + 6*E.p2*E.p^2 - 3*E.p^4 )   / s.p^4  
cat(sprintf("E(p^4) = %.3e; Kurtosis = %.3f\n", E.p4, Kurt))

moments <- c(mean = E.p,variance = s2.p,skewness = Skew, kurtosis = Kurt)
curve(dpearson(x, moments=moments), max(0.001,E.p-4*s.p), min(0.999,E.p+4*s.p),
      lwd=2, col='blue', xlab='p', ylab='f(p)')
pr <- rpearson(100000, moments = moments)
hist(pr , nc=100, freq=FALSE, col='cyan', add=TRUE)
cat(sprintf("\n\n mean    : %.4f\n", mean(pr))) 
cat(sprintf(" sigma: %.4f\n",sd(pr))) 

\end{verbatim}
}
\begin{center}
 \epsfig{file=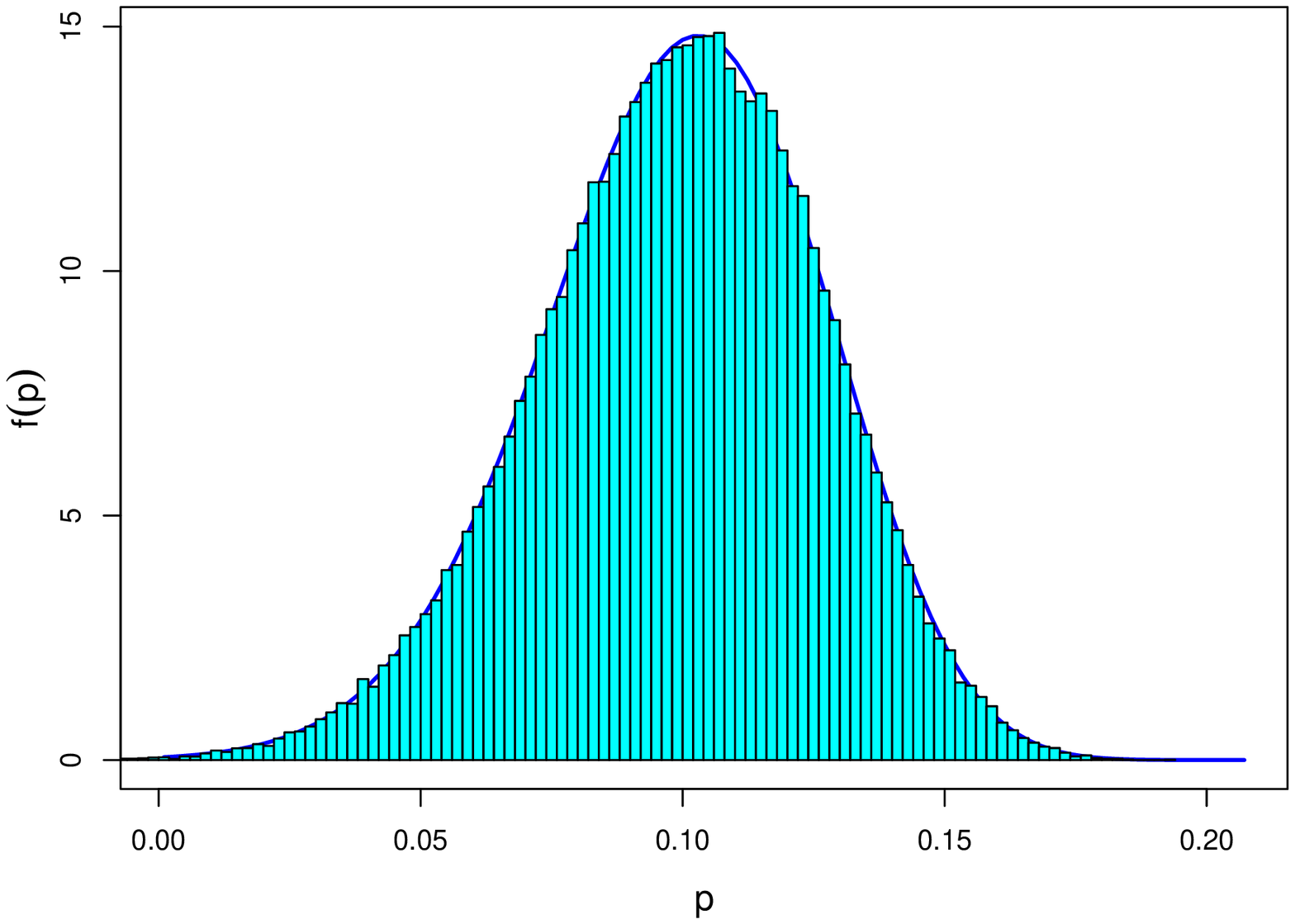,clip=,width=0.84\linewidth}
\end{center}

\end{document}